\documentclass[aps,prb,twocolumn,superscriptaddress,floatfix,showpacs,10pt,letterpaper]{revtex4-1}
\pdfoutput=1

\def\b#1{\bar{#1}}
\def\dd{\mathrm{d}}

\begin{document}
\begin{titlepage}

\title{
Symmetry-protected topological orders for interacting fermions\\
-- Fermionic topological nonlinear $\si$ models\\
and a special group supercohomology theory
}

\author{Zheng-Cheng Gu}
\affiliation{Perimeter Institute for Theoretical Physics, Waterloo, Ontario, N2L 2Y5 Canada}

\author{Xiao-Gang Wen}
\affiliation{Perimeter Institute for Theoretical Physics, Waterloo, Ontario, N2L 2Y5 Canada}
\affiliation{Department of Physics, Massachusetts Institute of
Technology, Cambridge, Massachusetts 02139, USA}

\begin{abstract}
Symmetry-protected topological (SPT) phases are gapped short-range-entangled
quantum phases with a symmetry $G$, which can all be smoothly connected to the
trivial product states if we break the symmetry.
It has been shown that a large class of \emph{interacting} bosonic SPT phases
can be systematically described by group cohomology theory.  In this paper, we
introduce a (special) group supercohomology theory which is a generalization
of the standard group cohomology theory.  We show that a large class of
\emph{short-range interacting} fermionic SPT phases can be described by the
group supercohomology theory.
Using the data of super cocycles,
we can obtain the ideal ground state wave function for the corresponding
fermionic SPT phase.  We can also obtain the bulk Hamiltonian that realizes the
SPT phase, as well as the anomalous (\ie non-on-site) symmetry for the boundary
effective Hamiltonian.  The anomalous symmetry on the boundary implies that the
\emph{symmetric} boundary must be gapless for 1+1D boundary, and must be
gapless or topologically ordered beyond 1+1D.
As an application of this general result, we construct a new SPT phase in 3D,
for interacting fermionic superconductors with coplanar spin order (which have
$T^2=1$ time-reversal $Z_2^T$ and fermion-number-parity $Z_2^f$ symmetries
described by a full symmetry group $Z_2^T\times Z_2^f$).  Such a fermionic SPT
state can neither be realized by free fermions nor by interacting bosons
(formed by fermion-pairs), and thus are not included in the K-theory
classification for free fermions or group cohomology description for
interacting bosons.  We also construct three interacting fermionic SPT phases
in 2D with a full symmetry group $Z_2\times Z_2^f$. Those 2D fermionic SPT
phases all have central-charge $c=1$ gapless edge excitations, if the symmetry
is not broken.

\end{abstract}


\maketitle

\vspace{2mm}

\end{titlepage}

{\small \setcounter{tocdepth}{1} \tableofcontents }

\section{Introduction}

\subsection{Short-range and long-range entangled states}

\begin{figure}[b]
\begin{center}
\includegraphics[scale=0.47]{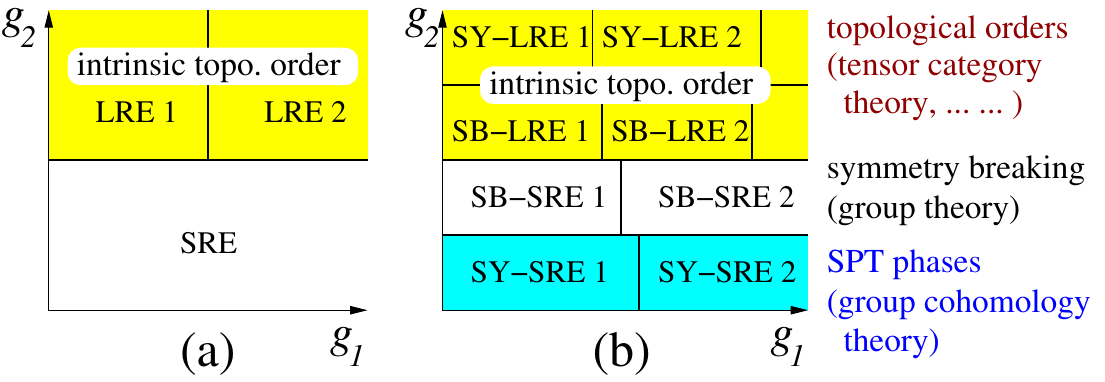}
\end{center}
\caption{
(Color online)
(a) The possible gapped phases for a class of Hamiltonians $H(g_1,g_2)$ without
any symmetry.  (b) The possible gapped phases for the class of
Hamiltonians $H_\text{symm}(g_1,g_2)$ with a symmetry.  The yellow regions in
(a) and (b) represent the phases with long range entanglement.  Each phase is
labeled by its entanglement properties and symmetry breaking properties.  SRE
stands for short range entanglement, LRE for long range entanglement, SB for
symmetry breaking, SY for no symmetry breaking.  SB-SRE phases are the Landau
symmetry breaking phases, which are understood by introducing group theory.
The SY-SRE phases are the SPT phases, which can be understood by introducing
group cohomology theory.
}
\label{topsymm}
\end{figure}

Recently, it was realized that highly entangled quantum states can give rise to
new kind of quantum phases beyond Landau symmetry
breaking,\cite{L3726,GL5064,LanL58} which include topologically ordered
phases,\cite{Wtop,WNtop,Wrig} and Symmetry Protected Topological (SPT)
phases\cite{GW0931,PBT0959} (see Fig. \ref{topsymm}).  The topologically
ordered phases contain long range entanglement\cite{CGW1038} as revealed by
topological entanglement entropy,\cite{LW0605,KP0604} and cannot be transformed
to product states via local unitary (LU)
transformations.\cite{LW0510,VCL0501,V0705}.  Fractional quantum Hall
states\cite{TSG8259,L8395}, chiral spin liquids,\cite{KL8795,WWZ8913} $Z_2$
spin liquids,\cite{RS9173,W9164,MS0181} non-Abelian fractional quantum Hall
states,\cite{MR9162,W9102,WES8776,RMM0899} \etc are examples of topologically
ordered phases.  The mathematical foundation of topological orders is closely
related to tensor category theory\cite{FNS0428,LW0510,CGW1038,GWW1017} and
simple current algebra.\cite{MR9162,LWW1024} Using this point of view, we have
developed a systematic and quantitative theory for topological
orders with gappable edge for 2+1D interacting boson and fermion
systems.\cite{LW0510,CGW1038,GWW1017} Also for 2+1D topological orders
with only Abelian statistics, we find that we can use integer $K$-matrices to
describe them.\cite{BW9045,R9002,FK9169,WZ9290}

The SPT states are short-range entangled (SRE) states with symmetry (\ie they
do not break the symmetry of the Hamiltonian), which can be transformed to
product states via LU transformations that break the symmetry.  However,
nontrivial SPT states cannot be transformed to product states via the LU
transformations that preserve the symmetry, and different  SPT states  cannot
be transformed to each other via the LU transformations that preserve the
symmetry.  The 1D Haldane phase for spin-1
chain\cite{H8364,AKL8877,GW0931,PBT0959} and topological
insulators\cite{KM0501,BZ0602,KM0502,MB0706,FKM0703,QHZ0824} are nontrivial
examples of SPT phases.  Some examples of 2D SPT phases protected by
translation and some other symmetries were discussed in
\Ref{YK1002,CGW1107,SPC1139}.

It turns out that there is no gapped bosonic LRE state in 1D.\cite{VCL0501} So
all 1D gapped bosonic states are either symmetry breaking states or SPT states.
This realization led to a complete classification of all 1+1D gapped bosonic
quantum phases.\cite{CGW1107,SPC1139,CGW1128}

In \Ref{CLW1141,CGL1172}, the result for 1D SPT phase is generalized to
any dimensions: \emph{For gapped bosonic systems in $d_{sp}$-spatial dimension
with an on-site symmetry $G$, we can construct distinct SPT phases that do not
break the symmetry $G$ from the distinct elements in $\cH^{d_{sp}+1}[G,
U_T(1)]$ -- the $(1+d_{sp})$-cohomology class of the symmetry group $G$ with
$G$-module $U_T(1)$ as coefficient.} Note that the above result does not
require the translation symmetry.  The results for some simple on-site symmetry
groups are summarized in Table \ref{tb}.  In particular, the interacting
bosonic topological insulators/superconductors are included in the table.

\subsection{The definition of fermionic SPT phases}

In \Ref{GWW1017}, the LU transformations for fermionic systems were introduced,
which allow us to define and study gapped quantum liquid phases and topological
orders in fermionic systems.  In particular, we developed a fermionic tensor
category theory to classify intrinsic fermionic topological orders with
gappable edge, as defined through the fermionic LU transformations without any
symmetry.  In this paper, we are going to use the similar line of thinking to
study fermionic gapped liquid phases with symmetries.  To begin, we will study
the simplest kind of them -- \emph{short-range-entangled fermionic phases with
symmetries}.  Those phases are called fermionic SPT phases:\\
(1)
They are $T=0$ gapped phases of fermionic Hamiltonians with
certain symmetries.\\
(2)
Those phases do not break any symmetry of the Hamiltonian.\\
(3)
Different SPT phases cannot be connected without phase transition
if we deform the Hamiltonian while preserving the symmetry
of the Hamiltonian.\\
(4)
All the  SPT phases can be  connected to the  trivial product states
without phase transition
through the deformation the Hamiltonian if we allow to break the symmetry
of the Hamiltonian.\\
{}[The term ``connected'' can also mean connected via fermionic LU
transformations (with or without symmetry)
defined in \Ref{GWW1017}.]

Note that in 0 spatial dimension, the trivial product states can have even or
odd numbers of fermions.  So the trivial product states in 0D can belong to two
different phases.  In higher dimensions, all the trivial product states belong
to one phase (if there is no symmetry).  This is why when there is no symmetry,
there are two fermionic SPT phases in 0D and only one trivial fermionic SPT
phase in higher dimensions.

\subsection{Group super-cohomology construction of fermionic SPT states}

After introducing the concept of fermionic SPT phase, we find that we can
generalize our construction of bosonic SPT orders\cite{CLW1141,CGL1172} to
fermion systems, by generalizing the group cohomology theory\cite{RS,CGL1172}
to group super-cohomology theory.  This allows us to systematically construct a
large class of fermionic SPT phases for interacting fermions in any dimensions.

In the group cohomology theory for bosonic SPT states, we find that the
different  topological terms in bosonic nonlinear $\si$ models in
\emph{discrete} space-time are described by different cocycles in group
cohomology theory (see section \ref{bSPT} for a detailed
review).  The different types of topological terms lead to different
bosonic SPT phases, and thus different cocycles describe different bosonic SPT
phases.  The idea behind our approach in this paper is similar: we show that
different topological terms in fermionic nonlinear $\si$ models in
\emph{discrete} space-time are described by different cocycles in group
super-cohomology theory.  The different types of fermionic topological terms
lead to different fermionic SPT phases, and thus different super-cocycles
describe different fermionic SPT phases.

So far, our construction only applies for a certain type of symmetries where
the fermions form an 1D representation of the symmetry group.  It cannot handle
the situation where fermions do not form an 1D representation of the symmetry
group in the fixed point wavefunctions.
For this reason, we call the current formulation of group super-cohomology
theory a special group super-cohomology theory.  On the other hand, the current
version of group super-cohomology theory can indeed systematically generate
a large class of fermionic SPT phases, and many of those examples are totally
new since they can neither be constructed from free fermions nor from
interacting bosons (that correspond to bound states of fermion pair).

\begin{table}[tb]
 \centering
 \begin{tabular}{ |c||c|c|c|c| }
 \hline
 Symm. $G$ & $d_{sp}=0$ & $d_{sp}=1$ & $d_{sp}=2$ & $d_{sp}=3$  \\
\hline
\hline
\red{$U(1)\rtimes Z_2^T$}  & \red{$\Z$} & \red{$\Z_2$} & \red{$\Z_2$} & \red{$\Z^2_2$}   \\
$U(1)\times Z_2^T$  & $\Z_1$ & $\Z^2_2$ & $\Z_1$ & $\Z^3_2$   \\
{$U(1)$} & {$\Z$}  & {$\Z_1$} & {$\Z$} & {$\Z_1$}    \\
$SO(3)$ & $\Z_1$  & $\Z_2$ & $\Z$ & $\Z_1$    \\
$SO(3)\times Z_2^T$ & $\Z_1$  & $\Z_2^2$ & $\Z_2$ & $\Z_2^3$    \\
\hline
\color{blue}{$Z_2^T$}  & \color{blue}{$\Z_1$} & \color{blue}{$\Z_2$} & \color{blue}{$\Z_1$} & \color{blue}{$\Z_2$}   \\
$Z_n$ & $\Z_n$  & $\Z_1$ & $\Z_n$ & $\Z_1$    \\
$Z_m \times Z_n$
& {\scriptsize $\Z_m\times \Z_n$} & $\Z_{(m,n)}$ & {\scriptsize $\Z_m\times \Z_n\times \Z_{(m,n)}$} & $ \Z^2_{(m,n)}$   \\
%
$Z_n \times Z_2^T$
& $\Z_{(2,n)}$ & {\scriptsize $\Z_2\times \Z_{(2,n)}$} & $\Z^2_{(2,n)}$ & {\scriptsize $\Z_2\times \Z^2_{(2,n)}$}   \\
\hline
 \end{tabular}
 \caption{
(Color online)
Bosonic SPT phases (from group cohomology theory) in $d_{sp}$-spatial dimensions protected by some simple
symmetries (represented by symmetry groups $G$).  Here $\Z_1$ means that our
construction only gives rise to the trivial phase.  $\Z_n$ means that the
constructed nontrivial SPT phases plus the trivial phase are labeled by the
elements in $\Z_n$.  $Z_2^T$ represents time-reversal symmetry, $U(1)$
represents $U(1)$ symmetry, $Z_n$ represents cyclic symmetry, \etc.  Also
$(m,n)$ is the greatest common divisor of $m$ and $n$.  The red rows are for
bosonic topological insulators and the blue rows bosonic topological
superconductors.
}
 \label{tb}
\end{table}
\begin{table}[tb]
 \centering
 \begin{tabular}{ |c||c|c|c|c|c| }
\hline
\multicolumn{6}{|c|} {Interacting fermionic SPT phases}   \\
 \hline
  $G_f\ \backslash d_{sp}$ & $0$ & $1$ & $2$ & $3$ & Example  \\
\hline
\red{``none''=$Z_2^f$}  & \blue{$\Z_2$} & \blue{$\Z_1$} & \blue{$\Z_1$}
& \blue{$\Z_1$} & superconductor  \\
$Z_2\times Z_2^f$
& $\blue{\Z_2^2}$ & \blue{$\Z_2$} & $\Z_4$ & $\Z_1$ & \\
\red{$Z_2^T\times Z_2^f$}
& \blue{$\Z_2$} & \blue{$\Z_4$} & $\Z_1$ & $\Z_2$ &
{\scriptsize $\bmm
\text{supercond. with}\\
\text{coplanar spin order}\\
\emm$}
\\
$Z_{2k+1}\times Z_2^f$
& \blue{$\Z_{4k+2}$} & \blue{$\Z_1$} & $\Z_{2k+1}$ & $\Z_1$ &  \\
$Z_{2k}\times Z_2^f$
& \blue{$\Z_{2k}\times Z_2$} & \blue{$\Z_2$} & $\Z_{4k}$ & $\Z_1$ &   \\
\hline
\hline
\multicolumn{6}{|c|} {Interacting fermionic gapped symmetric phases}   \\
 \hline
 $G_f\ \backslash d_{sp}$ & $0$ & $1$ & $2$ & $3$ & Example  \\
\hline
\red{``none''=$Z_2^f$}
& \blue{$\Z_2$} & \blue{$\Z_2$} & ?  & ? &
superconductor  \\
$Z_2\times Z_2^f$
& $\blue{\Z_2^2}$ & $\blue{\Z_4}$ & ? & ? & \\
\red{$Z_2^T\times Z_2^f$}
& \blue{$\Z_2$} & $\blue{\Z_8}$ & ? & ? &
{\scriptsize $\bmm
\text{supercond. with}\\
\text{coplanar spin order}\\
\emm$}
\\
$Z_{2k+1}\times Z_2^f$
& \blue{$\Z_{4k+2}$} & \blue{$\Z_2$} & ? & ? &  \\
$Z_{2k}\times Z_2^f$
& \blue{$\Z_{2k}\times Z_2$} & $\blue{\Z_4}$ & ? & ? &   \\
\hline
\hline
\multicolumn{6}{|c|} {Non-interacting fermionic SPT phases}   \\
 \hline
 $G_f\ \backslash d_{sp}$ & $0$ & $1$ & $2$ & $3$ & Example  \\
\hline
\red{``none''=$Z_2^f$}  & \blue{$\Z_2$} & \blue{$\Z_1$} & \blue{$\Z_1$} & \blue{$\ \Z_1\ $} & superconductor  \\
$Z_2\times Z_2^f$  & \blue{$\Z_2^2$} & \blue{$\Z_2$} & \blue{$\Z$} & \blue{$\Z_1$} &   \\
\red{$Z_2^T\times Z_2^f$}  & \blue{$\Z_2$} & \blue{$2\Z$} & \blue{$\Z_1$} & \blue{$\Z_1$} &
{\scriptsize $\bmm
\text{supercond. with}\\
\text{coplanar spin order}\\
\emm$}
\\
\hline
\hline
\multicolumn{6}{|c|} {Non-interacting fermionic gapped phases}   \\
 \hline
 $G_f\ \backslash d_{sp}$ & $0$ & $1$ & $2$ & $3$ & Example  \\
\hline
\red{``none''=$Z_2^f$}  & \blue{$\Z_2$} & \blue{$\Z_2$} & \blue{$\Z$} & \blue{$\ \Z_1\ $} & superconductor  \\
$Z_2\times Z_2^f$  & \blue{$\Z_2^2$} & \blue{$\Z_4$} & \blue{$\Z^2$} & \blue{$\Z_1$} &   \\
\red{$Z_2^T\times Z_2^f$}  & \blue{$\Z_2$} & \blue{$\Z$} & \blue{$\Z_1$} & \blue{$\Z_1$} &
{\scriptsize $\bmm
\text{supercond. with}\\
\text{coplanar spin order}\\
\emm$}
\\
\hline
 \end{tabular}
 \caption{
(Color online)
Fermionic SPT phases in $d_{sp}$-spatial dimensions constructed using group
super-cohomology  for some simple symmetries (represented by the full symmetry
groups $G_f$). The red symmetry groups can be realized by electron systems.
Here $\Z_1$ means that our construction only gives rise to the trivial phase.
$\Z_n$ means that the constructed nontrivial SPT phases plus the trivial phase
are labeled by the elements in $\Z_n$.  Note that fermionic SPT phases always
include the bosonic SPT phases with the corresponding bosonic symmetry group
$G_b\equiv G_f/Z_2^f$ as special cases.  The blue entries are complete
constructions which become classifications.  $Z_2^T$ represents time-reversal
symmetry, $U(1)$ represents $U(1)$ symmetry, \etc.  As a comparison, the
results for non-interacting fermionic gapped/SPT
phases,\cite{W1103,K0886,SRF0825} as well as the interacting symmetric phases
in 1D,\cite{FK1009,FK1103,TPB1102,CGW1128} are also listed.  Note that the
symmetric interacting and non-interacting fermionic gapped phases can be the
SPT phases or intrinsically topologically ordered phases.  This is why the
lists for gapped phases and for SPT phases are different.  $2\Z$ means that the
phases are labeled by even integers.  Note that all the phases listed above
respect the symmetry $G_f$.
}
\label{tbF}
\end{table}

\subsection{A summary of main results}

The constructed fermionic SPT phases for some simple symmetry groups are given
in Table \ref{tbF}, which lists the  special group super-cohomology class
$\fH^{d+1}[G_f, U(1)]$.  The rows correspond to different symmetries for the
fermion systems.  We note that, in literature, when we describe the symmetry of
a fermion system, sometimes we include the fermion-number-parity transformation
$P_f=(-)^{N}$ in the symmetry group, and sometimes we do not.  In this paper
and in Table \ref{tbF}, we always use the full symmetry group $G_f$ which
includes the fermion-number-parity transformation to describe the symmetry of
fermion systems.\cite{W1103}  So the full  symmetry group of a fermion system
with no symmetry is $G_f=Z_2^f$ generated by $P_f$.  The bosonic symmetry group
$G_b$ is given by $G_b=G_f/Z_2^f$, which correspond to the physical symmetry of
the fermion system that can be broken.  In fact, the full symmetry group is a
projective symmetry group discussed in \Ref{Wqoslpub}, which is a $Z_2^f$
extension of the physical symmetry group $G_b$.

The columns of the Table \ref{tbF} correspond to different spatial dimensions.
In 0D and 1D, our results reproduce the exact results obtained from previous
studies.\cite{FK1009,FK1103,TPB1102,CGW1128}  The results for 2D and 3D are
new.

Each entry indicates the number of nontrivial phases plus one trivial phase.
For example, $\Z_2$ means that there is one nontrivial SPT phase labeled by 1
and one trivial phase labeled by 0. Also, say, $\Z_8$ means that there are seven
nontrivial SPT phases and one trivial phase.  For each nontrivial fermionic
SPT phase, we can construct the ideal ground state wave function and the ideal
Hamiltonian that realizes the SPT phase, using group super-cohomology theory.


When $G_f=G_b\times Z_2^f$, $\fH^{d}[G_f, U(1)]$ can be calculated from the
following short exact sequence
\begin{align}
\label{HHH}
0 &\to \cH^d[G_b,U_T(1)]/\Ga \to  \fH^d[G_f,U_T(1)]
\nonumber\\
&
\to B\cH^{d-1}(G_b,\Z_2) \to 0.
\end{align}
In the above $B\cH^{d-1}(G_b,\Z_2)$ is a subgroup of $\cH^{d-1}(G_b,\Z_2)$,
which is formed by elements $n_{d-1}$ that satisfy $Sq^2(n_{d-1})=0$ in
$\cH^{d+1}(G_b,U(1)_T)$, where $Sq^2$ is the Steenrod square $Sq^2:
\cH^{d-1}(G_b,\Z_2)\to \cH^{d+1}(G_b,\Z_2) \subset \cH^{d+1}[G_b,U(1)_T]$ .
Also $\Ga$ is a subgroup of $\cH^d[G_b,U(1)_T]$ that is generated by $Sq^2
(n_{d-2})=f_d$, where $n_{d-2} \in \cH^{d-2}(G_b,\Z_2)$ and $f_d$ are viewed as
elements of $\cH^d[G_b,U_T(1)]$.

\subsection{The boundary of fermionic SPT states}

Topologically ordered states and SPT states have short-ranged correlations for
any local operators.  So, it is impossible to probe and distinguish the
different topological states by bulk linear response measurements.  However,
for chiral topological order (such as integer and fractional quantum Hall
states)\cite{Wedgerev,Wtoprev} and free fermion SPT states (such as topological
insulators/superconductors\cite{KM0501,BZ0602,KM0502,MB0706,FKM0703,QHZ0824}),
their boundary states are gapless if the symmetry is not broken.  In this case,
we can use boundary linear response measurements to probe those gapless
excitations, which allow us to indirectly measure the bulk topological phases.

Thus, it is very natural to ask the following: can we use the gapless boundary states to
probe the SPT order in interacting fermionic SPT states? It turns out that, in
the presence of interaction, the situation is much more complicated.  In fact,
the situation is already complicated even for non-interacting cases.  It is
well known that even trivial band insulator can have gapless boundary states.
So it is incorrect to say that topological insulators/superconductors are
characterized by gapless boundary states.  The situation gets worse for
interacting cases: regardless if the bulk is a trivial or nontrivial
bosonic/fermionic SPT phase, the interacting boundary can be symmetry breaking,
gapless, topological, \etc.  In fact, there can be infinite many different
boundary phases for every fixed 3+1D bulk phase.  So in order to use the
boundary states to characterize the bulk SPT order, we must identify the common
features among \emph{all} those infinite many different boundary phases of the
same bulk.  Only those common features characterize the bulk SPT order.  This
is a highly nontrivial task and has not been studied carefully in literature.

In this paper, just like the bosonic case,\cite{CLW1141,CGL1172} we propose the
boundary anomalous symmetry (\ie the boundary non-on-site symmetry) as one of
the common features that characterize the bulk SPT order.  The standard global
symmetry transformation (\ie on-site symmetry transformation) in the bulk,
$\hat U(g), g\in G$, has the following tensor product decomposition:
\begin{align}
\hat U(g) = \prod_i \hat U_i(g)
\end{align}
where $\hat U_i(g)$ acts on single site $i$.  However, although the boundary of
a SPT state has the same symmetry as the bulk, the symmetry transformation on
the boundary cannot be on-site if the bulk SPT order is nontrivial:
\begin{align}
\hat U_\text{bndry}(g) = w(\{g_i\}, g)\prod_i \hat U_i(g).
\end{align}
Here $g_i$ labels the effective degrees of freedom on the boundary site $i$,
and the $U(1)$ phase factor $w(\{g_i\}, g)$ makes the boundary symmetry
transformation $\hat U_\text{bndry}(g)$ non-on-site or anomalous.  In fact, the
$U(1)$ phase factor $w(\{g_i\}, g)$ can be constructed from the super-cocycle
in $\fH^d[G_f,U_T(1)]$ that describes the bulk SPT state (for details, see
Appendix \ref{symmedge}).

Since all the effective boundary Hamiltonians satisfy
\begin{align}
 \hat U^\dag_\text{bndry}(g) H_\text{bndry} \hat U_\text{bndry}(g)
 = H_\text{bndry},
\end{align}
and all the possible boundary types are described by the ground states of the
above boundary Hamiltonians, many low energy properties of boundary state are
determined by the anomalous (\ie non-on-site) symmetry  $\hat
U_\text{bndry}(g)$.  For example, for 1+1D boundary, an anomalous symmetry
makes the boundary state gapless if the symmetry is not broken.\cite{CLW1141}
For 2+1D boundary and beyond, an anomalous symmetry makes the symmetric
boundary state gapless or topologically ordered.

The above results can also be understood from space-time path integral point of
view.  Our discrete fermionic topological nonlinear $\si$ model, when defined
on a space-time with boundary, can be viewed as a ``non-local'' boundary
effective Lagrangian, which is a fermionic and discrete generalization of the
bosonic continuous Wess-Zumino-Witten (WZW) term.\cite{WZ7195,W8322}  As a
result of this ``non-local'' boundary effective Lagrangian, the action of
symmetry transformation on the low energy boundary degrees of freedom must be
non-on-site, and, we believe, the boundary excitations of a nontrivial SPT
phase are gapless or topologically ordered if the symmetry is not broken.

\subsection{Structure of the paper}

In the rest of this paper, we will first compare the results in Table \ref{tbF}
for a few interacting and free fermion systems. This will give us some physical
understanding of Table \ref{tbF}.  We then briefly review the topological
bosonic nonlinear $\si$ model on discretized space-time, which leads to the
group cohomology theory for the bosonic SPT states.  We start our development
of group super-cohomoloy theory for fermionic SPT phases by carefully defining
fermionic path integral for the fermionic nonlinear $\si$ model on discrete
space-time.  Next we discuss the conditions under which the fermionic path
integral becomes a fixed-point theory under the coarse-graining transformation
of the space-time complex.  Such a fixed-point theory is a fermionic
topological nonlinear $\si$ model.  The fixed-point path integral describes a
fermionic SPT phase. We then construct the ground state wave function from the
fixed-point path integral, as well as the exact solvable Hamiltonian that
realizes the SPT states.  In the Appendices, we develop a group super-cohomology
theory, and calculate the Table \ref{tbF} from the group super-cohomology
theory.

\section{Physical pictures of some generic results}

Before describing how to obtain the generic results \eqn{HHH}  and Table
\ref{tbF}, in this section, we will compare some of our results with known
results for 1D interacting systems and 2D/3D non-interacting systems.  The
comparison will give us a physical understanding for some of the interacting
fermionic SPT phases.

\subsection{Fermion systems with symmetry $G_f=Z_2^T\times Z_2^f$}

First, from the  Table \ref{tbF}, we find that interacting fermion systems with
$T^2=1$ time-reversal symmetry (or the full symmetry group $G_f=Z_2^T\times
Z_2^f$) can have three nontrivial fermionic SPT phases in $d_{sp}=1$ spatial
dimension and one nontrivial fermionic SPT phase in $d_{sp}=3$ spatial
dimensions. We would like to compare such results with those for free fermion
systems.

\subsubsection{0D case}

For 0D free or interacting fermion systems with $T^2=1$ time-reversal symmetry
$Z_2^T$, the possible symmetric gapped phases are the 1D representations of
$Z_2^T\times Z_2^f$. Since the time-reversal transformation $T$ is anti unitary,
$Z_2^T$ has only one trivial 1D representation.  $Z_2^f$ has two 1D
representations.  Thus 0D fermion systems with $T^2=1$ time-reversal symmetry
are classified by $\Z_2$ corresponding to even fermion states and odd fermion
states.

\subsubsection{1D case}

For 1D \emph{free} fermion systems with $T^2=1$ time-reversal symmetry $Z_2^T$,
the possible gapped phases are classified by $\Z$.\cite{K0886,SRF0825} For the
phase labeled by $n\in \Z$, it has $n$ Majorana zero modes at one end of the
chain.\cite{K0131} Those boundary states form a representation of $n$ Majorana
fermion operators $\eta_1,...,\eta_n$ which all transform in the same way under
the time-reversal transformation $T$: $\eta_a \to \eta_a$ for all $a$ or
$\eta_a \to -\eta_a$ for all $a$.  (If, say $\eta_1$ and $\eta_2$ transform
differently under $T$: $\eta_1 \to \eta_1$ and $\eta_2 \to -\eta_2$, $\imth
\eta_1\eta_2$ will be invariant under $T$ and such a term will gap out the
$\eta_1$ and $\eta_2$ modes.) Some of those 1D gapped states have intrinsic
topological orders.  Only those labeled by even integers (which have even
numbers of Majorana boundary zero modes) become the trivial phase after we
break the time-reversal symmetry $Z_2^T$.  Thus the free fermion SPT phases are
labeled by even integers $2\Z$.

For interacting 1D systems with $T^2=1$ time-reversal symmetry, there are eight
possible gapped phases that do not break the time-reversal
symmetry.\cite{FK1009,FK1103,TPB1102,CGW1128}  Four of them have intrinsic
topological orders (which break the fermion-number-parity symmetry in the
bosonized model)\cite{CGW1128} and the other four are fermionic SPT phases
given in Table \ref{tbF} (three of them are nontrivial fermionic SPT phases
and the fourth one is the trivial phase).

\subsubsection{3D case}

For 3D fermion systems with $T^2=1$ time-reversal symmetry, there is no
nontrivial gapped phase for non-interacting fermions,\cite{K0886,SRF0825} but
in contrast, there is (at least) one nontrivial SPT phase for strongly
interacting fermions.  Such a nontrivial 3D fermionic SPT phase cannot even be
viewed as a nontrivial bosonic SPT state formed by bounded fermion pairs.  To
see this point, we note that bounded fermion pairs behave like a boson system
with $Z_2^T$ time-reversal symmetry, which has one nontrivial bosonic SPT
state (see Table \ref{tb}).  Such a state cannot be smoothly deformed into
bosonic product state without breaking the $Z_2^T$ symmetry within the space of
\emph{many-boson} states.  But it can  be smoothly deformed into bosonic
product state without breaking the $Z_2^T$ symmetry within the space of
\emph{many-fermion} states.  Therefore, the discovered non-trival fermionic SPT
phase with $T^2=1$ time-reversal symmetry is totally new.

\subsection{Fermion systems with symmetry $G_f=Z_2\times Z_2^f$}

We also find that fermion systems with symmetry group $G_b=Z_2$ (or the full
symmetry group $G_f=Z_2\times Z_2^f$) can have one nontrivial fermionic SPT
phases in $d_{sp}=1$ spatial dimension and three nontrivial fermionic SPT
phases in $d_{sp}=2$ spatial dimensions.  (One of the nontrivial 3D fermionic
SPT phases is actually a bosonic SPT phase while the other two are
intrinsically fermionic.)

\subsubsection{Free fermion systems}

To compare the above result with the known free fermion result,
let us consider free fermion systems with symmetry group $G_b=Z_2$ (or the full
symmetry group $G_f=Z_2\times Z_2^f$), which contain two kinds of fermions: one
carries the $Z_2$-charge 0 and the other carries $Z_2$-charge 1.  For such free
fermion systems, their gapped phases are classified
by\cite{W1103}
\begin{align}
 \bmm d_{sp}:& 0 & 1 & 2 & 3 & 4 &  5 & 6& 7\\
  \text{gapped phases}: &\Z_2^2 & \Z_2^2 & \Z^2 & 0 & 0 &  0 & \Z^2 & 0  \\
\emm
\end{align}
The four $d_{sp}=0$ phases correspond to the ground state with even or odd
$Z_2$-charge-0 fermions and even or odd $Z_2$-charge-1 fermions.  The four
$d_{sp}=1$ phases correspond to the phases where the $Z_2$-charge-0 fermions
are in the trivial or nontrivial phases of Majorana chain\cite{K0131} and the
$Z_2$-charge-1 fermions are in the trivial or nontrivial phases of Majorana
chain.  A $d_{sp}=2$ phase labeled by two integers $(m,n)\in \Z^2$
corresponds to the phase where the $Z_2$-charge-0 fermions form $m$ $(p_x+\imth
p_y)$ states with $m$ right moving Majorana chiral modes and the $Z_2$-charge-1
fermions form $n$ $(p_x+\imth p_y)$ states with $n$ right moving Majorana
chiral modes.  (If $m$ and/or $n$ are negative, the fermions then form the
corresponding number of $(p_x-\imth p_y)$ states with the corresponding number
of left moving Majorana chiral modes.)

Some of the above non-interacting gapped phases have intrinsic fermionic
topological orders. (Those phases are stable and have intrinsic fermionic
topological orders even after we turn on interactions.) So only a subset of
them are non-interacting fermionic SPT phases:
\begin{align}
 \bmm d_{sp}:& 0 & 1 & 2 & 3 & 4 &  5 & 6& 7\\
  \text{SPT phases}: &\Z_2^2 & \Z_2 & \Z & 0 & 0 &  0 & \Z & 0  \\
\emm
\end{align}
The four $d_{sp}=0$ phases correspond to the ground states with even or odd
numbers of fermions and 0 or 1 $Z_2$-charges.  The two $d_{sp}=1$ phases
correspond to the phases where the $Z_2$-charge-0 fermions and the
$Z_2$-charge-1 fermions are both in the trivial or nontrivial phases of
Majorana chain.  The $d_{sp}=2$ phase labeled by one integer $n\in \Z$
corresponds to the phase where the $Z_2$-charge-0 fermions have $n$ right
moving Majorana chiral modes and the $Z_2$-charge-1 fermions have $n$ left
moving Majorana chiral modes.  We see that in 0D and 1D, the non-interacting
fermionic SPT phases are the same as the interacting fermionic SPT phases.

\subsubsection{2D case}

However, in two dimensions, the non-interacting fermionic SPT phases are quite
different from the interacting ones. In this paper, we are able to construct
three nontrivial interacting fermionic SPT phases.  Despite very different
phase diagram, it appears that the above three nontrivial interacting
fermionic SPT phases in 2D can be realized by free fermion systems.

In fact, it appears that there are seven nontrivial interacting 2D fermionic
SPT phases protected by $Z_2$ symmetry\cite{fduality}.  All the seven
nontrivial SPT phases can be represented by free fermion SPT phases.  This
suggests that our current construction is incomplete, since we only obtain four
fermionic SPT phases (including the trivial one) with the $G_f=Z_2\times Z_2^f$
symmetry.  One possible reason of the incompleteness may be due to our limiting
requirement that fermions only form 1D representations of $G_f$.

One way to understand why there can only be seven nontrivial interacting 2D
fermionic SPT phases protected by $Z_2$ symmetry is using the idea of duality
between intrinsic topological orders and SPT orders discovered
recently\cite{duality}.  The key observation in such a duality map is that for
any SPT orders associated with a (discrete) global symmetry $G$, we can always
promote the global symmetry to a local (gauge) symmetry. For different SPT
phases, the corresponding promoted (discrete) gauge models describe different
intrinsic topological orders. For fermionic SPT phases protected by $Z_2$
symmetry, we can let $Z_2$-charge-1 fermion couple to a $Z_2$ gauge field. In
this way, different intrinsic topological ordered phases can be characterized
by different statistics of the $Z_2$-flux. According to Kitaev's
classification\cite{KitaevSC} for different types of vortices in
superconductors\footnote{In a superconductor, the $U(1)$ symmetry is broken
down to $Z_2$ symmetry, hence the vortex of a superconductor can be regarded as
$Z_2$-flux.}, we know that there are seven nontrivial cases. Thus, we see that
although in free 2D fermionic systems, SPT phases protected by $Z_2$ symmetry
are classified by an integer (Chern number), the interactions dramatically
changes the classifications.

Following Kitaev's idea\cite{KitaevSC}, we can have a very simple way to
understand the seven nontrivial types of vortices by counting the number of
Majorana modes in the vertex core. In the corresponding free fermion model, the
number of Majorana modes $n$ corresponds to the Chern number of the
$Z_2$-charge-1 free fermion. The seven nontrivial SPT phases are described by
$n=1,2,\cdots,7$.

We see that an interacting 2D fermionic SPT state with $Z_2$ symmetry
($G_f=Z_2\times Z_2^f$) is characterized by having $n$ right-moving Majorana
chiral modes for the $Z_2$-charge-0 fermions and $n$ left-moving Majorana
chiral modes for the $Z_2$-charge-1 fermions, where $n\in \Z_8$.  Such kind of
edge states can be realized by free fermions.  Thus the interacting 2D
fermionic SPT states with $Z_2$ symmetry can be realized by free fermions.
When $n=$ even, the 2D fermionic SPT states with $Z_2$ symmetry can be realized
by 2D topological insulators with fermion number conservation, $S^z$ spin
rotations symmetry, and time reversal symmetry.  The $Z_2$ symmetry
transformation correspond to $\pi/2$ charge rotrations and $\pi$ spin rotation
\begin{align}
 U_{Z_2}=
\e^{\imth \pi N_F/2}
\e^{\imth \pi S^z},
\end{align}
where $N_F$ is the total fermion number and $S^z$ is the total $S^z$ spin.
Such a state is stable even if we break the fermion number conservation, $S^z$
spin rotations symmetry, and time reversal symmetry, as long as we keep the
above $Z_2$ symmetry.

Now, let us try to understand why four of the nontrivial fermionic SPT phases
protected by the $Z_2$ symmetry require fermions to form high dimensional
representations of $G_f=Z_2\times Z_2^f$. It is easy to see that when $n=even$,
the free fermion models are described by $n/2$ $Z_2$-charge-1 complex fermions
per site and these fermions form 1D representations of $G_f=Z_2\times Z_2^f$.
However, the situations are more complicated when $n=odd$. For example, when
$n=1$, the free fermion Hamiltonian describes one $Z_2$-charge-0 Majorana
fermion and one $Z_2$-charge-1 Majorana fermion per site, labeled as
$\gamma_{i,\uparrow}$ and $\gamma_{i,\downarrow}$. Under the $Z_2$ action,
$\gamma_{i,\uparrow}$ does not change while $\gamma_{i,\downarrow}$ changes to
$-\gamma_{i,\downarrow}$. Thus the symmetry operation $U_{Z_2}$ can be
constructed as $U_{Z_2}=\imth^{(N-1)N/2}\prod_{i=1}^N\gamma_{i,\uparrow}$,
where $N$ is the number of sites. It is easy to check that
$U_{Z_2}\gamma_{i,\uparrow} U_{Z_2}^\dagger=\gamma_{i,\uparrow}$,
$U_{Z_2}\gamma_{i,\downarrow} U_{Z_2}^\dagger=-\gamma_{i,\downarrow}$ and
$U_{Z_2}^2=1$. However, the symmetry operator $U_{Z_2}$ can be regarded as the
$Z_2$ symmetry transformation only when $N=$ even, in that case $U_{Z_2}$
contains an even number of fermion operators.  But in our construction, the
$Z_2$ symmetry can be realized as bosonic unitary transformation regardless
$N=$ even or $N=$ odd.
Thus, we can only construct  four fermionic SPT phases that are labeled by even
$n$.

Another example requiring fermions to carry high dimensional representations is
the well known time reversal symmetry with $T^2=P_f$. We believe that the
principle/framework developed in this paper should be applicable for all these
interesting cases and the results will be presented in our future work.

\section{Path integral approach to bosonic SPT phases}
\label{bSPT}

\begin{figure}[tb]
\begin{center}
\includegraphics[scale=0.15]{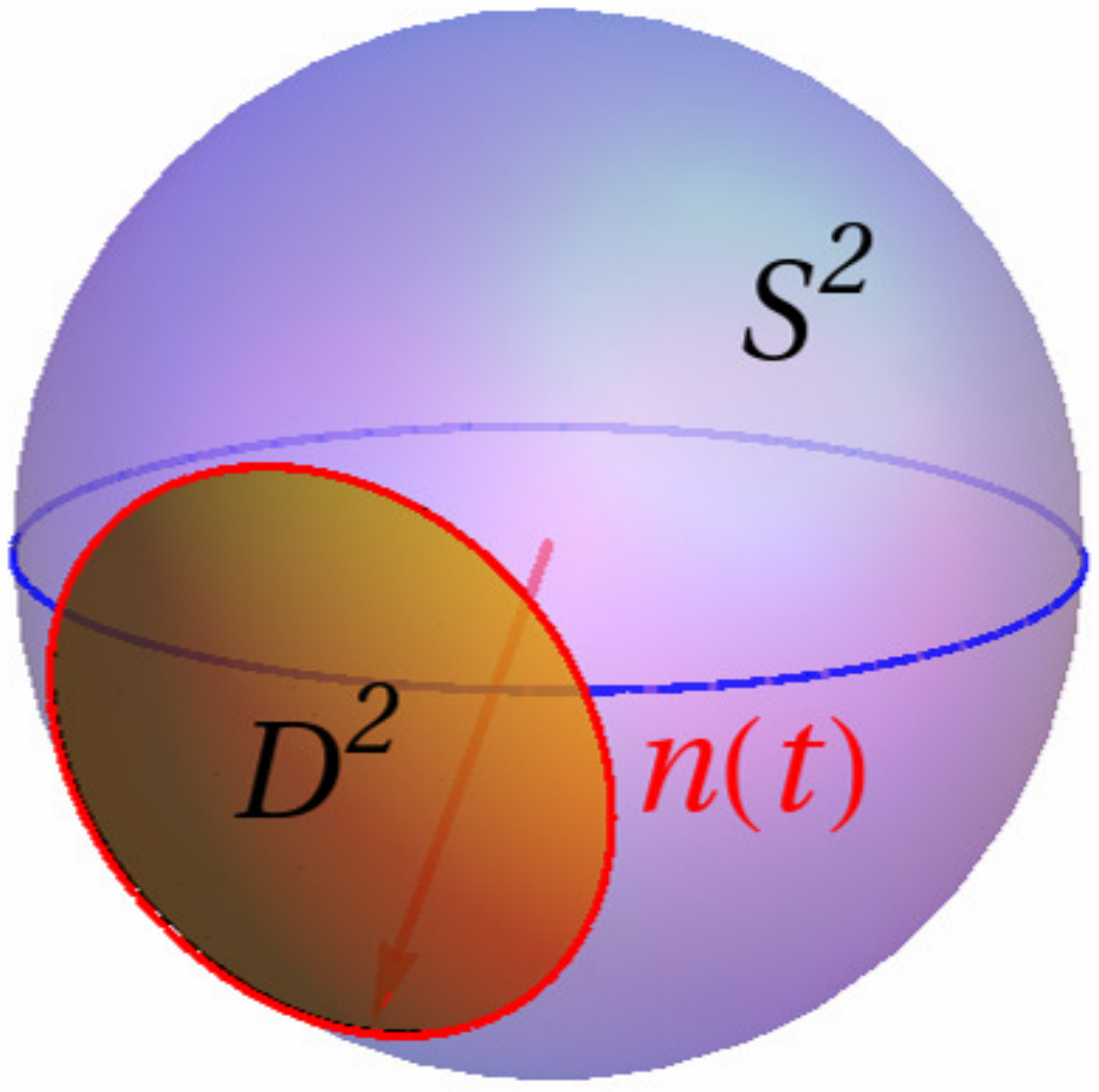}
\end{center}
\caption{
(Color online)
If we extend $\v n(t)$ that traces out a loop
to $\v n(t,\xi)$ that covers the shaded disk,
then the WZW term $\int_{D^2} \dd t\dd \xi\; \v n(t,\xi)\cdot
[\prt_t \v n(t,\xi) \times  \prt_\xi \v n(t,\xi)]$ corresponds to the area of
the disk.
}
\label{topterm1}
\end{figure}

In this paper, we are going develop a group super-cohomology theory, trying to
describe the SPT phase for interacting fermions.  Our approach is motivated by
the group cohomology theory for bosonic SPT phases, based on the path integral
approach.  In this section, we will briefly review such path integral approach
for the group cohomology description of bosonic SPT phases.  Those whose are
familar with bosonic SPT theory can go directly to the next section.

Here we are going to use the Haldane phase of spin-1 chain as an example.  It
has been pointed out that the Haldane phase (a nontrivial 1D SPT phase)
is described by a $2\pi$-quantized topological term in continuous nonlinear
$\si$ model.\cite{N9455} However, such kinds of $2\pi$-quantized topological
terms cannot describe more general 1D SPT phases.  We argue that to describe
SPT phases correctly, we must generalize the $2\pi$-quantized topological terms
to discrete space-time.  The generalized $2\pi$-quantized topological terms
turn out to be nothing but the cocycles of group cohomology (see
\Ref{CLW1141,CGL1172} for more details.)

\subsection{Path integral approach to a single spin}

Before considering a spin-1 chain, let us first consider a (0+1)D nonlinear
$\si$ model that describes a single spin, whose imaginary-time action is given
by
\begin{align}
\label{spins}
\oint \dd t  \frac{[\prt_t \v n(t)]^2}{2g}
+ \imth s \int_{D^2} \dd t\dd \xi\; \v n(t,\xi)\cdot
[\prt_t \v n(t,\xi) \times  \prt_\xi \v n(t,\xi)]
\end{align}
where $\v n(t)$ is an unit 3d vector and we have assumed that the time
direction forms a circle.  The second term is the Wess-Zumino-Witten (WZW)
term.\cite{WZ7195,W8322} We note that the WZW term cannot be calculated from
the field $\v n(t)$ on the time-circle.  We have to extend $\v n(t)$ to a disk
$D^2$ bounded by the time-circle: $\v n(t) \to \v n(t,\xi)$ (see Fig.
\ref{topterm1}).  Then the WZW term can be calculated from $\v n(t,\xi)$. When
$2s$ is an integer, WZW terms from different extensions only differ by a
multiple of $2\imth \pi$.  So $\e^{-S}$ is determined by $\v n(t)$ and is
independent of how we extend $\v n(t)$ to the disk $D^2$.

The ground states of the above nonlinear $\si$ model have $2s+1$ fold
degeneracy, which form the spin-$s$ representation of $SO(3)$.  The energy gap
above the ground state approaches to infinity as $g\to \infty$. Thus a pure WZW
term describes a pure spin-$s$ spin.

\begin{figure}[tb]
\begin{center}
\includegraphics[scale=0.42]{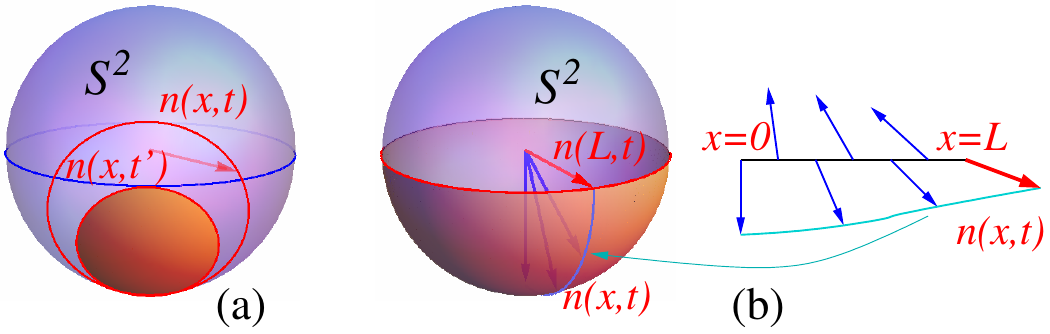}
\end{center}
\caption{
(Color online)
(a) The topological term $W$ describes the number of times that $\v n(x,t)$
wraps around the sphere (as we change $t$).  (b) On an open chain $x \in
[0,L]$, the  topological term $W$ in the (1+1)D bulk becomes the WZW term for
the end spin $ \v n_L(t)=\v n(L,t)$ (where the end spin at $x=0$ is hold
fixed).
}
\label{toptermChain}
\end{figure}

\subsection{Path integral approach to a spin-1 chain}

To obtain the action for the $SO(3)$ symmetric antiferromagnetic spin-1 chain,
we can assume that the spins $\v S_i$
are described
by a smooth unit vector field $\v n(x,t)$:
$\v S_i = (-)^i \v n(i a,t)$ (see Fig. \ref{toptermChain}b).
Putting the above single-spin
action for different spins together, we obtain the following (1+1)D nonlinear
$\si$ model\cite{H8559}
\begin{align}
\label{2DS}
S=\int \dd x\dd t \frac{1}{2g}(\prt \v n(x,t))^2 +  \imth\th W ,\ \ \
\th=2\pi,
\end{align}
where $W=(4\pi)^{-1} \int \dd t\dd x\; \v n(t,x)\cdot [\prt_t \v n(t,x) \times
\prt_x \v n(t,x)]$ and $ \imth \th  W$ is the topological term.\cite{H8559}  If
the space-time manifold has no boundary, then $\e^{-\imth \th W}=1$ when $\th=0
\text{ mod } 2\pi$.  We will call such a topological term -- a $2\pi$-quantized
topological term.
The above nonlinear $\si$ model describes a gapped phase
with short range correlation and the $SO(3)$ symmetry, which is the Haldane
phase.\cite{H8364} In the low energy limit, $g$ flows to infinity and the
fixed-point action contains only the $2\pi$-quantized topological term.  Such a
nonlinear $\si$ model will be called topological nonlinear $\si$ model.

It appears that the $2\pi$-quantized topological term has no contribution
to the path integral and can be dropped.  In fact, the $2\pi$-quantized topological
term has physical effects and cannot be dropped.  On an open chain, the
$2\pi$-quantized topological term $2\pi \imth W$ becomes a WZW term
for the boundary spin $\v n_L(t) \equiv \v n(x=L,t)$ (see Fig.
\ref{toptermChain}).\cite{N9455} The motion of $\v n_L$ is described by
\eqn{spins} with $s=1/2$.  So the Haldane phase of spin-1 chain has a spin-1/2
boundary spin at each chain end!\cite{HKA9081,GGL9114,N9455}

We see that the Haldane phase is described by a fixed point action which is a
topological nonlinear $\si$ model containing only the $2\pi$-quantized
topological term.  \emph{The non-trivialness of the Haldane phase is encoded
in the non trivially quantized topological term}.\cite{N9455,SRF0825}

\begin{figure}[tb]
\begin{center}
\includegraphics[scale=0.55]{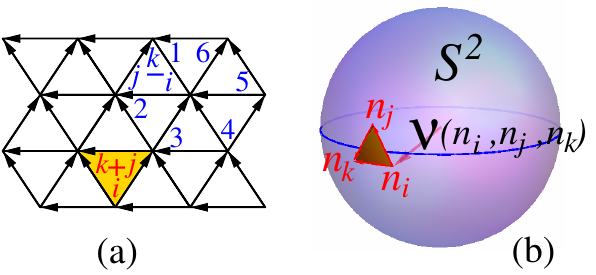}
\end{center}
\caption{
(Color online)
(a) A branched triangularization of space-time. Each edge has an orientation
and the orientations on the three edges of any triangle do not form a loop.
The orientations on the edges give rise to a natural order of the three
vertices of a triangle $(i,j,k)$ where the first vertex $i$ of a triangle has
two outgoing edges on the triangle and the last vertex $k$ of a triangle has
two incoming edges on the triangle.  $s(i,j,k)=\pm 1$ depending on the
orientation of $i\to j \to k$ to be clockwise or anti-clockwise.  (b) The phase
factor $\nu(\v n_i, \v n_j, \v n_k)$ depends on the image of a space-time
triangle on the sphere $S^2$.
}
\label{discreteST}
\end{figure}
\begin{figure}[tb]
\begin{center}
\includegraphics[scale=0.45]{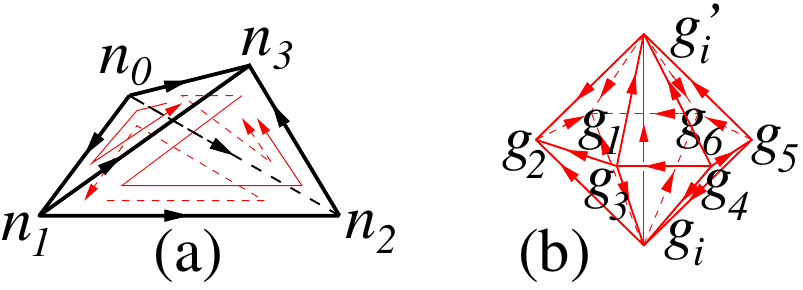}
\end{center}
\caption{
(Color online)
A tetrahedron -- the simplest discrete sphere.
$\prod \nu^{s(i,j,k)}( \v n_i,
\v n_j, \v n_k)=1$ on the tetrahedron becomes \eqn{2ccy1}.
Note that $s(1,2,3)=s(0,1,3)=1$ and
$s(0,2,3)=s(0,1,2)=-1$.
(b) The total action amplitude of
the topological nonlinear $\si$ model
on the complex gives rise to the phase factor in \eqn{Hi1}.
}
\label{n0123}
\end{figure}

\subsection{Topological term on discrete space-time}

From the above example,
one might guess that various SPT phases can be
described by various topological nonlinear $\si$ models, and thus by various
$2\pi$-quantized topological terms.
But such a guess is not correct.

This is because the fixed-point action (the topological nonlinear $\si$ model)
describes a short-range-correlated state.
Since the renormalized cut-off length scale of the  fixed-point action is always
larger than the correlation length,
the field $\v n( x,t)$ fluctuates strongly even at the
cut-off length scale.  Thus, the fixed-point action has no continuum limit, and
must be defined on discrete space-time.  On the other hand, in our fixed-point
action, $\v n(x,t)$ is assumed to be a continuous field in space-time. The very
existence of the continuum $2\pi$-quantized topological term depends on the
nontrivial mapping classes from the \emph{continuous} space-time manifold
$T^2$ to the \emph{continuous} target space $S^2$.  It is not self consistent
to use such a continuum topological term to describe the fixed-point action for
the
Haldane phase.

As a result, the continuum $2\pi$-quantized topological terms fail to properly
describe
bosonic SPT phases. For example, different possible continuum $2\pi$-quantized
topological terms in \eqn{2DS} are labeled by integers, while the integer spin
chain has only two gapped phases protected by spin rotation symmetry: all
even-integer topological terms give rise to the trivial phase and all
odd-integer topological terms give rise to the Haldane phase.  Also nontrivial
SPT phases may exist even when there are no continuum $2\pi$-quantized
topological terms (such as when the symmetry $G$ is discrete).

However, the general idea of using fixed-point actions
to describe
SPT phases is still correct.  But, to use $2\pi$-quantized topological terms
to describe bosonic SPT phases, we need to generalize  them to discrete
space-time.  In the following, we will show that this indeed can be done, using
the (1+1)D model \eq{2DS} as an example.

A discrete (1+1)D space-time is given by a branched triangularization\cite{C0527,CGL1172}
(see Fig.  \ref{discreteST}).  Since $S=\int \dd x\dd t L$, on triangularized
space-time, we can rewrite
\begin{align}
& e^{-S}=\prod \nu^{s(i,j,k)}( \v n_i, \v n_j, \v n_k),
\nonumber\\
& \nu^{s(i,j,k)}( \v n_i, \v n_j, \v n_k) =e^{-\int_\triangle \dd x\dd t\; L}
\in U(1),
\end{align}
where $\int_\triangle \dd x\dd t\; L$ is the action on a single triangle.
We see that, on discrete space-time, the action and the path integral are
described by a 3-variable function $\nu( \v n_i, \v n_j, \v n_k)$,
which is called action amplitude.
The $SO(3)$ symmetry requires that
\begin{align}
\label{2cch}
 \nu( g\v n_i, g\v n_j, g\v n_k)=\nu( \v n_i, \v n_j, \v n_k),
\ \ \ g\in SO(3).
\end{align}
In order to use the action amplitude $\nu^{s(i,j,k)}( \v n_i, \v n_j, \v n_k) $
to describe  a $2\pi$-quantized topological term, we must have $\prod
\nu^{s(i,j,k)}( \v n_i, \v n_j, \v n_k)=1$ on any sphere.  This can be
satisfied iff $\prod \nu^{s(i,j,k)}( \v n_i, \v n_j, \v n_k)=1$ on a
tetrahedron -- the simplest discrete sphere (See Fig. \ref{n0123}a):
\begin{align}
\label{2ccy1}
\frac{\nu( \v n_1, \v n_2, \v n_3) \nu( \v n_0, \v n_1, \v n_3)}
{\nu( \v n_0,  \v n_2, \v n_3) \nu( \v n_0, \v n_1, \v n_2 )}
 =1.
\end{align}
(Another way to define topological term on discretized space-time can be
found in \Ref{S8437}.)

A $\nu( \v n_0,  \v n_1, \v n_2)$ that satisfies \eqn{2cch} and \eqn{2ccy1} is
called a 2-cocycle.  If $\nu( \v n_0,  \v n_1, \v n_2)$ is a 2-cocycle, then
\begin{align}
 \nu'( \v n_0,  \v n_1, \v n_2)=
 \nu( \v n_0,  \v n_1, \v n_2)
\frac{\mu(\v n_1,\v n_2) \mu(\v n_0,\v n_1)}{\mu(\v n_0,\v n_2)}
\end{align}
is also a 2-cocycle, for any $\mu(\v n_0,\v n_1)$ satisfying $\mu(g\v n_0,g\v
n_1)=\mu(\v n_0,\v n_1)$, $g\in SO(3)$.  Since $\nu( \v n_0,  \v n_2, \v n_3)$
and $\nu'( \v n_0,  \v n_2, \v n_3)$ can continuously deform into each other,
they correspond to the same kind of $2\pi$-quantized topological term. So we
say that $\nu( \v n_0,  \v n_2, \v n_3)$ and $\nu'( \v n_0,  \v n_2, \v n_3)$
are equivalent.  The equivalent classes of the 2-cocycles $\nu( \v n_0,  \v
n_2, \v n_3)$ give us $\cH^2[S^2,U(1)]$ -- the 2-cohomology group of sphere
$S^2$ with $U(1)$ coefficient.  $\cH^2(S^2,U(1)$ classifies the
$2\pi$-quantized topological term in \emph{discrete} space-time and with $S^2$
as the target space.

\subsection{Maximum symmetric space and group cohomology classes}

Does $\cH^2[S^2,U(1)]$ classify the SPT phases with $SO(3)$ symmetry?  The
answer is no. We know that $S^2$ is just one of the symmetric spaces of
$SO(3)$. To classify the SPT phases, we need to replace the target space $S^2$
by the maximal symmetric space, which is the group itself $SO(3)$ (see
\Ref{CGL1172} for more discussions).  So we need to consider discrete nonlinear
$\si$ model described by $\nu(g_i,g_j,g_k)$, $g_i,g_j,g_k\in SO(3)$.  Now the
2-cocycle conditions becomes
\begin{align}
&
 \nu( gg_i, gg_j, gg_k)=\nu( g_i, g_j, g_k) \in U(1),
\nonumber\\
& \frac{\nu( g_1, g_2, g_3) \nu( g_0, g_1, g_3)}
{\nu( g_0,  g_2, g_3) \nu( g_0, g_1, g_2 )}
 =1,
\end{align}
which defines a ``group cohomology'' $\cH^2[SO(3),U(1)]$.  It
classifies the $2\pi$-quantized topological term for the maximal symmetric
space.  It also classifies the SPT phases with $SO(3)$ symmetry in (1+1)D.

The above 2-cocycle condition can be generalized to any group $G$, including
discrete groups!  We conclude that $\cH^2[G,U(1)]$ classifies the SPT phases
with on-site symmetry $G$ in (1+1)D.  The above discussion can also be
generalized to any dimensions by replacing the 2-cocycle condition on
$\nu(g_0,g_1,g_2)$ with $d$-cocycle condition on $\nu(g_0,g_1,...,g_d)$. For
example, the functions $\nu(g_0,...,g_3)$, $g_i \in G$, satisfying the
following conditions
\begin{align}
&
 \nu( gg_0, gg_1,gg_2, gg_3)= \nu( g_0, g_1,g_2, g_3) \in U(1),
\nonumber\\
& \frac{\nu( g_1, g_2, g_3,g_4) \nu( g_0, g_1, g_3,g_4)\nu( g_0, g_1, g_2 ,g_3)}
{\nu( g_0,  g_2, g_3,g_4) \nu( g_0, g_1, g_2 ,g_4)}
 =1.
\end{align}
are 3-cocycles which form $\cH^3[G,U(1)]$.  Using $\nu(g_0,...,g_{d_{sp}+1})\in
\cH^{d_{sp}+1}[G,U(1)]$ as the action amplitude on a simplex in $(d_{sp}+1)$D space-time,
we can construct the path integral of topological nonlinear $\si$ model, which
describes the corresponding SPT phase.  This way, we show that the SPT phases
with symmetry $G$ in $d$ spatial dimensions are described by the elements in
$\cH^{d_{sp}+1}[G,U(1)]$.

In conclusion, a quantized topological term in the path integral of a $(d_{sp}+1)D$ system can be written in terms of the $(d_{sp}+1)$-cocycles by (1) discretizing the $(d_{sp}+1)D$ space time with branched triangularization (a branching structure\cite{C0527,CGL1172} will induce a local order of vertices on each simplex); (2) putting group element labeled degrees of freedom onto the vertices (3) assigning action amplitude to each simplex with the corresponding cocycle. The path integral then takes the form
\begin{align}
Z =|G|^{-N_v}\sum_{\{g_i\}} \prod_{\{ij...k\}} \nu_{d_{sp}+1}^{s_{ij...k}}(g_i,g_j,...,g_k)
\label{topo_term}
\end{align}
where $s_{ij...k}=\pm 1$ depending on the orientation of the simplex $ij...k$. Similar to the $(1+1)D$ case, it can be shown that the path integral is symmetric under symmetries in group $G$, the action amplitude $e^{-S(\{g_i\})}=\prod_{\{ij...k\}} \nu_{d_{sp}+1}^{s_{ij...k}}(g_i,g_j,...,g_k)$ is in a fixed point form and is quantized to $1$ on a closed manifold.

On a space with boundary, the $2\pi$-quantized topological term obtained from a
nontrivial cocycle  $\nu(g_0,...,g_{d_{sp}+1})$ gives rise to a discretized
analog of the WZW term in the low energy effective theory for boundary
excitations.  We believe that such a discretized  WZW term will make the
boundary excitations gapless if the symmetry is not broken.  In $(2+1)$D, we
can indeed prove rigorously that a nontrivial SPT states must have gapless
edge modes if the symmetry is not broken.\cite{CLW1141} In additional, it has
been further pointed out that by "gauging" the global symmetry\cite{duality},
each SPT phase can be identified by the braiding statistics of the
corresponding gauge flux, which is known to be classified by $\cH^3[G,U(1)]$ in
$(2+1)$D\cite{DW}.

\subsection{Ground state wave function and Hamiltonian for a bosonic SPT phase}

{}From each element in $\cH^{d_{sp}+1}[G,U(1)]$ we can also construct
the $d_{sp}$-dimensional ground state wave function for the corresponding SPT phase.
In 2D, we can start with a triangle lattice model where the physical states on
site-$i$ are given by $|g_i\>$, $g_i\in G$ (see Fig.  \ref{discreteST}a).  The
ground state wave function can be obtained by viewing the 2D lattice (which is
a torus) as the surface of a solid torus. The evaluation of the path integral
of the topological nonlinear $\si$ model (which is given by
$\nu_3(g_0,g_1,g_2,g_3)$, an element in $\cH^3[G,U(1)]$) on the solid torus
gives rise to the following ground state wave function: $ \Phi(\{ g_i \}) =
\prod_{\bigtriangleup} \nu_3(1,g_i, g_j, g_k) \prod_{\bigtriangledown}
\nu_3^{-1}(1,g_i, g_j, g_k) $, where $ \prod_{\bigtriangleup} $ and $
\prod_{\bigtriangledown} $ multiply over all up- and down-triangles, and the
order of $ijk$ is clockwise for up-triangles and anti-clockwise for
down-triangles (see Fig.  \ref{discreteST}a). To construct exactly solvable
Hamiltonian $H$ that realizes the above wave function as the ground state, we
start with an exactly solvable Hamiltonian $H_0=-\sum_i |\phi_i\>\<\phi_i|$,
$|\phi_i\>=\sum_{g_i\in G} |g_i\>$, whose ground state is $\Phi_0(\{ g_i \})
=1$.  Then, using the local unitary transformation $U=\prod_{\bigtriangleup}
\nu_3(1,g_i, g_j, g_k) \prod_{\bigtriangledown} \nu_3^{-1}(1,g_i, g_j, g_k)$,
we find that $\Phi=U\Phi_0$ and $H=\sum_i H_i$, where $H_i=U |\phi_i\>\<\phi_i|
U^\dag$. $H_i$ acts on a seven-spin cluster labeled by $i$, 1 -- 6 in blue in
Fig. \ref{discreteST}a
(see Fig. \ref{n0123}b):
\begin{align}
\label{Hi1}
&
 H_i|g_i,g_1g_2g_3g_4g_5g_6\> =\sum_{g_i'}
|g_i',g_1g_2g_3g_4g_5g_6\> \times
\nonumber\\
&\ \ \ \ \ \ \
\frac{
\nu_3(g_4,g_5,g_i,g_i')
\nu_3(g_5,g_i,g_i',g_6)
\nu_3(g_i,g_i',g_6,g_1)
}{
\nu_3( g_i,g_i',g_2,g_1)
\nu_3(g_3,g_i,g_i',g_2)
\nu_3(g_4,g_3,g_i,g_i')
}
\end{align}
We see that $H$ has a short ranged interaction and has the symmetry $G$:
$|\{g_i\}\> \to |\{g g_i\}\>, \ g\in G$.\cite{CLW1141}

\section{Grassmann tensor network and fermionic path integral}

\begin{figure}[tb]
\begin{center}
\includegraphics[scale=0.6]{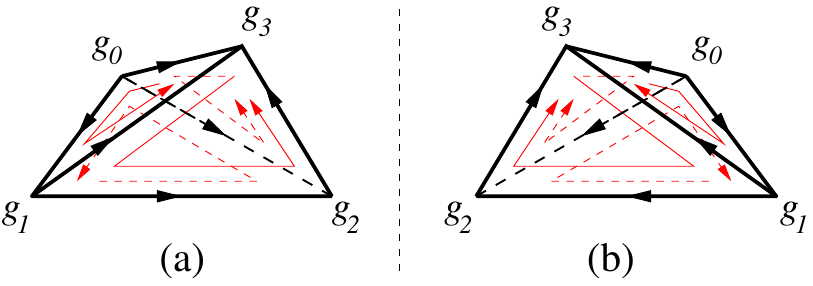}
\end{center}
\caption{
(Color online)
Two branched simplexes with opposite orientations.
(a) A branched simplex with positive orientation and (b)
a branched simplex with negative orientation.
}
\label{mir}
\end{figure}

After having some physical understanding of the fermionic SPT phases protected
by $Z_2^T\times Z_2^f$ or $Z_2\times Z_2^f$ symmetries and
after some understanding of the path integral approach to the
bosonic SPT phases, in the following, we
will discuss a generic construction that allows us to obtain fermionic SPT
phases protected by more general symmetries in any dimensions.

As discussed above,
one way to obtain a systematic construction of bosonic SPT phases is through a
systematic construction of \emph{topological} bosonic path
integral\cite{CGL1172} using a tensor network representation of the path
integral.\cite{LN0701,GW0931} Here we will use a similar approach to obtain a
systematic construction of fermionic SPT phases through \emph{topological}
fermionic path integral using a Grassmann tensor network\cite{GVW1063,GWW1017}
representation of the fermionic path integral.  The Grassmann tensor network
and the associated fermion path integral are defined on a discretized
space-time.  So first, let us describe the structure of a discretized
space-time.

\subsection{Discretized space-time}

A $d$-dimensional discretized space-time is a $d$-dimensional complex $\Si_d$
formed by many $d$-dimensional simplexes $\Si_{[i_0...i_d]}$.  We will use
$i,j,...$ to label the vertices of complex $\Si_d$. $\Si_{[i_0...i_d]}$ is
the simplex with vertices $i_0,...,i_d$.

However, in order to define Grassmann tensor network (and the associated
fermion path integral) on the discretized space-time $\Si_d$, the space-time
complex must have a so called branching structure.\cite{C0527,CGL1172} A
branching structure is a choice of orientation of each edge in the
$d$-dimensional complex so that there is no oriented loop on any triangle
(see Fig. \ref{mir}).

The branching structure induces a \emph{local order} of the vertices on each
$d$-simplex.  The first vertex of a $d$-simplex is the vertex with $d$ outgoing
edges, and the second vertex is the vertex with $d-1$ outgoing edges and 1 incoming
edge, \etc.  So the simplex in  Fig. \ref{mir}a has the following vertex
ordering: $0,1,2,3$.

The branching structure also gives the simplex (and its sub simplexes) an
orientation.  Fig. \ref{mir} illustrates two $3$-simplexes with opposite
orientations.  The red arrows indicate the orientations of the $2$-simplexes
which are the subsimplexes of the $3$-simplexes.  The black arrows on the edges
indicate the orientations of the $1$-simplexes.

\subsection{Physical variables on discretized space-time}

To define a path integral on a discretized space-time (\ie on a complex $\Si_d$
with a branching structure), we associate each vertex $i$ (a $0$-simplex) in
$\Si_d$ with a variable $g_i$.  So $g_i$ is a local physical dynamical variable
of our system.  The allowed values of $g_i$ form a space which is denoted as
$G_b$. At the moment, we treat $G_b$ as an arbitrary space.  Later, we will
assume that $G_b$ is the space of a group.

We also associate each $(d-1)$-simplex $(i_0..\hat i_i..i_d)$ of a
$d$-simplex $[i_0...i_d]$ with a Grassmann number. Here the $(d-1)$-simplex
$(i_0..\hat i_i..i_d)$ is a subsimplex of the  $d$-simplex $[i_0...i_d]$
that does not contain the vertex $i_i$.  Relative to the orientation of the
$d$-simplex $[i_0...i_d]$, the $(d-1)$-simplex $(i_0..\hat i_i..i_d)$ can
have a ``$+$'' or ``$-$'' orientation (see Fig. \ref{mir}).  We associate
$(d-1)$-simplex $(i_0..\hat i_i..i_{d-1})$ with a Grassmann number
$\th_{(i_0..\hat i_i..i_d)}$ if the  $(d-1)$-simplex has a ``$+$'' orientation
and with a Grassmann number $\bar\th_{(i_0..\hat i_i..i_d)}$ if the
$(d-1)$-simplex has a ``$-$'' orientation.
However, the Grassmann number $\th_{(i...k)}$ (or $\bar \th_{(i...k)}$) may or
may not present on the $(d-1)$-simplex $(i...k)$. So on each $(d-1)$-simplex
$(i...k)$ we also have a local physical dynamical variable $n_{i...k}=0,1$ or
$\bar n_{i...k}=0,1$, indicating
whether $\th_{(i...k)}$ or $\bar \th_{(i...k)}$
is present or not.  So each $(d-1)$-simplex $(i...k)$ is really associated
with a dynamical Grassmann variable $\th_{(i...k)}^{n_{i...k}}$  or
$\bar\th_{(i...k)}^{\bar n_{i...k}}$.

\begin{figure}[tb]
\begin{center}
\includegraphics[scale=0.6]{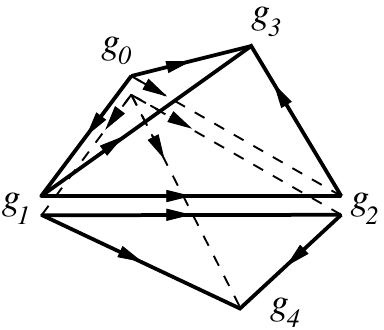}
\end{center}
\caption{
The $2$-simplex $(012)$ is associated with $\bar\th_{(012)}$ which
belongs to the top $3$-simplex $[0123]$.
The $2$-simplex $(012)$ is also associated with $\th_{(012)}$ which
belong to the bottom  $3$-simplex $[0124]$.
A rank $4$-tensor $\cV_3$ on the top tetrahedron is given by
$\cV^+_3( g_0, g_1, g_2, g_3)$
and on the lower tetrahedron by
$\cV^-_3( g_0, g_1, g_2, g_4)$.
}
\label{2tetra}
\end{figure}

We see that each  $(d-1)$-simplex $(i_0...i_{d-1})$ on the
surface of the $d$-complex $\Si_d$ is associated with two types of Grassmann
variables $\th_{(i_0...i_{d-1})}^{n_{i_0...i_{d-1}}}$ and $\bar
\th_{(i_0...i_{d-1})}^{\bar n_{i_0...i_{d-1}}}$, each belongs to the
$d$-simplex on each side of the $(d-1)$-simplex $(i_0...i_{d-1})$ (see Fig.
\ref{2tetra}).  We also see that each $d$-simplex has its own set of Grassmann
variables that do not overlap with the Grassmann variables of other simplexes.

\subsection{A constraint on physical variables}
\label{constraint}

In general, the physical dynamical variables $g_i$, $n_{i...k}$, and $\bar
n_{i...k}$ are independent.  We can formulate fermionic topological theory
and obtain fermionic topological phases by treating $g_i$, $n_{i...k}$, and
$\bar n_{i...k}$ as independent variables.  This is essentially the approach
used in \Ref{GWW1017} in (2+1)D, and we have obtained a system of nonlinear
algebraic equations, whose solutions describe various 2D fermionic topological
phases.  However, the nonlinear algebraic equations are very hard to solve,
despite they only describe 2D fermionic topological orders.  As a result, we
have obtained very few fermionic solutions.

In this paper, we are going to study a simpler and less general problem, by
putting a constraint between the physical dynamical variables $g_i$ and
$n_{i...k}$ ($\bar n_{i...k}$):
\begin{align}
 n_{i...k}=n(g_i,...,g_k), \ \ \ \
 \bar n_{i...k}=n(g_i,...,g_k).
\end{align}
That is when $n_{i...k}\neq n(g_i,...,g_k)$ or $\bar n_{i...k}\neq
n(g_i,...,g_k)$, there will be a huge energy cost.  So at low energies, the
system always stays in the subspace that satisfies the above constraint.
In this case $n_{i...k}$, and $\bar n_{i...k}$ are determined from $g_i$'s.

It turns out that the constraint indeed simplifies the mathematics a lot, which
allows us to systematically describe fermionic topological phases in any
dimensions.  The resulting nonlinear algebraic equations also have nice
structures which can be solved more easily.  So we can also find many solutions
in any dimensions and hence obtain many new examples of topological phases in
any dimensions.  However, the constraint limits us to describe only SPT phases.
To describe phases with intrinsic topological orders, we have to use
more general unconstrained formalism, which is discussed in \Ref{GWW1017}.  In
this paper, we will only consider the simple constraint cases.

\subsection{Grassmann tensor on a single $d$-dimensional simplex}

In the low energy constraint subspace, we can associate each $d$-simplex of
``$+$'' orientation with a Grassmann rank $(1+d)$-tensor $\cV_d^+$, and  each
$d$-simplex of ``$-$'' orientation with a different Grassmann rank
$(1+d)$-tensor $\cV_d^-$.

Let us first assume that a $d$-simplex $[0...d]$ has a ``$+$'' orientation, and
discuss the structure of the Grassmann tensor $\cV_d^+$.  The Grassmann tensor
$\cV_d^+$ is a map from space $G_b^{1+d} \to M_f$:
\begin{align}
 \cV_d^+(g_0,g_1,...,g_d) \in M_f, \ \ \ g_i\in G_b .
\end{align}
The order of the variables $g_0,g_1,...,g_d$
in $\cV_d$ is the same as the order of the branching structure: vertex-0 $<$
vertex-1 $<$ ... $<$ vertex-$d$.

Here an elements in $M_f$ is a complex number times an even number of
Grassmann numbers.  Thus $\cV_d^+(g_0,g_1,...,g_d)$ has a form
\begin{align}
 \cV_d^+&(g_0,g_1,...,g_d) = \nu^+_d(g_0,g_1,g_2,g_3,g_4,...,g_d) \times
\nonumber\\
&\ \
 \th_{(1234...d)}^{n_{d-1}(g_1,g_2,g_3,g_4,...,g_d)}
 \th_{(0134...d)}^{n_{d-1}(g_0,g_1,g_3,g_4,...,g_d)}... \times
\nonumber \\
&\ \
\bar \th_{(0234...d)}^{n_{d-1}(g_0,g_2,g_3,g_4,...,g_d)}
\bar \th_{(0124...d)}^{n_{d-1}(g_0,g_1,g_2,g_4...,g_d)}...
\end{align}
where $\nu^+_d(g_0,g_1,...,g_d)$ is a pure phase factor and $\th_{(ij...k)}$
or $\bar \th_{(ij...k)}$ is the Grassmann number associated with
$(d-1)$-dimensional subsimplex on $\Si^0_d$, $(ij...k)$.  The constraint is
implemented explicitly by writing, say, $\th_{(1234...d)}^{n_{1234...d}}$
as $\th_{(1234...d)}^{n_{d-1}(g_1,g_2,g_3,g_4,...,g_d)}$.

Note that the vertices on the branched simplex $\Si^0_d$ have a natural order:
vertex-0 $<$ vertex-1 $<$ ... $<$ vertex-$d$.  This leads to the ordering of
the  Grassmann numbers: the first Grassmann number is associated with the
subsimplex that does not contain the vertex-0, the second Grassmann number is
associated with the subsimplex that does not contain the vertex-2, \etc.  Then
it is followed by the Grassmann number associated with the subsimplex that does
not contain the vertex-1, the Grassmann number associated with the subsimplex
that does not contain the vertex-3, \etc.

The above example is for simplexes with a ``$+$'' orientation.
For branched simplex $\Si^0_d$ with a ``$-$'' orientation, we have
\begin{align}
 \cV_d^-&(g_0,g_1,...,g_d) = \nu^-_d(g_0,g_1,g_2,g_3,g_4,...,g_d) \times
\nonumber\\
&\ \
...
 \th_{(0124...d)}^{n_{d-1}(g_0,g_1,g_2,g_4...,g_d)}
 \th_{(0234...d)}^{n_{d-1}(g_0,g_2,g_3,g_4,...,g_d)}  \times
\nonumber \\
&\ \
... \bar \th_{(0134...d)}^{n_{d-1}(g_0,g_1,g_3,g_4,...,g_d)}
\bar \th_{(1234...d)}^{n_{d-1}(g_1,g_2,g_3,g_4,...,g_d)}
\end{align}
where the order of the Grassmann numbers is reversed and $\th$'s are switched
with $\bar\th$'s.

The number of the  Grassmann numbers on the $(d-1)$-dimensional simplex
$(ij...k)$ is given by
\begin{align}
n_{d-1}(g_i,g_j,...,g_k)=0,1 .
\end{align}
which depends on $d$ variables $g_i,g_j,...,g_k$, and must satisfy
\begin{align}
\label{ncond}
\sum_{i=0}^d n_{d-1}(g_0,...,\hat g_{i},...,g_d) =\text{ even}  ,
\end{align}
so that the Grassmann rank $(1+d)$-tensor always has an even number of
Grassmann numbers.  Here the sequence $g_0,...,\hat g_{i},...,g_d$ is the
sequence $g_0,...,g_d$ with $g_i$ removed.

All the solutions of \eqn{ncond} can be obtained using an integer function
$m_{d-2}(g_1,g_2,...,g_{d-1})=0,1$ with $d-1$ variables  (while $n_{d-1}(g_1,g_2,...,g_d)$ is
an integer function of $d$ variables):
\begin{align}
\label{hfrel}
&\ \ \ \ n_{d-1}(g_1,g_2,...,g_d)
\nonumber\\
&=\sum_{i=1}^d m_{d-2}(g_1,...,\hat g_i,...,g_d) \text{ mod } 2.
\end{align}

For example, a rank $1$-tensor has a form
\begin{align}
 \cV_0(g_0)= \nu_0(g_0)
\end{align}
which contains no Grassmann number.
A rank $2$-tensor has a form
\begin{align}
 \cV_1^+(g_0,g_1)= \nu_1^+(g_0,g_1)
\th_{(1)}^{n_0(g_1)} \bar\th_{(0)}^{n_0(g_0)}
\end{align}
where $n_0(g)$ has only two consistent choices
\begin{align}
 n_0(g) =0,\ \  \forall g \in G_b,\ \ \text{ and } \ \ n_0(g)=1,
\ \ \forall g \in G_b.
\end{align}
As another example,
a rank 4-tensor $\cV^+_3$ for the tetrahedron in Fig. \ref{mir}a
is given by
\begin{align}
&\cV^+_3( g_0, g_1, g_2, g_3)
=
\nu^+_3( g_0, g_1, g_2, g_3) \times
\nonumber\\
&\ \ \ \
\th_{(123)}^{n_2(g_1,g_2,g_3)}
\th_{(013)}^{n_2(g_0,g_1,g_3)}
\bar\th_{(023)}^{n_2(g_0,g_2,g_3)}
\bar\th_{(012)}^{n_2(g_0,g_1,g_2)} .
\end{align}
Note that the four triangles of the tetrahedron have different orientations:
the triangles $(123)$ and $(013)$ point outwards and the triangles $(023)$ and
$(012)$ point inwards.  We have used $\th$ and $\bar\th$ for triangles with
different orientations.

For the tetrahedron in Fig. \ref{mir}b, which has an opposite orientation to
that of the tetrahedron in Fig. \ref{mir}a, we have a rank 4-tensor $\cV^-_3$
which is given by
\begin{align}
&\cV^-_3( g_0, g_1, g_2, g_3)
=
\nu^-_3( g_0, g_1, g_2, g_3) \times
\nonumber\\
&\ \ \ \
\th_{(012)}^{n_2(g_0,g_1,g_2)}
\th_{(023)}^{n_2(g_0,g_2,g_3)}
\bar\th_{(013)}^{n_2(g_0,g_1,g_3)}
\bar\th_{(123)}^{n_2(g_1,g_2,g_3)}
\end{align}
Note the different order of the Grassmann numbers
$\th_{(012)}^{n_2(g_0,g_1,g_2)}   \th_{(023)}^{n_2(g_0,g_2,g_3)}
\bar\th_{(013)}^{n_2(g_0,g_1,g_3)} \bar\th_{(123)}^{n_2(g_1,g_2,g_3)} $.

\subsection{Evaluation of the Grassmann tensors on a complex}

Let $\Si_d$ be a $d$-complex with a branching structure which is formed by
several simplexes $[ab...c]$.  As discussed in the last section, a simplex is
associated with one of the two Grassmann tensors $\cV_d^\pm$.  Let us use
$\int_{\text{in}(\Si_d)} \prod_{[ab...c]} \cV^{s(a,b,...,c)}_d$ to represent
the evaluation of rank $(1+d)$ Grassmann tensors $\cV_d^\pm$ on the $d$-complex
$\Si_d$:
\begin{align}
\label{pVd}
& \ \ \ \
\int_{\text{in}(\Si_d)} \prod_{[ab...c]} \cV^{s(a,b,...,c)}_d
\\
& \equiv  \int
\prod_{(ij...k)}
\dd\th_{(ij...k)}^{n_{d-1}(g_i,g_j,...,g_k)}
\dd\bar\th_{(ij...k)}^{n_{d-1}(g_i,g_j,...,g_k)}
\times
\nonumber \\
&
\prod_{\{xy...z\}} (-)^{m_{d-2}(g_x,g_y,...,g_z)}
\prod_{[ab...c]} \cV_d^{s(a,b,...,c)}(g_a,g_b,...,g_c)
\nonumber
\end{align}
where $\prod_{[ab...,c]}$ multiplies over all the $d$-dimensional simplexes
$[ab...c]$ in $\Si_d$ and $\cV_d^{s(a,b,...,c)}(g_a,g_b,...,g_c)$ is a
Grassmann rank $(1+d)$-tensor associated with the simplex $[ab...c]$.
$s(a,b,...,c)=\pm $ depending on the orientation of the $d$-dimensional simplex
$[ab...c]$.  Also $\prod_{(ij...k)}$ multiplies over all the interior
$(d-1)$-dimensional simplexes, $(ij...k)$, of the complex $\Si_d$ [\ie those
$(d-1)$-dimensional simplexes that are not on the surface of $\Si_d$].
$\prod_{\{xy...z\}}$ multiplies over all the interior $(d-2)$-dimensional
simplexes [\ie those $(d-2)$-dimensional simplexes that are not on the surface
of $\Si_d$], $\{xy...z\}$, of the complex $\Si_d$.  We note that, when $\Si_d$
is a single simplex $[ab...c]$, $\int_{\text{in}(\Si_d)} \prod_{[ab...c]}
\cV_d^{s(a,b,...,c)}$ is given by $\cV_d^{s(a,b,...,c)}(g_a,g_b,...,g_c)$.

We see that in the formal notation, $\int_{\text{in}(\Si_d)} \prod_{[ab...c]}
\cV^{s(a,b,...,c)}_d$, $\int_{\text{in}(\Si_d)}$ represents a Grassmann integral
over the Grassmann numbers on all the interior $(d-1)$-simplexes in the
$d$-complex $\Si_d$.  So $\text{in}(\Si_d)$ in $\int_{\text{in}(\Si_d)}$
actually represents the collection of all those interior $(d-1)$-simplexes in
the $d$-complex $\Si_d$.  The Grassmann numbers on the $(d-1)$-simplexes on
the surface of the $d$-complex $\Si_d$ are not integrated over.  In fact
$\int_{\text{in}(\Si_d)}$ has the following explicit form
\begin{align}
\label{Gint}
\int_{\text{in}(\Si_d)}=  \int &
\prod_{(ij...k)}
\dd\th_{(ij...k)}^{n_{d-1}(g_i,g_j,...,g_k)}
\dd\bar\th_{(ij...k)}^{n_{d-1}(g_i,g_j,...,g_k)}
\times
\nonumber \\
&
\prod_{\{xy...z\}} (-)^{m_{d-2}(g_x,g_y,...,g_z)}
\end{align}
We note that $\dd \th$ always appears in front of $\dd \bar \th$.  We also note
that the integration measure contains a nontrivial sign factor
$(-)^{m_{d-2}(g_x,g_y,...,g_z)}$ on the interior $(d-2)$ simplexes
$\{xy...z\}$.

The sign factor $\prod_{\{xy...z\}} (-)^{m_{d-2}(g_x,g_y,...,g_z)}$ is included
to help us to define topological fermionic path integral later.  Choosing
orientation dependent tensors $\cV_d^\pm$ also help us to define topological
fermionic path integral.  Adding the sign factor $\prod_{\{xy...z\}}
(-)^{m_{d-2}(g_x,g_y,...,g_z)}$ and choosing orientation dependent tensors
$\cV_d^\pm$ appear to be very unnatural.  However, they are two extremely
important features of our approach.  We cannot obtain topological fermionic
path integral without these two features.  Realizing these two features is one
of a few breakthroughs that allows us to obtain topological fermionic path
integral on discrete space-time.  The two features are related to our choice of
the ordering convention of $\th$ and $\bar \th$ in the Grassmann tensors
$\cV^\pm_d$, and the ordering convention of $\dd \th$ and $\dd \bar \th$ in the
integration measure.

Now let us consider two $d$-complexes $\Si_d^1$ and $\Si_d^2$, which do not
overlap but may share part of their surfaces.  Let $\Si_d$ be the
union of the two  $d$-complexes.  From our definition of the Grassmann
integral, we find that the Grassmann integral on $\Si_d$ can be expressed as
\begin{align}
\label{glue}
&\int_{\text{in}(\Si_d)}
\prod_{[ab...c]} \cV^{s(a,b,...,c)}_d
\\
=&\int_{ \Si_d^1 \cap \Si_d^2 }
\Big(
\int_{\text{in}(\Si_d^1)}
\prod_{[ab...c]} \cV^{s(a,b,...,c)}_d
\int_{\text{in}(\Si_d^2)}
\prod_{[ab...c]} \cV^{s(a,b,...,c)}_d
\Big)
\nonumber
\end{align}
where $\Si_d^1 \cap \Si_d^2$ is the intersection of the two complexes, which
contains only $(d-1)$-simplexes on the shared surface of the two complexes
$\Si_d^1$ and $\Si_d^2$.
More precisely
\begin{align}
\int_{ \Si_d^1 \cap \Si_d^2 }=
\int &
\prod_{(ij...k)}
\dd\th_{(ij...k)}^{n_{d-1}(g_i,g_j,...,g_k)}
\dd\bar\th_{(ij...k)}^{n_{d-1}(g_i,g_j,...,g_k)}
\times
\nonumber \\
&
\prod_{\{xy...z\}} (-)^{m_{d-2}(g_x,g_y,...,g_z)}\label{integral}
\end{align}
where $\prod_{(ij...k)}$ is a product over all the $(d-1)$-simplexes in
$\Si_d^1 \cap \Si_d^2$ and $\prod_{\{xy...z\}}$ is a product over all the
interior $(d-2)$-simplexes in $\Si_d^1 \cap \Si_d^2$ (note that $\Si_d^1 \cap
\Si_d^2$ is a $(d-1)$-complex).  Eqn. (\ref{glue}) describes the process to
glue the two complexes $\Si_d^1$ and $\Si_d^2$ together.

\subsection{An example for evaluation of Grassmann tensor network}

Let us consider an example to help us to understand the complicated expression
\eq{pVd}.  When $\Si$ is formed by two tetrahedrons as in Fig. \ref{2tetra},
the top tetrahedron with a ``$+$'' orientation is associated with a rank
4-tensor $\cV_3^+$:
\begin{align}
&\cV^+_3( g_0, g_1, g_2, g_3)
=
\nu^+_3( g_0, g_1, g_2, g_3) \times
\nonumber\\
&\ \ \ \
\th_{(123)}^{n_2(g_1,g_2,g_3)}
\th_{(013)}^{n_2(g_0,g_1,g_3)}
\bar\th_{(023)}^{n_2(g_0,g_2,g_3)}
\bar\th_{(012)}^{n_2(g_0,g_1,g_2)}
\end{align}
The lower tetrahedron with a ``$-$'' orientation is associated with a rank
4-tensor $\cV_3^-$:
\begin{align}
&\cV^-_3( g_0, g_1, g_2, g_4)
=
\nu^-_3( g_0, g_1, g_2, g_4) \times
\nonumber\\
&\ \ \ \
\th_{(012)}^{n_2(g_0,g_1,g_2)}
\th_{(024)}^{n_2(g_0,g_2,g_4)}
\bar\th_{(014)}^{n_2(g_0,g_1,g_4)}
\bar\th_{(124)}^{n_2(g_1,g_2,g_4)}
\end{align}
Note that $\cV^+_3( g_0, g_1, g_2, g_3)$ and $\cV^-_3( g_0, g_1, g_2, g_4)$
have independent Grassmann numbers $\th_{(012)}$ and $\bar\th_{(012)}$ on the
shared face $(012)$.  $\bar\th_{(012)}$ belongs to $\cV^-_3( g_0, g_1, g_2,
g_4)$ and $\th_{(012)}$ belongs to $\cV^+_3( g_0, g_1, g_2, g_3)$.  Since
$\text{in}(\Si) =(012)$, the evaluation of $\cV_3^\pm$ on the complex $\Si$ is
given by
\begin{align}
& \int_{(012)}  \cV^+_3( g_0, g_1, g_2, g_3)
\cV^-_3( g_0, g_1, g_2, g_4)
\nonumber\\
=&\int
\dd\th_{(012)}^{n_2(g_0,g_1,g_2)}\dd\bar\th_{(012)}^{n_2(g_0,g_1,g_2)}
\times
\nonumber\\
&\ \
\cV^+_3( g_0, g_1, g_2, g_3)
\cV^-_3( g_0, g_1, g_2, g_4)
\end{align}
which is a special case of \eqn{pVd}.

\subsection{Fermionic path integral}
\label{path}

Given a closed space-time manifold, $M_{ST}$, with a time direction, its
triangularization $\Si$ is a complex with no boundary.  The time direction
gives rise to a local order of the vertices. So the complex $\Si$ has a
branching structure.

Since $\Si$ has no boundary, the evaluation of Grassmann tensors $\cV_d^\pm$
on it gives rise to a complex number:
\begin{align}
e^{-S}= \int_{\text{in}(\Si)} \prod_{[ab...c]} \cV_d^{s(a,b,...,c)}  .
\end{align}
Such a complex number can be viewed as the action-amplitude in the
imaginary-time path integral of a fermionic system.  The different choices of
the complex functions $\nu_d^\pm(g_0,...,g_d)$ and the integer functions
$m_{d-2}(g_0,...,g_{d-2})$ correspond to  different choices of Lagrangian of
the fermionic system.  The partition function of  the imaginary-time path
integral is given by
\begin{align}
Z=\sum_{\{g_i\}} \int_{\text{in}(\Si)} \prod_{[ab...c]} \cV_d^{s(a,b,...,c)} ,\label{fspt}
\end{align}
where each vertex of the complex $\Si$ is associated with a variable $g_i$.  We
note that
\begin{align*}
\frm{the fermionic partition function is determined from
two $(1+d)$-variable complex functions: $\nu_d^+(g_i,g_j,...)$, $\nu_d^-(g_i,g_j,...)$, and one $(d-1)$-variable integer function $m_{d-2}(g_i,g_j,...)$.
}
\end{align*}

\section{Topologically invariant Grassmann tensor network
and fermionic topological nonlinear $\si$ model}

\label{fixZ}

Now we would like to study the low energy effective theory of a gapped fermion
system.  The fixed point low energy effective action-amplitude of a gapped
system must describe a topological quantum field theory.  So the fixed point
action-amplitude must be invariant under a renormalization group (RG) flow
which generates coarse-graining transformation of the space-time complex $\Si$.
In fact, such a fixed-point requirement under RG flow completely fixes the form
of the fixed-point action-amplitude.  The fixed-point action-amplitude, in
turn, describes the possible fermionic SPT phases.

The detailed RG flow steps depend on the dimensions of the space-time.  We will
discuss different space-time dimensions and the corresponding fixed-point
action-amplitude separately.  Here we will only give a brief discussion. A more
detailed discussion will be given in Appendix \ref{topZ}.

\subsection{ (1+1)D case}

\begin{figure}[tb]
\begin{center}
\includegraphics[scale=0.6]{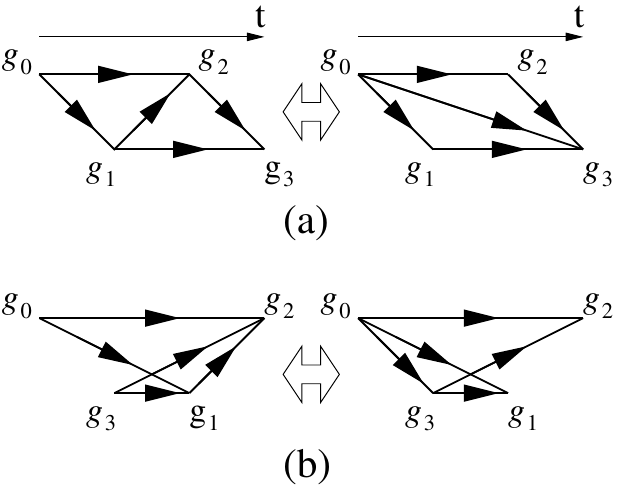}
\end{center}
\caption{
(a) Consider two triangles which are a part of a space-time complex $\Si$ with
a branching structure.  The first type of the RG flow step changes the two triangles
to other two triangles.  The branching structure leads to the following vertex
ordering $(0,1,2,3)$.  (b) A different space-time complex $\Si$ which
corresponds to a different ordered vertices $(0,3,1,2)$.  The corresponding
complex does have a valid branching structure because some triangles overlap.
}
\label{2to2}
\end{figure}

We first consider fermion systems in $(1+1)$ space-time dimension.
Its imaginary-time path integral is given by
\begin{align}
Z=\sum_{\{g_i\}} \int_{\text{in}(\Si_2)} \prod_{[abc]}\cV_2^{s(a,b,c)}  ,
\end{align}
where $\Si_2$ is the (1+1)D space-time complex with a branching structure (see
Fig. \ref{2to2}), and $\cV_2^\pm$ are rank-3 Grassmann tensors.

The  first type of the RG flow step that changes the space-time complex is
described by Fig. \ref{2to2}a.  If the Grassmann tensors $\cV_2^\pm$ describe a
fixed point action-amplitude, their evaluation on the two complexes in Fig.
\ref{2to2}a should be the same.  This leads to the following condition
\begin{align}
\label{nu2nu2}
&\ \ \ \
\nu_2 (g_0,g_1,g_3) \nu_2 (g_1,g_2,g_3)
\nonumber\\
&=\nu_2 (g_0,g_1,g_2) \nu_2 (g_0,g_2,g_3)
\end{align}
where $\nu_2$ is a complex phase related to $\nu_2^\pm$ and $m_0(g)$ through
\begin{align}
\label{nu2nupm}
 \nu_2^+ (g_0,g_1,g_2) &= \nu_2 (g_0,g_1,g_2),
\nonumber\\
 \nu_2^- (g_0,g_1,g_2) &=
(-)^{m_0(g_1)} /\nu_2 (g_0,g_1,g_2) .
\end{align}

Note that the above condition is obtained for a particular vertex ordering
$(0,1,2,3)$
(a particular branching structure)
as described in Fig.  \ref{2to2}a.
Different \emph{valid} branching structures,
in general, lead to other conditions on $\nu_2 (g_0,g_1,g_2)$.  It turns out
that all those conditions are equivalent to \eqn{nu2nu2}.  For example, for
vertex ordering  $(1,3,0,2)$, we obtain
\begin{align}
&\ \ \ \
\nu_2 (g_1,g_0,g_2) \nu_2 (g_1,g_3,g_0)
\nonumber\\
&=\nu_2 (g_1,g_3,g_2) \nu_2 (g_3,g_0,g_2)
\end{align}
which is the same as \eqn{nu2nu2} after replacing
$g_1\to g_0$,
$g_3\to g_1$,
$g_0\to g_2$, and
$g_2\to g_3$.

Some vertex orders do not correspond to valid branching structure due to the
overlap of the simplexes, and are not considered.  One of the invalid ordering
is described by Fig. \ref{2to2}b which has a vertex order $(0,3,1,2)$.  The
invalid orderings can give rise to conditions on $\nu_2 (g_0,g_1,g_2)$ that are
\emph{not} equivalent to \eqn{nu2nu2}.

\begin{figure}[tb]
\begin{center}
\includegraphics[scale=0.6]{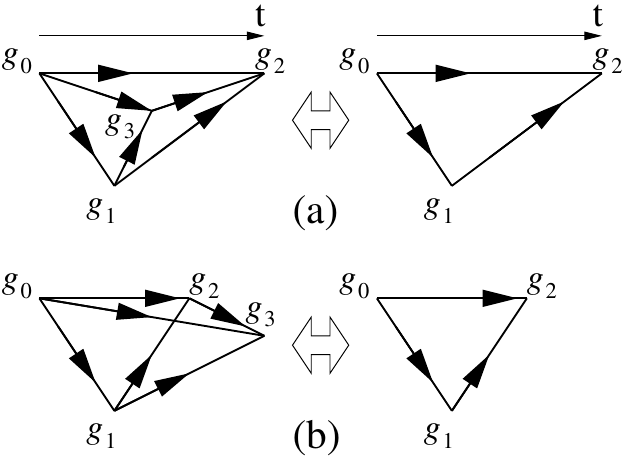}
\end{center}
\caption{
(a)
Consider three triangles which are a part of a space-time complex $\Si$ with a
branching structure.  The second type of the RG flow step changes the three triangles
to one triangle.  The branching structure leads to the following vertex ordering
$(0,1,3,2)$.
(b) The vertex order
$(0,1,2,3)$ does not correspond to a valid branching structure.
}
\label{3to1}
\end{figure}

\begin{figure}[tb]
\begin{center}
\includegraphics[scale=0.6]{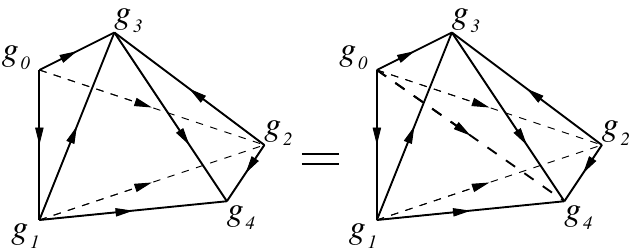}
\end{center}
\caption{
(a) A 3D complex formed by two 3-simplexes $(g_0,g_1,g_2,g_3)$ and
$(g_1,g_2,g_3,g_4)$.  The two 3-simplexes share a 2-simplex
$(g_1,g_2,g_3)$.  (b) A 3D complex formed by three 3-simplexes
$(g_0,g_2,g_3,g_4)$, $(g_0,g_1,g_3,g_4)$, and $(g_0,g_1,g_2,g_4)$.
}
\label{2to3}
\end{figure}
\begin{figure}[tb]
\begin{center}
\includegraphics[scale=0.6]{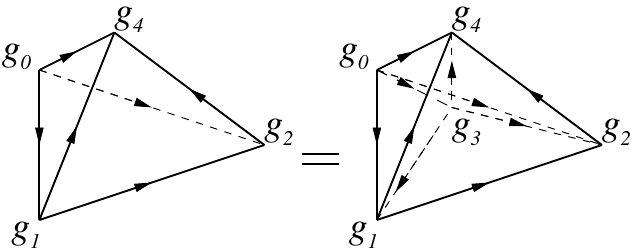}
\end{center}
\caption{
(a) A 3D complex formed by one 3-simplex $(g_0,g_1,g_2,g_4)$.  (b) A 3D complex
formed by four 3-simplexes $(g_1,g_2,g_3,g_4)$, $(g_0,g_2,g_3,g_4)$,
$(g_0,g_1,g_3,g_4)$, and $(g_0,g_1,g_2,g_3)$.
}
\label{1to4}
\end{figure}

The  second type of the RG flow step that changes the space-time complex is
described by Fig. \ref{3to1}.  Different \emph{valid} branching structures (see
Fig. \ref{3to1}a for one example), lead to different conditions on $\nu_2
(g_0,g_1,g_2)$ (which is defined in \eqn{nu2nupm}).  It turns out that all
those conditions are equivalent to \eqn{nu2nu2}.  (Some vertex orderings do not
correspond to valid branching structures.  One of the invalid ordering is
described by Fig.  \ref{3to1}b.)

Since $\nu_2(g_0,g_1,g_2)$ is related to $\nu_2^\pm(g_0,g_1,g_2)$ and $m_0(g)$
through \eqn{nu2nupm}, \eqn{nu2nu2} is actually a condition on
$\nu_2^\pm(g_0,g_1,g_2)$ and $m_0(g)$ that determines the fermion path integral.
The fermion path integral described by $\nu_2^\pm(g_0,g_1,g_2)$ and $m_0(g)$
that satisfies \eqn{nu2nu2} is a topological  fermion path integral which is a
fixed-point theory.

We would like to point out that \eqn{nu2nu2} is the standard
2-cocycle condition of group cohomology.\cite{RS,CGL1172}

\subsection{ (2+1)D case}

\begin{figure}[t]
\begin{center}
\includegraphics[scale=0.6]{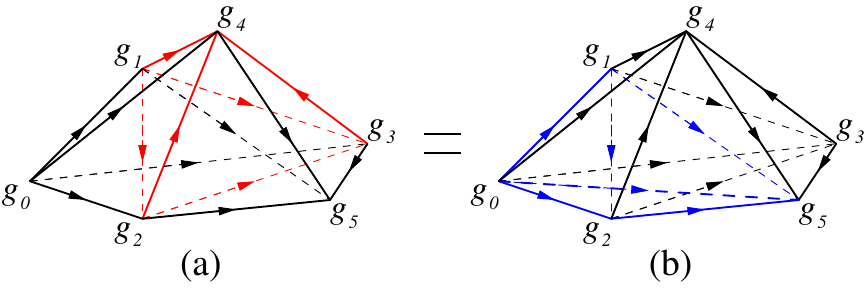}
\end{center}
\caption{
(Color online)
(a) A 4D complex formed by two 4-simplexes $(g_0,g_1,g_2,g_3,g_4)$ and
$(g_1,g_2,g_3,g_4,g_5)$.  The two 4-simplexes share a 3-simplex
$(g_1,g_2,g_3,g_4)$ (color red).  (b) A 4D complex formed by four 4-simplexes
$(g_0,g_2,g_3,g_4,g_5)$, $(g_0,g_1,g_3,g_4,g_5)$, $(g_0,g_1,g_2,g_4,g_5)$, and
$(g_0,g_1,g_2,g_3,g_5)$.  The two simplexes $(g_0,g_1,g_2,g_4,g_5)$, and
$(g_0,g_1,g_2,g_3,g_5)$ share a 3-simplex $(g_0,g_1,g_2,g_5)$ (color blue).
}
\label{4to2}
\end{figure}
\begin{figure}[t]
\begin{center}
\includegraphics[scale=0.6]{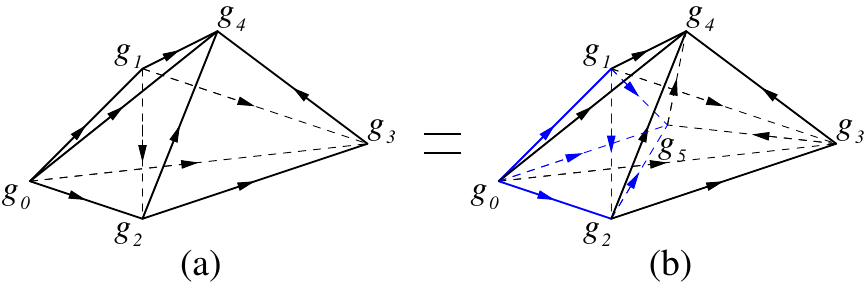}
\end{center}
\caption{
(Color online)
(a) A 4D complex formed by one 4-simplex $(g_0,g_1,g_2,g_3,g_4)$.  (b) A 4D
complex formed by five 4-simplexes $(g_1,g_2,g_3,g_5,g_4)$,
$(g_0,g_2,g_3,g_5,g_4)$, $(g_0,g_1,g_3,g_5,g_4)$, $(g_0,g_1,g_2,g_5,g_4)$, and
$(g_0,g_1,g_2,g_3,g_5)$.  The two simplexes $(g_0,g_1,g_2,g_5,g_4)$, and
$(g_0,g_1,g_2,g_3,g_5)$ share a 3-simplex $(g_0,g_1,g_2,g_5)$ (color blue).
}
\label{5to1}
\end{figure}

We next consider fermion systems in $(2+1)$ space-time dimension described by
imaginary-time path integral
\begin{align}
Z=\sum_{\{g_i\}} \int_{\text{in}(\Si_3)} \prod_{[abcd]}\cV_3^{s(a,b,c,d)}  ,\label{2Damplitude}
\end{align}
where $\Si_3$ is the (2+1)D space-time complex with a branching structure and
$\cV_3^\pm$ are rank-4 Grassmann tensors.

The  first type of the RG flow step that changes the space-time complex is
described by Fig. \ref{2to3}.  In order for the Grassmann tensors $\cV_3^\pm$
to describe a fixed point action-amplitude, their evaluations on the two
complexes in Fig. \ref{2to3} should be the same.  This leads to the following
condition
\begin{align}
\label{nu3nu3}
&
\nu_3 (g_1,g_2,g_3,g_4)
\nu_3(g_0,g_1,g_3,g_4)
\nu_3(g_0,g_1,g_2,g_3)=
\\
&(-)^{ n_2(g_0,g_1,g_2) n_2(g_2,g_3,g_4)}
\nu_3 (g_0,g_2,g_3,g_4)
\nu_3 (g_0,g_1,g_2,g_4)
\nonumber
\end{align}
where $\nu_3$ is a complex phase and is given by
\begin{align}
\label{nu3nupm}
 \nu_3^+ (g_0,g_1,g_2,g_3) &= (-)^{m_1(g_0,g_2)}\nu_3 (g_0,g_1,g_2,g_3),
\nonumber\\
 \nu_3^- (g_0,g_1,g_2,g_3) &=
(-)^{m_1(g_1,g_3)} /\nu_3 (g_0,g_1,g_2,g_3)
\end{align}

Different vertex orders give rise to different branching structures.
It turns out that all the different \emph{valid} branching
structures give rise to the same condition \eq{nu3nu3}.

The  second type of the RG flow step that changes the space-time complex is
described by Fig. \ref{1to4}.  If the Grassmann tensors $\cV_3^\pm$ describe a
fixed point action-amplitude, their evaluations on the two complexes in Fig.
\ref{1to4} should also be the same.  This leads to the same condition
\eq{nu3nu3}.

\subsection{ (3+1)D case}

Last, we consider fermion systems in $(3+1)$ space-time dimension, described
by imaginary-time path integral
\begin{align}
Z=\sum_{\{g_i\}} \int_{\text{in}(\Si_4)} \prod_{[abcde]}\cV_4^{s(a,b,c,d,e)}  ,
\end{align}
where $\Si_4$ is the (3+1)D space-time complex with a branching structure and
$\cV_4^\pm$ are rank-5 Grassmann tensors.

The  first type of the RG flow step that changes the space-time complex is
described by Fig. \ref{4to2}.  The  second type of the RG flow step that changes
the space-time complex is described by Fig. \ref{5to1}.  If the Grassmann
tensor $\cV_4$ describes a fixed point action-amplitude, its evaluation on the
two complexes in Fig. \ref{4to2} or Fig. \ref{5to1} should be the same.  This
leads to the following condition
\begin{widetext}
\begin{align}
\label{nu4nu4}
&\ \ \ \
\nu_4 (g_1,g_2,g_3,g_4,g_5)
\nu_4(g_0,g_1,g_3,g_4,g_5)
\nu_4(g_0,g_1,g_2,g_3,g_5)
\\
&=(-)^{
n_3(g_0,g_1,g_2,g_3)
n_3(g_0,g_3,g_4,g_5)
+
n_3(g_1,g_2,g_3,g_4)
n_3(g_0,g_1,g_4,g_5)
+
n_3(g_2,g_3,g_4,g_5)
n_3(g_0,g_1,g_2,g_5)
}\times
\nonumber\\
&\ \ \ \
\nu_4 (g_0,g_2,g_3,g_4,g_5)
\nu_4 (g_0,g_1,g_2,g_4,g_5)
\nu_4 (g_0,g_1,g_2,g_3,g_4)
\nonumber
\end{align}
\end{widetext}
where $\nu_4$ is a complex phase and is given by
\begin{align}
\label{nu4nupm}
&
 \nu_4^+ (g_0,g_1,g_2,g_3,g_4) =
\\
&
(-)^{
m_2(g_0,g_1,g_3)+
m_2(g_1,g_3,g_4)+
m_2(g_1,g_2,g_3)
}\nu_4 (g_0,g_1,g_2,g_3,g_4),
\nonumber\\
& \nu_4^- (g_0,g_1,g_2,g_3,g_4) =
(-)^{m_2(g_0,g_2,g_4)} /\nu_4 (g_0,g_1,g_2,g_3,g_4)
\nonumber
\end{align}
The sign factors in the above relations between $\nu_4^\pm$ and $\nu_4$, and the
sign factor in \eqn{nu4nu4} are very strange.  Obtaining those strange sign
factors is another break through that allows us to obtain topological fermionic
path integral beyond (1+1) dimensions.  It appears that those sign factors are
related to a deep mathematical structure called Steenrod
squares.\cite{S4790,S5313} Different vertex orders give rise to different
branching structures.  It turns out that all the valid branching structures give
rise to the same condition \eq{nu4nu4}.

\subsection{Fixed-point action on a closed complex}

We know that the fixed-point theory of a bosonic SPT phase is a discrete
bosonic topological nonlinear $\si$ model.\cite{CGL1172}  The  bosonic
topological nonlinear $\si$ models are characterized by action-amplitudes that
is equal to 1 on any closed complex.  The action-amplitude is equal to 1 on
any closed complex is the key reason why the models describe bosonic SPT
phases without intrinsic topological orders.

In the above, we have developed a fixed-point theory of fermionic SPT
phases, and defined discrete fermionic topological nonlinear $\si$ model.  A
discrete fermionic topological nonlinear $\si$ model is described by a pair of
functions $[\nu_d(g_0,...,g_d), m_{d-2}(g_0,...,g_{d-2})]$, that satisfy the
conditions \eqn{nu2nu2}, \eqn{nu3nu3}, or \eqn{nu4nu4}.
According to the branching moves discussed above, the evaluation on any closed
manifold is reduced to an equivalent evaluation
\begin{align}
\int \cV_d^+ \cV_d^-
\end{align}
where the symbol $\int$ means integrating out all the Grassmann variables on the
corresponding $d-1$ simplexes (see \eqn{Gint}).  We note that the sign factor
$\prod_{\{x,y,...,z\}} (-)^{m_{d-2}(g_x,g_y,...,g_z)}$ should be included in
the fermionic path integral.  In Appendix \ref{close} we will show that $\int \cV_d^+
\cV_d^-$ is always equal to 1.

\subsection{Symmetry and stability of the fixed-point action amplitude}
\label{symmstb}

We have seen that a fermionic system can be described by a fermionic path
integral on the discretized space-time.  In $d$ dimensional space-time, the
fermionic path integral is determined by two $(d+1)$-variable complex functions
$\nu_d^\pm(g_0,...,g_d)$ and one $(d-1)$-variable integer function
$m_{d-2}(g_0,...,g_{d-2})$.  In the last few sections, we have shown that if
$\nu_d^\pm(g_0,...,g_d)$ and $m_{d-2}(g_0,...,g_{d-2})$ satisfy some
conditions, \eqn{nu2nu2}, \eqn{nu3nu3}, and \eqn{nu4nu4}, the fermionic path
integral determined from $\nu_d^\pm(g_0,...,g_d)$ and
$m_{d-2}(g_0,...,g_{d-2})$ is actually a fixed-point theory under the
coarse-graining transformation of the space-time complex.

A fixed-point theory can be used to describe a phase if the fixed-point theory
is stable. To see whether a fixed-point theory is stable or not, we perturb the
fixed-point action amplitude $\nu_d^\pm(g_0,...,g_d)$ by a small amount
$\nu_d^\pm(g_0,...,g_d) \to \nu_d^\pm(g_0,...,g_d)+ \del \nu_d^\pm(g_0,...,g_d)
$.  If $\nu_d^\pm(g_0,...,g_d)+ \del \nu_d^\pm(g_0,...,g_d) $ flows back to
$\nu_d^\pm(g_0,...,g_d)$ under the coarse-graining transformation of space-time
complex, the fixed-point theory is stable.

{}From this point of view, our constructed fixed-point theories described by
$\nu_d^\pm(g_0,...,g_d)$ and $m_{d-2}(g_0,...,g_{d-2})$ are not stable.  If we
add a perturbation of form $\del \nu_d^\pm(g_0,...,g_d)=\eps \prod_{i=0}^d
\del_{g_i,\bar g}$ to increase the weight of $\bar g$ state in the action
amplitude, we believe that the action amplitude will flow to the simple form
$\nu_d^\pm(g_0,...,g_d)=\prod_{i=0}^d \del_{g_i,\bar g}$ where only the $\bar g$
state has a non-zero weight.

However, if we require the path integral to have a symmetry, then our
constructed fixed-point theories can be stable.  To impose a symmetry, we
assume that our fermion system has full symmetry group $G_f$.  Since $G_f$
contains a normal subgroup $Z^f_2$, we can view $G_f$ as a fiber bundle with
fiber $Z^f_2$ and base space $G_b=G_f/Z_2^f$.  Thus there is natural projection
map $p:\ G_f\to G_b$.

In general, $g_i$ in $\nu_d^\pm(g_0,...,g_d)$ and $m_{d-2}(g_0,...,g_{d-2})$
can be a group element in $G_f$.
Here we will assume that
 $\nu_d^\pm(g_0,...,g_d)$ and $m_{d-2}(g_0,...,g_{d-2})$
satisfy
\begin{align}
 \nu_d^\pm(g_0,...,g_d) &= \nu_d^\pm(\t g_0,...,\t g_d) ,
\nonumber\\
m_{d-2}(g_0,...,g_{d-2}) &=
m_{d-2}(\t g_0,...,\t g_{d-2}) ,
\nonumber\\
g_i \in G_f, &\ \ \ \t g_i =p(g_i) \in G_b.
\end{align}
Therefore, we can view
$g_i$ in $\nu_d^\pm(g_0,...,g_d)$ and $m_{d-2}(g_0,...,g_{d-2})$
as a group element in $G_b$.

From Section \ref{constraint}, we see that different degrees of freedom are
attached to the $(d-1)$-simplexes for different choices of $g_i,...,g_k$ in
$G_b$.  The degrees of freedom on a $(d-1)$-simplex form a 1D representation
$u_{d-1}^{g}$ of the \emph{full} symmetry group $g\in G_f$, which may depend on
$g_i,...,g_k$:
\begin{align}
& u_{d-1}^{g_1g_2}(g_i,g_j,...,g_k)
=
 u_{d-1}^{g_1}(g_i,g_j,...,g_k)
 u_{d-1}^{g_2}(g_i,g_j,...,g_k)
\nonumber\\
&
 u_{d-1}^{g}(p(g'g_i),...,p(g'g_k))= u_{d-1}^{g}(g_i,...,g_k),
\end{align}
where $p$ is the projection map $G_f \to G_b$.
Here we would like to stress that, in $u_{d-1}^{g}(g_i,g_j,...,g_k)$, $g\in
G_f$ and $g_i,g_j,... \in G_b$.  We also require the variables with bar and
without bar to carry opposite quantum number (\ie their 1D representations are
inverse of each other).  In this case, the symmetry of the path integral (\ie
the invariance of the action amplitude) can be implemented by requiring:
\begin{align}
 \th_{(ij...k)} \to & {(u^g_{d-1})}^{-1}(g_i,g_j,...,g_k) \th_{(ij...k)},
\nonumber\\
 \bar \th_{(ij...k)} \to &  u^g_{d-1}(g_i,g_j,...,g_k) \bar \th_{(ij...k)},
\end{align}
and
\begin{align}
&\ \ \ \ (\nu_d^\pm(gg_0,..,gg_d))^{s(g)} \nonumber\\
&=
\nu_d^\pm(g_0,..,g_d)
\prod_i (u_{d-1}^{g})^{(-)^i}(g_0,..,\hat g_i,..,g_d) ,
\nonumber \\
& m_{d-2}(gg_0,..,gg_{d-2})=m_{d-2}(g_0,..,g_{d-2}), \
g\in G_f .
\end{align}
where $s(g)=-1$ if $g$ contains the anti unitary time-reversal transformation
and $s(g)=1$ otherwise. If we view $\mathcal{V}^\pm_d(g_0,..,g_d))$ as a whole object, it is invariant
under the symmetry action $g$:
\begin{align}
\mathcal{V}^\pm_d(gg_0,..,gg_d))^{s(g)}
=\mathcal{V}^\pm_d(g_0,..,g_d)
\end{align}

Indeed, the above symmetric $\nu_d^\pm(g_0,...,g_d)$ and
$m_{d-2}(g_0,...,g_{d-2})$ describe a fermion system with $G_f$ symmetry. If
$\nu_d^\pm(g_0,...,g_d)$ and $m_{d-2}(g_0,...,g_{d-2})$ further satisfy
\eqn{nu2nu2}, \eqn{nu3nu3}, or \eqn{nu4nu4}, we believe, they even
describe a fixed-point theory that is stable against perturbations which do not
break the symmetry.  But further consideration suggests that such a symmetry
condition is too strong. The fermionic systems that are described by the
symmetric $\nu_d^\pm(g_0,...,g_d)$ and $m_{d-2}(g_0,...,g_{d-2})$
appear to be essentially bosonic. At least, their
description is the same as the description of bosonic
SPT phases.

So in this paper, we will use a weaker symmetry condition.  We only require
$\nu_d(g_0,...,g_d)$ and $n_{d-1}(g_0,...,g_{d-1})$ (which are certain
combinations of $\nu_d^\pm(g_0,...,g_d)$ and $m_{d-2}(g_0,...,g_{d-2})$) to
be symmetric
\begin{align}
\label{symmcnd}
&\ \ \ \ \nu_d^{s(g)}(gg_0,..,gg_d)
\nonumber\\
&
 =\nu_d(g_0,..,g_d)
\prod_i (u_{d-1}^{g})^{(-)^i}(g_0,..,\hat g_i,..,g_d) ,
\nonumber\\
&
n_{d-1}(gg_0,..,gg_{d-1})=n_{d-1}(g_0,..,g_{d-1}), \
g\in G_f .
\end{align}
Since the fermion fields always change sign under the fermion-number-parity
transformation $P_f\in G_f$,
the 1D representations satisfy
\begin{align}
 u_{d-1}^{P_f}(g_0,..,g_{d-1})=(-)^{n_{d-1}(g_0,...,g_{d-1})}.
\end{align}

Note that we only require $n_{d-1}(g_0,...,g_{d-1})$ to be symmetric.  In
general,  $m_{d-2}(g_0,...,g_{d-2})$ is not symmetric
$m_{d-2}(gg_0,...,gg_{d-2})\neq m_{d-2}(g_0,...,g_{d-2})$.  Although the
fermionic path integral appears not to be symmetric under the weaker symmetry
condition, it turns out that the symmetry breaking is only a boundary effect
and the symmetry can be restored by adding an additional boundary action to the
space-time path integral.  In fact, we will show that the ground state
wave function and the Hamiltonian obtained from the path integral with the
weaker symmetry condition are indeed symmetric. So a fermion path integral
determined by $\nu_d(g_0,...,g_d)$ and $n_{d-1}(g_0,...,g_{d-1})$ that satisfy
\eqn{symmcnd} as well as \eqn{nu2nu2}, \eqn{nu3nu3}, or \eqn{nu4nu4} does
describe a symmetric stable fixed-point theory.  Such $\nu_d(g_0,...,g_d)$,
$n_{d-1}(g_0,...,g_{d-1})$, and $u_{d-1}^{g}(g_0,..,g_{d-1})$ describe a
fermionic topological phase with symmetry.

Let us introduce
\begin{align}
\label{fdn}
 f_1(g_0,g_1) &= 0;
\nonumber\\
 f_2(g_0,g_1,g_2) &= 0;
\nonumber\\
 f_3(g_0,g_1,...,g_3) &= 0;
\\
 f_4(g_0,g_1,...,g_4) &=
n_2(g_0,g_1,g_2)
n_2(g_2,g_3,g_4)
;
\nonumber\\
 f_5(g_0,g_1,...,g_5) &=\ \ \
 n_3(g_0,g_1,g_2,g_3) n_3(g_0,g_3,g_4,g_5)
\nonumber\\ & \ \ \
+n_3(g_1,g_2,g_3,g_4) n_3(g_0,g_1,g_4,g_5)
\nonumber\\ & \ \ \
+n_3(g_2,g_3,g_4,g_5) n_3(g_0,g_1,g_2,g_5)
.
\nonumber
\end{align}
Using $f_d$, we can define a mapping from $(d+1)$-variable functions
$\nu_d(g_0,...,g_d)$ to $(d+2)$-variable functions
$(\del\nu_d)(g_0,...,g_{d+1})$:
\begin{align}
\label{delnudef}
&\ \ \ \ (\del\nu_d)(g_0,...,g_{d+1})
\\
& \equiv (-)^{f_{d+1}(g_0,...,g_{d+1})}
\prod_{i=0}^{d+1}
\nu_d^{(-)^i}(g_0,..,\hat g_i,..,g_{d+1}) .
\nonumber
\end{align}
Then using $(\del\nu_d)$, we can
rewrite the conditions,
\eqn{nu2nu2}, \eqn{nu3nu3}, and \eqn{nu4nu4},
in a uniform way:
\begin{align}
\label{fccnd}
(\del\nu_d)(g_0,...,g_{d+1})
 =1.
\end{align}

Here $[\nu_d(g_0,...,g_d), n_{d-1}(g_0,...,g_{d-1}),
u_{d-1}^{g}(g_0,...,g_{d-1})]$ that satisfies \eqn{symmcnd},  \eqn{ncond}, and
\eqn{fccnd}, will be called a $d$-cocycle.  The space of $d$-cocycle is
denoted as $\fZ^d[G_f,U_T(1)]$.  The fermionic path integral obtained from a
$d$-cocycle $(\nu_d,n_{d-1},u_{d-1}^{g})$ (via $\nu_d^\pm(g_0,...,g_d)$ and
$m_{d-2}(g_0,...,g_{d-2})$)
will be called a fermionic topological nonlinear $\si$ model.

In Appendix \ref{fcoh}, we study the fermionic cocycles $(\nu_{d+1},n_d,u^g_d)$
systematically.  In particular, we will study the equivalence relation between
them.  This will lead to the notion of group super-cohomology class
$\fH^{d+1}[G_f,U_T(1)]$.

Similar to the bosonic case, we argue that each element in group super-cohomology
class $\fH^{d_{sp}+1}[G_f,U_T(1)]$ correspond to
a fermionic topological nonlinear $\si$ model defined via the fermion path
integral in imaginary time (see section \ref{path}).  The fermion path integral
over a $(d_{sp}+1)$-dimensional complex $\Si$ will give rise to a ground state
on its $d_{sp}$ boundary.  Such a ground state represents the corresponding
fermionic SPT state.  So each element in group super-cohomology class
$\fH^{d_{sp}+1}[G_f,U_T(1)]$ corresponds to a fermionic SPT state.

In Appendix \ref{eqwav}, we will show the elements in $\fH^{d_{sp}+1}[G_f,U_T(1)]$
correspond to distinct fermionic SPT phases.
We will calculate $\fH^{d_{sp}+1}[G_f,U_T(1)]$ for some simple symmetry groups $G_f$
in Appendix \ref{calfcoh}.  This allows us to construct several new fermionic
SPT phases.  The results are summarized in Table \ref{tbF}.

\section{General scheme of calculating the group super-cohomology classes $\fH^d[G_f,U_T(1)]$ -- an outline }
\label{gen}
In this section, we will give a general description
on how to calculate the group super-cohomology classes.
More concrete calculations will be given in Appendix \ref{calfcoh}.
From the discussions in Appendix \ref{fcoh}, we see that to calculate
$\fH^d[G_f,U_T(1)]$ (which is formed by the equivalent classes of the fermionic
cocycles $[\nu_d(g_0,...,g_d),
n_{d-1}(g_0,...,g_{d-1}),u^g_{d-1}(g_0,...,g_{d-1})]$), we need to go through
the following steps.

\noindent
\textbf{First step}: Calculate $\cH^{d-1}(G_b,\Z_2)$ which give us different
classes of $(d-1)$D-graded structure $n_{d-1}(g_0,...,g_{d-1})$.  For each
$n_{d-1}(g_0,...,g_{d-1})$, find an allowed $u^g_{d-1}(g_0,...,g_{d-1})$.

\noindent
\textbf{Second step}: Next we need to find out $B\cH^{d-1}(G_b,\Z_2)$ from
$\cH^{d-1}(G_b,\Z_2)$.  We note that
$B\cH^{d-1}(G_b,\Z_2)$ denote the subset of $\cH^{d-1}(G_b,\Z_2)$, such that for
each graded structure $n_{d-1}(g_0,...,g_{d-1})$ in $B\cH^{d-1}(G_b,\Z_2)$,  the
corresponding $ (-)^{f_{d+1}}$ is a $(d+1)$-coboundary in $\cB^{d+1}[G_b,
U_T(1)]$.  We note that for $n_{d-1}(g_0,...,g_{d-1}) \notin
B\cH^{d-1}(G_b,\Z_2)$, the corresponding $(-)^{f_{d+1}}$ will not be a
coboundary in $\cB^{d+1}[G_b, U_T(1)]$ and the super cocycle condition
\eqn{fccnd} will not have any solution.  Mathematically, such an inconsistency
is called as an \emph{obstruction}, and the group $B\cH^{d-1}(G_b,\Z_2)$ is
called as the obstruction-free subgroup of $\cH^{d-1}(G_b,\Z_2)$. We note that
for $d\leq 2$, $B\cH^{d-1}(G_b,\Z_2)\equiv\cH^{d-1}(G_b,\Z_2)$ since
$f_{d+1}\equiv 0$ when $d\leq 2$.

\noindent
\textbf{Third step}: We need to calculate $\nu_d(g_0,...,g_d)$ from
\eqn{fccnd}.  For each fixed $n_{d-1}(g_0,...,g_{d-1}) \in
B\cH^{d-1}(G_b,\Z_2)$, the solutions $\nu_d$ of \eqn{fccnd} have an one-to-one
correspondence with the standard group cocycles, provided that
$u^g_{d-1}(g_0,...,g_{d-1})=1$ for $g\in G_b$.  We see that, for each element
in $ B\cH^{d-1}(G_b,\Z_2)$, there is a class of solutions. For two different
elements in $ B\cH^{d-1}(G_b,\Z_2)$, their classes of solutions have a
one-to-one correspondence.  Therefore, we have an exact sequence
\begin{align}
  \fH^d[G_f,U_T(1)]\to B\cH^{d-1}(G_b,\Z_2) \to 0.
\end{align}

\noindent
\textbf{Fourth step}: From the equivalence relation of supercohomology class
defined in Appendix \ref{fcoh}, we see that, for each fixed element
$n_{d-1}(g_0,...,g_{d-1})$ in $ B\cH^{d-1}(G_b,\Z_2)$, the equivalent classes of the
super cocycle solutions can be labeled by $\cH^d[G_b,U_T(1)]$.  However, the
labeling is not one-to-one, since in obtaining  $\cH^d[G_b,U_T(1)]$, we only
used the equivalence relation for the standard group cohomology class.  The
labeling is one-to-one only when $(-)^{f_d}$ happen to be a coboundary in
$\cB^d[G_b,U_T(1)]$.  But when  $(-)^{f_d}$ is not a coboundary in
$\cB^d[G_b,U_T(1)]$, we need to consider the more general equivalence relation
including all possible shifts $(-)^{f_d}$, where $f_d$ is generated by
$n_{d-2}(g_0,...,g_{d-2})\in \cH^{d-2}(G_b,\Z_2)$, e.g., see Eq.(\ref{fdn}) for
the precise definition for $f_4$ and $f_5$ in terms of $n_2(g_0,g_1,g_2)$ and
$n_3(g_0,g_1,g_2,g_3)$.  So the group cohomology description of bosonic SPT
phases will collapse into a smaller quotient group
$\cH^d_{\rm{rigid}}[G_b,U_T(1)]=\cH^d[G_b,U_T(1)]/\Ga$, where $\Ga$ is a
subgroup of $\cH^d[G_b,U_T(1)]$ generated by $(-)^{f_d}$.  Physically, such a
result implies that when we embed interacting bosons systems into interacting
fermion systems by viewing bosons as tightly bounded fermion pairs, some times,
a non-trival bosonic SPT state, described by a cocyle $\nu_d$ in
$\cH^d[G_b,U_T(1)]$, may correspond to a trivial fermionic SPT state, since
$\nu_d$ corresponds to the trivial element in $\cH^d_{\rm{rigid}}[G_b,U_T(1)]$.
We call $\cH^d_{\rm{rigid}}[G_b,U_T(1)]$ as a \emph{rigid} center, which is a
normal subgroup of the standard group cohomology class $\cH^d[G_b,U_T(1)]$.
Fortunately, such an additional complication only happens when $d\geq 4$ since
$f_d\equiv0$ for all $d<4$, and we have $\cH^d_{\rm{rigid}}[G_b,U_T(1)]\equiv
\cH^d[G_b,U_T(1)]$ for all $d<4$.
We see that, for each element (such as the trivial element)
in $ B\cH^{d-1}(G_b,\Z_2)$, the class of solutions is described by
$\cH^d_{\rm{rigid}}[G_b,U_T(1)]$.
This leads to the following short exact sequence
\begin{align}
0
&\to \cH^d_{\rm{rigid}}[G_b,U_T(1)]
\to  \fH^d[G_f,U_T(1)]
\nonumber\\
&
\to B\cH^{d-1}(G_b,\Z_2)
\to 0,
\end{align}
which completely determines $\fH^d[G_f,U_T(1)]$.
Roughly speaking
\begin{align}
 \fH^d[G_f,U_T(1)]= \cH^d_{\rm{rigid}}[G_b,U_T(1)] \times  B\cH^{d-1}(G_b,\Z_2).
\end{align}

We see that, through the above four steps and combining with results from the
standard group cohomology, we can calculate the group super-cohomology classes
$\fH^d[G_f,U_T(1)]$. In Appendix \ref{group}, we will further prove the group
structure of the super-cohomology classes $\fH^d[G_f,U_T(1)]$.  In the
following, we summarize how to describe a minimal set of fermionic SPT phases
by using (special) group supercohomology class in $0,1,2$ and $3$ spatial
dimensions with arbitrary $G_b=G_f/Z_2^f$.

\begin{table*}[tb]
\centering
\begin{tabular}{|c|c|}
\hline
$d_{sp}$  & short exact sequence   \\
\hline
$0$ &  $0\rightarrow \cH^1[G_b,U_T(1)]\rightarrow \fH^{1}[G_f,U_T(1)]\rightarrow \Z_2  \rightarrow 0$ \\
\hline
$1$ &  $0\rightarrow \cH^2[G_b,U_T(1)]\rightarrow \fH^{2}[G_f,U_T(1)]\rightarrow \cH^{1}(G_b,\Z_2)  \rightarrow 0$ \\
\hline
$2$ & $0\rightarrow \cH^3[G_b,U_T(1)] \rightarrow \fH^{3}[G_f,U_T(1)]\rightarrow B\cH^{2}(G_b,\Z_2) \rightarrow 0$    \\
\hline
$3$ & $0\rightarrow \cH^4_{\rm{rigid}}[G_b,U_T(1)] \rightarrow \fH^{4}[G_f,U_T(1)]\rightarrow B\cH^{3}(G_b,\Z_2) \rightarrow 0$  \\
\hline
\end{tabular}
\caption{Computing (special) group supercohomology class by using short exact sequence. }\label{supercohomology}
\end{table*}

In $d_{sp}=0$ spatial dimension, the elements in $\fH^1[G_f,U_T(1)]$ can always
have the trivial graded structure $n_0(g_0)=0$, or the nontrivial graded
structure $n_0(g_0)=1$.  The corresponding fermionic gapped states can have
even or odd numbers of fermions.  So we can have two different fermionic SPT
phases in $d_{sp}=0$ spatial dimension even without symmetry. In $d_{sp}=1$
spatial dimension, $\fH^2[G_f,U_T(1)]$ is just an extension of the graded
structure  $\cH^{1}(G_b,\Z_2)$ by the standard group cohomology class
$\cH^2[G_b,U_T(1)]$.  In $d_{sp}=2$ spatial dimension, $\fH^3[G_f,U_T(1)]$ is
just an extension of the graded structure $B\cH^{2}(G_b,\Z_2)$ by the standard
group cohomology class $\cH^3[G_b,U_T(1)]$.  In $d_{sp}=3$ spatial dimension,
$\fH^4[G_f,U_T(1)]$ is just an extension of the graded $B\cH^{3}(G_b,\Z_2)$ by
$\cH^4_{\rm{rigid}}[G_b,U_T(1)]$ -- the rigid center of the standard group
cohomology class $\cH^4[G_b,U_T(1)]$ (see Table \ref{supercohomology}).
\begin{widetext}
\begin{align}
0\rightarrow \fH^3[G_f,U_T(1)] \rightarrow \fH^{3}_{\rm{general}}[G_f,U_T(1)]\rightarrow H^{1}(G_b,\Z_2) \rightarrow 0
\end{align}
\begin{align}
0\rightarrow \fH^{d+1}[G_f,U_T(1)] \rightarrow \fH^{d+1}_{\rm{general}}[G_f,U_T(1)]\rightarrow  ? \rightarrow 0
\end{align}
\end{widetext}

\section{Ideal ground state wave function}
\label{idwav}
In the following, we will show that we can construct an exactly solvable local
fermionic Hamiltonian in $d$ spatial dimensions from each $(d+1)$-cocycle
$(\nu_{d+1},n_d,u_d^{g})$ in $\fZ^{d+1}[G_f,U_T(1)]$.  The Hamiltonian has a symmetry
$G_f$.  The ground state wave function of the constructed Hamiltonian can also
be obtained exactly from the $(d+1)$-cocycle.  Such a ground state does not
break the symmetry $G_f$ and describes a fermionic SPT phase.

We have shown that from each element $(\nu_{d+1},n_d,u_d^g)$ of
$\fZ^{d+1}[G_f,U_T(1)]$, we can define a fermionic topological nonlinear
$\si$ model in $(d+1)$ space-time dimensions.  The action amplitude
$\cV_{d+1}^\pm$ of the model [obtained from $(\nu_{d+1},n_d,u_d^g)$] is a
fixed-point action amplitude under the coarse-graining transformation of the space-time
complex.  The fermionic path integral is supposed to give us a quantum ground
state.  We claim that such a quantum ground state is a fermionic SPT phase
described by $(\nu_{d+1},n_d,u_d^g)$. In the section, we will construct the
ground state wave function, in (2+1)D as an example.

\subsection{Construction of 2D wave function}
\label{2dwav}

\begin{figure}[tb]
\centerline{
\includegraphics[scale=0.9]{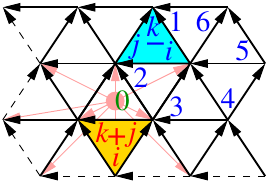}
}
\caption{
A 9-site triangular lattice on a torus, where each site $i$ has physical states
$|g_i\>$ labeled by $g_i \in G_b$, and each triangle has fermionic states
$|n_{ijk}\>$ where $n_{ijk}=0,1$ is the fermion occupation number.  The
orientations on the edges give rise to a natural order of the three vertices of
a triangle $(i,j,k)$ where the first vertex $i$ of a triangle has two outgoing
edges on the triangle and the last vertex $k$ of a triangle has two incoming
edges on the triangle.  The triangular lattice can be viewed as the surface of
solid torus.  A discretization of the solid torus can be obtained by adding a
vertex-0 inside the solid torus. The branching structure of the resulting
complex is indicated by the arrows on the edges.
}
\label{trilatt}
\end{figure}

We will assume that our 2D system lives on a triangular lattice which forms a 2D
torus (see Fig.  \ref{trilatt}).  On each lattice site $i$, we have physical
states $|g_i\>$ labeled by $g_i\in G_b$.  On each triangle $(ijk)$, we have two
states: no fermion state $|0\>$ and one fermion state $|1\>$.

The ideal ground state wave function can be obtained by viewing the 2D torus as
the surface of a 3D solid torus.  The fermionic path integral on the 3D solid
torus with the action amplitude $\cV_{d+1}^\pm$ obtained from the cocycle
$(\nu_{d+1},n_d,u^g_d)$ should give us the ground state wave function for the
corresponding fermionic SPT state.  To do the fermionic path integral, we need
to divide the 3D solid torus into a 3D complex with a branching structure.  Due
to the topological invariance of our constructed action amplitude, the
resulting wave function should not depend on how we divide the 3D solid torus
into 3D complex.  So we choose a very simple 3D complex which is formed by the
triangular lattice on the surface of the 3D solid torus and one additional
vertex-$0$ inside the 3D solid torus.  The resulting 3D complex is formed by
simplexes $[0ijk]$, where $ijk$ is a triangle on the surface (see Fig.
\ref{trilatt}). The branching structure of the 3D complex is given by the
orientations on the edges.  Those orientations for the edges on the surface are
given in Fig.  \ref{trilatt}.  For the edges inside the 3D solid torus, the
orientation is always pointing from vertex-0 to the vertex on the surface.  We
note that the simplexes associated with the down triangles in Fig.
\ref{trilatt} have a ``$+$'' orientation while the simplexes associated with
the up triangles have a ``$-$'' orientation.

{}From each 3-cocycle $(\nu_{3},n_2,u^g_2)$ in $\fZ^3[G_f,U_T(1)]$, we can
construct a fixed-point action amplitude.  For simplexes with ``$+$''
orientation, the fixed-point action amplitude is given by
\begin{align}
&\cV^+_3( g_0, g_i, g_j, g_k)
=
\nu^+_3( g_0, g_i, g_j, g_k) \times
\nonumber\\
&\ \ \ \
\th_{(123)}^{n_2(g_i,g_j,g_k)}
\th_{(013)}^{n_2(g_0,g_i,g_k)}
\bar\th_{(023)}^{n_2(g_0,g_j,g_k)}
\bar\th_{(012)}^{n_2(g_0,g_i,g_j)}
\end{align}
and
for simplexes with ``$-$'' orientation given by
\begin{align}
&\cV^-_3( g_0, g_i, g_j, g_k)
=
\nu^-_3( g_0, g_i, g_j, g_k) \times
\nonumber\\
&\ \ \ \
\th_{(012)}^{n_2(g_0,g_i,g_j)}
\th_{(023)}^{n_2(g_0,g_j,g_k)}
\bar\th_{(013)}^{n_2(g_0,g_i,g_k)}
\bar\th_{(123)}^{n_2(g_i,g_j,g_k)} ,
\end{align}
where
\begin{align}
 \nu_3^+ (g_0,g_i,g_j,g_k) &= (-)^{m_1(g_0,g_j)}\nu_3 (g_0,g_i,g_j,g_k),
\nonumber\\
 \nu_3^- (g_0,g_i,g_j,g_k) &=
(-)^{m_1(g_i,g_k)} /\nu_3 (g_0,g_i,g_j,g_k) ,
\end{align}
Since $n_2(g_0,g_1,g_2)$ is in $\cH^2(G,\Z_2)$, we can always write
$n_2(g_0,g_1,g_2)$ in terms of $m_1(g_0,g_1)$:
\begin{align}
 n_2(g_0,g_1,g_2) &=
m_1(g_1,g_2)+
m_1(g_0,g_2)+
m_1(g_0,g_1) \text{ mod } 2,
\nonumber\\
 m_1(g_0,g_1) &=0,1.
\end{align}
However, in general $m_1(gg_0,gg_1)\neq m_1(g_0,g_1)$, even though
$n_2(g_0,g_1,g_2)$ satisfies $n_2(gg_0,gg_1,gg_2)=n_2(g_0,g_1,g_2)$.

Now, the wave function is given by
\begin{align}
& \ \ \ \ \Psi( \{g_i\}, \{\th_{(ijk)}\}, \{\bar\th_{(ijk)}\})
\nonumber\\
&= \int \prod_{(0ij)}
\dd \th_{(0ij)}^{n_2(g_0,g_i,g_j)}
\dd \bar\th_{(0ij)}^{n_2(g_0,g_i,g_j)}
\prod_{\{0i\}} (-)^{m_1(g_0,g_i)}
\times
\nonumber\\
&\ \ \ \
\prod_{\bigtriangleup} \cV_3^-(g_0,g_i, g_j, g_k) \prod_{\bigtriangledown}
\cV_3^+(g_0,g_i, g_j, g_k) ,
\end{align}
where
$\prod_{(0ij)}$ is the product over all links of the triangular lattice,
$\prod_{\{0i\}}$ is the product over all sites of the triangular lattice,
$\prod_{\bigtriangleup} $ is the product over all up-triangles, and
$\prod_{\bigtriangledown} $ is the product over all down-triangles.
Clearly, the above wave function depends on
$g_i$ through $\nu_3$, $n_2$ and $m_1$.
The wave function also appears to depend on $g_0$,
the variable that we assigned to the vertex-0 inside the
solid torus. In fact, the $g_0$ dependence cancels out, and
the wave function is independent of $g_0$.
We can simply set $g_0=1$.

If all the $m_1$ dependence also cancels out, the wave function
will have a symmetry described by $G_f$:
$\Psi( \{gg_i\}, \{\th_{(ijk)}\}, \{\bar\th_{(ijk)}\})=
\Psi^{s(g)}( \{g_i\}, \{\th_{(ijk)}\}, \{\bar\th_{(ijk)}\})$.
But does the $m_1$ dependence cancel out?
Let us only write down the $m_1$ dependence of the
wave function:
\begin{align}
& \ \ \ \ \Psi( \{g_i\}, \{\th_{(ijk)}\}, \{\bar\th_{(ijk)}\})
\nonumber \\
&=
\prod_{\{0i\}} (-)^{m_1(g_0,g_i)}
\prod_{\bigtriangleup} (-)^{m_1(g_i,g_k)} \prod_{\bigtriangledown} (-)^{m_1(g_0,g_j)}
... ...
\nonumber\\
&=\prod_{\bigtriangleup} (-)^{m_1(g_i,g_k)} ... ...
\end{align}
where $ijk$ around the up- and down-triangles are arranged in a way as
illustrated in Fig. \ref{trilatt}. We also have used the relation
$\prod_{\{0i\}} (-)^{m_1(g_0,g_i)} \prod_{\bigtriangledown} (-)^{m_1(g_0,g_j)}
=1$.  We see that $m_1$ does not cancel out and the wave function is not
symmetric, since $m_1(gg_0,gg_1)\neq m_1(g_0,g_1)$.

This is a serious problem.  But the symmetry breaking is only on the surface
and it can be easily fixed: we simply redefine the wave function by including
an extra factor $\prod_\text{up-left} (-)^{m_1(g_i,g_k)}$
on the surface:
\begin{align}
\label{fWav1}
& \ \ \ \ \Psi( \{g_i\}, \{\th_{(ijk)}\}, \{\bar\th_{(ijk)}\})
\\
&=
\prod_\text{up-left} (-)^{m_1(g_i,g_k)}
\int \prod_{(0ij)}
\dd \th_{(0ij)}^{n_2(g_0,g_i,g_j)}
\dd \bar\th_{(0ij)}^{n_2(g_0,g_i,g_j)}
\times
\nonumber\\
&
\prod_{\{0i\}} (-)^{m_1(g_0,g_i)}
\prod_{\bigtriangleup} \cV_3^-(g_0,g_i, g_j, g_k) \prod_{\bigtriangledown}
\cV_3^+(g_0,g_i, g_j, g_k)
\nonumber\\
&=
\int \prod_{(0ij)}
\dd \th_{(0ij)}^{n_2(g_0,g_i,g_j)}
\dd \bar\th_{(0ij)}^{n_2(g_0,g_i,g_j)}
\times
\nonumber\\
&\ \ \ \ \ \ \ \
\prod_{\bigtriangleup} \nu_3^{-1}(g_0,g_i, g_j, g_k)
\prod_{\bigtriangledown} \nu_3(g_0,g_i, g_j, g_k)
\times
\nonumber\\
&\ \ \ \ \ \ \ \
\prod_{\bigtriangleup}
\th_{(0ij)}^{n_2(g_0,g_i,g_j)}
\th_{(0jk)}^{n_2(g_0,g_j,g_k)}
\bar\th_{(0ik)}^{n_2(g_0,g_i,g_k)}
\bar\th_{(ijk)}^{n_2(g_i,g_j,g_k)}
\times
\nonumber\\
&\ \ \ \ \ \ \ \
\prod_{\bigtriangledown}
\th_{(ijk)}^{n_2(g_i,g_j,g_k)}
\th_{(0ik)}^{n_2(g_0,g_i,g_k)}
\bar\th_{(0jk)}^{n_2(g_0,g_j,g_k)}
\bar\th_{(0ij)}^{n_2(g_0,g_i,g_j)}
\nonumber
\end{align}
where $\prod_\text{up-left}$ is a product over all the links with the up-left
orientation (note that $\prod_\text{up-left}
(-)^{m_1(g_i,g_k)}=\prod_{\bigtriangleup} (-)^{m_1(g_i,g_k)}$ and see Fig.
\ref{trilatt}).  The redefined wave function is indeed symmetric, and is the
wave function that corresponds to the fermionic SPT state described by the
cocycle $(\nu_3,n_2) \in \fZ^3[G_f,U_T(1)]$.

We note that the wave function $\Psi( \{g_i\}, \{\th_{(ijk)}\},
\{\bar\th_{(ijk)}\})$ described above contains Grassmann numbers. Indeed,
$\Psi( \{g_i\}, \{\th_{(ijk)}\}, \{\bar\th_{(ijk)}\})$ can be regarded as the
wave function in the fermion coherent state basis. After we expand the wave
function in power of the Grassmann numbers, we obtain
\begin{align}
& \ \ \ \ \Psi( \{g_i\}, \{\th_{(ijk)}\}, \{\bar\th_{(ijk)}\})
\nonumber\\
&= \sum_{n_{ijk}=0,1}\Phi( \{g_i\}, \{ n_{ijk}\})
\prod_{\bigtriangleup} \bar\th_{(ijk)}^{n_{ijk}}
\prod_{\bigtriangledown} \th_{(ijk)}^{n_{ijk}} .
\end{align}
Then $\Phi( \{g_i\}, \{ n_{ijk}\})$ is the amplitude of the ground state on the
fermion number basis $ \otimes_i |g_i\> \otimes_{(ijk)} |n_{ijk}\> $, where $|g_i\>$ is
the state on site-$i$ and $|n_{ijk}\>$ is the state on triangle $(ijk)$ (where
$n_{ijk}=0,1$ is the fermion occupation number on the  triangle $(ijk)$).  Note
that the sign of $\Phi( \{g_i\}, \{ n_{ijk}\})$ will depend on how the
Grassmann numbers are ordered in $\prod_{\bigtriangleup} \bar\th_{(ijk)}^{n_{ijk}}
\prod_{\bigtriangledown} \th_{(ijk)}^{n_{ijk}}$.
We also note that the wave function vanishes
if $n_{ijk}\neq n_2(g_i,g_j,g_k)$.

\subsection{No intrinsic topological orders}
\label{notop}

In this section, we are going to show that the wave function constructed in the
last section $\Phi( \{g_i\}, \{ n_{ijk}\})$ contain no intrinsic topological
orders as a fermion system.  In other words, starting with the following pure
bosonic direct product state
\begin{align}
 |\Phi_0\>=\otimes_i |\phi_i\>,\ \ \ \
|\phi_i\>=|G_b|^{-1/2} \sum_{g_i\in G_b} |g_i\>,
\end{align}
we can obtain the fermionic state $|\Phi\>$ constructed in the last section after a
fermionic LU transformation defined in \Ref{GWW1017}.

\begin{figure}[tb]
\begin{center}
\includegraphics[scale=1.4]{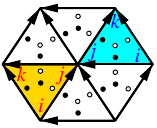}
\end{center}
\caption{
(Color online) A triangular lattice with four fermionic orbitals
in each triangle $(ijk)$. The fermions on the two solid dots in the yellow
triangle are described by operators $c_{(ijk)}$ (the center one)
and $ c_{(0ik)}$ (the side one), and the fermions on the two open dots are
described by operators $\bar c_{(0ij)}$ and $\bar c_{(0jk)}$.  The
fermions on the two open dots in the blue triangle are described by
operators $\bar c_{(ijk)}$ (the center one) and $\bar c_{(0ik)}$ (the side
one), and the fermions on the two solid dots are described by
operators $ c_{(0ij)}$ and $ c_{(0jk)}$.
Each vertex has bosonic states described by $|g_i\>$, $g_i\in G_b$.
}
\label{trifermU}
\end{figure}

To show this, we start with an expanded fermionic Hilbert space, where we have
four fermionic orbitals within each triangles. We also have bosonic state
$|g_i\>$ on each vertex (see Fig. \ref{trifermU}).  In the Grassmann number
form, the constructed fermionic wave function is given by \eqn{fWav1}:
\begin{align}
\label{fWav}
& \ \ \ \ \Psi( \{g_i\}, \{\th_{(ijk)}\}, \{\bar\th_{(ijk)}\})
\\
&=
\int \prod_{(0ij)}
\dd \th_{(0ij)}^{n_2(g_0,g_i,g_j)}
\dd \bar\th_{(0ij)}^{n_2(g_0,g_i,g_j)}
\times
\nonumber\\
&\ \ \ \ \ \ \ \
\prod_{\bigtriangleup} \nu_3^{-1}(g_0,g_i, g_j, g_k)
\prod_{\bigtriangledown} \nu_3(g_0,g_i, g_j, g_k)
\times
\nonumber\\
&\ \ \ \ \ \ \ \
\prod_{\bigtriangleup}
\th_{(0ij)}^{n_2(g_0,g_i,g_j)}
\th_{(0jk)}^{n_2(g_0,g_j,g_k)}
\bar\th_{(0ik)}^{n_2(g_0,g_i,g_k)}
\bar\th_{(ijk)}^{n_2(g_i,g_j,g_k)}
\times
\nonumber\\
&\ \ \ \ \ \ \ \
\prod_{\bigtriangledown}
\th_{(ijk)}^{n_2(g_i,g_j,g_k)}
\th_{(0ik)}^{n_2(g_0,g_i,g_k)}
\bar\th_{(0jk)}^{n_2(g_0,g_j,g_k)}
\bar\th_{(0ij)}^{n_2(g_0,g_i,g_j)}
\nonumber
\end{align}
We would like to point out that although the above expression contains $g_0$,
the topological invariance of the fermion path integral ensures that the
Grassmann wave function on the left-hand-side does not depend on $g_0$.

If we treat $\th$'s and $\bar\th$'s as the following complex fermion operators
\begin{align}
& \th_{(abc)}=c^\dag_{(abc)}, \ \ \
 \bar\th_{(abc)}=\bar c^\dag_{(abc)},
\\
& \dd\th_{(abc)}=c_{(abc)}, \ \ \
 \dd\bar\th_{(abc)}=\bar c_{(abc)},
\nonumber
\end{align}
the expression \eqn{fWav} can be viewed as an operator:
\begin{align}
\label{hatU}
& \hat U=
\prod_{(0ij)}
 c_{(0ij)}^{n_2(g_0,g_i,g_j)}
\bar c_{(0ij)}^{n_2(g_0,g_i,g_j)}
\times
\\
&\ \ \ \ \
\prod_{\bigtriangleup} \nu_3^{-1}(g_0,g_i, g_j, g_k)
\prod_{\bigtriangledown} \nu_3(g_0,g_i, g_j, g_k)
\times
\nonumber\\
&\ \ \ \ \
\prod_{\bigtriangleup}
 c_{(0ij)}^{\dag n_2(g_0,g_i,g_j)}
 c_{(0jk)}^{\dag n_2(g_0,g_j,g_k)}
\bar c_{(0ik)}^{\dag n_2(g_0,g_i,g_k)}
\bar c_{(ijk)}^{\dag n_2(g_i,g_j,g_k)}
\times
\nonumber\\
&\ \ \ \ \
\prod_{\bigtriangledown}
 c_{(ijk)}^{\dag n_2(g_i,g_j,g_k)}
 c_{(0ik)}^{\dag n_2(g_0,g_i,g_k)}
\bar c_{(0jk)}^{\dag n_2(g_0,g_j,g_k)}
\bar c_{(0ij)}^{\dag n_2(g_0,g_i,g_j)} .
\nonumber
\end{align}
Again $\hat U$ is independent of $g_0$, despite the
appearance of $g_0$ on the right-hand-side.
Then the fermionic state constructed in the last section
can be obtained from the bosonic product state $|\Phi_0\>$:
\begin{align}
 |\Psi\>=\hat U |\Phi_0\>.
\end{align}
Note that $ |\Phi_0\>$ is a ``no-fermion'' state satisfying
\begin{align}
 c_{(abc)} |\Phi_0\>=0.
\end{align}

Now we would like to point out that $\hat U$ itself is formed by several layers of
\emph{fermionic} LU transformations.  Since $\nu_3$ is a pure $U(1)$ phase,
thus $\prod_{\bigtriangleup} \nu_3^{-1}(g_0,g_i, g_j, g_k)
\prod_{\bigtriangledown} \nu_3(g_0,g_i, g_j, g_k)$ represents layers of bosonic
LU transformations.  Also the $\hat U$ has a property that when acting on
$|\Phi_0\>$, $c^\dag$ and $\bar c^\dag$ always act on states with no fermion
and $c$ and $\bar c$ always act on states with one fermion.  So those operators
map a set of orthogonal states to another set of orthogonal states.  In this
case, an even numbers of $ c$'s and $\bar c$'s correspond to a fermionic LU
transformation.  Therefore, the terms in $\hat U$, such as $ c_{(ijk)}^{\dag
n_2(g_i,g_j,g_k)} c_{(0ik)}^{\dag n_2(g_0,g_i,g_k)} \bar c_{(0jk)}^{\dag
n_2(g_0,g_j,g_k)} \bar c_{(0ij)}^{\dag n_2(g_0,g_i,g_j)}$ and $
c_{(0ij)}^{n_2(g_0,g_i,g_j)} \bar c_{(0ij)}^{n_2(g_0,g_i,g_j)}$, all represent
fermionic LU transformation as defined in \Ref{GWW1017}.  Note that none of the
above transformations changes $g_i$.  So we can show those transformations to be
unitary within each fixed set of $\{g_i\}$.  Therefore, the state $|\Psi\>$ has
no fermionic long range entanglement as defined in \Ref{GWW1017} (\ie no
fermionic intrinsic topological order).  $|\Psi\>$ is the fermionic SPT state
described by a cocycle $(\nu_3,n_2,u^g_2)$.

The fermionic LU transformation \eq{hatU} that maps the fermionic SPT state to
a trivial product state is one of the most important results in this paper.  All
the properties of the fermionic SPT state as well as the classification of the
fermionic SPT states can be described in terms of the fermionic LU
transformation.  The fact that the fermionic LU transformation is expressed in
terms of group super-cohomology gives us a systematic understanding of
fermionic SPT states in terms of group super-cohomology.

\section{Ideal Hamiltonians that realize the fermionic SPT states}
\label{idHam}

After obtain the wave function $|\Psi\>=\hat U|\Phi_0\>$ for the fermionic SPT
state, it is easy to construct a Hamiltonian $H$ such that $|\Psi\>$ is its
ground state.  We start with an nonnegative definite hermitian operator $H_0$
that satisfies $H_0 |\Phi_0\>=0$.
For example, we may choose
\begin{align}
 H_0=\sum_i (1-|\phi_i\>\<\phi_i|) .
\end{align}
The Hamiltonian $H$ can then be obtained as
\begin{align}
\label{HiHp}
H&=VH'+\sum_i H_i
,
\nonumber\\
H_i&= \hat U (1-|\phi_i\>\<\phi_i|) \hat U^\dag
\nonumber\\
H' &=
\sum_\bigtriangledown [ c^\dag_{(ijk)} c_{(ijk)}-n_2(g_i,g_j,g_k)]^2
\nonumber\\
&\ \ \ \
+\sum_\bigtriangleup [ \bar c^\dag_{(ijk)} \bar c_{(ijk)}-n_2(g_i,g_j,g_k)]^2 .
\end{align}
When $V$ is positive and very large, the $H'$ enforces that the fermion number on each triangle $(ijk)$ is given by
$n_2(g_i,g_j,g_k)$.  Since $\hat U^\dag \hat U|\Phi_0\>=|\Phi_0\>$, one can easily show that $H$ is
nonnegative definite and $H |\Psi\>=H \hat U|\Phi_0\>=0$.

Similar as the bosonic case, $H_i$ acts on site $i$ as well as its $6$
neighbor. However, since there are $6$ more fermionic degrees of freedom in the
$6$ triangles surround $i$, $H_i$ also acts on these $6$ triangles. Moreover,
when $V$ is positive and very large, the low states are the zero energy
subspace of $H^\prime$. Within such a low energy subspace, all the $H_i$ are
hermitian (unconstrained)commuting projectors satisfying $H_i^2=H_i$ and
$H_iH_j=H_jH_i$.(We note that in the zero energy subspace of $H^\prime$,
$H_i^2=\hat U (1-|\phi_i\>\<\phi_i|) \hat U^\dag \hat U (1-|\phi_i\>\<\phi_i|)
\hat U^\dag=\hat U (1-|\phi_i\>\<\phi_i|) \hat U^\dag=H_i$ and $H_iH_j=\hat U
(1-|\phi_i\>\<\phi_i|) (1-|\phi_j\>\<\phi_j|) \hat U^\dag=\hat U
(1-|\phi_j\>\<\phi_j|) (1-|\phi_i\>\<\phi_i|) \hat U^\dag=H_jH_i$.) Such a nice
property(frustration free) make it exact solvable, with $\Psi( \{g_i\},
\{\th_{(ijk)}\}, \{\bar\th_{(ijk)}\})$ as its unique ground state.

In Appendix \ref{Hamiltonian}, we provide an alternative way to construct the parent Hamiltonian from path integral formulism, which is equivalent to the construction here in the infinity $V$ limit.

\section{An example of 2D fermionic SPT states with a $Z_2$ symmetry}
\subsection{Ideal Hamiltonians that realize the fermionic
2D $Z_2$ SPT states}

\begin{table*}[tb]
 \centering
 \begin{tabular}{ |c|c|c|c|}
\hline
$ g_1 g_2 g_3 g_4 g_5 g_6 $ & $v( g_1, g_2, g_3, g_4, g_5, g_6)$
&
$ g_1 g_2 g_3 g_4 g_5 g_6 $ & $v( g_1, g_2, g_3, g_4, g_5, g_6)$
\\
\hline
000000 & $\bpm 1 ,& -1\epm$ &
000001 & $\bpm 1 ,& -1\epm$ \\
000010 & $\bpm 1 ,& -1\epm$ &
000011 & $\bpm -1 ,& -1\epm$ \\
000100 & $\bpm 1 ,& -1\epm$ &
000101 & $\bpm 1 ,& -1\epm$ \\
000110 & $\bpm 1 ,& 1\epm$ &
000111 & $\bpm -1 ,& 1\epm$ \\
\hline
001000 & $\bpm 1 ,& -1\epm$ &
001001 & $\bpm 1 ,& -1\epm$ \\
001010 & $\bpm 1 ,& -1\epm$ &
001011 & $\bpm -1 ,& -1\epm$ \\
001100 & $\bpm 1 ,& 1\epm$ &
001101 & $\bpm 1 ,& 1\epm$ \\
001110 & $\bpm 1 ,& -1\epm$ &
001111 & $\bpm -1 ,& -1\epm$ \\
\hline
010000 & $\bpm 1 ,& -1\epm$ &
010001 & $\bpm 1 ,& -1\epm$ \\
010010 & $\bpm 1 ,& -1\epm$ &
010011 & $\bpm -1 ,& -1\epm$ \\
010100 & $\bpm 1 ,& -1\epm$ &
010101 & $\bpm 1 ,& -1\epm$ \\
010110 & $\bpm 1 ,& 1\epm$ &
010111 & $\bpm -1 ,& 1\epm$ \\
\hline
011000 & $\bpm -1 ,& -1\epm$ &
011001 & $\bpm -1 ,& -1\epm$ \\
011010 & $\bpm -1 ,& -1\epm$ &
011011 & $\bpm 1 ,& -1\epm$ \\
011100 & $\bpm -1 ,& 1\epm$ &
011101 & $\bpm -1 ,& 1\epm$ \\
011110 & $\bpm -1 ,& -1\epm$ &
011111 & $\bpm 1 ,& -1\epm$ \\
\hline
\end{tabular}
 \caption{
The 1 by 2 matrices $ v( g_1, g_2, g_3, g_4, g_5, g_6)$ for the bosonic SPT
state \eq{n2Z2B1}.
}
\label{HiTableB}
\end{table*}

In this section, we will apply the above general discussion to a particular
example -- the fermionic  2D $Z_2$ SPT states.  The 3-cocycles that describe
the three nontrivial fermionic  2D $Z_2$ SPT states are given in section
\ref{2DZ2fSPT}.  All those SPT phases can be realized on a triangle lattice as
described in Fig.  \ref{trilatt}.  Each site has two bosonic states $|g_i\>$,
$g_i=0,1$, and each triangle has a fermionic orbital which can occupied
$|n_{ijk}=1\>$ or unoccupied $|n_{ijk}=0\>$.  The bulk Hamiltonian
\begin{align}
H=H'+\sum_i H_i
\end{align}
that realizes the SPT phase on such a triangle lattice can constructed from the
data in the 3-cocycles $[\nu_3(g_0,g_1,g_2,g_3),n_2(g_0,g_1,g_2)]$ as discussed
in the above section (see \eqn{HiHp}).

$H_i$ in the Hamiltonian acts on the bosonic states on site-$i$ and the six
sites 1,2,3,4,5,6 around the site-$i$ (see Fig.  \ref{trilatt}).  $H_i$ also
acts on the fermionic states on the six triangles, $(i21)$, $(3i2)$, $(43i)$,
$(45i)$, $(5i6)$, $(i61)$, around the site-$i$.  $H_i$ has a property that it
does not change $g_1$, $g_2$, $g_3$, $g_4$, $g_5$, $g_6$, but it can change the
bosonic state on site-$i$: $|g_i\> \to |g_i'\>$ and the fermionic states on the
triangles $(ijk)$: $|n_{ijk}\> \to |n'_{ijk}\>$.  So we can express $H_i$ as
operator-valued 2 by 2 matrix $M( g_1, g_2, g_3, g_4, g_5, g_6)$ where the
matrix elements are given by
\begin{align}
&\ \ \ \
 M_{g_i'g_i}( g_1, g_2, g_3, g_4, g_5, g_6)
\nonumber\\
&=
\<  g_1 g_2 g_3 g_4 g_5 g_6 g_i'|H_i |  g_1 g_2 g_3 g_4 g_5 g_6 g_i\> .
\end{align}
It turns out that $M( g_1, g_2, g_3, g_4, g_5, g_6)$
are always 2 by 2 projection matrices
\begin{align}
&\ \ \ \
 M( g_1, g_2, g_3, g_4, g_5, g_6)
\\
&=\frac 12
v^\dag( g_1, g_2, g_3, g_4, g_5, g_6)
 v( g_1, g_2, g_3, g_4, g_5, g_6) ,
\nonumber
\end{align}
where $v( g_1, g_2, g_3, g_4, g_5, g_6)$ are operator-valued 1 by 2 matrices
that satisfy the $Z_2$ symmetry condition
\begin{align}
&\ \ \ \
 v( g_1, g_2, g_3, g_4, g_5, g_6)
\\
&=v( 1-g_1, 1-g_2, 1-g_3, 1-g_4, 1-g_5, 1-g_6)\si^x .
\nonumber
\end{align}

$H'$ in the Hamiltonian is given by
\begin{align}
H' &= U
\sum_{(ijk)} [ c^\dag_{(ijk)} c_{(ijk)}-n_2(g_i,g_j,g_k)]^2
\end{align}
It enforces the constraints that the fermion number on each triangle $(ijk)$
is given by $n_2(g_i,g_j,g_k)$ in the ground state.  The $H_i$ mentioned above
preserve the constraints: $[H_i,H']=0$.  We see that $n_2(g_i,g_j,g_k)$
describes the fermionic character of the SPT phases.  For the SPT phases
described by \eqn{n2Z2B}, $n_2(g_i,g_j,g_k)=0$. So there is no fermion in those
SPT states.  Those SPT states are actually bosonic SPT states.  For the
SPT phases described by \eqn{n2Z2F}, $n_2(g_i,g_j,g_k)\neq 0$. The
corresponding SPT states are fermionic SPT states.

\begin{table*}[tb]
 \centering
 \begin{tabular}{ |c|c|c|c|}
\hline
$ g_1 g_2 g_3 g_4 g_5 g_6 $ & $v( g_1, g_2, g_3, g_4, g_5, g_6)$
&
$ g_1 g_2 g_3 g_4 g_5 g_6 $ & $v( g_1, g_2, g_3, g_4, g_5, g_6)$
\\
\hline
000000 & $\bpm 1 ,& - \bar c_{(5i6)} c_{(3i2)}\epm$ &
000001 & $\bpm  c_{(i61)} ,& - c_{(3i2)}\epm$ \\
000010 & $\bpm  c_{(45i)} ,& - c_{(3i2)}\epm$ &
000011 & $\bpm  \imth  c_{(i61)} \bar c_{(5i6)} c_{(45i)} ,& - %
c_{(3i2)}\epm$ \\
000100 & $\bpm 1 ,& - \bar c_{(5i6)} c_{(45i)} \bar c_{(43i)} %
c_{(3i2)}\epm$ &
000101 & $\bpm  c_{(i61)} ,& - c_{(45i)} \bar c_{(43i)} %
c_{(3i2)}\epm$ \\
000110 & $\bpm 1 ,&  \imth  \bar c_{(43i)} c_{(3i2)}\epm$ &
000111 & $\bpm  \imth  c_{(i61)} \bar c_{(5i6)} ,&  \imth  \bar %
c_{(43i)} c_{(3i2)}\epm$ \\
\hline
001000 & $\bpm  \bar c_{(43i)} ,& - \bar c_{(5i6)}\epm$ &
001001 & $\bpm  c_{(i61)} \bar c_{(43i)} ,& -1\epm$ \\
001010 & $\bpm  c_{(45i)} \bar c_{(43i)} ,& -1\epm$ &
001011 & $\bpm  \imth  c_{(i61)} \bar c_{(5i6)} c_{(45i)} \bar %
c_{(43i)} ,& -1\epm$ \\
001100 & $\bpm 1 ,& - \imth  \bar c_{(5i6)} c_{(45i)}\epm$ &
001101 & $\bpm  c_{(i61)} ,& - \imth  c_{(45i)}\epm$ \\
001110 & $\bpm 1 ,& -1\epm$ &
001111 & $\bpm  \imth  c_{(i61)} \bar c_{(5i6)} ,& -1\epm$ \\
\hline
010000 & $\bpm  \bar c_{(i21)} ,& - \bar c_{(5i6)}\epm$ &
010001 & $\bpm  c_{(i61)} \bar c_{(i21)} ,& -1\epm$ \\
010010 & $\bpm - \bar c_{(i21)} c_{(45i)} ,& -1\epm$ &
010011 & $\bpm  \imth  c_{(i61)} \bar c_{(i21)} \bar c_{(5i6)} %
c_{(45i)} ,& -1\epm$ \\
010100 & $\bpm  \bar c_{(i21)} ,& - \bar c_{(5i6)} c_{(45i)} \bar %
c_{(43i)}\epm$ &
010101 & $\bpm  c_{(i61)} \bar c_{(i21)} ,& - c_{(45i)} \bar %
c_{(43i)}\epm$ \\
010110 & $\bpm  \bar c_{(i21)} ,&  \imth  \bar c_{(43i)}\epm$ &
010111 & $\bpm - \imth  c_{(i61)} \bar c_{(i21)} \bar c_{(5i6)} ,&  %
\imth  \bar c_{(43i)}\epm$ \\
\hline
011000 & $\bpm - \imth  \bar c_{(i21)} \bar c_{(43i)} c_{(3i2)} ,& - %
\bar c_{(5i6)}\epm$ &
011001 & $\bpm - \imth  c_{(i61)} \bar c_{(i21)} \bar c_{(43i)} %
c_{(3i2)} ,& -1\epm$ \\
011010 & $\bpm  \imth  \bar c_{(i21)} c_{(45i)} \bar c_{(43i)} %
c_{(3i2)} ,& -1\epm$ &
011011 & $\bpm  c_{(i61)} \bar c_{(i21)} \bar c_{(5i6)} c_{(45i)} %
\bar c_{(43i)} c_{(3i2)} ,& -1\epm$ \\
011100 & $\bpm  \imth  \bar c_{(i21)} c_{(3i2)} ,& - \imth  \bar %
c_{(5i6)} c_{(45i)}\epm$ &
011101 & $\bpm  \imth  c_{(i61)} \bar c_{(i21)} c_{(3i2)} ,& - \imth  %
c_{(45i)}\epm$ \\
011110 & $\bpm  \imth  \bar c_{(i21)} c_{(3i2)} ,& -1\epm$ &
011111 & $\bpm  c_{(i61)} \bar c_{(i21)} \bar c_{(5i6)} c_{(3i2)} ,& %
-1\epm$ \\
\hline
\end{tabular}
 \caption{
The 1 by 2 matrices $ v( g_1, g_2, g_3, g_4, g_5, g_6)$
for the fermionic SPT state \eq{n2Z2F1}.
}
\label{HiTable}
\end{table*}

We also would like to mention that the constructed Hamiltonian $H$
has an on-site $Z_2$ symmetry generated by
\begin{align}
 \hat W(g) = \prod_{\v i} |g_{\v i}\>\<gg_{\v i}|,\ \ \ \
g,g_{\v i}\in Z_2.
\end{align}

For the trivial phase described by
\begin{align}
\label{n2Z2B0}
& n_2(g_0,g_1,g_2)=0,
\nonumber\\
& \nu_3(0,1,0,1)=\nu_3(1,0,1,0)=1, \ \ \ \ \text{other } \nu_3 = 1,
\end{align}
we find that $v( g_1, g_2, g_3, g_4, g_5, g_6)=(1,-1)$ and $M( g_1, g_2, g_3,
g_4, g_5, g_6)=\frac 12\bpm 1 & -1\\ -1 & 1\epm=H_i$.  Note that $H_i$ is a
projection operator on the site-$i$. Thus the ground state of $H$ is a product
state $\otimes_i |\phi_i\>$, $|\phi_i\>\propto |0\>+|1\>$.

One of the nontrivial SPT phases is described by
\begin{align}
\label{n2Z2B1}
& n_2(g_0,g_1,g_2)=0,
\nonumber\\
& \nu_3(0,1,0,1)=\nu_3(1,0,1,0)=1, \ \ \ \ \text{other } \nu_3 = 1,
\end{align}
It is a bosonic SPT state since $n_2(g_0,g_1,g_2)=0$. Such a bosonic SPT state
can also be viewed as a fermionic SPT state forming tightly bounded fermion
pairs.  The Hamiltonian for such a bosonic SPT state is given by $v( g_1, g_2,
g_3, g_4, g_5, g_6)$ in the Table \ref{HiTableB}.  The corresponding $H_i=M(
g_1, g_2, g_3, g_4, g_5, g_6)$ are either $\frac 12\bpm 1 & -1\\ -1 & 1\epm $
or $\frac 12\bpm 1 & 1\\ 1 & 1\epm $.  $H_i$ are still projection operators.
But now $H_i$ can be two different projection operators depending on the values
of $g_1, g_2, g_3, g_4, g_5, g_6$ on the neighboring sites.  Such a
bosonic SPT phase is nothing but the $Z_2$ SPT state studied in \Ref{CLW1141}.

The 3-cocycle
\begin{align}
\label{n2Z2F1}
 n_2(0,1,0)=n_2(1,0,1)=1, \ \ \ \ &\text{other } n_2 = 0,
\nonumber\\
 \nu_3(0,1,0,1)=\nu_3(1,0,1,0)= \imth, \ \ \ \ &\text{other } \nu_3 = 1,
\end{align}
describes a nontrivial fermionic SPT state.  The Hamiltonian for such a
fermionic SPT state is given by $v( g_1, g_2, g_3, g_4, g_5, g_6)$ in the Table
\ref{HiTableB}.  We see that $H_i$ are still projection operators.  But now
$H_i$ can be many different projection operators depending on the values of
$g_1, g_2, g_3, g_4, g_5, g_6$ on the neighboring sites.  Also the
projection operators mix the bosonic and fermionic states.

\subsection{Edge excitations of
2D fermionic SPT state with $Z_2$ symmetry
}
\label{2DZ2fSPT}

In this section, we will discuss 2D fermionic SPT states with $Z_2$ symmetry in
more detail.  In particular, the nontrivial realization of the $Z_2$ symmetry
on the edge states and its protection of gapless edge excitations against
interactions.

In Appendix \ref{calfcoh}, we have calculated the group
super-cohomology classes $\fH^3[Z_2\times Z_2^f,U(1)]=\Z_4$. This means that
interacting fermion systems with a $Z_2$ symmetry can have (at least) 4
different SPT phases: a trivial one plus three nontrivial ones.  This result
is described by the $Z_2\times Z_2^f$ row and $d_{sp}=2$ column of Table
\ref{tbF}.

\subsubsection{3-cocycles --  data that characterize the fermionic
2D $Z_2$ SPT states}

The fermionic 2D $Z_2$ SPT phases are characterized by the data
$[\nu_3(g_0,g_1,g_2,g_3),n_2(g_0,g_1,g_2)]$ where $\nu_3(g_0,g_1,g_2,g_3)$ is a
complex function and $n_2(g_0,g_1,g_2)$ an integer function with variables
$g_i\in G_b=G_f/Z_2^f=Z_2$.  The data
$[\nu_3(g_0,g_1,g_2,g_3),n_2(g_0,g_1,g_2)]$ are called a fermionic 3-cocycle,
which is an element in $\fH^3[Z_2\times Z_2^f,U(1)]$.  The
first two SPT phases are given by
\begin{align}
\label{n2Z2B}
& n_2(g_0,g_1,g_2)=0,
\nonumber\\
& \nu_3(0,1,0,1)=\nu_3(1,0,1,0)=\pm 1, \ \ \ \ \text{other } \nu_3 = 1,
\end{align}
and the next two SPT phases are given by
\begin{align}
\label{n2Z2F}
 n_2(0,1,0)=n_2(1,0,1)=1, \ \ \ \ &\text{other } n_2 = 0,
\nonumber\\
 \nu_3(0,1,0,1)=\nu_3(1,0,1,0)=\pm \imth, \ \ \ \ &\text{other } \nu_3 = 1,
\end{align}
where we have assumed that the elements in $G_b=Z_2$ are described by $\{0,1\}$
with $0$ being the identity element.  The 3-cocycle $n_2(g_0,g_1,g_2)=0$,
$\nu_3(g_0,g_1,g_2,g_3)=1$ corresponds to the trivial fermionic SPT phase.

Using the above  3-cocycle data $[\nu_3(g_0,g_1,g_2,g_3), n_2(g_0,g_1,g_2)]$ we
can construct the ideal wave functions that realize the above fermionic SPT
phases (see section \ref{idwav}).  We can also construct local Hamiltonian (see
section \ref{idHam}) such that the above ideal wave functions are the exact
ground states.

\subsubsection{Low energy edge excitation -- their effective symmetry and
effective Hamiltonian}

The nontrivial 2D $Z_2$ SPT states described by the above three 3-cocycles
have symmetry-protected gapless edge excitations.  The detailed discussions of
those gapless edge excitations are presented in Appendix
\ref{edgestate} and \ref{symmedge}.  Here we just present the results.

The low energy edge excitations of the SPT phase can be described by an
effective Hamiltonian $H_\text{eff}=\sum_{i=1}^L H_\text{eff}(i)$, where $L$ is
the number of sites on the edge and the states on each edge-site are described
by $|g_i\>$, $g_i\in G_b$, $i=1,2,...,L$.  We use $|\{g_i\}_\text{edge},g_0\>$
to denote an low energy edge state (where $g_0$ takes a fixed value, say
$g_0=0$).  Since the edge is 1D, we can use a purely bosonic model to describe
the edge states of 2D fermion system (see section \ref{edgestate}).

For the nontrivial SPT phase described by 3-cocycles
$[\nu_3(g_0,g_1,g_2,g_3),n_2(g_0,g_1,g_2)]$, the low energy effective edge
Hamiltonian $H_\text{eff}$ satisfies an unusual symmetry condition (see Appendix
\ref{symmedge}):
\begin{align}
&\ \ \ \
 W^\dag_\text{eff}(g) H_\text{eff}(i) W_\text{eff}(g) = H_\text{eff}(i) ,
\nonumber\\
&\ \ \ \
W_\text{eff}(g) |\{g_i\}_\text{edge},g_0\>
\nonumber\\
&=
(-)^{
n_2(g^{-1}g_0,g_0,g_1)
\sum_{i=1}^L [n_2(g_{0},g_{i+1},g_{i})+1] }
\nonumber\\
&\ \ \ \
\prod_{i} w^*_{i,i+1}
|\{g g_i\}_\text{edge},g_0\>,
\end{align}
where
\begin{align}
& w_{i,i+1} =
\imth^{
n_2(g^{-1}g_{0},g_{i+1},g_{i})
-n_2(g_{0},g_{i+1},g_{i})
+2n_2(g^{-1}g_0,g_0,g_i)
}
\times \nonumber \\ & \ \ \ \
\nu_3(g^{-1}g_0,g_0,g_{i+1}, g_i) 
.
\end{align}
We see that the effective edge symmetry $ W_\text{eff}(g)$ on the low energy
edge states is determined by the cocycle data
$[\nu_3(g_0,g_1,g_2,g_3),n_2(g_0,g_1,g_2)]$.  Due to the $ w_{i,i+1}$ factor,
the symmetry is not an on-site symmetry.  Such a symmetry can protect the
gaplessness of the edge excitations if the symmetry is not spontaneously broken
on the edge.

For the trivial SPT phase \eq{n2Z2B0}, we find that the symmetry action on the
edge states is given by $ W_\text{eff}(g) |\{g_i\}_\text{edge},g_0\>
=|\{gg_i\}_\text{edge},g_0\>$ which is an on-site symmetry.  Such an on-site
symmetry cannot protect the gapless edge excitations: the edge excitations can
be gapped without breaking the symmetry.

For the nontrivial bosonic SPT phase \eq{n2Z2B1}, we find that the symmetry
action on the edge states to be
\begin{align}
 W_\text{eff}(g) |\{g_i\}_\text{edge},g_0\>
=\prod_{i} w^*_{i,i+1}
 |\{gg_i\}_\text{edge},g_0\>
\end{align}
which is not an on-site symmetry, with $w_{i,i+1}(g_i,g_{i+1})$ given by
\begin{align}
 w_{i,i+1}(0,0)&=1, &
 w_{i,i+1}(0,1)&=-1,
\nonumber\\
 w_{i,i+1}(1,0)&=1, &
 w_{i,i+1}(1,1)&=1.
\end{align}
Such an non-on-site symmetry can protect the gapless edge excitations: the edge
excitations must be gapless without breaking the symmetry.\cite{CLW1141}

If we identify $|0\>$ as $|\up\>$ and $|1\>$ as $|\down\>$,
we may rewrite the above in an operator form
\begin{align}
 w_{i,i+1} &=\e^{\imth \pi [\frac 14
(\si_i^z+1)(\si_{i+1}^z-1)
]}
,
\\
W_\text{eff}(1) &=\Big[\prod_i \si_i^x\Big]\Big[ \prod_i
\e^{\imth \pi [-\frac 14
-\frac 14 \si^z_{i}
+\frac 14 \si^z_{i+1}
+\frac 14 \si_i^z\si_{i+1}^z
]} \Big]
\nonumber\\
&=
\Big[\prod_i \si_i^x\Big]\Big[ \prod_i
\e^{\imth \pi [-\frac 14
+\frac 14 \si_i^z\si_{i+1}^z
]} \Big]
.
\nonumber
\end{align}
We find
\begin{align}
[W_\text{eff}(1)]^2=
  \prod_{i=1}^L
\e^{\imth \pi [-\frac 12
+\frac 12 \si_i^z\si_{i+1}^z
]}
=
\prod_{i=1}^L
 \si_i^z\si_{i+1}^z
=1.
\end{align}
So $W_\text{eff}(0)=1$ and $W_\text{eff}(1)$ indeed form a $Z_2$
representation.  From
\begin{align}
&
 W_\text{eff}^\dag(1) \si_i^x  W_\text{eff}(1)
= - \si_{i-1}^z \si_i^x  \si_i^z
\end{align}
we find that the following edge Hamiltonian
\begin{align}
 H_\text{eff}=
\sum_{i} \Big[
J_z \si^z_i\si^z_{i+1} +
h_x (\si^x_i- \si^z_{i-1}\si^x_i\si^z_{i+1} )
\Big]
\end{align}
respects the non-on-site $Z_2$ symmetry on the edge.  Such a system either
spontaneously breaks the $Z_2$ symmetry on the edge or has gapless edge
excitations.\cite{CLW1141,duality}

For the nontrivial fermionic SPT phase \eq{n2Z2F1}, we find the symmetry
action on the edge states to be
\begin{align}
&\ \ \ \
 W_\text{eff}(g) |\{g_i\}_\text{edge},g_0\>
\nonumber\\
&=
(-)^{
n_2(g^{-1}g_0,g_0,g_1)
\sum_{i=1}^L [n_2(g_{0},g_{i+1},g_{i})+1] }
\times \nonumber\\
&\ \ \ \
\prod_{i} w^*_{i,i+1}
 |\{gg_i\}_\text{edge},g_0\>
\end{align}
which is also not an on-site symmetry, with $w_{i,i+1}(g_i,g_{i+1})$ given by
\begin{align}
 w_{i,i+1}(0,0)&=1, &
 w_{i,i+1}(0,1)&=1,
\nonumber\\
 w_{i,i+1}(1,0)&=-\imth, &
 w_{i,i+1}(1,1)&=-1.
\end{align}
Again, we may rewrite the above in an operator form
\begin{align}
 w_{i,i+1} &=\e^{\imth \pi [-\frac 38
+\frac 18 \si^z_{i+1}
+\frac 38 \si^z_{i}
-\frac 18 \si_i^z\si_{i+1}^z
]}
.
\end{align}
When $g=1$ and $g_0=0$, we find that $(-)^{n_2(g^{-1}g_0,g_0,g_1)}=\si^z_1$.
Also since
\begin{align}
n_2(g_0,0,0)&=0, &
n_2(g_0,0,1)&=0,
\nonumber\\
n_2(g_0,1,0)&=1, &
n_2(g_0,1,1)&=0,
\end{align}
we find that
\begin{align}
\sum_{i=1}^L
n_2(g_0,g_i,g_{i+1})
=
\frac 14 \sum_{i=1}^L (1-\si_i^z\si_{i+1}^z)
.
\end{align}
Therefore,
\begin{align}
\label{Weff1}
&
 W_\text{eff}(1) =
\Big[\prod_i \si_i^x\Big]\Big[ \prod_i\e^{\imth \pi [\frac 38
-\frac 18 \si^z_{i+1}
-\frac 38 \si^z_{i}
+\frac 18 \si_i^z\si_{i+1}^z
]} \Big]
\times \nonumber \\ &
\ \ \ \
\ \ \ \
\ \ \ \
\ \
[\si_1^z]^{\sum_{i=1}^L [
\frac 14  (1-\si_i^z\si_{i+1}^z)+1]
}
\\
=&
\Big[\prod_i \si_i^y\Big]\Big[ \prod_i\e^{\imth \pi [
\frac 18 \si_i^z\si_{i+1}^z
-\frac 58
]} \Big]
[\si_1^z]^{\sum_{i=1}^L
\frac 14  (5-\si_i^z\si_{i+1}^z)
}
.
\nonumber
\end{align}
We note that
\begin{align}
[W_\text{eff}(1)]^2
=&
\e^{\imth\frac{\pi}{4}\sum_{i=1}^L[5-\si_i^z\si_{i+1}^z]}
\prod_i\e^{\imth \pi [
\frac 14 \si_i^z\si_{i+1}^z
-\frac 54
]} =1.
\end{align}
We see that $W_\text{eff}(g)$ forms a $Z_2$ representation.

{}From the expression \eqn{edgeWavS} of the low energy edge state, we see that
the total number of fermions in the low energy edge states
$|\{g_i\}_\text{edge},g_0\>$ is given by
\begin{align}
N_F= \sum_i n_2(g_a,g_{i+1},g_i)
\end{align}
(which is independent of $g_a$).  If we choose $g_a=0$, we see that only the
link $(g_i,g_{i+1})=(0,1)$ has one fermion, while other links have no fermion.
If we choose $g_a=1$ instead, then only the link $(g_i,g_{i+1})=(1,0)$ has one
fermion, while other links have no fermion.  Let us call a link $(i,i+1)$ with
$\si^z_i\si^z_{i+1}=-1$ a domain wall.  We see that either only the step-up
domain wall has a fermion or the step-down domain wall has a fermion.  Since
the numbers of step-up domain walls and the step-down domain walls are equal on
a ring, the expression $N_F= \sum_i n_2(g_a,g_{i+1},g_i)$ does not depend on
$g_a$.  The number of fermions is equal to the half of the number of the domain
walls.  So the fermion parity operator $P_f = (-)^{N_F}$ is given by
\begin{align}
 P_f=&(-)^{
\frac 14 \sum_{i=1}^L (1-\si_i^z\si_{i+1}^z)
}
=
\prod_i \e^{\imth
\frac {\pi}{4}  (1-\si_i^z\si_{i+1}^z)
}
\nonumber\\
=&
\prod_i
 \e^{\imth
\frac {\pi}{4}}
[
\cos(\frac {\pi}{4})-\imth \sin(\frac {\pi}{4})\si_i^z\si_{i+1}^z
]
.
\end{align}
The effective edge Hamiltonian must be invariant under both $P_f$ and
$W_\text{eff}(1)$ transformations.

The edge effective Hamiltonian must be invariant under both the fermion-number
parity $P_f$ and the $Z_2$ symmetry $W_\text{eff}(1)$ transformations.  One
example is given by the following (for infinite long edge):
\begin{align}
\label{Hef}
& H_\text{edge}   =\sum_i
 \Big[-J \si^z_i\si^z_{i+1} + h_0 \si^x_i (\si^z_{i-1}-\si^z_{i+1})
\nonumber\\ &\ \ \ \
+
h_1\si^y_i (1- \si^z_{i-1} \si^z_{i+1})
\\ & \
+
\imth
h_2
(\si^+_{i} \si^+_{i+2} -\si^-_{i} \si^-_{i+2})
(1+ \si^z_{i-1} \si^z_{i+1})
(1+ \si^z_{i+1} \si^z_{i+3})
\Big]
,
\nonumber
\end{align}
where
\begin{align}
 \si^-=\bpm 0 & 0\\ 1&0 \epm,\ \ \ \ \
 \si^+=\bpm 0 & 1\\ 0&0 \epm.
\end{align}

To understand the behavior of such a Hamiltonian,
we note that the $J$,
$h_0$ and $h_1$ terms in the above $H_\text{eff}$ cannot change the number of
domain walls.
The $h_0$ and $h_1$ terms can only induce domain wall hopping.
So if $h_2=0$, the model will have an effective $U(1)$ symmetry.
It has two phases: a gapped phase for large $J$ where $\si^z=\pm 1$ and there
are no domain walls, and a gapless phase for large $h_0,h_1$ where the domain
walls form a ``superfluid''.  The gapped phase breaks the $Z_2$ symmetry while
the gapless phase is described by a central-charge $c=1$ non-chiral Luttinger liquid
theory.

The $h_2$ term can only change the domain wall number by $\pm 4$ since the
number of fermions on the edge is given by half of the domain wall number. So,
the fermion-number-parity conservation only allows the domain wall number to
change by a multiple of 4.  The $h_2$ term can be irrelevant. So, the edge
excitations still can be gapless.

Now, let us explain why the effective edge Hamiltonian \eq{Hef}
respects the non-on-site $Z_2$ symmetry \eq{Weff1}.
We note that $\frac12 \sum_{i=1}^L(1-{\sigma}^z_i {\sigma}^z_{i+1})$ counts the
number of domain walls.  Therefore $J$, $h_0$ and $h_1$ terms in \eqn{Hef}
commute with $\exp\left(\frac{\imth\pi}{4}\frac 12\sum_{i=1}^L(1-{\sigma}^z_i
{\sigma}^z_{i+1})\right)$ since those terms do not change the domain wall
number, while the $h_2$ term in \eqn{Hef} anti-commute with
$\exp\left(\frac{\imth\pi}{4}\frac 12\sum_{i=1}^L(1-{\sigma}^z_i
{\sigma}^z_{i+1})\right)$ since that term changes the domain wall number by $\pm
4$. We can also show that the $J$, $h_0$ and $h_1$ terms commute with
$\prod_{i=1}^L {\sigma}^y_i$, while the $h_2$ terms anti-commute with
$\prod_{i=1}^L {\sigma}^y_i$.

For the terms in  \eqn{Hef} that do not act on the site-1, we can treat the
$\si^z_1$ in $W_\text{eff}(1)$ (see \eqn{Weff1}) as a c-number. In this case
$[\si_1^z]^{\sum_{i=1}^L \frac 14  (5-\si_i^z\si_{i+1}^z) } $ are either $\pm
1$ or $\pm P_f$.  So $[\si_1^z]^{\sum_{i=1}^L \frac 14  (5-\si_i^z\si_{i+1}^z)
} $ always commute with the terms in  \eqn{Hef} (as long as they do not act on
the site-1).  After dropping the term $[\si_1^z]^{\sum_{i=1}^L \frac 14
(5-\si_i^z\si_{i+1}^z) } $ we find that $W_\text{eff}(1)$ and
$\exp\left(\frac{\imth\pi}{4}\frac 12\sum_{i=1}^L(1-{\sigma}^z_i
{\sigma}^z_{i+1})\right)\prod_{i=1}^L {\sigma}^y_i$ only differ by a phase.
This way, we show that the edge Hamiltonian is invariant under $P_f$ and
$W_\text{eff}(1)$.

\subsubsection{The stability and instability of edge theory}
Here we make use the method developed in \Ref{duality} to study the stability and instability of the edge theory for the nontrivial fermionic SPT phase. As we know, the $c=1$ non-chiral Luttinger liquid can be described as:
\begin{eqnarray}
\mathcal{L} = \frac{1}{4\pi} (\partial_x \theta \partial_t \phi + \partial_x
\phi \partial_t \theta)
- \frac{v}{8\pi} \left[ K (\partial_x \theta)^2 +
\frac{4}{K} (\partial_x \phi)^2 \right]\nonumber\\\label{edgeth}
\end{eqnarray}
with Luttinger parameter $K $ and velocity $v $.

The key step for understanding the stability and instability of edge theory is to figure out how the low energy fields $\theta$ and $\phi$ transform under the non-on-site $Z_2$ symmetry.
Let us first introduce the domain wall representation:
\begin{align}
\tau_{i}^z &={\sigma}^z_i
{\sigma}^z_{i+1}
\end{align}
In principle, we can re-express everything in terms of the ${\tau}$'s.
However, the above duality transformation does not quite work for a system
with periodic boundary conditions, since the ${\tau}^z_i$ variables
obey the global constraint $\prod_{i=1}^L {\tau}^z_i = 1$, and therefore only
describe $L-1$ independent degrees of freedom.

In order to incorporate the missing degree of freedom
and make the dual description complete, we
introduce an additional $Z_2$ gauge field
$\mu^z_{i-1,i}$ that lives on the links connecting neighboring
boundary sites $i-1,i$. We then
define the duality transformation between ${\sigma}$ and ${\tau}, \mu$ by
the relation
\begin{equation}
\mu^x_{i-1,i} = {\sigma}^z_i
\end{equation}
together with the gauge invariance constraint
\begin{equation}
\mu^x_{i-1,i} \mu^x_{i,i+1} {\tau}^z_i = 1
\label{giconst}
\end{equation}
It is easy to check that this duality transformation is complete: there
is a one-to-one correspondence between configurations of ${\sigma}^z_i = \pm 1$
and configurations of $\mu^x_i = \pm 1, {\tau}^z_i = \pm 1$ obeying the
constraint (\ref{giconst}). Similarly, there is a one-to-one correspondence between
physical operators written in terms of the ${\sigma}$'s and gauge invariant
combinations of $\mu, {\tau}$ (i.e. operators that commute with the left hand side of
(\ref{giconst})). In particular, the operators ${\sigma}^x, {\sigma}^y,
{\sigma}^z$ are given by
\begin{eqnarray}
{\sigma}^x_i &=& {\tau}^x_{i-1} {\tau}^x_i \mu^y_{i-1,i} \nonumber \\
{\sigma}^y_i &=& {\tau}^x_{i-1} {\tau}^x_i \mu^z_{i-1,i} \nonumber \\
{\sigma}^z_i &=& \mu^x_{i-1,i}
\label{dualform}
\end{eqnarray}
while the symmetry transformation $S$ is given by:
\begin{eqnarray}
W_\text{eff}(1) &\sim&
\prod_{i=1}^L {\sigma}^y_i\exp\left(\frac{\imth\pi}{4}\frac 12\sum_{i=1}^L(1-{\sigma}^z_i
{\sigma}^z_{i+1})\right)\nonumber\\
 &=& \prod_{i=1}^L \mu^z_{i-1,i} \cdot \exp\left(\frac{i\pi}{8}\sum_{i=1}^L (1-{\tau}^z_i)\right)
\label{symmtau}
\end{eqnarray}

In the long wavelength limit, the domain wall density is given by
\begin{equation}
\frac{{\tau}^z_i}{2} \sim \frac{1}{\pi} \partial_x \phi
\label{domwall}
\end{equation}

We also note that (\ref{domwall}) implies that
\begin{equation}
\exp\left(-\frac{\pi i}{8}\sum_{i=1}^L \tau^z_i\right) =
\exp\left(-\frac{i}{4} \int \partial_x \phi dx \right)
\end{equation}
Similarly, we have:
\begin{equation}
\prod_{i=1}^L \mu^z_{i-1,i} = \exp\left(\frac{i}{2} \int \partial_x \theta dx \right)
\end{equation}
This equality follows from the observation that the periodic/anti-periodic
sectors $\prod_{i=1}^L \mu^z_{i-1,i} = \pm 1$ correspond to the two boundary conditions
$\theta(L) - \theta(0) = 4m \pi, (4m+2)\pi$ respectively.

Combining these two results, we see that our expression (\ref{symmtau}) for $S$
becomes
\begin{equation}
W_\text{eff}(1)\sim \exp\left(\frac{i}{2} \int \partial_x \theta dx - \frac{i}{4} \int
\partial_x \phi dx \right)
\end{equation}
Using the commutation relation $[\theta(x), \partial_x \phi(y)] =
2\pi i \delta(x-y)$, we deduce that
\begin{eqnarray}
W_\text{eff}(1)^{-1} \theta W_\text{eff}(1) &=& \theta + \frac{\pi}{2} \nonumber\\ W_\text{eff}(1)^{-1} \phi W_\text{eff}(1) &=& \phi + \pi
\label{symmedgeth}
\end{eqnarray}

For the fermion parity symmetry $P_f$, a similar calculation gives out:
\begin{eqnarray}
P_f &=&\exp\left(\frac{i\pi}{4}\sum_{i=1}^L (1-{\tau}^z_i)\right)\nonumber\\
&=& \exp\left(- \frac{i}{2} \int
\partial_x \phi dx \right)
\end{eqnarray}

\begin{equation}
P_f^{-1} \theta P_f = \theta + \pi ;\quad P_f^{-1} \phi P_f = \phi
\label{parity}
\end{equation}

The above transformation laws Eq.(\ref{symmedgeth},\ref{parity}) together with the
action (\ref{edgeth}) gives a complete description of the
low energy edge physics.

It is easy to see terms like $\cos(4\theta-\alpha(x))$ or $\cos(2\phi-\alpha(x))$ are allowed by both the non-on-site $Z_2$ symmetry and fermion parity $P_f$. Obviously, the $Z_2$ symmetry will be broken if a mass gap is generated by these terms. By performing a simple scaling dimension calculation, we find that both terms are irrelevant and the edge theory remains gapless when $2<K<8$. To this end, we see that in contrast to the bosonic $Z_2$ SPT phase\cite{duality}, the fermionic $Z_2$ SPT phases have a stable gapless edge.

\section{An example of $3$D fermionic SPT states with $T^2=1$ time-reversal symmetry}

In Appendix \ref{calfcoh}, we have also calculated the group super-cohomology
classes $\fH^4[Z_2^T\times Z_2^f,U(1)]=\Z_2$. So interacting fermion systems
with a $T^2=1$ time-reversal symmetry can have (at least) one nontrivial SPT phase.  This is described by the
$Z_2^T\times Z_2^f$ row and $d_{sp}=3$ column of Table \ref{tbF}.

Let us list the fermionic 4-cocycles
$[\nu_4(g_0,g_1,g_2,g_3,g_4),n_3(g_0,g_1,g_2,g_3)]$ that describe the intrinsic fermionic SPT phase that can
neither be realized by free fermion models nor by interacting boson models.
\begin{align}
\label{3DSPT2}
& n_3(0,1,0,1)=n_3(1,0,1,0)=1, \ \ \ \ \text{other } n_3 = 0,
\\
& \nu_4(0,1,0,1,0)=-\nu_4(1,0,1,0,1)=\pm \imth, \ \ \ \ \text{other } \nu_4 = 1.
\nonumber
\end{align}
Using the above cocycles, we can write the corresponding wave functions
and exactly solvable Hamiltonians.  However, the explicit Hamiltonian is very
complicated.  We wonder if there exists a better basis, in which the Hamiltonian
will have a simpler form. We will address this problem in future publications.

Moreover, we note that the above two fermionic 4-cocycles only differ by the following
bosonic 4-cocycle:
\begin{align}
\label{3DSPT1}
& n_3(g_0,g_1,g_2,g_3)=0,
\\
& \nu_4(0,1,0,1,0)=\nu_4(1,0,1,0,1)=\pm 1, \ \ \ \ \text{other }\nu_4 = 1,
\nonumber
\end{align}
which describes a $T^2=1$ (nontrivial) bosonic SPT phase.(We note that in the
limit with tightly bounded fermion pairs, e.g., $n_3(g_0,g_1,g_2,g_3)=0$, a
fermionic system can always be viewed as a bosonic system.) Surprisingly, in
Appendix \ref{calfcoh}, we found such a solution can be generated by a
fermionic coboundary.  Physically, such a statement indicates that the $T^2=1$
(nontrivial) bosonic SPT phase can be connected to a trivial product state or
an atomic insulator state without breaking the corresponding $T^2=1$ time
reversal symmetry through interacting fermion systems. In addition, such a
result also implies the two fermionic 4-cocycles in Eq.(\ref{3DSPT2}) actually
describe the same fermionic SPT phase.

\section{Summary}

It was shown in \Ref{CGL1172} that generalized topological nonlinear
$\si$ models with symmetry can be constructed from group cohomology theory of
the symmetry group.  This leads to a systematic construction of bosonic SPT
phases in any dimensions and for any symmetry groups.  This result allows us to
construct new topological insulators (with symmetry group $U(1)\times Z_2^T$)
and new topological superconductors (with symmetry group $Z_2^T$) for
interacting boson systems (or qubit systems).  It also leads to a complete
classification of all gapped phases in 1D interacting boson/fermion systems.

In this paper, we introduce a special group super-cohomology theory which is a
generalization of the standard group cohomology theory.  Using the special group
super-cohomology theory, we can construct new discrete fermionic topological
nonlinear $\si$ models with symmetry.  This leads to a systematic construction
of fermionic SPT phases in any dimensions and for certain symmetry groups $G_f$
where the fermions form an 1D representation.  The discrete fermionic
topological nonlinear $\si$ model, when defined on a space-time with boundary,
can be viewed as a ``non-local'' boundary effective Lagrangian, which is a
fermionic and discrete generalization of the bosonic continuous
Wess-Zumino-Witten term.  Thus the boundary excitations of a nontrivial SPT
phase are described by a ``non-local'' boundary effective Lagrangian, which, we
believe, implies that the boundary excitations are gapless or topologically
ordered if the symmetry is not broken.

As a simple application of our group supercohomology theory, we constructed a
nontrivial SPT phase in 3D, for interacting fermionic superconductors with
coplanar spin order.  Such a topological superconductor has time-reversal
$Z_2^T$ and fermion-number-parity $Z_2^f$ symmetries described by the full
symmetry group $G_f=Z_2^T\times Z_2^f$.\cite{W1103}  The nontrivial SPT phase
should have gapless or fractionalized excitations on the 2D surface if the
time-reversal symmetry is not broken.  It is known that such a nontrivial 3D
gapped topological superconductor does not exist if the fermions are
non-interacting. In addition, such a nontrivial SPT phase can not be realized
in any interacting bosonic models either. So the constructed nontrivial
fermionic SPT phase is totally new.

We also constructed three nontrivial SPT phases in 2D, for interacting
fermionic systems with the full symmetry group $G_f=Z_2\times Z_2^f$. We show
that the three nontrivial SPT phases indeed have gapless excitations on the 1D
edge which are described by central-charge $c=1$ conformal field theory, if the
$Z_2$ symmetry is not broken. In several recent
works\cite{duality,fduality,CG,ftoric}, it has been further shown that by
"gauging" the $Z_2$ global symmetry, each of the three nontrivial SPT phases
can be uniquely identified by the braiding statistics of the corresponding
gauge flux.

Clearly, more work is needed to generalize the special group super-cohomology
theory to the yet-to-be-defined full group super-cohomology theory, so that we
can handle the cases when the fermions do not form an 1D representation in the
fixed point wavefunctions.  This will allow us to construct more general
interacting fermionic SPT phases.

\section{acknowledgements}

This research is supported by NSF Grant No.  DMR-1005541, NSFC 11074140, and
NSFC 11274192.  It is also supported by BMO financial group through the Newton
chair and by the John Templeton Foundation.  Research at Perimeter Institute is
supported by the Government of Canada through Industry Canada and by the
Province of Ontario through the Ministry of Research.

\appendix

\section{Topological invariance of the partition amplitudes}
\label{topZ}

In this appendix, we will prove the topological invariance of the
partition amplitudes Eq. (\ref{fspt}) in $1+1$D and $2+1$D under
their corresponding fermionic group cocycle condition. The $3+1$D
case is much more complicated but can still be checked by computer.

\subsection{$1+1$D}

\begin{figure}[b]
\begin{center}
\includegraphics[scale=0.4]{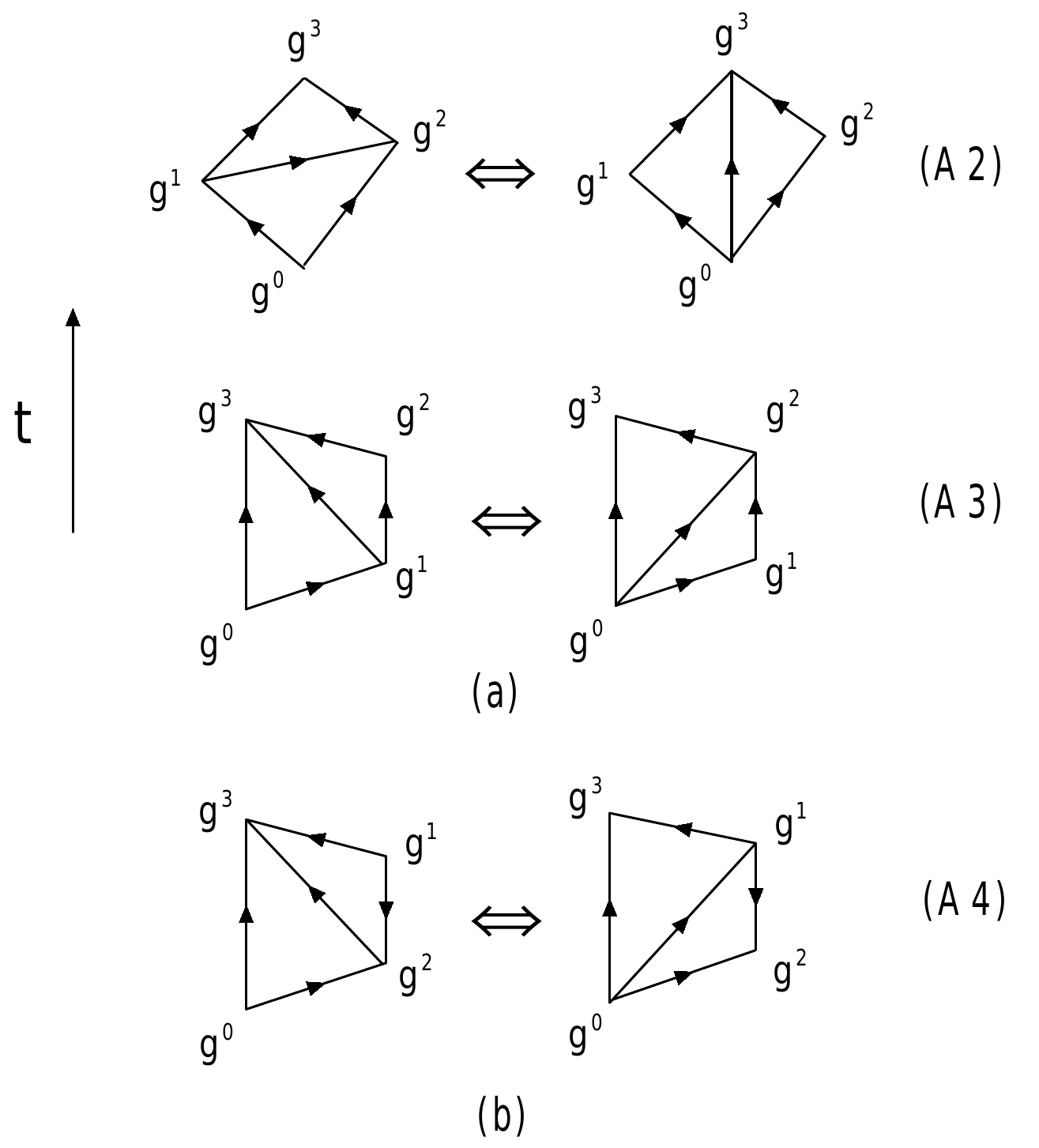}
\end{center}
\caption{The admissible branching $2 \leftrightarrow 2$ moves.
(a)Branching moves that can be induced by a global time ordering.
(b)Branching moves that can not be induced by a global time
ordering.} \label{1D22}
\end{figure}

\begin{figure}[tb]
\begin{center}
\includegraphics[scale=0.35]{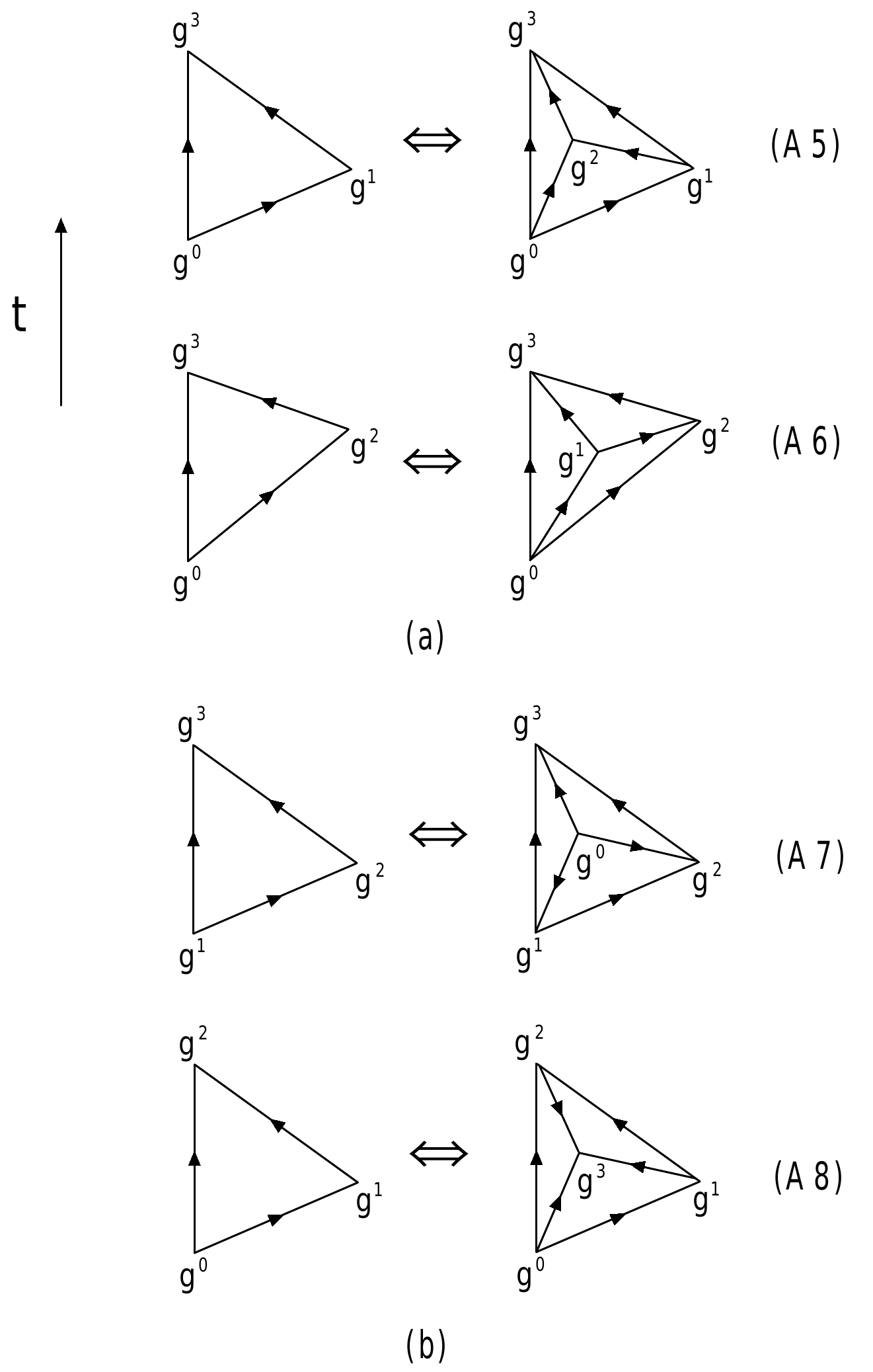}
\end{center}
\caption{The admissible branching $1 \leftrightarrow 3$ moves.
(a)Branching moves that can be induced by a global time ordering.
(b)Branching moves that can not be induced by a global time
ordering.} \label{1D13}
\end{figure}
Similarly as we prove the topological invariance of the partition
amplitudes that describe bosonic SPT phases\cite{CGL1172}, we can
check all the admissible branching $2 \leftrightarrow 2$ and $1
\leftrightarrow 3$ moves for the Grassmann graded 2 cocycle
function:
\begin{align}
\mathcal{V}^{\pm}_2(g_i,g_j,g_k)=\nu^{\pm}_2(g_i,g_j,g_k)
&\theta_{(i,j)}^{n_1(g_i,g_j)} \theta_{(j,k)}^{n_1(g_j,g_k)}{\b
\theta_{(i,k)}}^{n_1(g_i,g_k)}.
\end{align}

For admissible branching $2 \leftrightarrow 2$ moves, we can have
three different equations (up to the orientation conjugate
$+\rightarrow -;-\rightarrow +$):
\begin{align}
&\int \dd\theta_{(12)}^{n_1(g_1,g_2)} \dd
\b\theta_{(12)}^{n_1(g_1,g_2)}
\mathcal{V}^+_2(g_1,g_2,g_3)\mathcal{V}^-_2(g_0,g_1,g_2)\nonumber\\=&\int
\dd\theta_{(03)}^{n_1(g_0,g_3)} \dd \b\theta_{(03)}^{n_1(g_0,g_3)}
\mathcal{V}^+_2(g_0,g_2,g_3)\mathcal{V}^-_2(g_0,g_1,g_3)\\
&\int\dd\theta_{(13)}^{n_1(g_1,g_3)} \dd
\b\theta_{(13)}^{n_1(g_1,g_3)}
\mathcal{V}^+_2(g_1,g_2,g_3)\mathcal{V}^+_2(g_0,g_1,g_3)\nonumber\\=&\int
\dd\theta_{(02)}^{n_1(g_0,g_2)} \dd \b\theta_{(02)}^{n_1(g_0,g_2)}
\mathcal{V}^+_2(g_0,g_2,g_3)\mathcal{V}^+_2(g_0,g_1,g_2),
\end{align}
and
\begin{align}
&\int\dd\theta_{(23)}^{n_1(g_2,g_3)} \dd
\b\theta_{(23)}^{n_1(g_2,g_3)}
\mathcal{V}^-_2(g_1,g_2,g_3)\mathcal{V}^+_2(g_0,g_2,g_3)\nonumber\\=&\int
\dd\theta_{(01)}^{n_1(g_0,g_1)} \dd \b\theta_{(01)}^{n_1(g_0,g_1)}
\mathcal{V}^+_2(g_0,g_1,g_3)\mathcal{V}^-_2(g_0,g_1,g_2).
\label{G1D22move}
\end{align}
For admissible branching $1 \leftrightarrow 3$ moves, we can have
four different equations (up to the orientation conjugate
$+\rightarrow -;-\rightarrow +$):
\begin{align}
&\mathcal{V}^+_2(g_0,g_1,g_3)\nonumber\\=&\int
\dd\theta_{(02)}^{n_1(g_0,g_2)} \dd
\b\theta_{(02)}^{n_1(g_0,g_2)}\dd\theta_{(12)}^{n_1(g_1,g_2)} \dd
\b\theta_{(12)}^{n_1(g_1,g_2)}\nonumber\\
&\times\dd\theta_{(23)}^{n_1(g_2,g_3)} \dd
\b\theta_{(23)}^{n_1(g_2,g_3)}(-)^{m_0(g_2)} \nonumber\\
&\times
\mathcal{V}^-_2(g_1,g_2,g_3)\mathcal{V}^+_2(g_0,g_2,g_3)\mathcal{V}^+_2(g_0,g_1,g_2)
\\
&\mathcal{V}^+_2(g_0,g_2,g_3)\nonumber\\=&\int
\dd\theta_{(01)}^{n_1(g_0,g_1)} \dd
\b\theta_{(01)}^{n_1(g_0,g_1)}\dd\theta_{(12)}^{n_1(g_1,g_2)} \dd
\b\theta_{(12)}^{n_1(g_1,g_2)}\nonumber\\
&\times\dd\theta_{(13)}^{n_1(g_1,g_3)} \dd
\b\theta_{(13)}^{n_1(g_1,g_3)}(-)^{m_0(g_1)} \nonumber\\
&\times
\mathcal{V}^+_2(g_1,g_2,g_3)\mathcal{V}^+_2(g_0,g_1,g_3)\mathcal{V}^-_2(g_0,g_1,g_2),
\end{align}
and
\begin{align}
&\mathcal{V}^+_2(g_1,g_2,g_3)\nonumber\\=&\int
\dd\theta_{(01)}^{n_1(g_0,g_1)} \dd
\b\theta_{(01)}^{n_1(g_0,g_1)}\dd\theta_{(02)}^{n_1(g_0,g_2)} \dd
\b\theta_{(02)}^{n_1(g_0,g_2)}\nonumber\\
&\times\dd\theta_{(03)}^{n_1(g_0,g_3)} \dd
\b\theta_{(03)}^{n_1(g_0,g_3)}(-)^{m_0(g_0)} \nonumber\\
&\times
\mathcal{V}^+_2(g_0,g_2,g_3)\mathcal{V}^-_2(g_0,g_1,g_3)\mathcal{V}^+_2(g_0,g_1,g_2)\\
&\mathcal{V}^+_2(g_0,g_1,g_2)\nonumber\\=&\int
\dd\theta_{(03)}^{n_1(g_0,g_3)} \dd
\b\theta_{(03)}^{n_1(g_0,g_3)}\dd\theta_{(13)}^{n_1(g_1,g_3)} \dd
\b\theta_{(13)}^{n_1(g_1,g_3)}\nonumber\\
&\times\dd\theta_{(23)}^{n_1(g_2,g_3)} \dd
\b\theta_{(23)}^{n_1(g_2,g_3)}(-)^{m_0(g_3)} \nonumber\\
&\times
\mathcal{V}^+_2(g_1,g_2,g_3)\mathcal{V}^-_2(g_0,g_2,g_3)\mathcal{V}^+_2(g_0,g_1,g_3).\label{G1D13move1}
\end{align}
Note that the first two equations of $2 \leftrightarrow 2$ and $1
\leftrightarrow 3$ moves can be induced by a global time
ordering (see in Fig. \ref{1D22} (a) and Fig. \ref{1D13} (a)) while
the rest can not (see in Fig. \ref{1D22} (b) and Fig. \ref{1D13}
(b)). Here $g^0$ is defined on the vertex with no incoming edge,
$g^1$ with one incoming edge, etc.

Let us use the definition of $\mathcal{V}_2^\pm$ and integral out
the Grassmann variables. For the branching $2 \leftrightarrow 2$
moves, we have:
\begin{align}
\nu^+_2(g_1,g_2,g_3)\nu^-_2(g_0,g_1,g_2)=&
\nu^+_2(g_0,g_2,g_3)\nu^-_2(g_0,g_1,g_3)\\
\nu^+_2(g_1,g_2,g_3)\nu^+_2(g_0,g_1,g_3)=&
\nu^+_2(g_0,g_2,g_3)\nu^+_2(g_0,g_1,g_2),
\end{align}
and
\begin{align}
 &\nu^-_2(g_1,g_2,g_3)\nu^+_2(g_0,g_2,g_3)\nonumber\\=& (-)^{n_1(g_1,g_2)}
\nu^+_2(g_0,g_1,g_3)\nu^-_2(g_0,g_1,g_2).
\end{align}
Similarly, for the branching $1\leftrightarrow 3$ moves, we have:
\begin{align}
&\nu^+_2(g_0,g_1,g_3)\\=&(-)^{m_0(g_2)}
\nu^-_2(g_1,g_2,g_3)\nu^+_2(g_0,g_2,g_3)\nu^+_2(g_0,g_1,g_2)\nonumber\\
 &\nu^+_2(g_0,g_2,g_3)\\=&(-)^{m_0(g_1)}
\nu^+_2(g_1,g_2,g_3)
\nu^+_2(g_0,g_1,g_3)\nu^-_2(g_0,g_1,g_2),\nonumber
\end{align}
and
\begin{align}
 &\nu^+_2(g_1,g_2,g_3)\\=&(-)^{m_0(g_1)}
\nu^+_2(g_0,g_2,g_3)\nu^-_2(g_0,g_1,g_3)\nu^+_2(g_0,g_1,g_2)\nonumber\\
 &\nu^+_2(g_0,g_1,g_2)\\=&(-)^{m_0(g_2)} \nu^+_2(g_1,g_2,g_3)
\nu^-_2(g_0,g_2,g_3)\nu^+_2(g_0,g_1,g_3).\nonumber\label{G1D13move}
\end{align}
If we use the definition of $\nu_2^\pm$:
\begin{align}
\nu_2^+ (g_0,g_1,g_2)&= \nu_2 (g_0,g_1,g_2),
\nonumber\\
\nu_2^- (g_0,g_1,g_2)&= (-)^{m_0(g_1)} /\nu_2 (g_0,g_1,g_2),
\end{align}
All the above admissible branching moves will be equivalent to a
single fermionic 2 cocycle equation of $\nu_2$
\begin{align}
\nu_2 (g_0,g_1,g_3) \nu_2 (g_1,g_2,g_3)  =\nu_2 (g_0,g_1,g_2) \nu_2
(g_0,g_2,g_3).
\end{align}
We note that this equation is the same as the 2 cocycle equation in
bosonic systems.

\subsection{$2+1$D}
In $2+1$D, there are in total 10 admissible branching $2
\leftrightarrow 3$ moves and 5 admissible branching $1
\leftrightarrow 4$ moves. Let us show all theses moves will lead to
the same fermionic 3 cocycle condition.

For example, the admissible $2 \leftrightarrow 3$ in Fig. \ref{2to3}
represents the following equation for $\mathcal{V}^\pm_3$:
\begin{widetext}
\begin{align}
&\int \dd \th_{(123)}^{n(g_1,g_2,g_3)}\dd
\b\th_{(123)}^{n(g_1,g_2,g_3)}
\mathcal{V}_3^+(g_0,g_1,g_2,g_3)\mathcal{V}_3^+(g_1,g_2,g_3,g_4)\nonumber\\
= & \int \dd \th_{(014)}^{n(g_0,g_1,g_4)}\dd
\b\th_{(014)}^{n(g_0,g_1,g_4)}\dd
\th_{(024)}^{n(g_0,g_2,g_4)}\dd \b\th_{(024)}^{n(g_0,g_2,g_4)}
\dd \th_{(034)}^{n(g_0,g_3,g_4)}\dd
\b\th_{(034)}^{n(g_0,g_3,g_4)}(-)^{m_1(g_0,g_4)}\nonumber\\
\times&\mathcal{V}_3^+(g_0,g_1,g_2,g_4)
\mathcal{V}_3^+(g_0,g_2,g_3,g_4)\mathcal{V}_3^-(g_0,g_1,g_3,g_4)
\end{align}
\end{widetext}
Note that in the above expression, we put the sign factor
$(-)^{m_1(g_0,g_4)}$ on the interior link and integral out the
Grassmann variables on the interior faces. To simplify the
representation, we can formally rewrite the above equation as (see \eqn{Gint}):
\begin{align}
&\int
\mathcal{V}_3^+(g_0,g_1,g_2,g_3)\mathcal{V}_3^+(g_1,g_2,g_3,g_4)\\
= & \int \mathcal{V}_3^+(g_0,g_1,g_2,g_4)
\mathcal{V}_3^+(g_0,g_2,g_3,g_4)\mathcal{V}_3^-(g_0,g_1,g_3,g_4)\nonumber
\end{align}

In such a way, we can formally write down all admissible branching
$2 \leftrightarrow 3$ moves in terms of $\mathcal{V}_3^{\pm}$(up to
the orientation conjugate $+\rightarrow -;-\rightarrow +$):
\begin{align}
& \int{\mathcal{V}_3^-(g_0,g_1,g_2,g_4)}
{\mathcal{V}_3^-(g_0,g_2,g_3,g_4 )}\\=& \int
{\mathcal{V}_3^-(g_0,g_1,g_2,g_3)}{\mathcal{V}_3^-(g_0,g_1,g_3,g_4 )}{\mathcal{V}_3^-(g_1,g_2,g_3,g_4)}\nonumber
\end{align}
\begin{align}
&\int \mathcal{V}_3^-(g_0,g_1,g_2,g_4)\mathcal{V}_3^+(g_1,g_2,g_3,g_4)\\
= & \int \mathcal{V}_3^-(g_0,g_1,g_2,g_3)\mathcal{V}_3^-(g_0,g_1,g_3,g_4)\mathcal{V}_3^+(g_0,g_2,g_3,g_4)\nonumber
\end{align}
\begin{align}
&\int \mathcal{V}_3^-(g_0,g_1,g_2,g_3)\mathcal{V}_3^+(g_0,g_2,g_3,g_4)\\
= & \int \mathcal{V}_3^-(g_0,g_1,g_2,g_4)\mathcal{V}_3^+(g_0,g_1,g_3,g_4)\mathcal{V}_3^+(g_1,g_2,g_3,g_4)\nonumber
\end{align}
\begin{align}
&\int \mathcal{V}_3^+(g_0,g_1,g_3,g_4)\mathcal{V}_3^+(g_1,g_2,g_3,g_4)\\
= & \int \mathcal{V}_3^-(g_0,g_1,g_2,g_3)\mathcal{V}_3^+(g_0,g_1,g_2,g_4)\mathcal{V}_3^+(g_0,g_2,g_3,g_4)\nonumber
\end{align}
\begin{align}
&\int
\mathcal{V}_3^+(g_0,g_1,g_2,g_3)\mathcal{V}_3^+(g_0,g_1,g_3,g_4)\\
= &\int
\mathcal{V}_3^+(g_0,g_1,g_2,g_4)\mathcal{V}_3^+(g_0,g_2,g_3,g_4)
\mathcal{V}_3^-(g_1,g_2,g_3,g_4)\nonumber
\end{align}
\begin{align}
&\int
\mathcal{V}_3^-(g_0,g_1,g_2,g_4)\mathcal{V}_3^+(g_0,g_1,g_3,g_4)\\
= & \int \mathcal{V}_3^-(g_0,g_1,g_2,g_3)\mathcal{V}_3^+(g_0,g_2,g_3,g_4)\mathcal{V}_3^-(g_1,g_2,g_3,g_4)\nonumber
\end{align}
\begin{align}
&\int \mathcal{V}_3^-(g_0,g_1,g_3,g_4)\mathcal{V}_3^+(g_0,g_2,g_3,g_4)\\
=&\int \mathcal{V}_3^+(g_0,g_1,g_2,g_3)\mathcal{V}_3^-(g_0,g_1,g_2,g_4)\mathcal{V}_3^+(g_1,g_2,g_3,g_4)\nonumber
\end{align}
\begin{align}
&\int
\mathcal{V}_3^+(g_0,g_1,g_2,g_3)\mathcal{V}_3^+(g_1,g_2,g_3,g_4)\\
= & \int \mathcal{V}_3^+(g_0,g_1,g_2,g_4)
\mathcal{V}_3^+(g_0,g_2,g_3,g_4)\mathcal{V}_3^-(g_0,g_1,g_3,g_4)\nonumber
\end{align}
and
\begin{align}
&\int\mathcal{V}_3^+(g_0,g_2,g_3,g_4)\mathcal{V}_3^-(g_1,g_2,g_3,g_4)\\
= & \int
\mathcal{V}_3^+(g_0,g_1,g_2,g_3)\mathcal{V}_3^-(g_0,g_1,g_2,g_4)
)\mathcal{V}_3^+(g_0,g_1,g_3,g_4)\nonumber
\end{align}
\begin{align}
&\int \mathcal{V}_3^-(g_0,g_1,g_2,g_3)\mathcal{V}_3^+(g_0,g_1,g_2,g_4)\\
= & \int
\mathcal{V}_3^-(g_0,g_2,g_3,g_4)\mathcal{V}_3^+(g_0,g_1,g_3,g_4)\mathcal{V}_3^+(g_1,g_2,g_3,g_4)\nonumber
\end{align}

\begin{figure}[tb]
\begin{center}
\includegraphics[scale=0.3]{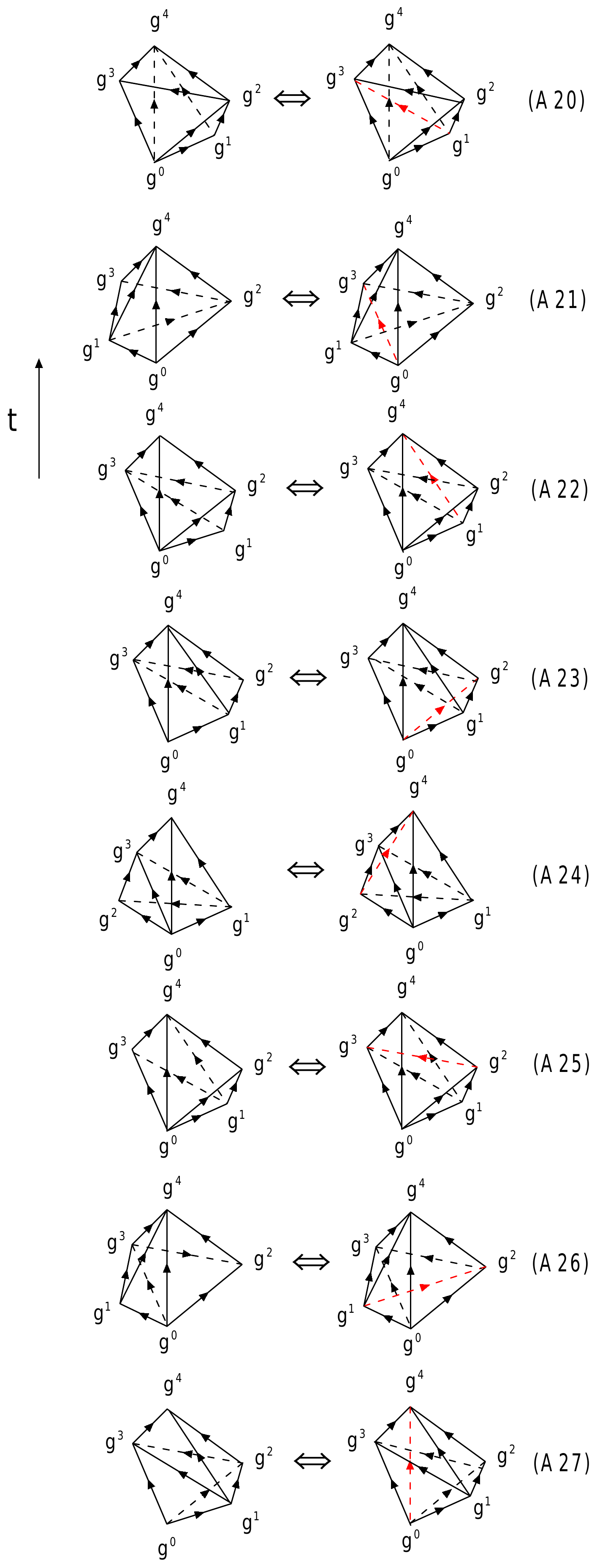}
\end{center}
\caption{(Color online) The admissible branching $2 \leftrightarrow
3$ moves that can be induced by a global time ordering. }
\label{2DT23}
\end{figure}

\begin{figure}[tb]
\begin{center}
\includegraphics[scale=0.3]{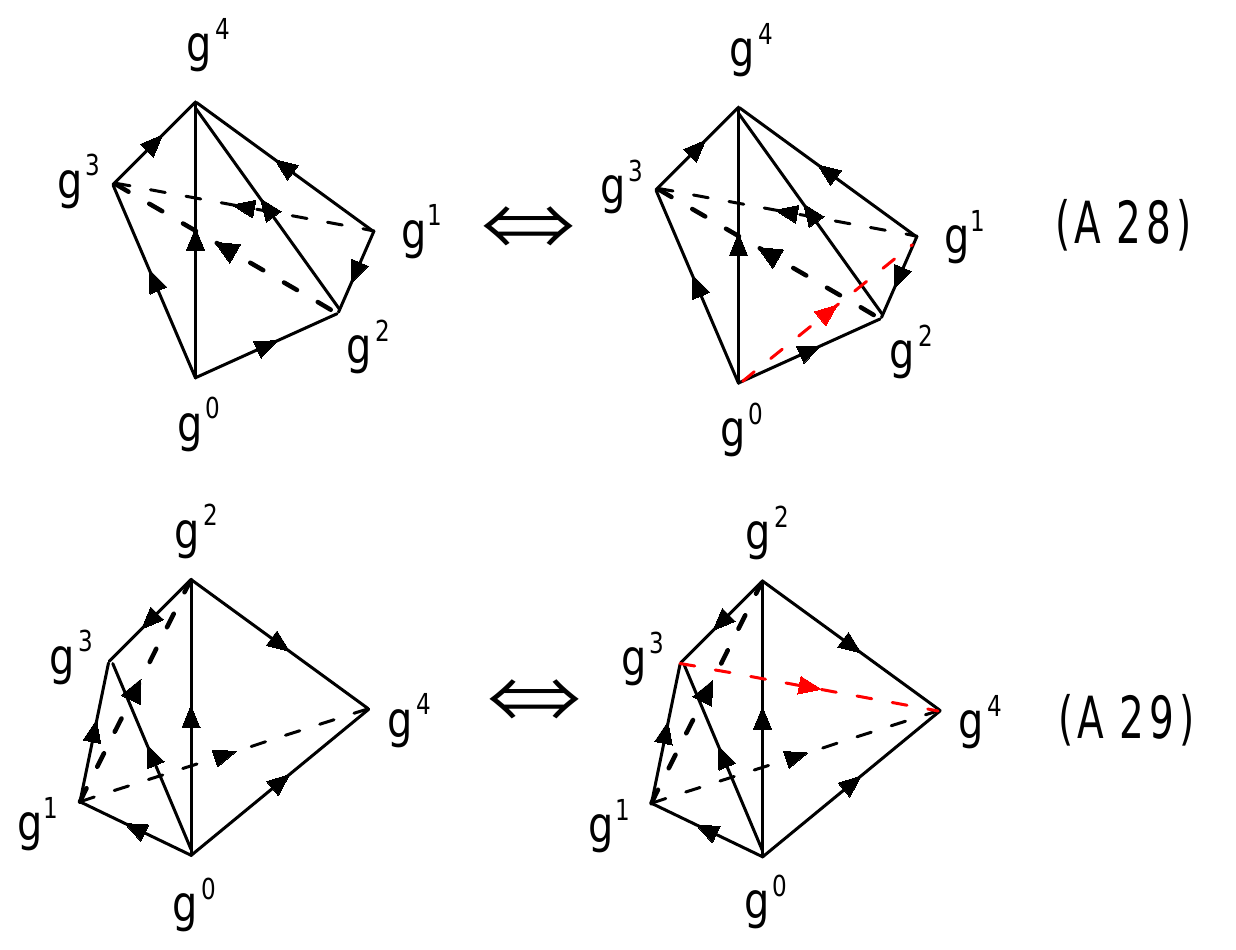}
\end{center}
\caption{(Color online) The admissible branching $2 \leftrightarrow
3$ moves that can not be induced by a global time ordering.}
\label{2DnonT23}
\end{figure}

\begin{figure}[tb]
\begin{center}
\includegraphics[scale=0.26]{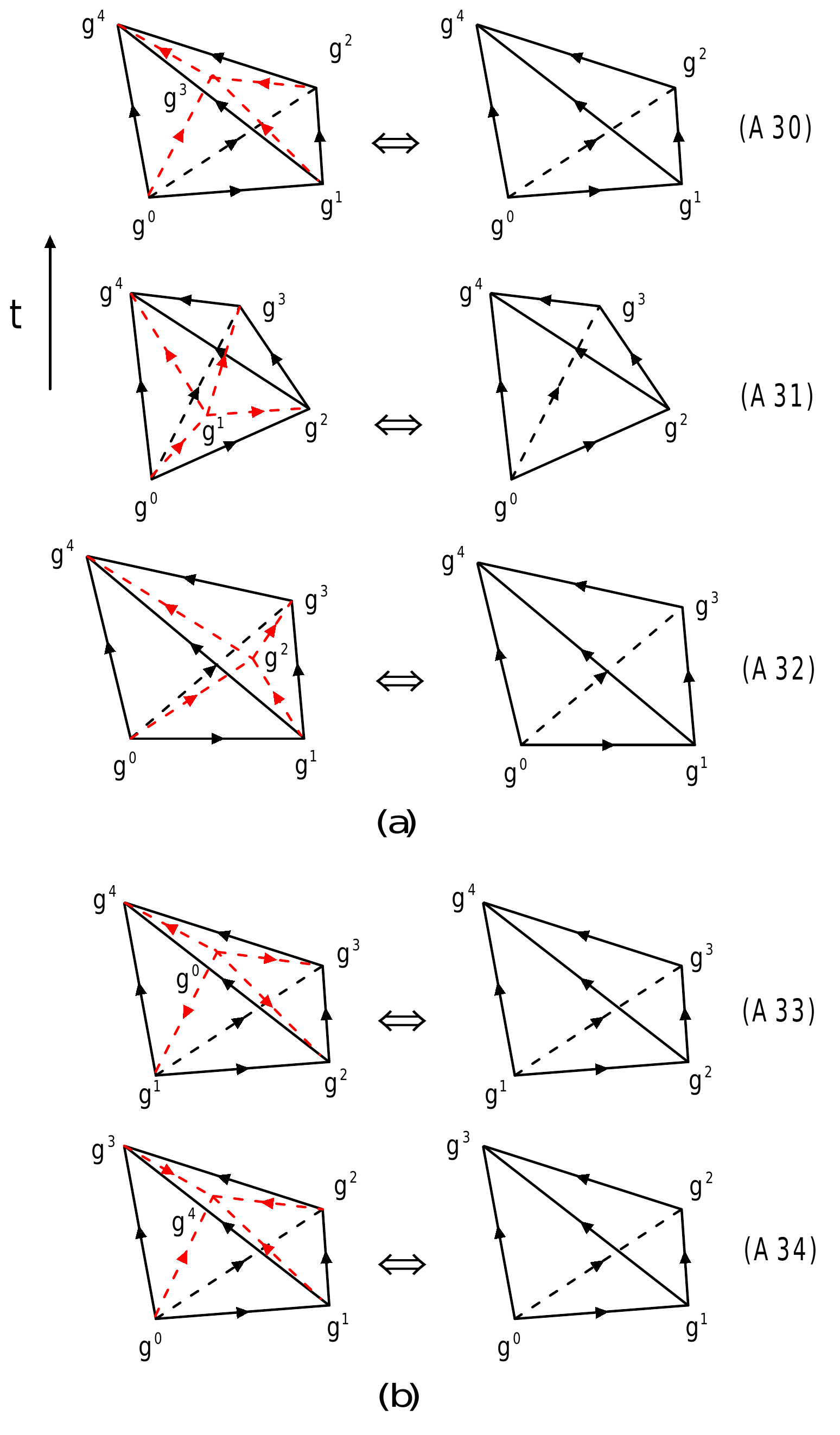}
\end{center}
\caption{(Color online) The admissible branching $1 \leftrightarrow
4$ moves that (a) can be induced by a global time ordering and (b)
can not be induced by a global time ordering } \label{2DT14}
\end{figure}

Similarly, we can formally write all the admissible branching
$1 \leftrightarrow 4$ moves in terms of $\mathcal{V}_3^{\pm}$(up to
the orientation conjugate $+\rightarrow -;-\rightarrow +$):
\begin{align}
\mathcal{V}_3^+ & (g_0,g_1,g_2,g_4) =  \int
\mathcal{V}_3^+(g_0,g_1,g_2,g_3)\mathcal{V}_3^+(g_0,g_1,g_3,g_4)
\nonumber\\ &\times
\mathcal{V}_3^-(g_0,g_2,g_3,g_4)\mathcal{V}_3^+(g_1,g_2,g_3,g_4)
\end{align}
\begin{align}
\mathcal{V}_3^+&(g_0,g_2,g_3,g_4) =  \int
\mathcal{V}_3^+(g_0,g_1,g_2,g_3)\mathcal{V}_3^-(g_0,g_1,g_2,g_4)
\nonumber\\ &\times
\mathcal{V}_3^+(g_0,g_1,g_3,g_4)\mathcal{V}_3^+(g_1,g_2,g_3,g_4)
\end{align}
\begin{align}
\mathcal{V}_3^+&(g_0,g_1,g_3,g_4) =
\int\mathcal{V}_3^-(g_0,g_1,g_2,g_3)\mathcal{V}_3^+(g_0,g_1,g_2,g_4)
\nonumber\\ &\times
\mathcal{V}_3^+(g_0,g_2,g_3,g_4)\mathcal{V}_3^-(g_1,g_2,g_3,g_4)
\end{align}
and
\begin{align}
\mathcal{V}_3^+ &(g_1,g_2,g_3,g_4)=
\int\mathcal{V}_3^-(g_0,g_1,g_2,g_3)\mathcal{V}_3^+(g_0,g_1,g_2,g_4)
\nonumber\\ &\times
\mathcal{V}_3^-(g_0,g_1,g_3,g_4)\mathcal{V}_3^+(g_0,g_2,g_3,g_4)
\end{align}
\begin{align}
\mathcal{V}_3^+&(g_0,g_1,g_2,g_3) =
 \int\mathcal{V}_3^+(g_0,g_1,g_2,g_4)\mathcal{V}_3^-(g_0,g_1,g_3,g_4) \nonumber\\ &\times
\mathcal{V}_3^+(g_0,g_2,g_3,g_4)\mathcal{V}_3^-(g_1,g_2,g_3,g_4)
\end{align}
Here the symbol $\int$ means that we put the sign factors
$(-)^{m_1(g_i,g_j)}$ on all four interior links and integrate over
the Grassmann variables on all six interior faces. Note that the
first 8 $2 \leftrightarrow 3$ moves (Fig. \ref{2DT23}) and first 3 $1
\leftrightarrow 4$ can be induced by a global time ordering while
the rest can not. Again here $g^0$ is defined on the vertex with no
incoming edge, $g^1$ with one incoming edge, etc.

Expressing $\mathcal{V}_3^{\pm}$ in terms of $\nu_3^{\pm}$ and
integrating out all the Grassmann variables, the admissible branching
$2 \leftrightarrow 3$ moves lead to the following equations:
\begin{align}
& {\nu_3^-(g_0,g_1,g_2,g_4)} {\nu_3^-(g_0,g_2,g_3,g_4
)}\\=& (-)^{n_2(g_0,g_1,g_2)n_2(g_2,g_3,g_4)+m_1(g_1,g_3)} \nonumber\\
\times &
{\nu_3^-(g_0,g_1,g_2,g_3)}{\nu_3^-(g_0,g_1,g_3,g_4 )}{\nu_3^-(g_1,g_2,g_3,g_4)}
\nonumber
\end{align}
\begin{align}
&\nu_3^-(g_0,g_1,g_2,g_4)\nu_3^+(g_1,g_2,g_3,g_4)\\ = &
(-)^{n_2(g_0,g_1,g_2)n_2(g_2,g_3,g_4)+m_1(g_0,g_3)}\nonumber \\
\times &\nu_3^-(g_0,g_1,g_2,g_3)\nu_3^-(g_0,g_1,g_3,g_4)\nu_3^+(g_0,g_2,g_3,g_4)
\nonumber
\end{align}
\begin{align}
&\nu_3^-(g_0,g_1,g_2,g_3)\nu_3^+(g_0,g_2,g_3,g_4)\\ = &
(-)^{n_2(g_0,g_1,g_2)n_2(g_2,g_3,g_4)+m_1(g_1,g_4)}\nonumber \\
\times & \nu_3^-(g_0,g_1,g_2,g_4)\nu_3^+(g_0,g_1,g_3,g_4)\nu_3^+(g_1,g_2,g_3,g_4)
\nonumber
\end{align}
\begin{align}
&\nu_3^+(g_0,g_1,g_3,g_4)\nu_3^+(g_1,g_2,g_3,g_4)\\ = &
(-)^{n_2(g_0,g_1,g_2)n_2(g_2,g_3,g_4)+m_1(g_0,g_2)}\nonumber \\
\times &\nu_3^-(g_0,g_1,g_2,g_3)\nu_3^+(g_0,g_1,g_2,g_4)\nu_3^+(g_0,g_2,g_3,g_4)
\nonumber
\end{align}
\begin{align}
&\nu_3^+(g_0,g_1,g_2,g_3)\nu_3^+(g_0,g_1,g_3,g_4)\\ = &
(-)^{n_2(g_0,g_1,g_2)n_2(g_2,g_3,g_4)+m_1(g_2,g_4)}\nonumber\\
\times & \nu_3^+(g_0,g_1,g_2,g_4)\nu_3^+(g_0,g_2,g_3,g_4)
\nu_3^-(g_1,g_2,g_3,g_4)
\nonumber
\end{align}
\begin{align}
&\nu_3^-(g_0,g_1,g_2,g_4)\nu_3^+(g_0,g_1,g_3,g_4)\\ = &
(-)^{n_2(g_0,g_1,g_2)n_2(g_2,g_3,g_4)
+n_2(g_1,g_2,g_3)+n_2(g_1,g_2,g_4)+m_1(g_2,g_3)}\nonumber\\&
\times\nu_3^-(g_0,g_1,g_2,g_3)\nu_3^+(g_0,g_2,g_3,g_4)\nu_3^-(g_1,g_2,g_3,g_4)
\nonumber
\end{align}
\begin{align}
&\nu_3^-(g_0,g_1,g_3,g_4)\nu_3^+(g_0,g_2,g_3,g_4)\\
=&(-)^{n_2(g_0,g_1,g_2)n_2(g_2,g_3,g_4)+n_2(g_0,g_1,g_2)+n_2(g_0,g_1,g_3)+m_1(g_1,g_2)}\nonumber\\&
\times\nu_3^+(g_0,g_1,g_2,g_3)\nu_3^-(g_0,g_1,g_2,g_4)\nu_3^+(g_1,g_2,g_3,g_4)
\nonumber
\end{align}
\begin{align}
&\nu_3^+(g_0,g_1,g_2,g_3)\nu_3^+(g_1,g_2,g_3,g_4)\\ = &
(-)^{n_2(g_0,g_1,g_2)n_2(g_2,g_3,g_4)+n_2(g_0,g_3,g_4)+n_2(g_1,g_3,g_4)+m_1(g_0,g_4)}\nonumber\\
&\times\nu_3^+(g_0,g_1,g_2,g_4)
\nu_3^+(g_0,g_2,g_3,g_4)\nu_3^-(g_0,g_1,g_3,g_4)
\nonumber
\end{align}
and
\begin{align}
&\nu_3^+(g_0,g_2,g_3,g_4)\nu_3^-(g_1,g_2,g_3,g_4)\\ = &
(-)^{n_2(g_0,g_1,g_2)n_2(g_2,g_3,g_4)+n_2(g_0,g_1,g_2)+n_2(g_1,g_2,g_4)+m_1(g_0,g_1)}\nonumber\\
&\times\nu_3^+(g_0,g_1,g_2,g_3)\nu_3^-(g_0,g_1,g_2,g_4)
)\nu_3^+(g_0,g_1,g_3,g_4)
\nonumber
\end{align}
\begin{align}
&\nu_3^-(g_0,g_1,g_2,g_3)\nu_3^+(g_0,g_1,g_2,g_4)\\
= &(-)^{n_2(g_0,g_1,g_2)n_2(g_2,g_3,g_4)+n_2(g_0,g_2,g_4)+n_2(g_0,g_3,g_4)+m_1(g_3,g_4)}\nonumber\\
&\times
\nu_3^-(g_0,g_2,g_3,g_4)\nu_3^+(g_0,g_1,g_3,g_4)\nu_3^+(g_1,g_2,g_3,g_4)
\nonumber
\end{align}
We note that the sign factor $(-)^{m_1(g_i,g_j)}$ comes from the
definition of the fermionic path integral. Later we will see such a
factor is very important to make all the branching moves to be self
consistent.

Similarly, all the admissible branching $1 \leftrightarrow 4$ moves
lead to the following equations for $\nu_3^{\pm}$:
\begin{align}
&\nu_3^+(g_0,g_1,g_2,g_4)\\ = &
(-)^{n_2(g_0,g_1,g_2)n_2(g_2,g_3,g_4)+n_2(g_2,g_3,g_4)}\nonumber\\
\times&
(-)^{m_1(g_0,g_3)+m_1(g_1,g_3)+m_1(g_2,g_3)+m_1(g_3,g_4)}\nu_3^+(g_0,g_1,g_2,g_3)
\nonumber\\ \times &
\nu_3^+(g_0,g_1,g_3,g_4)\nu_3^-(g_0,g_2,g_3,g_4)\nu_3^+(g_1,g_2,g_3,g_4)
\nonumber
\end{align}
\begin{align}
&\nu_3^+(g_0,g_2,g_3,g_4)\\ = &
(-)^{n_2(g_0,g_1,g_2)n_2(g_2,g_3,g_4)+n_2(g_0,g_1,g_2)}\nonumber\\
\times&
(-)^{m_1(g_0,g_1)+m_1(g_1,g_2)+m_1(g_1,g_3)+m_1(g_1,g_4)}\nu_3^+(g_0,g_1,g_2,g_3)
\nonumber\\ \times
&\nu_3^-(g_0,g_1,g_2,g_4)\nu_3^+(g_0,g_1,g_3,g_4)\nu_3^+(g_1,g_2,g_3,g_4)
\nonumber
\end{align}
\begin{align}
&\nu_3^+(g_0,g_1,g_3,g_4)\\ = &
(-)^{n_2(g_0,g_1,g_2)n_2(g_2,g_3,g_4)+n_2(g_1,g_2,g_3)}\nonumber\\
\times&
(-)^{m_1(g_0,g_2)+m_1(g_1,g_2)+m_1(g_2,g_3)+m_1(g_2,g_4)}\nu_3^-(g_0,g_1,g_2,g_3)
\nonumber\\ \times
&\nu_3^+(g_0,g_1,g_2,g_4)\nu_3^+(g_0,g_2,g_3,g_4)\nu_3^-(g_1,g_2,g_3,g_4)
\nonumber
\end{align}
and
\begin{align}
&\nu_3^+(g_1,g_2,g_3,g_4)\\ = &
(-)^{n_2(g_0,g_1,g_2)n_2(g_2,g_3,g_4)+n_2(g_0,g_1,g_4)}\nonumber\\
\times&
(-)^{m_1(g_0,g_1)+m_1(g_0,g_2)+m_1(g_0,g_3)+m_1(g_0,g_4)}\nu_3^-(g_0,g_1,g_2,g_3)
\nonumber\\ \times
&\nu_3^+(g_0,g_1,g_2,g_4)\nu_3^-(g_0,g_1,g_3,g_4)\nu_3^+(g_0,g_2,g_3,g_4)
\nonumber
\end{align}
\begin{align}
&\nu_3^+(g_0,g_1,g_2,g_3)\\ = &
(-)^{n_2(g_0,g_1,g_2)n_2(g_2,g_3,g_4)+n_2(g_0,g_3,g_4)}\nonumber\\
\times&
(-)^{m_1(g_0,g_4)+m_1(g_1,g_4)+m_1(g_2,g_4)+m_1(g_3,g_4)}\nu_3^+(g_0,g_1,g_2,g_4)
\nonumber\\ \times
&\nu_3^-(g_0,g_1,g_3,g_4)\nu_3^+(g_0,g_2,g_3,g_4)\nu_3^-(g_1,g_2,g_3,g_4)
\nonumber
\end{align}
Amazingly, if we define:
\begin{align}
 \nu_3^+ (g_0,g_1,g_2,g_3) &= (-)^{m_1(g_0,g_2)}\nu_3 (g_0,g_1,g_2,g_3),
\\
 \nu_3^- (g_0,g_1,g_2,g_3) &=
(-)^{m_1(g_1,g_3)} /\nu_3 (g_0,g_1,g_2,g_3)
\nonumber
\end{align}
we find all the above equations are equivalent to the following
single equation, which is the fermionic 3 cocycle condition of
$\nu_3$.
\begin{align}
& \nu_3 (g_1,g_2,g_3,g_4) \nu_3(g_0,g_1,g_3,g_4)
\nu_3(g_0,g_1,g_2,g_3)=
\nonumber\\
&(-)^{ n_2(g_0,g_1,g_2) n_2(g_2,g_3,g_4)} \nu_3 (g_0,g_2,g_3,g_4)
\nu_3 (g_0,g_1,g_2,g_4)
\end{align}

\subsection{$3+1$D}

The admissible branching moves for $3+1$D case is much more
complicated. There are in total 10 $3 \leftrightarrow 3$ moves, 15
$2 \leftrightarrow 4$ moves and 6 $1 \leftrightarrow 5$ moves.

For example, Fig. \ref{4to2} represents one admissible branching $2
\leftrightarrow 4$ move, which leads to the following equation for
$\mathcal{V}_4^{\pm}$:
\begin{widetext}
\begin{align}
&\int \dd \th_{(1234)}^{n(g_1,g_2,g_3,g_4)}\dd
\b\th_{(1234)}^{n(g_1,g_2,g_3,g_4)}{\mathcal{V}_4^+(g_1,g_2,g_3,g_4,g_5)}{\mathcal{V}_4^-(g_0,g_1,g_2,g_3,g_4)}
\nonumber\\=& \int \dd\th_{(0125)}^{n(g_0,g_1,g_2,g_5)}\dd
\b\th_{(0125)}^{n(g_0,g_1,g_2,g_5)}\dd\th_{(0135)}^{n(g_0,g_1,g_3,g_5)}\dd
\b\th_{(0135)}^{n(g_0,g_1,g_3,g_5)} \dd
\th_{(0145)}^{n(g_0,g_1,g_4,g_5)}\dd
\b\th_{(0145)}^{n(g_0,g_1,g_4,g_5)}\dd\th_{(0235)}^{n(g_0,g_2,g_3,g_5)}\dd
\b\th_{(0235)}^{n(g_0,g_2,g_3,g_5)}\nonumber\\
& \times \dd\th_{(0245)}^{n(g_0,g_2,g_4,g_5)}\dd
\b\th_{(0245)}^{n(g_0,g_2,g_4,g_5)}\dd\th_{(0345)}^{n(g_0,g_3,g_4,g_5)}\dd
\b\th_{(0345)}^{n(g_0,g_3,g_4,g_5)}(-)^{m_2(g_0,g_1,g_5)+m_2(g_0,g_2,g_5)+m_2(g_0,g_3,g_5)+m_2(g_0,g_4,g_5)}\nonumber\\
& \times
{\mathcal{V}_4^+(g_0,g_2,g_3,g_4,g_5)}{\mathcal{V}_4^+(g_0,g_1,g_2,g_4,g_5)}
{\mathcal{V}_4^-(g_0,g_1,g_3,g_4,g_5)}{\mathcal{V}_4^-(g_0,g_1,g_2,g_3,g_5)}
\end{align}
Here we integrate out the Grassmann variables on the interior
tetrahedra and put the sign factor on the interior surfaces. We can
formally denote the above equation as (see \eqn{Gint}):
\begin{align}
&\int{\mathcal{V}_4^+(g_1,g_2,g_3,g_4,g_5)}{\mathcal{V}_4^-(g_0,g_1,g_2,g_3,g_4)}
\nonumber\\
=& \int
{\mathcal{V}_4^+(g_0,g_2,g_3,g_4,g_5)}{\mathcal{V}_4^+(g_0,g_1,g_2,g_4,g_5)}
{\mathcal{V}_4^-(g_0,g_1,g_3,g_4,g_5)}{\mathcal{V}_4^-(g_0,g_1,g_2,g_3,g_5)}
\end{align}
Similarly, Fig. \ref{5to1} represents one admissible branching $1
\leftrightarrow 5$ move and can be formally written down in terms of
$\mathcal{V}_4^{\pm}$ as:
\begin{align}
&{\mathcal{V}_4^-(g_0,g_1,g_2,g_3,g_4)}\nonumber\\ =& \int
{\mathcal{V}_4^+(g_0,g_2,g_3,g_4,g_5)}
{\mathcal{V}_4^+(g_0,g_1,g_2,g_4,g_5)}
{\mathcal{V}_4^-(g_0,g_1,g_3,g_4,g_5)}
{\mathcal{V}_4^-(g_0,g_1,g_2,g_3,g_5)}
{\mathcal{V}_4^-(g_1,g_2,g_3,g_4,g_5)}
\end{align}
We note that here the symbol $\int$ means integrating over
all Grassmann variables on 10 interior tetrahedra and put the sign
factor $(-)^{m_2(g_i,g_j,g_k)}$ on 10 interior surfaces.

In addition, there are $3 \leftrightarrow 3$ moves. For example, one of
such moves gives rise the following formal equation for
$\mathcal{V}_4^{\pm}$:
\begin{align}
&\int{\mathcal{V}_4^+(g_1,g_2,g_3,g_4,g_5)}{\mathcal{V}_4^+(g_0,g_1,g_3,g_4,g_5)}{\mathcal{V}_4^+(g_0,g_1,g_2,g_3,g_5)}\nonumber\\=&
\int
{\mathcal{V}_4^+(g_0,g_1,g_2,g_3,g_4)}{\mathcal{V}_4^+(g_0,g_2,g_3,g_4,g_5)}{\mathcal{V}_4^+(g_0,g_1,g_2,g_4,g_5)}\nonumber\\
\end{align}
In this case, the symbol $\int$ on both sides mean
integrating over all Grassmann variables on 3 interior tetrahedra
and put the sign factor $(-)^{m_2(g_i,g_j,g_k)}$ on 1 interior
surface.

After integrating over all the Grassmann variables, we can express the
above three equations in terms of $\nu_4^{\pm}$:
\begin{align}
&\nu_4^+ (g_1,g_2,g_3,g_4,g_5)\nu_4^- (g_0,g_1,g_2,g_3,g_4)
\nonumber\\
=&(-)^{ n_3(g_0,g_1,g_2,g_3) n_3(g_0,g_3,g_4,g_5) +
n_3(g_1,g_2,g_3,g_4) n_3(g_0,g_1,g_4,g_5) + n_3(g_2,g_3,g_4,g_5)
n_3(g_0,g_1,g_2,g_5)+n_3(g_0,g_1,g_4,g_5)
}\nonumber\\
&\ \ \ \times
(-)^{m_2(g_0,g_1,g_5)+m_2(g_0,g_2,g_5)+m_2(g_0,g_3,g_5)+m_2(g_0,g_4,g_5)}\nonumber\\
&\ \ \ \times \nu_4^+ (g_0,g_2,g_3,g_4,g_5) \nu_4^+ (g_0,g_1,g_2,g_4,g_5)
\nu_4^-(g_0,g_1,g_3,g_4,g_5) \nu_4^-(g_0,g_1,g_2,g_3,g_5),
\\
&\nu_4^- (g_0,g_1,g_2,g_3,g_4)
\nonumber\\
=&(-)^{ n_3(g_0,g_1,g_2,g_3) n_3(g_0,g_3,g_4,g_5) +
n_3(g_1,g_2,g_3,g_4) n_3(g_0,g_1,g_4,g_5) + n_3(g_2,g_3,g_4,g_5)
n_3(g_0,g_1,g_2,g_5)+n_3(g_0,g_1,g_4,g_5)+n_3(g_1,g_2,g_4,g_5)+n_3(g_2,g_3,g_4,g_5)
}\nonumber\\
\times
&(-)^{m_2(g_0,g_1,g_5)+m_2(g_0,g_2,g_5)+m_2(g_0,g_3,g_5)+m_2(g_0,g_4,g_5)
+m_2(g_1,g_2,g_5)+m_2(g_1,g_3,g_5)+m_2(g_1,g_4,g_5)+m_2(g_2,g_3,g_5)+m_2(g_2,g_4,g_5)+m_2(g_3,g_4,g_5)}\nonumber\\
&\ \ \ \times \nu_4^+ (g_0,g_2,g_3,g_4,g_5) \nu_4^+ (g_0,g_1,g_2,g_4,g_5)
\nu_4^-(g_0,g_1,g_3,g_4,g_5) \nu_4^-(g_0,g_1,g_2,g_3,g_5)\nu_4^-
(g_1,g_2,g_3,g_4,g_5) ,
\\
&\nu_4^+ (g_1,g_2,g_3,g_4,g_5)\nu_4^+(g_0,g_1,g_3,g_4,g_5)
\nu_4^+(g_0,g_1,g_2,g_3,g_5)(-)^{m_2(g_1,g_3,g_5)}
\nonumber\\
=&(-)^{ n_3(g_0,g_1,g_2,g_3) n_3(g_0,g_3,g_4,g_5) +
n_3(g_1,g_2,g_3,g_4) n_3(g_0,g_1,g_4,g_5)+n_3(g_2,g_3,g_4,g_5)
n_3(g_0,g_1,g_2,g_5)+m_2(g_0,g_2,g_4) }\nonumber\\
&\ \ \ \times \nu_4^+
(g_0,g_1,g_2,g_3,g_4)\nu_4^+ (g_0,g_2,g_3,g_4,g_5) \nu_4^+
(g_0,g_1,g_2,g_4,g_5) .
\end{align}

Surprisingly, if we define $\nu_4^{\pm}$ as:
\begin{align}
\nu_4^+ (g_0,g_1,g_2,g_3,g_4) =
& (-)^{ m_2(g_0,g_1,g_3)+ m_2(g_1,g_3,g_4)+ m_2(g_1,g_2,g_3) }\nu_4
(g_0,g_1,g_2,g_3,g_4),
\nonumber\\
 \nu_4^- (g_0,g_1,g_2,g_3,g_4) = & (-)^{m_2(g_0,g_2,g_4)} /\nu_4
(g_0,g_1,g_2,g_3,g_4),
\end{align}
we find all the above three equations will be equivalent to the
following single equation of $\nu_4$, which is the fermionic 4
cocycle condition:
\begin{align}
&\ \ \ \ \nu_4 (g_1,g_2,g_3,g_4,g_5) \nu_4(g_0,g_1,g_3,g_4,g_5)
\nu_4(g_0,g_1,g_2,g_3,g_5)
\\
&=(-)^{ n_3(g_0,g_1,g_2,g_3) n_3(g_0,g_3,g_4,g_5) +
n_3(g_1,g_2,g_3,g_4) n_3(g_0,g_1,g_4,g_5) + n_3(g_2,g_3,g_4,g_5)
n_3(g_0,g_1,g_2,g_5) }\times
\nonumber\\
&\ \ \ \ \nu_4 (g_0,g_2,g_3,g_4,g_5) \nu_4 (g_0,g_1,g_2,g_4,g_5)
\nu_4 (g_0,g_1,g_2,g_3,g_4) .
\nonumber
\end{align}
Indeed, it can be verified by computer that all other admissible  $3
\leftrightarrow 3$, $2 \leftrightarrow 4$ and $1 \leftrightarrow 5$
branching moves will give rise to the same fermionic 4 cocycle
condition! Again, the phase factor $(-)^{m_2(g_i,g_j,g_k)}$ in the
fermionic path integral is crucial for the self consistency of all
admissible branching moves.

\section{The fixed-point action on a closed complex and the unitary
condition}
\label{close}
In the following we will show that $\int \cV_d^+
\cV_d^-$ is always equal to 1.

In $1+1$D, we have:
\begin{align}
&\int \cV_2^+(g_0,g_1,g_2) \cV_2^-(g_0,g_1,g_2) \nonumber\\=&\int \dd
\th_{(01)}^{n_1(g_0,g_1)}\dd \b \th_{(01)}^{n_1(g_0,g_1)} \dd
\th_{(02)}^{n_1(g_0,g_2)}\dd \b \th_{(02)}^{n_1(g_0,g_2)}\dd
\th_{(12)}^{n_1(g_1,g_2)}\dd \b \th_{(12)}^{n_1(g_1,g_2)}(-)^{m_0(g_0)+m_0(g_1)+m_0(g_2)}\nonumber\\
&\times
\nu_2^+(g_0,g_1,g_2)\th_{(12)}^{n_1(g_1,g_2)}\th_{(01)}^{n_1(g_0,g_1)}\b\th_{(02)}^{n_1(g_0,g_2)}
\nu_2^-(g_0,g_1,g_2)\th_{(02)}^{n_1(g_0,g_2)}\b\th_{(01)}^{n_1(g_0,g_1)}\b\th_{(12)}^{n_1(g_1,g_2)}\nonumber\\
=&\int  \dd \th_{(01)}^{n_1(g_0,g_1)}\dd \b
\th_{(01)}^{n_1(g_0,g_1)} \dd \th_{(02)}^{n_1(g_0,g_2)}\dd \b
\th_{(02)}^{n_1(g_0,g_2)}\dd
\th_{(12)}^{n_1(g_1,g_2)}\dd \b \th_{(12)}^{n_1(g_1,g_2)}(-)^{n_1(g_0,g_2)}\nonumber\\
&\times
\th_{(12)}^{n_1(g_1,g_2)}\th_{(01)}^{n_1(g_0,g_1)}\b\th_{(02)}^{n_1(g_0,g_2)}
\th_{(02)}^{n_1(g_0,g_2)}\b\th_{(01)}^{n_1(g_0,g_1)}\b\th_{(12)}^{n_1(g_1,g_2)}\nonumber\\
=&1
\end{align}
In the above calculation, we can move the pair $\dd
\th_{(01)}^{n_1(g_0,g_1)}\dd \b \th_{(01)}^{n_1(g_0,g_1)}$ to the front of
$\b\th_{(02)}^{n_1(g_0,g_2)} \th_{(02)}^{n_1(g_0,g_2)}$ without generating any
signs.  We then can evaluate $\int \dd \th_{(02)}^{n_1(g_0,g_1)}\dd \b
\th_{(02)}^{n_1(g_0,g_1)} \b\th_{(02)}^{n_1(g_0,g_2)} \th_{(02)}^{n_1(g_0,g_2)}
=1$.  We next move $\dd \th_{(01)}^{n_1(g_0,g_1)}\dd \b
\th_{(01)}^{n_1(g_0,g_1)}$ to the front of $ \th_{(01)}^{n_1(g_0,g_1)} \b
\th_{(01)}^{n_1(g_0,g_1)}$ without generating any signs.  We then evaluate
$\int \dd \th_{(01)}^{n_1(g_0,g_1)}\dd \b \th_{(01)}^{n_1(g_0,g_1)}
\th_{(01)}^{n_1(g_0,g_1)} \b \th_{(01)}^{n_1(g_0,g_1)} =(-)^{n_1(g_0,g_1)}$,
which has a sign factor.
The Grassmann integral
$\int \dd \th_{(12)}^{n_1(g_1,g_2)}\dd \b \th_{(12)}^{n_1(g_1,g_2)}
 \th_{(12)}^{n_1(g_1,g_2)} \b \th_{(12)}^{n_1(g_1,g_2)}=
(-)^{n_1(g_1,g_2)}$
also generates a sign factor.
Such two sign factors
$(-)^{n_1(g_0,g_1)} (-)^{n_1(g_1,g_2)}$
cancel the sign factor
$(-)^{n_1(g_0,g_2)}$
due to the condition \eqn{ncond} on $n_d$.

Similarly, in $(2+1)$D and in $(3+1)$D, we find that
the sign factors in the integration measure \eq{Gint} and in the $\nu_d^\pm$
expressions \eq{nu4nupm} or \eq{nu4nupm},
when combined with those from exchanging the
Grassmann numbers, just cancel each other:
\begin{align}
&\int \cV_3^+(g_0,g_1,g_2,g_3) \cV_3^-(g_0,g_1,g_2,g_3)
\nonumber\\=&\int \dd \th_{(012)}^{n_2(g_0,g_1,g_2)}\dd \b
\th_{(012)}^{n_2(g_0,g_1,g_2)} \dd
\th_{(013)}^{n_2(g_0,g_1,g_3)}\dd
\b\th_{(013)}^{n_2(g_0,g_1,g_3)}\dd
\th_{(023)}^{n_2(g_0,g_2,g_3)}\dd \b
\th_{(023)}^{n_2(g_0,g_2,g_3)}\dd
\th_{(123)}^{n_2(g_1,g_2,g_3)}\dd
\b\th_{(123)}^{n_2(g_1,g_2,g_3)}\nonumber\\
&\times(-)^{m_1(g_0,g_1)+m_1(g_0,g_2)+m_1(g_0,g_3)+m_1(g_1,g_2)+m_1(g_1,g_3)+m_1(g_2,g_3)}\nonumber\\
&\times
\nu_3^+(g_0,g_1,g_2,g_3)\th_{(123)}^{n_2(g_1,g_2,g_3)}\th_{(013)}^{n_2(g_0,g_1,g_3)}\b\th_{(023)}^{n_2(g_0,g_2,g_3)}
\b\th_{(012)}^{n_2(g_0,g_1,g_2)}\nonumber\\ &\times
\nu_3^-(g_0,g_1,g_2,g_3)\th_{(012)}^{n_2(g_0,g_1,g_2)}\th_{(023)}^{n_2(g_0,g_2,g_3)}
\b\th_{(013)}^{n_2(g_0,g_1,g_3)}\b\th_{(123)}^{n_2(g_1,g_2,g_3)}\nonumber\\
=&\int \dd \th_{(012)}^{n_2(g_0,g_1,g_2)}\dd \b
\th_{(012)}^{n_2(g_0,g_1,g_2)} \dd
\th_{(013)}^{n_2(g_0,g_1,g_3)}\dd
\b\th_{(013)}^{n_2(g_0,g_1,g_3)}\dd
\th_{(023)}^{n_2(g_0,g_2,g_3)}\dd \b
\th_{(023)}^{n_2(g_0,g_2,g_3)}\dd
\th_{(123)}^{n_2(g_1,g_2,g_3)}\dd
\b\th_{(123)}^{n_2(g_1,g_2,g_3)}\nonumber\\
&\times
(-)^{n_2(g_0,g_1,g_3)+n_2(g_1,g_2,g_3)}\th_{(123)}^{n_2(g_1,g_2,g_3)}\th_{(013)}^{n_2(g_0,g_1,g_3)}\b\th_{(023)}^{n_2(g_0,g_2,g_3)}
\b\th_{(012)}^{n_2(g_0,g_1,g_2)}
\th_{(012)}^{n_2(g_0,g_1,g_2)}\th_{(023)}^{n_2(g_0,g_2,g_3)}
\b\th_{(013)}^{n_2(g_0,g_1,g_3)}\b\th_{(123)}^{n_2(g_1,g_2,g_3)}\nonumber\\
=& 1 ;
\end{align}
\begin{align}
&\int \cV_4^+(g_0,g_1,g_2,g_3,g_4) \cV_4^-(g_0,g_1,g_2,g_3,g_4)
\nonumber\\=&\int \dd \th_{(0123)}^{n_3(g_0,g_1,g_2,g_3)}\dd \b
\th_{(0123)}^{n_3(g_0,g_1,g_2,g_3)} \dd
\th_{(0124)}^{n_3(g_0,g_1,g_2,g_4)}\dd \b
\th_{(0124)}^{n_3(g_0,g_1,g_2,g_4)} \dd
\th_{(0134)}^{n_3(g_0,g_1,g_3,g_4)}\dd
\b\th_{(0134)}^{n_3(g_0,g_1,g_3,g_4)} \nonumber\\
&\times\dd \th_{(0234)}^{n_3(g_0,g_2,g_3,g_4)}\dd \b
\th_{(0234)}^{n_3(g_0,g_2,g_3,g_4)}\dd
\th_{(1234)}^{n_3(g_1,g_2,g_3,g_4)}\dd
\b\th_{(1234)}^{n_3(g_1,g_2,g_3,g_4)}(-)^{m_2(g_0,g_1,g_2)+m_2(g_0,g_1,g_3)}\nonumber\\
&\times  (-)^{m_2(g_0,g_1,g_4)+m_2(g_0,g_2,g_3)+m_2(g_0,g_2,g_4)+
m_2(g_0,g_3,g_4)+m_2(g_1,g_2,g_3)+m_2(g_1,g_2,g_4)+m_2(g_1,g_3,g_4)+m_2(g_2,g_3,g_4)}\nonumber\\
&\times
\nu_4^+(g_0,g_1,g_2,g_3,g_4)\th_{(1234)}^{n_3(g_1,g_2,g_3,g_4)}\th_{(0134)}^{n_3(g_0,g_1,g_3,g_4)}
\th_{(0123)}^{n_3(g_0,g_1,g_2,g_3)}\b\th_{(0234)}^{n_3(g_0,g_2,g_3,g_4)}
\b\th_{(0124)}^{n_3(g_0,g_1,g_2,g_4)}\nonumber\\ &\times
\nu_4^-(g_0,g_1,g_2,g_3,g_4)\th_{(0124)}^{n_3(g_0,g_1,g_2,g_4)}\th_{(0234)}^{n_3(g_0,g_2,g_3,g_4)}
\b\th_{(0123)}^{n_3(g_0,g_1,g_2,g_3)}\b\th_{(0134)}^{n_3(g_0,g_1,g_3,g_4)}\b\th_{(1234)}^{n_3(g_1,g_2,g_3,g_4)}\nonumber\\
=&\int \dd \th_{(0123)}^{n_3(g_0,g_1,g_2,g_3)}\dd \b
\th_{(0123)}^{n_3(g_0,g_1,g_2,g_3)} \dd
\th_{(0124)}^{n_3(g_0,g_1,g_2,g_4)}\dd \b
\th_{(0124)}^{n_3(g_0,g_1,g_2,g_4)} \dd
\th_{(0134)}^{n_3(g_0,g_1,g_3,g_4)}\dd
\b\th_{(0134)}^{n_3(g_0,g_1,g_3,g_4)} \nonumber\\
&\times\dd \th_{(0234)}^{n_3(g_0,g_2,g_3,g_4)}\dd \b
\th_{(0234)}^{n_3(g_0,g_2,g_3,g_4)}\dd
\th_{(1234)}^{n_3(g_1,g_2,g_3,g_4)}\dd
\b\th_{(1234)}^{n_3(g_1,g_2,g_3,g_4)}(-)^{n_3(g_0,g_1,g_2,g_4)+n_3(g_0,g_2,g_3,g_4)}\nonumber\\
&\times
\th_{(1234)}^{n_3(g_1,g_2,g_3,g_4)}\th_{(0134)}^{n_3(g_0,g_1,g_3,g_4)}
\th_{(0123)}^{n_3(g_0,g_1,g_2,g_3)}\b\th_{(0234)}^{n_3(g_0,g_2,g_3,g_4)}
\b\th_{(0124)}^{n_3(g_0,g_1,g_2,g_4)}\nonumber\\&\times\th_{(0124)}^{n_3(g_0,g_1,g_2,g_4)}\th_{(0234)}^{n_3(g_0,g_2,g_3,g_4)}
\b\th_{(0123)}^{n_3(g_0,g_1,g_2,g_3)}\b\th_{(0134)}^{n_3(g_0,g_1,g_3,g_4)}\b\th_{(1234)}^{n_3(g_1,g_2,g_3,g_4)}\nonumber\\
=& 1 .
\end{align}
\end{widetext}

We also note that the  bosonic topological nonlinear $\si$ models are
characterized by action-amplitudes that are pure $U(1)$ phases
$|\nu_d(g_0,...,g_d)|=1$ (where the action-amplitude on ``$+$'' oriented simplexes
is given by $\nu_d(g_0,...,g_d)$, and the action-amplitude on ``$-$'' oriented
simplexes is given by $\nu_d^{-1}(g_0,...,g_d)=\nu_d^*(g_0,...,g_d)$).  The
action-amplitude being a pure $U(1)$ phase ensures that the model defined by the
path integral to be unitary theory.

But what is the analog of the pure $U(1)$ phase condition on the Grassmann
amplitude $\cV^\pm_d$ (or on the  pair of functions $[\nu_d(g_0,...,g_d),
m_{d-2}(g_0,...,g_{d-2})]$)?  Clearly, the Grassmann amplitude $\cV^\pm_d$ is
not even a complex numbers.  It is hard to say when $\cV^\pm_d$ behaves like
$U(1)$ phases.

To address this issue, let us introduce the
complex conjugate of a quantity that contains Grassmann numbers:
\begin{align}
 \Big(
\nu \th_1\bar \th_2 \th_3...
\Big)^*
\equiv
\nu^*
...
\bar\th_3
 \th_2
\bar \th_1
.
\end{align}
We see that under complex conjugate a) the complex coefficients are complex
conjugated, b) the order of the Grassmann number is reversed, and c) $\th$ and
$\bar \th$ are exchanged.  With this definition, one is tempting to require
$\cV^+_d (\cV^+_d)^*=1$, in order for $\cV^+_d$ to be a $U(1)$ phase.  But,
$\cV^+_d (\cV^+_d)^*$ still contains Grassmann numbers and we cannot require it
to be 1.  So we try to require $\int \cV^+_d (\cV^+_d)^*=1$ where the Grassmann
integral is defined in \eqn{Gint}.  Now  $\int \cV^+_d (\cV^+_d)^*$ is a
complex number.  But $\int \cV^+_d (\cV^+_d)^*$ is not non-negative.  So we
cannot treat $\int \cV^+_d (\cV^+_d)^*$ as a norm-square of $ \cV^+_d$.  After
some considerations, we find that we need to define a different complex
conjugate:
\begin{align}
[ \cV^\pm_1(g_0,g_1)]^\dag &= [\cV^\pm_1(g_0,g_1)]^* ,
\nonumber\\
[ \cV^\pm_2(g_0,g_1,g_2)]^\dag &= (-)^{m_0(g_1)} [\cV^\pm_2(g_0,g_1,g_2)]^* ,
\\
[ \cV^\pm_3(g_0,..,g_3)]^\dag &= (-)^{m_1(g_0,g_2)+m_1(g_1,g_3)} [\cV^\pm_3(g_0,..,g_3)]^* ,
\nonumber\\
[ \cV^\pm_4(g_0,...,g_4)]^\dag &=
[\cV^\pm_4(g_0,...,g_4)]^* \times
\nonumber\\
&\hskip -0.5in
(-)^{
 m_1(g_0,g_1,g_3)
+m_1(g_1,g_3,g_4)
+m_1(g_1,g_2,g_3)
+m_1(g_0,g_2,g_4)
}
\nonumber
\end{align}
Using the new complex conjugate and the evaluation of $\int \cV_d^+ \cV_d^-$
described above, we find that
\begin{align}
 \int \cV_d^+ [\cV_d^+]^\dag
 =\int \cV_d^- [ \cV_d^-]^\dag =\nu_d \nu_d^*.
\end{align}
Thus
we would like to require the
Grassmann amplitude $\cV^\pm_d$ to satisfy
\begin{align}
\label{U1cnd}
\int \cV_d^\pm [\cV_d^\pm]^\dag =1.
\end{align}
which is to require $|\nu_d(g_0,...,g_d)|=1$.
Eqn. (\ref{U1cnd}) is the analog of the pure $U(1)$ phase condition on the
Grassmann amplitude $\cV^\pm_d$.

For the Grassmann amplitude
$\cV_d^\pm$ that satisfies
the pure $U(1)$ phase condition \eq{U1cnd},
we have
\begin{align}
[ \cV_d^\pm]^\dag =\cV_d^\mp.
\end{align}

\section{Group super-cohomology}
\label{fcoh}

In Section \ref{fixZ}, we studied the condition on $\nu_d(g_0,g_1,...)$ so that
the fermionic path integral constructed from $\nu_d(g_0,g_1,...)$ correspond to
a fermionic topological nonlinear $\si$ model which is a fixed point theory
under the RG flow. Those conditions on $\nu_d(g_0,g_1,...)$ actually define a
generalization of group cohomology.\cite{RS,CGL1172} We will call such a
generalization group super-cohomology. In this section, we will study the group
super-cohomology in detail.

A  $d$-cohomology class of group super-cohomology is a set that depend on a
full symmetry group $G_f$
and a fermionic $G$-module $M$.  We will denote the fermionic $d$-cohomology
class as $\fH^d(G_f,M)$.

We note that a fermion system always has a $Z_2$ symmetry which corresponds to
the conservation of fermion-number parity.  Such a symmetry group is denoted as
$Z_2^f$ which is generated by $P_f=(-)^{N_f}$ where $N_f$ is the fermion
number.  The full symmetry group $G_f$ always contain $Z_2^f$ as a subgroup.
We will define $G_b=G_f/Z_2^f$.

\subsection{Graded structure of a group $G_b$}

First, let us introduce the graded structure of a group $G_b$.  A $d$D-graded
structure of a group $G_b$ is an integer function $n_{d-1}(g_1,...,g_d)$ of $d$
variables, whose values are $0,1$.  Here we choose $g_i$ to be an element of
$G_b$, rather than an element of the full symmetry group $G_f$.  This is
because $g_i$ corresponds to the \emph{bosonic} fields in the path integral, and
the bosonic field is invariant under $Z_2^f$.

The function $n_{d-1}(g_1,...,g_d)$ satisfies
\begin{align}
\label{ncond1}
& n_{d-1}(gg_1,...,gg_d) = n_{d-1}(g_1,...,g_d),\ \ \ \forall g \in G_b,
\\
& \sum_{i=0}^d n_{d-1}(g_0,..,g_{i-1},g_{i+1},..,g_d) =\text{ even}
,\ \ \ \forall g_0,..,g_d \in G
.
\nonumber
\end{align}
We see that a $d$D-graded structure of a group $G_b$ is a $(d-1)$-cocycle
$n_{d-1} \in \cZ^{d-1}(G_b,\Z_2)$. If
a function $n_{d-1}$ only satisfies the first condition in
\eqn{ncond1}, then it will be called a $(d-1)$-cochain.
The space of all $(d-1)$-cochains is denoted as
$\cC^{d-1}(G_b,\Z_2)$.

The coboundary is given by
\begin{widetext}
\begin{align}
\cB^{d-1}(G_b,\Z_2)
=\Big\{ & n^\prime_{d-1}(g_1,...,g_d)=
\sum_i  m_{d-2}^\prime (g_1,...,g_{i-1},g_{i+1},...,g_d)\text{ mod } 2\Big|
\nonumber\\
&  m^\prime_{d-2}(gg_1,...,gg_{d-1}) =
 m^\prime_{d-2}(g_1,...,g_{d-1}) ,
\ \ \forall g,g_1,...,g_{d-1} \in G_b \Big\}.
\nonumber
\end{align}
\end{widetext}
We say that two graded structures differ by an above coboundary
are equivalent. Thus different classes of
$d$D-graded structures are given by
$ \cH^{d-1}(G_b,\Z_2)= \cZ^{d-1}(G_b,\Z_2)/ \cB^{d-1}(G_b,\Z_2) $.

\subsection{$G$-module}

For a group $G_f$, let $M$ be a $G_f$-module, which is an Abelian group (with
multiplication operation) on which $G_f$ acts compatibly with the multiplication
operation (\ie the Abelian group structure):
\begin{align}
\label{gm}
 g\cdot (ab)=(g\cdot a)(g\cdot b),\ \ \ \ g\in G_f,\ \ \ \ a,b\in M.
\end{align}

For the most cases studied in this paper, $M$ is simply the $U(1)$ group and
$a$ an $U(1)$ phase.  The multiplication operation $ab$ is the usual
multiplication of the $U(1)$ phases.  The group action is trivial: $g\cdot
a=a$, $g\in G_f$, $a\in U(1)$.  We will denote such a trivial $G_f$-module as
$M=U(1)$.

For a group $G_f$ that contain time-reversal operation, we can define a
nontrivial $G_f$-module which is denoted as $U_T(1)$.  $U_T(1)$ is also a
$U(1)$ group whose elements are the $U(1)$ phases.  The multiplication
operation $ab$, $a,b\in U_T(1)$, is still the usual multiplication of the
$U(1)$ phases.  However, the group action is nontrivial now: $g\cdot
a=a^{s(g)}$, $g\in G_f$, $a\in U_T(1)$. Here $s(g)=1$ if the number of
time-reversal operations in the group operation $g$ is even and $s(g)=-1$ if
the number of time-reversal operations in $g$ is odd.

\subsection{Fermionic $d$-cochain}
A fermionic $d$-cochain is described by a set of three functions
$\nu_d(g_0,...,g_d)$, $n_{d-1}(g_1,...,g_d)$, and $u^g_{d-1}(g_1,...,g_d)$.
$n_{d-1}(g_1,...,g_d)$ is $(d-1)$-cochain in $ \cC^{d-1}(G_b,\Z_2)$ which has
been discussed above.  $\nu_d(g_0,...,g_d)$ is a function of $1+d$ variables
whose value is in a $G$-module $M$, $\nu_d: G_b^{1+d} \to M$.  Again, note that
$g_i$ is an elements of $G_b$ rather than the full symmetry group $G_f$.  For
cases studied in this paper, $M$ is always a $U(1)$ group (\ie
$\nu_d(g_0,...,g_d)$ is a complex phase).

Since the action amplitude is invariant under the symmetry transformation in
$G_f$, $\nu_d(g_0,...,g_d)$ must satisfy certain conditions.  Note that a
transformation $g$ in $G_f$ will generate a transformation in $G_b$, $g G_b \to
G_b$, since $G_b=G_f/Z_2^f$.  So $\nu_d(g_0,...,g_d)$ transform as
$\nu_d(g_0,...,g_d) \to \nu_d^{s(g)}(gg_0,...,gg_d)$.

The invariance under the symmetry transformation in $G_f$ is discussed in
Sections \ref{symmstb} and \ref{2dwav}.  This motivates us to require that the
fermionic $d$-cochain $\nu_d(g_0,...,g_d)$ to satisfy the following symmetry
condition
\begin{align}
\label{gnu}
&\ \ \ \ \nu_d^{s(g)}(gg_0,...,gg_d)
\nonumber\\
 &=\nu_d(g_0,...,g_d)
\prod_i [u_{d-1}^{g}(g_0,..,\hat g_i,..,g_d)]^{(-)^i} ,
\\
&
n_{d-1}(gg_0,...,gg_{d-1})=n_{d-1}(g_0,...,g_{d-1}), \ \ \
g\in G_f ,
\nonumber
\end{align}
where the sequence $g_0,..,\hat g_i,..,g_d$
is the sequence $g_0,...,g_d$ with $g_i$ removed,
and $u_{d-1}^{g}(g_i,g_j,...,g_k)$ is an 1D representation of $G_f$
\begin{align}
\label{1Drep}
& u_{d-1}^{g_a}(g_i,...,g_k) [u_{d-1}^{g_b}(g_i,...,g_k)]^{s(g_a)}
=u_{d-1}^{g_ag_b}(g_i,...,g_k),
\nonumber\\
& u_{d-1}^{g_a}(gg_i,gg_j,...,gg_k)=[u_{d-1}^{g_a}(g_i,g_j,...,g_k)]^{s(g)},
\end{align}
such that
$u_{d-1}^{P_f}(g_i,g_j,...,g_k)=(-)^{n_{d-1}(g_i,g_j,...,g_k)}$.
The triple functions $(\nu_d, n_{d-1}, u_{d-1}^{g})$ that satisfy \eqn{gnu} and
\eqn{1Drep} are
called fermionic cochains.  We will denote the collection of all those
fermionic cochains $(\nu_d, n_{d-1}, u_{d-1}^{g})$ as $\fC^d[G_f,U_T(1)]$.

\subsection{Fermionic $d$-cocycle}
With the above setup, now we are ready to define fermionic $d$-cocycle.  The
fermionic $d$-cocycles are fermionic $d$-cochains $[\nu_d, n_{d-1},
u_{d-1}^{g}]$ in $\fC^d[G_f,U_T(1)]$ that satisfy some additional conditions.

First, we require $n_{d-1}(g_1,..,g_d)$ to be $(d-1)$-cocycles in
$\cZ^{d-1}(G_f,\Z_2)$.  Note that when $d-1$, we have $n_0(g_1)=0$ or
$n_0(g_1)=1$ for any $G_f$.  We also require $\nu_d(g_0,..,g_d)$ to satisfy
\begin{align}
\label{1ccy}
 \nu_1(g_1,g_2)
 \nu_1^{-1}(g_0,g_2)
 \nu_1(g_0,g_1) =1,
\end{align}
\begin{align}
\label{2ccy}
 \nu_2(g_1,g_2,g_3)
 \nu_2^{-1}(g_0,g_2,g_3)
 \nu_2(g_0,g_1,g_3)
 \nu_2^{-1}(g_0,g_1,g_2)
=1 ,
\end{align}
\begin{align}
\label{3ccy}
&(-)^{
n_2(g_0,g_1,g_2)
n_2(g_2,g_3,g_4)}
 \nu_3(g_1,g_2,g_3,g_4)
 \nu_3^{-1}(g_0,g_2,g_3,g_4)\times
\nonumber\\
&
 \nu_3(g_0,g_1,g_3,g_4)
 \nu_3^{-1}(g_0,g_1,g_2,g_4)
 \nu_3(g_0,g_1,g_2,g_3)
=1 ,
\end{align}
\begin{widetext}
\begin{align}
\label{4ccy}
&(-)^{
n_3(g_0,g_1,g_2,g_3)
n_3(g_0,g_3,g_4,g_5)
+
n_3(g_1,g_2,g_3,g_4)
n_3(g_0,g_1,g_4,g_5)
+
n_3(g_2,g_3,g_4,g_5)
n_3(g_0,g_1,g_2,g_5)
}
\nu_4(g_1,g_2,g_3,g_4,g_5)
\times \nonumber\\ &\ \ \
 \nu_4^{-1}(g_0,g_2,g_3,g_4,g_5)
 \nu_4(g_0,g_1,g_3,g_4,g_5)
 \nu_4^{-1}(g_0,g_1,g_2,g_4,g_5)
 \nu_4(g_0,g_1,g_2,g_3,g_5)
 \nu_4^{-1}(g_0,g_1,g_2,g_3,g_4)
=1.
\end{align}
\end{widetext}
We will denote the collections of all $d$-cocycles $[\nu_d(g_0,...,g_d),
u_{d-1}^{g}(g_1,...,g_d)]$ that satisfy the above
conditions as $\fZ^d[G_f,U_T(1)]$.

\subsection{An ``linearized'' representation of group super-cohomology}

Since the coefficient of the group super-cohomology is always $U(1)$, we can
map the fermionic cochains $[\nu_d, n_{d-1},u^{g}_{d-1}]$ into two real vectors
and an integer vector $(\v v_d,\v n_{d-1},\v h^{g}_{d-1})$, where the
components of $\v v_d$ and $\v h^{g}_{d-1}$ are in $[0,1)$ and the components
of $\v n_{d-1}$ are $0,1$.  Then, we can rewrite the cochain and the cocycle
conditions on $[\nu_d(g_0,...,g_d),
n_{d-1}(g_1,...,g_d),u^{g}_{d-1}(g_1,...,g_d)]$ as ``linear equations'' on
$(\v v_d,\v n_{d-1},\v h^{g}_{d-1})$.  Those linear equations can be solved
numerically more easily.

To map $\nu_d(g_0,...,g_d)$ into a real vector $\v v_d$, we can introduce the
log of $\nu_d(g_0,...,g_d)$:
\begin{align}
 v_d(g_0,g_1,...,g_d)=\frac{1}{2\pi \imth} \log \nu_d(g_0,g_1,...,g_d)
\text{ mod } 1,
\end{align}
So we define $\v v_d$ as a $|G_b|^{d+1}$ dimensional vector, whose
components are given by
\begin{align}
(\v v_d)_{(g_0,...,g_d)}=
 v_d(g_0,...,g_d).
\end{align}
Similarly, $\v n_{d-1}$ is a $|G_b|^{d}$ dimensional integer vector
\begin{align}
 (\v n_{d-1})_{(g_0,...,g_{d-1})}=
n_d(g_0,...,g_{d-1}),
\end{align}
and $\v h^{g}_{d-1}$ is a $|G_b|^{d}$ dimensional vector
\begin{align}
& (\v h^{g}_{d-1})_{(g_0,...,g_{d-1})} =
h^{g}_{d-1}(g_0,...,g_{d-1}),
\\
&
h^{g}_{d-1}(g_0,...,g_{d-1})=\frac{1}{2\pi\imth} \ln
u_{d-1}^{g}(g_0,...,g_{d-1})  \text{ mod } 1.
\nonumber
\end{align}

Now, the cochain conditions (\ie the symmetry condition), \eqn{gnu},
\eqn{ncond1}, and \eqn{1Drep}, can be written in a ``linearized'' form
\begin{align}
\label{scnd}
 s(g) S_d^{g} \v v_d  &= \v v_d + D_{d-1} \v h^{g}_{d-1} \text{ mod } 1,
\nonumber\\
S_{d-1}^{g} \v n_{d-1} &= \v n_{d-1},
\nonumber\\
s(g) S_{d-1}^{g} \v h^{g_1}_{d-1} &= \v h^{g_1}_{d-1},
\nonumber\\
\v h^{g_1g_2}_{d-1}
&
 =
\v h^{g_1}_{d-1} + s(g_1) \v h^{g_2}_{d-1}
\text{ mod } 1,
\nonumber\\
\v h^{P_f}_{d-1} & = \frac 12 \v n_{d-1} \text{ mod } 1,
\nonumber\\
 g,g_1,g_2 & \in G_f,
\end{align}
where $S_d^{g}$ is a $|G_b|^{d+1}\times |G_b|^{d+1}$ dimensional matrix, whose
elements are
\begin{align}
&\ \ \ \
 (S_d^{g})_{(g_0,...,g_{d}),(g'_0,...,g'_{d})}
\\
&=
\del_{(g'_0,...,g'_d),(gg_0,...,gg_{d+1})}.
\nonumber
\end{align}
Also $D_d$ is a
$|G_b|^{d+2}\times |G_b|^{d+1}$ dimensional integer matrix, whose elements are all 0 or
$\pm1$:
\begin{align}
&\ \ \ \
 (D_d)_{(g_0,...,g_{d+1}),(g'_0,...,g'_{d})}
\\
&=
(-)^i \del_{(g'_0,...,g'_d),(g_0,...,\hat g_i, ...,g_{d+1})},
\nonumber
\end{align}
where $(g_0,...,\hat g_i, ...,g_{d+1})$ is the sequence obtained from  the
sequence  $(g_0,...,g_{d+1})$ with the element $g_i$ removed.
The matrices $D_d$ satisfy
\begin{align}
 D_{d+1} D_d =0 .
\end{align}
A triple $(\v v_d,\v n_{d-1}, \v h^{g}_{d-1})$ that satisfies
\eqn{scnd} corresponds to a fermionic $d$-cochain
in $\fC^d[G_f,U_T(1)]$.

We can also rewrite the cocycle conditions \eqn{fccnd} as a ``linear'' equation:
\begin{align}
\label{vccy}
D_{d-1} \v n_{d-1}  &= 0 \text{ mod } 2,
\nonumber\\
 D_d \v v_d + \frac12 \v  f_{d+1} &= 0 \text{ mod } 1
.
\end{align}
Here $\v f_d$ is a $|G_b|^{d+1}$ dimensional integer vector whose components
are given by
\begin{align}
 (\v f_d)_{(g_0,...,g_d)}=
f_d(g_0,...,g_d) \text{ mod } 2.
\end{align}
where $f_d(g_0,...,g_d)$ is obtained from $n_{d-2}(g_0,...,g_{d-2})$ according
to the relation \eqn{fdn}.  A triple $(\v v_d,\v n_{d-1}, \v h^{g}_{d-1})$ that
satisfies \eqn{scnd} and \eqn{vccy} corresponds to a fermionic $d$-cocycle in
$\fZ^d[G_f,U_T(1)]$.

\subsection{Fermionic $d$-cohomology class}

As discussed in section \ref{fixZ}, each $(d+1)$-cocycle will define a
fermionic topological nonlinear $\si$ model in $d$ spatial dimensions which is
a fixed point of the RG flow.  Those $(d+1)$-cocycles give rise to new
fermionic SPT phases.  However, different  $(d+1)$-cocycles may give rise to
the same fermionic SPT phase.  Those $(d+1)$-cocycles are said to be equivalent.
To find (or guess) the equivalence relation between cocycles, in this appendix,
we will discuss some natural choices of possible equivalence relations.  A more
physical discussion of this issue is given in Appendix \ref{eqwav}.

First, we would like to show that (for $d\leq 4$)
\begin{align}
D_d \v f_d=0 \text{ mod } 2\label{coboundary},
\end{align}
where $\v f_d$ is determined from $n_{d-1}(g_1,...,g_d)$ in
$\cZ^{d-1}(G_f,\Z_2)$ (see \eqn{fdn}).  For $1\leq d\leq 3$, the above is
trivially satisfied since the corresponding $f_d=0$.  To show the above for
$d=4$, we note that $n_{d-2}(g_0,..,g_{d-2})$ (which give rise to $\v f_d$) is
a $(d-2)$-cocycle in $\cH^{d-2}(G_b,\Z_2)$.  Let us assume that a $2$-cocycle
$n_2(g_0,g_1,g_2)$ gives rise to $\v f_4$.  The expression, $f_4(g_0 ,g_1 ,g_2
,g_3 ,g_4) = n_2(g_0,g_1,g_2) n_2(g_2,g_3,g_4) $ implies that the 4-cochain
$f_4(g_0 ,g_1 ,g_2 ,g_3 ,g_4)$ is the cup product of two 2-cocycles
$n_2(g_0,g_1,g_2)$ and $n_2(g_0,g_1,g_2)$ in $\cZ^2(G_f,\Z_2)$.  So $f_4(g_0
,g_1 ,g_2 ,g_3 ,g_4)$ is also a cocycle in $\cZ^4(G_f,\Z_2)$.  Therefore $D_4\v
f_4=0\mod 2$.  We can also show that $D_5\v f_5=0\mod 2$ by an
explicit calculation using Mathematica.
Therefore, for $d\leq 5$,
\begin{align}
 D_d [D_{d-1} \v v_{d-1} + \frac12 \v  f_d]=0 \text{ mod } 1.
\end{align}
for any fermionic $(d-1)$-cochains $(\v v_{d-1}, \v  n_{d-1}, \v h^{g}_{d-1})$.

There is another important relation.  Let us assume that a $d$-cocycle
$n_d(g_0,g_1,...,g_d)$ gives rise to $\v f_{d+2}$, and another $d$-cocycle  $\t
n_2(g_0,g_1,...,g_d)$ gives rise to $\t{\v f}_{d+2}$ (assuming $d\leq 3$).  If
$n_d(g_0,g_1,...,g_d)$ and $\t n_d(g_0,g_1,...,g_d)$ are related by a
coboundary in $\cB^d(G_b,\Z_2)$
\begin{align}
&\ \ \ \
n_d(g_0,g_1,...,g_d) + \sum_i m_{d-1}^\prime(g_0,...,\hat g_i,...,g_d)
\nonumber\\
&=
n_d(g_0,g_1,...,g_d) + (\del m_{d-1}^\prime)(g_0,...,g_d)
\nonumber\\
&=
\t n_d(g_0,g_1,...,g_d)
\text{ mod } 2,
\end{align}
then what is the relation between $\v f_{d+2}$ and $\t{\v f}_{d+2}$?  In the
Appendix \ref{f5equivalent}, we will show that $\v f_{d+2}$ and $\t{\v f}_{d+2}$
also differ by a coboundary $D_{d+1} \v g_{d+1}$ in $\cB^{d+2}(G_b,\Z_2)$:
\begin{align}
\label{ffg}
\t{\v f}_{d+2}=\v f_{d+2} +D_{d+1} \v g_{d+1} \text{ mod } 2,
\end{align}
where $\v g_{d+1}$ is determined from $\v n_d$ and $\v m_{d-1}$ (see
Appendix \ref{f5equivalent}).

Using the above two relations, we can define the first equivalence relation
between two fermionic $d$-cocycles.  Starting from a fermionic $d$-cocycle,
$(\v v_{d}, \v n_{d-1}, \v h^{g}_{d-1})$, we can use a  fermionic
$(d-1)$-cochain $(\v v_{d-1}, \v n_{d-2}, \v h^{g}_{d-2})$ to transform the
above $d$-cocycle to an equivalent $d$-cocycle $(\t {\v v}_{d}, \t{\v n}_{d-1},
\t{\v h}^{g}_{d-1})$:
\begin{align}
\label{lccyeq}
\t {\v v}_{d} =& \v v_{d} -\frac12 \v  g_d +D_{d-1}\v v_{d-1},
\nonumber\\
\t {\v n}_{d-1} =& \v n_{d-1} + D_{d-2} \v n_{d-2}
\nonumber\\
\t {\v h}^{g}_{d-1} =& \v h^{g}_{d-1} + D_{d-2} \v h^{g}_{d-2}
,
\end{align}
where $\v g_d$ is determined from $\v n_{d-1}$  and $\v n_{d-2}$(see Appendix
\ref{f5equivalent} Eq.(\ref{gd})).

First, we would like to show that if $(\v v_{d}, \v n_{d-1}, \v h^{g}_{d-1})$
satisfies the cochain condition \eqn{scnd}, then $(\t {\v v}_{d}, \t{\v
n}_{d-1}, \t{\v h}^{g}_{d-1})$ obtained above also satisfies the same cochain
condition.  Let us first consider the first equation in \eqn{scnd}:
\begin{align}
\label{eq1}
& s(g)S^{g}_d \t {\v v}_{d}
= s(g)S^{g}_d \v v_{d} -\frac12 \v  g_d
+ s(g)S^{g}_d D_{d-1} \v v_{d-1}
\nonumber\\
=& \v v_{d}
+D_{d-1} \v h^{g}_{d-1}
-\frac12 \v  g_d
+ D_{d-1} s(g)S^{g}_{d-1} \v v_{d-1}
\text{ mod } 1
\nonumber\\
=& \t {\v v}_d +D_{d-1} (\v h^{g}_{d-1}+D_{d-2}\v h^{g}_{d-2})
\text{ mod } 1 ,
\nonumber\\
=& \t {\v v}_d +D_{d-1} \t{\v h}^{g}_{d-1}
\text{ mod } 1 ,
\end{align}
where we have used
\begin{align}
 S^{g}_d \v  g_d=\v  g_d;\quad  S^{g}_d \v  f_d=\v  f_d,
\end{align}
and the relation
\begin{align}
 S^{g}_d D_{d-1} = D_{d-1} S^{g}_{d-1}.
\end{align}
It is easy to show that $(\t {\v v}_{d}, \t{\v
n}_{d-1}, \t{\v h}^{g}_{d-1})$ also satisfies
other equations in \eqn{scnd}.

Now we like to show that if $(\v v_{d}, \v n_{d-1}, \v h^{g}_{d-1})$ satisfies
the cocycle condition \eqn{vccy}, then $(\t {\v v}_{d}, \t{\v n}_{d-1}, \t{\v
h}^{g}_{d-1})$ obtained above also satisfies the same cocycle condition:
\begin{align}
&\ \ \ \ D_d \t{\v v}_d + \frac12 \t{\v f}_{d+1}
\nonumber\\
&=
D_d \v v_d -\frac12 D_d \v g_d+
\frac12 (\v f_{d+1}+D_d \v g_d)
\nonumber\\
&= 0 \text{ mod } 1
.
\end{align}

We can also define the second equivalence relation between two fermionic
$d$-cocycles.  Let $\bar h^{g}_{d-1}(g_0,...,g_{d-1})$ be representations of
$G_b$ that satisfy
\begin{align}
\bar h^{g_a}_{d-1}(g_0,g_1,..)
 +&s(g_a)\bar h^{g_b}_{d-1}(g_0,g_1,..)
 =\bar h^{g_a g_b}_{d-1}(g_0,g_1,..) ,
\nonumber\\
 s(g) \bar h^{g_a}_{d-1}&(gg_0,...,gg_{d-1})
 =\bar h^{g_a }_{d-1}(g_0,...,g_{d-1}),
\nonumber\\
 g,g_a,g_b & \in G_b.
\end{align}
Then, starting from a fermionic $d$-cocycle, $(\v v_{d},
\v n_{d-1}, \v h^{g}_{d-1})$, we can use such a
$\bar h^{g}_{d-1}(g_0,...,g_{d-1})$
to transform the above $d$-cocycle to an
equivalent $d$-cocycle $(\t {\v v}_{d}, \t{\v n}_{d-1}, \t{\v h}^{g}_{d-1})$:
\begin{align}
\label{eq2}
&\t v_d(g_0,g_1,...,g_d) =
 v_d(g_0,g_1,...,g_d)+ s(g_1)\bar h^{g_1}_{d-1}(g_1,...,g_d)
\nonumber\\
&\ \ \ \ \ \ \
+s(g_0)\sum_{i=1}^d (-)^i \bar h^{g_0}_{d-1}(g_0,..,\hat g_i,..,g_d),
\nonumber\\
&\t n_{d-1}(g_0,g_1,...,g_{d-1}) = n_{d-1}(g_0,g_1,...,g_{d-1}) ,
\\
&\t h^{g}_{d-1}(g_0,..,g_{d-1}) =
h^{g}_{d-1}(g_0,..,g_{d-1}) +
s(gg_0)\bar h^{g}_{d-1}(g_0,..,g_{d-1}) .
\nonumber
\end{align}
The above can also be written in a more compact form:
\begin{align}
 \t{\v v}_d & = {\v v}_d + D_{d-1} \bar {\v h}_{d-1},
\nonumber\\
 \t{\v n}_{d-1} & = {\v n}_{d-1},
\nonumber\\
 \t{\v h}^g_{d-1} & = {\v h}^g_{d-1} + \bar {\v h}^g_{d-1}
,
\end{align}
where $\bar {\v h}_{d-1}$ and $\bar {\v h}^g_{d-1}$
are the $|G_b|^d$ dimensional vectors:
\begin{align}
 (\bar {\v h}_{d-1})_{(g_0,...,g_{d-1})} &= s(g_0)h^{g_0}_{d-1}(g_0,...,g_{d-1}),
\nonumber\\
 (\bar {\v h}^g_{d-1})_{(g_0,...,g_{d-1})} &= s(gg_0)h^g_{d-1}(g_0,...,g_{d-1}).
\end{align}

Let us first show that
$(\t {\v v}_{d}, \t{\v n}_{d-1}, \t{\v h}^{g}_{d-1})$ is a $d$-cochain:
\begin{align}
& s(g) \t v_d(gg_0,gg_1,...,gg_d)
= s(g) v_d(gg_0,gg_1,...,gg_d)
\nonumber\\
&\ \ \
+ s(g)s(gg_1) \bar h^{gg_1}_{d-1}(gg_1,..,gg_d)
\nonumber\\
&\ \ \
+ s(g)s(gg_0)\sum_{i=1}^d (-)^i \bar h^{gg_0}_{d-1}(gg_0,..,g\hat g_i,..,gg_d)
\nonumber\\
&=
 v_d(g_0,g_1,...,g_d)+
\sum_{i=0}^d (-)^i h^{g}_{d-1}(g_0,..,\hat g_i,..,g_d)
\\
&
+ s(gg_1)\bar h^{gg_1}_{d-1}(g_1,..,g_d)
+ s(gg_0)\sum_{i=1}^d (-)^i \bar h^{gg_0}_{d-1}(g_0,..,\hat g_i,..,g_d)
\nonumber \\
&=
 v_d(g_0,g_1,...,g_d)+
\sum_{i=0}^d (-)^i h^{g}_{d-1}(g_0,..,\hat g_i,..,g_d)
\nonumber \\
&
+ s(gg_1)\bar h^{g}_{d-1}(g_1,..,g_d)
+ s(gg_0)\sum_{i=1}^d (-)^i \bar h^{g}_{d-1}(g_0,..,\hat g_i,..,g_d)
\nonumber \\
&
+ s(g_1)\bar h^{g_1}_{d-1}(g_1,..,g_d)
+ s(g_0)\sum_{i=1}^d (-)^i \bar h^{g_0}_{d-1}(g_0,..,\hat g_i,..,g_d)
\nonumber \\
&=
\t v_d(g_0,g_1,...,g_d)+
\sum_{i=0}^d (-)^i\t h^{g}_{d-1}(g_0,..,\hat g_i,..,g_d)
.
\nonumber
\end{align}
Second, since $\t {\v v}_d= \v v_d + D_{d-1} \bar {\v h}_{d-1}$, it also clear
that $(\t {\v v}_{d}, \t{\v n}_{d-1}, \t{\v h}^{g}_{d-1})$ is a
$d$-cocycle.

The equivalence classes of fermionic $d$-cocycles obtained from the above two
kinds of equivalence relations, \eqn{eq1} and \eqn{eq2}, form the group
$d$-super-cohomology $\fH^d[G_f,U_T(1)]$.

We note that, when ${\v n}_{d-1}=0$,  the representations of $G_f$, ${\v
h}^{g}_{d-1}$, are also representations of $G_b$.  So the second equivalence
relation implies that a $d$-cocycle $(\v v_{d}, \v n_{d-1}=0, \v h^g_{d-1})$ is
always equivalent to a simpler $d$-cocycle $(\t{\v v}_{d}, \v n_{d-1}=0, \v
h^g_{d-1}=0)$.  Such kinds of  $d$-cocycles are described by group cohomology
and form group cohomology class $\cH^d[G_b,U_T(1)]$.

We also note that, due to the requirement $\v h^{P_f}_{d-1}  = \frac 12 \v
n_{d-1} \text{ mod } 1$, for each fixed $\v n_{d-1}$, all allowed choices of
$\v h^{g}_{d-1}$ belong to one class under the second equivalence relation.  So
we just need to choose one allowed $\v h^{g}_{d-1}$ for each $\v n_{d-1}$.

\section{The Abelian group structure of the group super-cohomology classes}
\label{group}

Similar to the standard group cohomology, the elements of super-cohomology
classes $\fH^d[G_f,U_T(1)]$ should form an Abelian group.  This Abelian group
structure has a very physical meaning.  Let us consider two SPT phases labeled
by $a,b$.  If we stack the phase-$a$ and the phase-$b$ on top of each other, we
still have a gapped SPT phase which should be labeled by $c$.  Such a stacking
operation $a+b \to c$ generates the Abelian group structure of the group
super-cohomology classes.  So if we our construction of the fermionic SPT
phases is in some sense complete, stacking the phase-$a$ and the phase-$b$ in
$\fH^d[G_f,U_T(1)]$ will result in a SPT phase-$c$ still in
$\fH^d[G_f,U_T(1)]$.  In other words, $\fH^d[G_f,U_T(1)]$ should be an Abelian
group.

The key step to showing $\fH^d[G_f,U_T(1)]$ to be an Abelian group is to show that
all the valid graded structures labeled by the elements in
$B\cH^{d-1}(G_b,\Z_2)$ form an Abelian group.  Since $\cH^{d-1}(G_b,\Z_2)$ is
an Abelian group, we only need to show that for any two elements
$n_{d-1},n_{d-1}^\prime \in B\cH^{d-1}(G_b,\Z_2)$, their summation
$n_{d-1}^{\prime\prime}=n_{d-1}+n_{d-1}^\prime \in B\cH^{d-1}(G_b,\Z_2)$.  In
Appendix \ref{additive}, we prove that
$f_{d+1}^{\prime\prime}-f_{d+1}-f_{d+1}^{\prime}$
is a coboundary in  $\cB^{d+1}(G_b,\Z_2)$ (given by $a_d$ below),
where $f_d$ is obtained from $n_d$, $f_d^\prime$ is obtained from $n_d^\prime$, and $f_d^{\prime\prime}$ is obtained from $n_d^{\prime\prime}$, respectively. Therefore, if $f_{d+1}$ and $f_{d+1}^\prime$ are $(d+1)$-coboundaries in
$\cB^{d+1}[G_b, U_T(1)]$, $f_{d+1}^{\prime\prime}$ is also a $(d+1)$-coboundary in
$\cB^{d+1}[G_b, U_T(1)]$. On the other hand, $B\cH^{d-1}(G_b,\Z_2)$ is a subset of $\cH^{d-1}(G_b,\Z_2)$, hence $B\cH^{d-1}(G_b,\Z_2)$ forms an (Abelian) subgroup of $\cH^{d-1}(G_b,\Z_2)$.

Using the above result, we can show that if $(\nu_{d},n_{d-1},u_{d-1}^g)$ and
$(\nu_{d}^\prime,n_{d-1}^\prime,{u^\prime}_{d-1}^g)$ are elements in
$\fH^d[G_f,U_T(1)]$,
then using an additive operation,
we can produce a
$(\nu_{d}^{\prime\prime},n_{d-1}^{\prime\prime},{u^{\prime\prime}}_{d-1}^g)$
\begin{align}
\label{fadd}
n_{d-1}^{\prime\prime} & = n_{d-1}+n_{d-1}^\prime
\nonumber\\
{u^{\prime\prime}}_{d-1}^g & =u_{d-1}^g{u^\prime}_{d-1}^g
\nonumber\\
\nu_{d}^{\prime\prime} &=
\nu_{d}\nu_{d}^\prime (-)^{a_d}
\end{align}
such that it is also an element in $\fH^d[G_f,U_T(1)]$.
Here the phase factor $(-)^{a_d}$ is given by
\begin{align}
& a_0 =a_1=a_2=0,
\\
&a_3={n_2(g_0,g_1,g_2)n_2^\prime(g_0,g_2,g_3)+n_2(g_1,g_2,g_3)n_2^\prime(g_0,g_1,g_3)}
\nonumber\\
&a_4 =
n_3(g_0,g_1,g_2,g_3)n_3^\prime(g_0,g_1,g_3,g_4)+
\nonumber\\ &\ \ \ \ \ \ \
n_3^\prime(g_0,g_1,g_2,g_4)n_3(g_0,g_2,g_3,g_4) +
\nonumber \\
&\ \ \ \ \ \ \
 n_3(g_0,g_1,g_3,g_4)n_3^\prime(g_1,g_2,g_3,g_4)+
\nonumber\\ &\ \ \ \ \ \ \
n_3(g_0,g_1,g_2,g_3)n_3^\prime(g_1,g_2,g_3,g_4) .
\nonumber
\end{align}

\section{Calculate $\cH^d(G_b,\Z_2)$ -- the graded structure}

In this appendix, we will discuss several methods that allow us
to calculate the graded structure $\cH^d(G_b,\Z_2)$ in general.

\subsection{Calculate $\cH^d(Z_n,\Z_2)$}

The cohomology group $\cH^d(Z_n,M)$ has a very simple form.  To describe the
simple form in a more general setting, let us define Tate cohomology groups
$\hat\cH^d(G,M)$.

For $d$ to be $0$ or $-1$, we have
\begin{align}
 \hat\cH^0(G,M) &= M^{G}/\text{Img}(N_{G},M),
\nonumber\\
 \hat\cH^{-1}(G,M) &= \text{Ker}(N_{G},M)/I_{G}M.
\end{align}
Here $M^{G}$, $\text{Img}(N_{G},M)$, $\text{Ker}(N_{G},M)$, and
$I_{G}M$ are submodule of $M$.
$M^{G}$ is the maximal submodule that is invariant under the
group action. Let us define a
map $N_{G}: M\to M$ as
\begin{align}
 a\to \sum_{g\in G} g\cdot a, \ \ \ a\in M.
\end{align}
$\text{Img}(N_{G},M)$ is the image of the map
and $\text{Ker}(N_{G},M)$ is the
kernel of the map.
The submodule $I_{G}M$ is given by
\begin{align}
 I_{G}M=
\{ \sum_{g\in G} n_g g\cdot a | \sum_{g\in G} n_g=0,\ a\in M\}
\end{align}
In other words, $I_{G}M$ is generated by $g\cdot a - a$,
$\forall\ g\in G,\ a \in M$.

For $d$ other than $0$ and $-1$,
the Tate cohomology groups $\hat\cH^d(G,M)$ is given by
\begin{align}
 \hat\cH^d(G,M) &=\cH^d(G,M), \ \text{ for } d>0
\nonumber\\
 \hat\cH^d[G,M] &=\cH_{-d-1}[G,M], \ \text{ for } d<-1 .
\end{align}

For cyclic group $Z_n$, its (Tate) group cohomology over a generic $Z_n$-module $M$
is given by\cite{DJ,RS}
\begin{align}
\label{ZnCoh}
\hat \cH^d(Z_n,M)=
\begin{cases}
\hat\cH^0(Z_n,M) & \text{ if } d=0 \text{ mod } 2,\\
\hat\cH^{-1}(Z_n,M) & \text{ if } d=1 \text{ mod } 2.\\
\end{cases}
\end{align}
where
\begin{align}
 \hat\cH^0(Z_n,M) &= M^{Z_n}/\text{Img}(N_{Z_n},M),
\nonumber\\
 \hat\cH^{-1}(Z_n,M) &= \text{Ker}(N_{Z_n},M)/I_{Z_n}M.
\end{align}

For example, when the group action is trivial, we have $M^{Z_n}=M$ and $
I_{Z_n}M=\Z_1$.  The map $N_{Z_n}$ becomes $N_{Z_n}: a\to n a$.  For $M=\Z_2$,
we have $\text{Img}(N_{Z_n},\Z_2)=n \Z_{2}$ and
$\text{Ker}(N_{Z_n},\Z_2)=\Z_{(n,2)}$
(where $(n,m)$ is the greatest common divisor of $n$ and $m$).  So we have
\begin{align}
\label{ZnZ2}
 \cH^d(Z_n,\Z_2)=
\begin{cases}
\Z_2  & \text{ if } d=0, \\
\Z_{(n,2)}  & \text{ if } d>0 .
\end{cases}
\end{align}

What is the nontrivial cocycle in $\cH^d[Z_2,\Z_2]$?  In \Ref{CGL1172}, it is
shown that a $d$-cocycle can be chosen to satisfy
\begin{align}
\label{gauge}
&\nu_d({\bf{g,g}},g_2,g_3,...,g_{d-2},g_{d-1},g_d)=1
\nonumber \\
&\nu_d(g_0,{\bf g,g},g_3,...,g_{d-2},g_{d-1},g_d)=1
\nonumber \\
&\ \ \ \ \ \ \ \ \ \ \ ...\ \ \ ...\ \ \ ...
\nonumber\\
&\nu_d(g_0,g_1,g_2,g_3,..., g_{d-2},{\bf g,g})=1.
\end{align}
by adding a proper coboundary.

Let us denote the two elements in $Z_2$ as $\{E,\si\}$ with $\si^2=E$.  If we
choose the cocycle $\nu_d(g_0,g_1,g_2,...)$ in $\cH^d(Z_{2},\Z_2)$ to satisfy
the above condition, then only $\nu_d(E,\si,E,...)=\nu_d(\si,E,\si,...)$ can be
non zero.  Since $\cH^d(Z_{2},\Z_2)=\Z_2$, So
$\nu_d(E,\si,E,...)=\nu_d(\si,E,\si,...)=0$ must correspond to the trivial
class and $\nu_d(E,\si,E,...)=\nu_d(\si,E,\si,...)=1$ must correspond to the
nontrivial class in  $\cH^d(Z_{2},\Z_2)$.

\subsection{Calculate $\cH^d(G,\Z_2)$ from $\cH^d(G,\Z)$ using K\"unneth
formula for group cohomology}

We can also calculate $\cH^d(G,\Z_2)$ from $\cH^d(G,\Z)$ using the K\"unneth
formula for group cohomology.  Let $M$ (resp. $M'$) be an arbitrary $G$-module
(resp. $G'$-module) over a principal ideal domain $R$.  We also assume that
either $M$ or $M'$ is $R$-free. Then we have a K\"unneth formula for group
cohomology\cite{G0691,Kunn}
\begin{align}
\label{kunn}
\Z_1 & \rightarrow
\prod_{p=0}^d \cH^p(G,M)\otimes_R \cH^{d-p}(G',M')
\nonumber\\
& \rightarrow
\cH^d(G\times G',M\otimes_R M')
\nonumber\\
& \rightarrow
\prod_{p=0}^{d+1}
\text{Tor}_1^R(\cH^p(G,M),\cH^{d-p+1}(G',M'))
\rightarrow \Z_1
\end{align}
If both  $M$ and $M'$ are $R$-free, then the sequence splits and we have
\begin{align}
&\ \ \ \ \cH^d(G\times G',M\otimes_R M')
\nonumber\\
&=\Big[\prod_{p=0}^d \cH^p(G,M)\otimes_R \cH^{d-p}(G',M')\Big]\times
\nonumber\\
&\ \ \ \ \ \
\Big[\prod_{p=0}^{d+1}
\text{Tor}_1^R(\cH^p(G,M),\cH^{d-p+1}(G',M'))\Big]  .
\end{align}
If $R$ is a field $K$, we have
\begin{align}
&\ \ \ \ \cH^d(G\times G',M\otimes_K M')
\nonumber\\
&=\prod_{p=0}^d \Big[\cH^p(G,M)\otimes_K \cH^{d-p}(G',M')\Big] .
\end{align}

Let us choose $R=\Z$, $M=\Z$ and $M'=\Z_2$.  Note that $M=\Z$ is a free module
over $R=\Z$ but $M'=\Z_2$ is not a free module over $R=\Z$.  We also note that
$M\otimes_\Z M' = \Z \otimes_\Z \Z_2=\Z_2$.  Let us choose $G'=Z_1$ the trivial
group with only identity.  We have
\begin{align}
\label{Z1Z2}
 \cH^d(Z_1,\Z_2)=
\begin{cases}
\Z_2  & \text{ if } d=0, \\
\Z_1  & \text{ if } d>0 .
\end{cases}
\end{align}
Thus we obtain
\begin{align}
\Z_1 & \rightarrow \cH^d(G,\Z)\otimes_\Z \Z_2 \rightarrow \cH^d(G,\Z_2)
\nonumber\\
& \rightarrow
\text{Tor}_1^\Z(\cH^{d+1}(G,\Z),\Z_2)
\rightarrow \Z_1 .
\end{align}

The above can be calculated using the following simple properties of the tensor
product $\otimes_\Z$ and $ \text{Tor}_1^\Z$ functor:
\begin{align}
& \Z \otimes_\Z M = M \otimes_\Z \Z =M ,
\nonumber\\
& \Z_n \otimes_\Z M = M \otimes_\Z \Z_n = M/nM ,
\nonumber\\
& \Z_m \otimes_\Z \Z_n  =\Z_{(m,n)} ,
\nonumber\\
&  (A\times B)\otimes_R M = (A \otimes_R M)\times (B \otimes_R M)   ,
\nonumber\\
& M \otimes_R (A\times B) = (M \otimes_R A)\times (M \otimes_R B)   ;
\end{align}
\begin{align}
& \text{Tor}_1^\Z(\Z, M) = \text{Tor}_1^\Z(M, \Z) = \Z_1,
\nonumber\\
& \text{Tor}_1^\Z(\Z_n, M) = \text{Tor}_1^\Z(M, \Z_n) = M/nM,
\nonumber\\
& \text{Tor}_1^\Z(\Z_m, \Z_n) = \Z_{(m,n)} ,
\nonumber\\
& \text{Tor}_1^R(A\times B,M) = \text{Tor}_1^R(A, M)\times\text{Tor}_1^R(B, M),
\nonumber\\
& \text{Tor}_1^R(M,A\times B) = \text{Tor}_1^R(M,A)\times\text{Tor}_1^R(M,B)
.
\end{align}
Here $(m,n)$ is the greatest common divisor of $m$ and $n$.

Using
\begin{align}
\label{ZnZ}
 \cH^d(Z_n,\Z)=
\begin{cases}
\Z  & \text{ if } d=0, \\
\Z_n  & \text{ if } d=0 \text{ mod } 2,\\
\Z_1  & \text{ if } d=1 \text{ mod } 2,
\end{cases}
\end{align}
we find that
for $d=$ even
\begin{align}
& \Z_1  \rightarrow \Z_n\otimes_\Z \Z_2 \rightarrow \cH^d(\Z_n,\Z_2)
\rightarrow \text{Tor}_1^\Z(\Z_1,\Z_2) \rightarrow \Z_1 ,
\nonumber\\
\text{or }
& \Z_1  \rightarrow \Z_{(n,2)}\rightarrow \cH^d(\Z_n,\Z_2)
\rightarrow \Z_1 \rightarrow \Z_1 ;
\end{align}
and for $d=$odd
\begin{align}
& \Z_1  \rightarrow \Z_1\otimes_\Z \Z_2 \rightarrow \cH^d(\Z_n,\Z_2)
\rightarrow \text{Tor}_1^\Z(\Z_n,\Z_2) \rightarrow \Z_1 ,
\nonumber\\
\text{or }
& \Z_1  \rightarrow \Z_1 \rightarrow \cH^d(\Z_n,\Z_2)
\rightarrow \Z_{(n,2)} \rightarrow \Z_1 .
\end{align}
This allows us to obtain \eqn{ZnZ2} using a different approach.

Using
\begin{align}
\label{Z2rZnZ}
 \cH^d[ Z_n\rtimes Z_2,\Z] \leq
\begin{cases}
\Z,  &  d=0 ,\\
\Z_2\times \Z_n\times \Z_{(2,n)}^{\frac {d-2} 2},  &  d=0 \text{ mod } 4,\\
\Z_{(2,n)}^{\frac{d-1}{2}},  &  d=1 \text{ mod } 2 ,\\
\Z_2\times \Z_{(2,n)}^{\frac d 2},  &  d=2 \text{ mod } 4,\\
\end{cases}
\end{align}
we can show that for $d=0$ mod 4, $d>0$,
\begin{align}
& \Z_1  \rightarrow (\Z_2\times \Z_n\times \Z_{(2,n)}^{\frac {d-2} 2}) \otimes_\Z \Z_2 \rightarrow \cH^d[Z_n\rtimes Z_2,\Z_2]
\nonumber\\
&\ \ \ \,
\rightarrow \text{Tor}_1^\Z(\Z_{(2,n)}^{\frac d 2},\Z_2) \rightarrow \Z_1 ,
\\
\text{or }
& \Z_1  \rightarrow \Z_2\times \Z_{(2,n)}^{\frac d 2} \rightarrow \cH^d[Z_n\rtimes Z_2,\Z_2]
\rightarrow \Z_{(2,n)}^{\frac d 2} \rightarrow \Z_1 ;
\nonumber
\end{align}
for $d=1$ mod 4, $d>0$,
\begin{align}
& \Z_1  \rightarrow \Z_{(2,n)}^{\frac {d-1} 2} \otimes_\Z \Z_2 \rightarrow \cH^d[Z_n\rtimes Z_2,\Z_2]
\nonumber\\
&\ \ \ \,
\rightarrow
\text{Tor}_1^\Z(\Z_2\times \Z_{(2,n)}^{\frac {d+1} 2},\Z_2) \rightarrow \Z_1 ,
\\
\text{or }
& \Z_1  \rightarrow \Z_{(2,n)}^{\frac {d-1} 2} \rightarrow \cH^d[Z_n\rtimes Z_2,\Z_2]
\rightarrow \Z_2\times \Z_{(2,n)}^{\frac {d+1} 2} \rightarrow \Z_1 ;
\nonumber
\end{align}
for $d=2$ mod 4, $d>0$,
\begin{align}
& \Z_1  \rightarrow (\Z_2\times \Z_{(2,n)}^{\frac d 2}) \otimes_\Z \Z_2 \rightarrow \cH^d[Z_n\rtimes Z_2,\Z_2]
\nonumber\\
&\ \ \ \,
\rightarrow \text{Tor}_1^\Z(\Z_{(2,n)}^{\frac d 2},\Z_2) \rightarrow \Z_1 ,
\\
\text{or }
& \Z_1  \rightarrow \Z_2\times \Z_{(2,n)}^{\frac d 2} \rightarrow \cH^d[Z_n\rtimes Z_2,\Z_2]
\rightarrow \Z_{(2,n)}^{\frac d 2} \rightarrow \Z_1 ;
\nonumber
\end{align}
for $d=3$ mod 4, $d>0$,
\begin{align}
& \Z_1  \rightarrow \Z_{(2,n)}^{\frac {d-1} 2} \otimes_\Z \Z_2 \rightarrow \cH^d[Z_n\rtimes Z_2,\Z_2]
\nonumber\\
&\ \ \ \,
\rightarrow
\text{Tor}_1^\Z(\Z_2\times \Z_n \times \Z_{(2,n)}^{\frac {d-1} 2},\Z_2) \rightarrow \Z_1 ,
\\
\text{or }
& \Z_1  \rightarrow \Z_{(2,n)}^{\frac {d-1} 2} \rightarrow \cH^d[Z_n\rtimes Z_2,\Z_2]
\rightarrow \Z_2\times \Z_{(2,n)}^{\frac {d+1} 2} \rightarrow \Z_1 .
\nonumber
\end{align}
Combining the above results, we find that, for $d>0$,
\begin{align}
 \cH^d[Z_n\rtimes Z_2,\Z_2]= \Z_2\times \Z_{(2,n)}^d .
\end{align}

Using
\begin{align}
\label{UZ}
 \cH^d[U(1),\Z]=
\begin{cases}
\Z  & \text{ if } d=0 \text{ mod } 2,\\
\Z_1  & \text{ if } d=1 \text{ mod } 2,
\end{cases}
\end{align}
we find that
for $d=$ even
\begin{align}
& \Z_1  \rightarrow \Z\otimes_\Z \Z_2 \rightarrow \cH^d[U(1),\Z_2]
\rightarrow \text{Tor}_1^\Z(\Z_1,\Z_2) \rightarrow \Z_1 ,
\nonumber\\
\text{or }
& \Z_1  \rightarrow \Z_2 \rightarrow \cH^d[U(1),\Z_2]
\rightarrow \Z_1 \rightarrow \Z_1 ;
\end{align}
and for $d=$odd
\begin{align}
& \Z_1  \rightarrow \Z_1\otimes_\Z \Z_2 \rightarrow \cH^d[U(1),\Z_2]
\rightarrow \text{Tor}_1^\Z(\Z,\Z_2) \rightarrow \Z_1 ,
\nonumber\\
\text{or }
& \Z_1  \rightarrow \Z_1 \rightarrow \cH^d(U(1),\Z_2)
\rightarrow \Z_1 \rightarrow \Z_1 .
\end{align}
This allows us to obtain
\begin{align}
\label{UZ2}
 \cH^d[U(1),\Z_2]=
\begin{cases}
\Z_2  & \text{ if } d=0 \text{ mod } 2,\\
\Z_1  & \text{ if } d=1 \text{ mod } 2,
\end{cases}
\end{align}
Note that in $\cH^d[U(1),\Z]$ and $\cH^d[U(1),\Z_2]$, the cocycles $\nu_d:
[U(1)]^{d+1} \to \Z$ or $\nu_d: [U(1)]^{d+1} \to \Z_2$ are Borel measurable
functions.

Let us choose $R=M=M'=\Z_2$.  Since $R$ is a field and $\Z_2\otimes_{\Z_2}
\Z_2$, we have
\begin{align}
\cH^d(G\times G',\Z_2)
=\prod_{p=0}^d \Big[\cH^p(G,\Z_2)\otimes_{\Z_2} \cH^{d-p}(G',\Z_2)\Big] .
\end{align}
Note that
\begin{align}
 \Z_2 \otimes_{\Z_2} \Z_2 =\Z_2, \ \ \ \
 \Z_2 \otimes_{\Z_2} \Z_1 =\Z_1.
\end{align}
We find
\begin{align}
\cH^d[Z_2\times Z_2,\Z_2] &= \Z_2^{d+1} .
\end{align}

\section{Calculate group super-cohomology class}
\label{calfcoh}

In this appendix we will demonstrate how to calculate group super-cohomology
classes, following the general outline described in section \ref{gen}. We will
do so by performing explicit calculations from some simple groups.

\subsection{Calculate $\fH^d[Z_1\times Z_2^f,U(1)]$}

Here we choose $G_b=Z_1$ (the trivial group) which corresponds to fermion
systems with no symmetry.  Since $\cH^{d-1}(Z_1, \Z_2)=\Z_2$ for $d=1$ and
$\cH^{d-1}(Z_1, \Z_2)=\Z_1=0$ for $d>1$.  We find that there are two 1D-graded
structures and only one trivial $d$D-graded structure for $d>1$.  For each
$d$D-graded structure, we can choose a $u^g_{d-1}(g_0,...,g_{d-1})$ that
$u^{P_f}_{d-1}(g_0,...,g_{d-1})=(-)^{n_{d-1}(g_0,...,g_{d-1})}$.

For $d>1$ and for the only trivial graded structure, the equivalent classes of
$\v v_d$ are described by $\cH^d[Z_1,U(1)]=\Z_1$. Therefore $\fH^d[Z_1\times
Z_2^f,U(1)]=\Z_1$ for $d>1$.

For $d=1$, $\cH^0(Z_1, \Z_2)=\Z_2$.  The trivial graded structure is given by
$n^{(0)}(E)=0$ and the other nontrivial graded structure is given by
$n^{(1)}(E)=1$. The corresponding $f_2^{(a)}(g_0,...,g_d)$ are given by
$f_2^{(a)}=0$, $a=0,1$. Thus for the each graded structure, the equivalent
classes of $\v v_d$ are described by $\cH^1[Z_1,U(1)]=\Z_1$.  Therefore
$\fH^1[Z_1\times Z_2^f,U(1)]=\Z_2$. To summarize
\begin{align}
\label{fHZ1U1}
 \fH^d[ Z_1\times Z_2^f,U(1)] =
\begin{cases}
\Z_2,  &  d=1 ,\\
 \Z_1,  &  d>1 .\\
\end{cases}
\end{align}

$\fH^1[ Z_1\times Z_2^f,U(1)]=\Z_2$ implies that there are two possible
fermionic SPT phases in $d_{sp}=0$ spatial dimension if there is no symmetry.
One has an even number of fermions and the other has an odd number of fermions.

$\fH^2[ Z_1\times Z_2^f,U(1)]=\Z_1$ implies that there is only one trivial gapped
fermionic SPT phases in $d_{sp}=1$ spatial dimension if there is no symmetry.
It is well known that there are two gapped fermionic phases in $d_{sp}=1$
spatial dimension if there is no symmetry, the trivial phase and the phase with
boundary Majorana zero mode.\cite{K0131} One way to see this result is to
use the Jordan-Wigner transformation to map the 1D fermion system with no
symmetry to an 1D boson system with $Z_2^f$ symmetry.  1D boson system with
$Z_2^f$ symmetry can have only two gapped phases: the symmetry unbroken one
(which corresponds to the trivial fermionic phase) and the symmetry breaking one
(which corresponds to the Majorana chain).\cite{CGW1128}  Both gapped phases can be realized by
non-interacting fermions (see Table \ref{tbF}).  However, the nontrivial
gapped phase (the bosonic $Z_2^f$ symmetry breaking phase) has
nontrivial intrinsic topological orders. So the nontrivial gapped phase is not
a fermionic SPT phase.  This is why we only have one trivial fermionic SPT phase
in 1D if there is no symmetry.

$\fH^3[ Z_1\times Z_2^f,U(1)]=\Z_1$ implies that there is only one trivial gapped
fermionic SPT phases in $d_{sp}=2$ spatial dimension if there is no symmetry.
It is well known that even non-interacting fermions have infinite gapped phases
labeled by $\Z$ in $d_{sp}=2$ spatial dimensions if there is no symmetry (see
Table \ref{tbF}).  Those phases correspond to integer quantum Hall states
and/or $p+\imth p$, $d+\imth d$ superconductors.  Again, all those gapped
phases have nontrivial intrinsic topological orders.  So we only have one
trivial fermionic SPT phase in $d_{sp}=2$ spatial dimensions if there is no
symmetry.

\subsection{Calculate $\fH^d[Z_{2k+1}\times Z_2^f,U(1)]$}

Next we choose $G_f=Z_{2k+1}\times Z_2^f$.  Again, since $\cH^{d-1}(Z_{2k+1},
\Z_2)=\Z_2$ for $d=1$ and $\cH^{d-1}(Z_{2k+1}, \Z_2)=\Z_1$ for $d>1$,  there
are two 1D-graded structures and only one trivial $d$D-graded structure for
$d>1$.  For each $d$D-graded structure, we can choose a
$u^g_{d-1}(g_0,...,g_{d-1})$ that satisfies $u^g_{d-1}(g_0,...,g_{d-1})=1$ for
$g\in G_b$
and $u^{P_f}_{d-1}(g_0,...,g_{d-1})=(-)^{n_{d-1}(g_0,...,g_{d-1})}$.

For $d>1$ and for the only trivial graded structure, the equivalent classes of
$\v v_d$ are described by $\cH^d[Z_{2k+1},U(1)]=\Z_1$ for $d$=even and
$\cH^d[Z_{2k+1},U(1)]=\Z_{2k+1}$ for $d$=odd.  Therefore $\fH^d[Z_1\times
Z_2^f,U(1)]=\Z_1$ for $d$=even and $\fH^d[Z_1\times Z_2^f,U(1)]=\Z_{2k+1}$ for
$d$=odd.

For $d=1$, $\cH^0(Z_{2k+1}, \Z_2)=\Z_2$.  The trivial graded structure is given
by $n^{(0)}(g_0)=0$ and the other nontrivial graded structure is given by
$n^{(1)}(g_0)=1$. The corresponding $f_2^{(a)}(g_0,...,g_d)$ are given by
$f_2^{(a)}=0$, $a=0,1$. Thus for the each graded structure, the equivalent
classes of $\v v_d$ are described by $\cH^1[Z_{2k+1},U(1)]=\Z_{2k+1}$.
Therefore $\fH^1[Z_{2k+1}\times Z_2^f,U(1)]=\Z_2\times \Z_{2k+1}=\Z_{4k+2}$. To
summarize
\begin{align}
\label{fHZoddU1}
 \fH^d[ Z_{2k+1}\times Z_2^f,U(1)] =
\begin{cases}
\Z_{4k+2},  &  d=1 ,\\
 \Z_1,  &  d=\text{even}, d>0 ,\\
 \Z_{2k+1},  &  d=\text{odd} .\\
\end{cases}
\end{align}

$\fH^1[ Z_{2k+1}\times Z_2^f,U(1)]=\Z_{4k+2}$ implies that there are
$4k+2$  possible fermionic SPT phases in $d_{sp}=0$ spatial dimension if there
is a $Z_{2k+1}$ symmetry.  $2k+1$ of them have even number of fermions and the
other $2k+1$ of them have odd number of fermions.  The $2k+1$ phases are
separated by the $2k+1$ possible $Z_{2k+1}$ quantum numbers.

$\fH^2[ Z_{2k+1}\times Z_2^f,U(1)]=\Z_1$ implies that there is no nontrivial
gapped fermionic SPT phases in $d_{sp}=1$ spatial dimension if there is a
$Z_{2k+1}$ symmetry. We can use the Jordan-Wigner transformation to map 1D
fermion systems with $Z_{2k+1}$ symmetry to 1D boson systems with
$Z_{2k+1}\times Z_2^f= Z_{4k+2}$ symmetry.  Since $\cH^2[Z_{4k+2}, U(1)]=\Z_1$, 1D
boson systems have only two phases that do not break the $Z_{2k+1}$ symmetry:
one does not break the $Z_2^f$ symmetry and one breaks the $Z_2^f$ symmetry.
The fermionic SPT phase has no intrinsic topological order and corresponds to
the bosonic phase that does not break the $Z_2^f$ symmetry (and the $Z_{2k+1}$
symmetry).  Thus 1D fermion systems with $Z_{2k+1}$ symmetry indeed have only
one trivial fermionic SPT phase.

$\fH^3[ Z_{2k+1}\times Z_2^f,U(1)]=\Z_{2k+1}$ implies that our construction
gives rise to $2k+1$ gapped fermionic SPT phases in $d_{sp}=2$ spatial
dimension if there is a $Z_{2k+1}$ symmetry.  The constructed $2k+1$ gapped
fermionic SPT phases actually correspond to the $2k+1$ gapped bosonic SPT
phases with $Z_{2k+1}$ symmetry [since $\cH^2(Z_{2k+1},\Z_2)=\Z_1$ and there is no
nontrivial graded structure].

\subsection{Calculate $\fH^d[Z_{2}\times Z_2^f,U(1)]$}

Now, let us choose $G_f=Z_{2}\times Z_2^f$.  Following the first step in
section \ref{gen}, we find that $\cH^{d-1}(Z_{2}, \Z_2)=\Z_2$ and there are
two graded structures in all dimensions.  For each $d$D-graded structure, we
can choose a $u^g_{d-1}(g_0,...,g_{d-1})$ that satisfies
$u^g_{d-1}(g_0,...,g_{d-1})=1$ for $g\in G_b$
and $u^{P_f}_{d-1}(g_0,...,g_{d-1})=(-)^{n_{d-1}(g_0,...,g_{d-1})}$.

Following the second step in section \ref{gen}, we note that the trivial graded
structure is given by $n_{d-1}(g_0,...,g_{d-1})=0$.  The nontrivial graded
structure is given by
\begin{align}
 n_{d-1}(e,o,e,...)= n_{d-1}(o,e,o,...)= 1, \ \ \
\text{ others}=0,
\end{align}
where $e$ represents the identity element in $Z_{2}$ and $o$ represents the
other element in $Z_{2}$.

For $d=1,2$, $\cH^{d-1}(Z_{2}, \Z_2)=\Z_2$.  For each graded structure
$n_{d-1}(g_0,...,g_{d-1})$, the corresponding $f_{d+1}(g_0,...,g_{d+1}) =0$.  Thus
$B\cH^{d-1}(Z_{2}, \Z_2)=\Z_2$.

For $d=3$, $\cH^2(Z_{2}, \Z_2)=\Z_2$.
For the nontrivial 3D-graded structure,
$ n_{2}(e,o,e)= n_2(o,e,o)= 1$ and others $n_2=0$,
the corresponding $f_4$ has a form
\begin{align}
 f_4(e,o,e,o,e)= f_4(o,e,o,e,o)= 1, \ \ \
\text{ others}=0.
\end{align}
One can show that $(-)^{f_4(g_0,...,g_4)}$ is a cocycle in $\cZ^4[Z_{2},
U(1)]$.  Since $\cH^4[Z_{2}, U(1)]=\Z_1$, $(-)^{f_4(g_0,...,g_4)}$ is also a
coboundary in $\cB^4[Z_{2}, U(1)]$.  Therefore, $B\cH^{2}(Z_{2}, \Z_2)=\Z_2$.

For $d=4$, $\cH^3(Z_{2}, \Z_2)=\Z_2$.  For the nontrivial 4D-graded
structure, $ n_3(e,o,e,o)= n_3(o,e,o,e)= 1$ and others $n_3=0$, the
corresponding $f_5$ has a form
\begin{align}
 f_5(e,o,e,o,e,o)= f_5(o,e,o,e,o,e)= 1, \ \ \
\text{ others}=0.
\end{align}
One can show that $(-)^{f_5(g_0,...,g_5)}$ is a cocycle in $\cZ^5[Z_{2},
U(1)]$.  Since $\cH^5[Z_{2}, U(1)]=\Z_{2}$ and $(-)^{f_5(g_0,...,g_5)}$
corresponds to a nontrivial cocycle in $\cH^5[Z_{2}, U(1)]=\Z_{2}$,
we find that $B\cH^{3}(Z_{2}, \Z_2)=\Z_1$.

Following the third step in section \ref{gen}, we can show that the elements in
$\fH^d[Z_{2}\times Z_2^f,U(1)]$ can be labeled by $\cH^d[Z_{2}\times
Z_2^f,U(1)]\times B\cH^{d-1}(Z_{2}, \Z_2)$.

Following the fourth step in section \ref{gen}, we would like to show that the above
labeling is one-to-one.  For $d=1,2,3$, $f_{d}(g_0,...,g_{d})=0$.  Thus the
labeling is one-to-one.  For $d=4$, a 3D-graded structure in
$B\cH^2(Z_{2},\Z_2)$ gives rise to a $f_4(g_0,...,g_4)$.  Since
$(-)^{f_4(g_0,...,g_4)}$ is a cocycle in $\cZ^4[Z_{2},U(1)]$ and since
$\cH^4[Z_{2},U(1)]=\Z_1$, we find that $(-)^{f_4(g_0,...,g_4)}$ is also a
coboundary in $\cB^4[Z_{2},U(1)]$ and the labeling is one-to-one.

So, using
\begin{align}
 \cH^d[ Z_n,U(1)] =
\begin{cases}
\Z_1,  &  d=0 \text{ mod }2,\ d>0,\\
\Z_n, &  d=1 \text{ mod }2,\\
\end{cases}
\end{align}
we find that\cite{CGL1172}
\begin{align}
\label{fHZevenU1}
 \fH^d[ Z_{2}\times Z_2^f,U(1)] =
\begin{cases}
\Z_2^2,  &  d=1 ,\\
\Z_2,  &  d=2 ,\\
\Z_4,  &  d=3 ,\\
\Z_1,  &  d=4 .\\
\end{cases}
\end{align}
Here the $\Z_4$ group structure for $2+1$D case comes from the fact that two copies of a fermionic SPT phase gives rise to the nontrival bosonic SPT phase, see Eq.(\ref{n2Z2B}) and Eq.(\ref{n2Z2F}).

$\fH^1[ Z_{2}\times Z_2^f,U(1)]=\Z_2^2$ implies that there are $4$  possible
fermionic SPT phases in $d_{sp}=0$ spatial dimension if there is a $Z_{2}$
symmetry.  $2$ of them have an even number of fermions and the other $2$ of them
have an odd number of fermions.  The $2$ phases with an even number of fermions are
separated by the $2$ possible $Z_{2}$ quantum numbers.  The $2$ phases with odd
number of fermions are also separated by the $2$ possible $Z_{2}$ quantum
numbers.

$\fH^2[ Z_{2}\times Z_2^f,U(1)]=\Z_2$ implies that there are two gapped
fermionic SPT phases in $d_{sp}=1$ spatial dimension if there is a $Z_{2}$
symmetry. We can use the Jordan-Wigner transformation to map 1D fermion
systems with $Z_{2}$ symmetry to 1D boson systems with $Z_{2}\times Z_2^f$
symmetry.  Since $\cH^2[Z_{2}\times Z_2, U(1)]=\Z_2$, 1D  boson systems have
two phases that do not break the $Z_{2}\times Z_2^f$ symmetry.  1D  boson
systems also have another phase that does not break the $Z_{2}$ symmetry: the
phase that breaks the $Z_2^f$ symmetry.  The fermionic SPT phases have no
intrinsic topological order and correspond to the bosonic phases that does not
break the $Z_2^f$ symmetry (and the $Z_{2}$ symmetry).  Thus 1D fermion
systems with $Z_{2}$ symmetry indeed have two fermionic SPT phases.

$\fH^3[ Z_{2}\times Z_2^f,U(1)]=\Z_4$ implies that our
construction gives rise to $4$ gapped fermionic SPT phases in $d_{sp}=2$
spatial dimension if there is a $Z_{2}$ symmetry.

\subsection{Calculate $\fH^d[Z_{2k}\times Z_2^f,U(1)]$}

Similarly, let us choose $G_f=Z_{2k}\times Z_2^f$.  Following the first step in
section \ref{gen}, we find that $\cH^{d-1}(Z_{2k}, \Z_2)=\Z_2$ and there are
two graded structures in all dimensions.  For each $d$D-graded structure, we
can choose a $u^g_{d-1}(g_0,...,g_{d-1})$ that satisfies
$u^g_{d-1}(g_0,...,g_{d-1})=1$ for $g\in G_b$
and $u^{P_f}_{d-1}(g_0,...,g_{d-1})=(-)^{n_{d-1}(g_0,...,g_{d-1})}$.

Following the second step in section \ref{gen}, we note that the trivial graded
structure is given by $n_{d-1}(g_0,...,g_{d-1})=0$.  The nontrivial graded
structure is given by
\begin{align}
 n_{d-1}(e,o,e,...)= n_{d-1}(o,e,o,...)= 1, \ \ \
\text{ others}=0,
\end{align}
where $e$ represents the even elements in $Z_{2k}$ and $o$ represents the odd
elements in $Z_{2k}$.

For $d=1,2$, $\cH^{d-1}(Z_{2k}, \Z_2)=\Z_2$.  For each graded structure
$n_{d-1}(g_0,...,g_{d-1})$, the corresponding $f_{d+1}(g_0,...,g_{d+1}) =0$.  Thus
$B\cH^{d-1}(Z_{2k}, \Z_2)=\Z_2$.

For $d=3$, $\cH^2(Z_{2k}, \Z_2)=\Z_2$.
For the nontrivial 3D-graded structure,
$ n_{2}(e,o,e)= n_2(o,e,o)= 1$ and others $n_2=0$,
the corresponding $f_4$ has a form
\begin{align}
 f_4(e,o,e,o,e)= f_4(o,e,o,e,o)= 1, \ \ \
\text{ others}=0.
\end{align}
One can show that $(-)^{f_4(g_0,...,g_4)}$ is a cocycle in $\cZ^4[Z_{2k},
U(1)]$.  Since $\cH^4[Z_{2k}, U(1)]=\Z_1$, $(-)^{f_4(g_0,...,g_4)}$ is also a
coboundary in $\cB^4[Z_{2k}, U(1)]$.  Therefore, $B\cH^{2}(Z_{2k}, \Z_2)=\Z_2$.

For $d=4$, $\cH^3(Z_{2k}, \Z_2)=\Z_2$.  For the nontrivial 4D-graded
structure, $ n_3(e,o,e,o)= n_3(o,e,o,e)= 1$ and others $n_3=0$, the
corresponding $f_5$ has a form
\begin{align}
 f_5(e,o,e,o,e,o)= f_5(o,e,o,e,o,e)= 1, \ \ \
\text{ others}=0.
\end{align}
One can show that $(-)^{f_5(g_0,...,g_5)}$ is a cocycle in $\cZ^5[Z_{2k},
U(1)]$.  Since $\cH^5[Z_{2k}, U(1)]=\Z_{2k}$ and $(-)^{f_5(g_0,...,g_5)}$
corresponds to a nontrivial cocycle in $\cH^5[Z_{2k}, U(1)]=\Z_{2k}$,
we find that $B\cH^{3}(Z_{2k}, \Z_2)=\Z_1$.

Following the third step in section \ref{gen}, we can show that the elements in
$\fH^d[Z_{2k}\times Z_2^f,U(1)]$ can be labeled by $\cH^d[Z_{2k},U(1)]\times B\cH^{d-1}(Z_{2k}, \Z_2)$.

Following the fourth step in section \ref{gen}, we would like to show that the above
labeling is one-to-one.  For $d=1,2,3$, $f_{d}(g_0,...,g_{d})=0$.  Thus the
labeling is one-to-one.  For $d=4$, a 3D-graded structure in
$B\cH^2(Z_{2k},\Z_2)$ gives rise to a $f_4(g_0,...,g_4)$.  Since
$(-)^{f_4(g_0,...,g_4)}$ is a cocycle in $\cZ^4[Z_{2k},U(1)]$ and since
$\cH^4[Z_{2k},U(1)]=\Z_1$, we find that $(-)^{f_4(g_0,...,g_4)}$ is also a
coboundary in $\cB^4[Z_{2k},U(1)]$ and the labeling is one-to-one.

So, using
\begin{align}
 \cH^d[ Z_n,U(1)] =
\begin{cases}
\Z_1,  &  d=0 \text{ mod }2,\ d>0,\\
\Z_n, &  d=1 \text{ mod }2,\\
\end{cases}
\end{align}
we find that\cite{CGL1172}
\begin{align}
\label{fHZevenU1a}
 \fH^d[ Z_{2k}\times Z_2^f,U(1)] =
\begin{cases}
\Z_{2k}\times \Z_2 ,  &  d=1 ,\\
\Z_2,  &  d=2 ,\\
\Z_{4k},  &  d=3 ,\\
\Z_1,  &  d=4 .\\
\end{cases}
\end{align}
Here the group structure $\Z_{4k}$ could be derived from a topological field
theory approach.\cite{CG}
We note that it is a central extension of $\Z_{2k}$(which classifies bosonic SPT phases) over a $\Z_2$ graded structure.

$\fH^1[ Z_{2k}\times Z_2^f,U(1)]=\Z_{2k}\times \Z_2$ implies that there are
$4k$  possible fermionic SPT phases in $d_{sp}=0$ spatial dimension if there is
a $Z_{2k}$ symmetry.  $2k$ of them have an even number of fermions and the other
$2k$ of them have an odd number of fermions.  The $2k$ phases are separated by the
$2k$ possible $Z_{2k}$ quantum numbers.

$\fH^2[ Z_{2k}\times Z_2^f,U(1)]=\Z_2$ implies that there are two gapped
fermionic SPT phases in $d_{sp}=1$ spatial dimension if there is a $Z_{2k}$
symmetry. We can use the Jordan-Wigner transformation to map 1D fermion systems
with $Z_{2k}$ symmetry to 1D boson systems with $Z_{2k}\times Z_2^f$ symmetry.
The fermionic SPT phases have no intrinsic topological order and correspond to
the bosonic phases that does not break the $Z_2^f$ symmetry (and the $Z_{2k}$
symmetry).  Since $\cH^2[Z_{2k}\times Z_2, U(1)]=\Z_2$, 1D  boson systems have
two phases that do not break the $Z_{2k}\times Z_2^f$ symmetry.  Thus 1D
fermion systems with $Z_{2k}$ symmetry indeed have two fermionic SPT phases.

$\fH^3[ Z_{2k}\times Z_2^f,U(1)]=\Z_{4k}$ implies that our
construction gives rise to $4k$ gapped fermionic SPT phases in $d_{sp}=2$
spatial dimension if there is a $Z_{2k}$ symmetry.

\subsection{Calculate $\fH^d[Z_2^T\times Z_2^f,U_T(1)]$}

Now, let us choose $G_f=Z_2^T\times Z_2^f$.  Since $\cH^{d-1}(Z_2^T,
\Z_2)=\Z_2$, there are two graded structures in all dimensions.  The trivial
graded structure is given by $n_{d-1}(g_0,...,g_{d-1})=0$.  The nontrivial
graded structure is given by
\begin{align}
 n_{d-1}(e,o,e,...)= n_{d-1}(o,e,o,...)= 1, \ \ \
\text{ others}=0,
\end{align}
where $e$ is the identity element in $Z_2^T$ and $o$ is the nontrivial element
in $Z_2^T$.  For each $d$D-graded structure, we can choose a
$u^g_{d-1}(g_0,...,g_{d-1})$ that satisfies $u^g_{d-1}(g_0,...,g_{d-1})=1$ for
$g\in G_b$.

For $d=1,2$, $\cH^{d-1}(Z_2^T, \Z_2)=\Z_2$.  For each graded structure
$n_{d-1}(g_0,...,g_{d-1})$, the corresponding $f_{d+1}(g_0,...,g_{d+1}) =0$,
and the equivalent classes of $\v v_d$ are described by $\cH^d[Z_2^T,U_T(1)]$,
Therefore $\fH^1[Z_2^T\times Z_2^f,U_T(1)]=\Z_2$ and $\fH^2[Z_2^T\times
Z_2^f,U_T(1)]=\Z_4$.

For $d=3$, $\cH^2(Z_2^T, \Z_2)=\Z_2$.  For the nontrivial 3D-graded structure,
$ n_{2}(e,o,e)= n_2(o,e,o)= 1$ and others $n_2=0$, the corresponding $f_4$ has
a form
\begin{align}
 f_4(e,o,e,o,e)= f_4(o,e,o,e,o)= 1, \ \ \
\text{ others}=0.
\end{align}
One can show that $(-)^{f_4(g_0,...,g_4)}$ is a cocycle in $\cZ^4[Z_2^T,
U_T(1)]$.  Since $\cH^4[Z_2^T, U_T(1)]=\Z_2$ and $(-)^{f_4(g_0,...,g_4)}$ is a
nontrivial cocycle in $\cH^4[Z_2^T, U_T(1)]$,  so for the nontrivial
3D-graded structure, \eqn{vccy} for $\v v_d$ has no solutions.  For the trivial
3D-graded structure, the equivalent classes of $\v v_d$ are described by
$\cH^3[Z_2^T,U_T(1)]$.  Thus $\fH^3[Z_2^T\times Z_2^f,U_T(1)]=\Z_1$.

For $d=4$, $\cH^3(Z_2^T, \Z_2)=\Z_2$.  For the nontrivial 4D-graded structure,
$ n_3(e,o,e,o)= n_3(o,e,o,e)= 1$ and others $n_3=0$, the corresponding $f_5$
has a form
\begin{align}
 f_5(e,o,e,o,e,o)= f_5(o,e,o,e,o,e)= 1, \ \ \
\text{ others}=0.
\end{align}
One can show that $(-)^{f_5(g_0,...,g_5)}$ is a cocycle in $\cZ^5[Z_2^T,
U_T(1)]$.  Since $\cH^5[Z_2^T, U_T(1)]=\Z_1$, $(-)^{f_5(g_0,...,g_5)}$ is also a
coboundary in $\cB^5[Z_2^T, U_T(1)]$. So for the nontrivial 4D-graded
structure, the equivalent classes of $\v v_d$ are labeled by
$\cH^4[Z_2^T,U_T(1)]=\Z_2$. On the other hand, if $(-)^{f_4(g_0,...,g_4)}$ is a
nontrivial cocycle in $\cB^4[Z_2^T,U_T(1)]$, then the labeling is \emph{not} one-to-one.
In fact, $(-)^{f_4(g_0,...,g_4)}$ is a nontrivial cocycle in $\cB^4[Z_2^T,U_T(1)]$ which implies that the
nontrivial bosonic SPT phase protected by $T^2=1$ time reversal symmetry can be connected to a
trivial direct product state or an atomic insulator state via interacting fermion systems. (We note
that such a "collapsing" can happen because local unitary transformations in interacting fermion
systems have a much more general meaning\cite{GWW1017}.)
Thus $\fH^4[Z_2^T\times Z_2^f,U_T(1)]=\Z_2$. To summarize
\begin{align}
\label{fHZTU1}
 \fH^d[ Z_2^T\times Z_2^f,U_T(1)] =
\begin{cases}
\Z_2,  &  d=1 ,\\
\Z_4,  &  d=2 ,\\
\Z_1,  &  d=3 ,\\
\Z_2,  &  d=4 .\\
\end{cases}
\end{align}

$\fH^1[ Z_2^T\times Z_2^f,U_T(1)]=\Z_2$ implies that there are two  possible
fermionic SPT phases in $d_{sp}=0$ spatial dimension if there is a $Z_2^T$
symmetry.  One has even number of fermions and the other one has odd number of
fermions.

$\fH^2[ Z_2^T\times Z_2^f,U_T(1)]=\Z_4$ implies that there are four gapped
fermionic SPT phases in $d_{sp}=1$ spatial dimension if there is a $Z_2^T$
symmetry. We can use the Jordan-Wigner transformation to map 1D fermion systems
with $Z_2^T$ symmetry to 1D boson systems with $Z_2^T\times Z_2^f$ symmetry.
Since $\cH^2[Z_2^T\times Z_2, U_T(1)]=\Z_2\times \Z_2$, 1D  boson systems have
four phases that do not break the $Z_2^T\times Z_2^f$ symmetry.  These four
bosonic phases correspond to the four gapped fermionic SPT phases described by
$\fH^2[ Z_2^T\times Z_2^f,U_T(1)]=\Z_4$.

Note that the stacking operation of 1D systems and the Jordan-Wigner
transformation of the 1D systems do not commute.  The state obtained by
stacking two 1D fermion systems then performing the  Jordan-Wigner
transformation is different from the state obtained by performing the
Jordan-Wigner transformation on each 1D fermion systems then stacking the two
resulting boson systems.  This is why, although $\fH^2[ Z_2^T\times
Z_2^f,U_T(1)]$ and  $\cH^2[Z_2^T\times Z_2, U_T(1)]$ classify the same four
states, their Abelian group structure is different. The Abelian group
multiplication operation in  $\fH^2[ Z_2^T\times Z_2^f,U_T(1)]$ corresponds to
stacking 1D fermion systems, while the  Abelian group multiplication operation
in $\cH^2[Z_2^T\times Z_2, U_T(1)]$ corresponds to stacking 1D boson systems.
Both types of group multiplication operations give rise to Abelian group
structures, but they give rise to different Abelian groups.

1D boson systems also have four phases that do break the $Z_2^f$ symmetry: two
do not break the time-reversal $T$ symmetry and two do not break the $TP_f$
symmetry.  The fermionic SPT phases have no intrinsic topological order and
correspond to the bosonic phases that do not break the $Z_2^f$ symmetry (and
the $Z_2^T$ symmetry).  Thus 1D fermion systems with $Z_2^T$ symmetry indeed
have only four fermionic SPT phases.

$\fH^3[ Z_2^T\times Z_2^f,U_T(1)]=\Z_1$ implies that our construction only
gives rise to one trivial gapped fermionic SPT phases in $d_{sp}=2$ spatial
dimension if there is a $Z_2^T$ symmetry. $\fH^4[ Z_2^T\times
Z_2^f,U_T(1)]=\Z_2$ implies that our construction gives rise to one
nontrivial gapped fermionic SPT phases in $d_{sp}=3$ spatial dimension if
there is a $Z_2^T$ symmetry.  We note
that, using non-interacting fermions, we cannot construct any nontrivial
gapped phases with  $Z_2^T$ symmetry (the $T^2=1$ time-reversal symmetry).

\section{Properties of the constructed SPT states}

\subsection{Equivalence between ground state wave functions}
\label{eqwav}

In Section \ref{idwav}, we have demonstrated how to construct an ideal ground
state wave function $\Phi( \{g_i\}, \{ n_{ij...k}\})$ in $d$ spatial
dimension from a $(d+1)$-cocycle $(\nu_{d+1},n_d,u^g_d)$. However, a
natural question is as follows: How do we know these wave functions describe
the same fermionic SPT phases or not? Here we will address this important
question based on the new mathematical structure - group
super-cohomology class invented in Appendix \ref{fcoh}.

\subsubsection{A special equivalent relation}

Let us choose arbitrary $(\nu_d,n_{d-1},u^g_{d-1})$ in one lower dimension.
Here $(\nu_d,n_{d-1},u^g_{d-1})$ is not a $d$-cocycle. They just satisfy the
symmetry condition \eqn{symmcnd}.  In Appendix \ref{fcoh}, we have shown that
$(\t \nu_{d+1},\t n_d,\t u^g_d)$ obtained from $(\nu_{d+1},n_d,u^g_d)$ through
\eqn{lccyeq} is also a $(d+1)$-cocycle.  Here we will assume that $n_{d-1}=0$
and $u^g_{d-1}=1$, and the resulting
\begin{align}
 \t \nu_{d+1}=\nu_{d+1} \del \nu_d, \ \ \
\t n_d=n_d,\ \ \  \t u^g_d=u^g_d,
\end{align}
is a $(d+1)$-cocycle.  The new  $(d+1)$-cocycle
$(\t \nu_{d+1},\t n_d,\t u^g_d)$ gives rise to a new ideal wave function $\Phi'(
\{g_i\}, \{ n_{ij...k}\})$.  Should we view the two wave functions
$\Phi( \{g_i\}, \{ n_{ij...k}\})$ and $\Phi'( \{g_i\}, \{
n_{ij...k}\})$ as wave functions for different SPT phases?

{}From the definition of $\del\nu_d$ \eq{delnudef}, we can show that the two wave
functions $\Phi( \{g_i\}, \{ n_{ijk}\})$ and $\Phi'( \{g_i\}, \{ n_{ijk}\})$
are related through
\begin{align}
&
 \Phi'( \{g_i\}, \{ n_{ijk}\})
=\Phi( \{g_i\}, \{ n_{ijk}\}) \times
\nonumber\\
&\ \ \ \
\prod_{(ij...k)}
(-)^{f_{d+1}(g_0,g_i,g_j,...,g_k)}
\nu_d^{s_{ij...k}}(g_i,g_j,...,g_k),
\end{align}
where $\prod_{(ij...k)}$ is the product over all the $d$-simplexes
that form the $d$ dimensional space [in our (2+1)D example,
$\prod_{(ij...k)}$ is the product over all triangles]. Also
$s_{ij...k}=\pm 1$ depending on the orientation of the $d$-simplex
$(ij...k)$.  Since $\nu_d^{s_{ij...k}}(g_i,g_j,...,g_k)$ is a pure
$U(1)$ phase that satisfies
$\nu_d^{s_{ij...k}}(gg_i,gg_j,...,gg_k)=\nu_d^{s_{ij...k}}(g_i,g_j,...,g_k)$,
it represents a bosonic \emph{symmetric} local unitary (LU)
transformation. Therefore the new wave function $\Phi'$ induced by
$(\nu_d\neq 0,n_{d-1}=0,u^g_{d-1}=1)$ belongs to the same SPT phase as the
original wave function $\Phi$.

Also since $f_{d+1}$ for $d<3$ is always zero.  Therefore, below 3 spatial
dimensions, the new wave function $\Phi'$ belongs to the same SPT phase as the
original wave function $\Phi$.

In 3 spatial dimensions and above ($d\geq 3$), if we choose
$(\nu_d=0,n_{d-1}\neq 0)$, the new induced wave function $\Phi'$
will differ from the original wave function $\Phi$ by a phase factor
$ \prod_{(ij...k)} (-)^{f_{d+1}(g_0,g_i,g_j,...,g_k)} $.
$(-)^{f_{d+1}(g_0,g_i,g_j,...,g_k)}$ represents a LU transformation,
and the two wave functions, $\Phi'$ and $\Phi$, have the same
intrinsic topological order. But since
$(-)^{f_{d+1}(g_0,gg_i,gg_j,...,gg_k)} \neq
(-)^{f_{d+1}(g_0,g_i,g_j,...,g_k)}$,
$(-)^{f_{d+1}(g_0,g_i,g_j,...,g_k)}$ does not represent a bosonic
symmetric LU transformation.  So we do not know whether $\Phi'$ and
$\Phi$ belong to the same fermionic SPT phase or not.

We would like to point out that in order for $\Phi'$ and $\Phi$ to
belong to the same \emph{fermionic} SPT phase, the two wave
functions can only differ by a \emph{fermionic} symmetric LU
transformation (which is defined in \Ref{GWW1017}).  The
\emph{bosonic} symmetric LU transformations are a subset of
\emph{fermionic} symmetric LU transformations.  So even though
$(-)^{f_{d+1}(g_0,g_i,g_j,...,g_k)}$ is not a \emph{bosonic}
symmetric LU transformation, we do not know whether it is a
\emph{fermionic} symmetric LU transformation or not.
The structure of group super-cohomology theory developed in
Appendix \ref{fcoh} shows that $f_{d+1}(g_0,g_i,g_j,...,g_k)$ is a
d+1-cocycle with $\mathbb{Z}_2$ coefficient, suggesting the sign
factor $(-)^{f_{d+1}(g_0,g_i,g_j,...,g_k)}$ can be generated by
\emph{fermionic} symmetric LU transformation.

\subsubsection{Generic equivalent relations}

In the above, we discussed a special case of $(\t \nu_{d+1},\t n_d, \t u^g_d)$
where $\t n_d$ and $\t u^g_d$ are the same as $n_d$ and $u^g_d$.  However, the
equivalent class of group super-cohomology has a more complicated
structure.  In Appendix \ref{fcoh}, we have shown that $(\t \nu_{d+1},\t n_d,\t
u^g_d)$ obtained from:
\begin{align}
 \t \nu_{d+1}=\nu_{d+1} (-)^{g_d}, \
\t n_d=n_d+\delta m_{d-1}^\prime,\
\t u^g_d= u^g_d
,
\end{align}
(with $g_d(g_0,\cdots,g_d)={(\v g_d)}_{(g_0,\cdots,g_d)}$ mod $2$) also belongs
to the same equivalent class of group super-cohomology. We note that here the graded
structure $n_d$ changes into $\t n_d$. It would be much harder to understand
why the corresponding new wave function still describes the same fermionic SPT
phase. Let us consider the $2+1$D wave function on a sphere as a simple
example. Since the wave function derived from the fixed point amplitude Eq.
(\ref{2Damplitude}) is a fixed point wave function with zero correlation
length, it is enough to only consider a minimal wave function with 4 sites on a
sphere(labeled as $1,2,3,4$). Such a minimal wave functions will contain 4 triangles $123,124,134$ and $234$. From the
definition of $\t \nu_d$, we can show that the two wave functions $\Phi(
\{g_i\}, \{ n_{ijk}\})$ and $\t \Phi( \{g_i\}, \{ \t n_{ijk}\})$ are related
through $\t\Phi( \{g_i\}, \{ n^\prime_{ijk}\})=(-)^{g_4(g_1,g_2,g_3,g_4)}\Phi(
\{g_i\}, \{ n_{ijk}\})$. To see this explicitly, let us construct
the ground state wave function by adding one more point $0$ inside the sphere. In this way, the
wave function can be formally expressed as:
\begin{align}
&\Psi(g_1,g_2,g_3,g_4,\theta_{234},\theta_{124},\bar\theta_{134},\bar\theta_{123})
\nonumber\\=& (-)^{m_1(g_1,g_3)}\int\mathcal{V}_3^-(g_0,g_1,g_2,g_3)\mathcal{V}_3^+(g_0,g_1,g_2,g_4)
\nonumber\\ &\times
\mathcal{V}_3^-(g_0,g_1,g_3,g_4)\mathcal{V}_3^+(g_0,g_2,g_3,g_4)
\nonumber\\=& (-)^{m_1(g_1,g_3)}\mathcal{V}_3^+(g_1,g_2,g_3,g_4)
\nonumber\\=&\nu_3(g_1,g_2,g_3,g_4)\times\nonumber\\ &\theta_{234}^{n_2(g_2,g_3,g_4)}\theta_{124}^{n_2(g_1,g_2,g_4)}
\bar\theta_{234}^{n_2(g_2,g_3,g_4)}\bar\theta_{123}^{n_2(g_1,g_2,g_3)}.
\end{align}
Again, the symbol $\int$ is defined as integrating over the internal Grassmann variables with respecting to the weights $m_1$ on the
internal links(see Eq.(\ref{Gint})).
Similarly, the wave function corresponding to a different solution $(\t \nu_{d+1},\t n_d,\t
u^g_d)$ takes a form:
\begin{align}
&\t\Psi(g_1,g_2,g_3,g_4,\theta_{234},\theta_{124},\bar\theta_{134},\bar\theta_{123})
\nonumber\\=&\nu_3(g_1,g_2,g_3,g_4)(-)^{g_4(g_1,g_2,g_3,g_4)}\times\nonumber\\ &\theta_{234}^{n'_2(g_2,g_3,g_4)}\theta_{124}^{n'_2(g_1,g_2,g_4)}
\bar\theta_{234}^{n'_2(g_2,g_3,g_4)}\bar\theta_{123}^{n'_2(g_1,g_2,g_3)}.
\end{align}

We note that $g_4(gg_1,gg_2,gg_3,gg_4)=g_4(g_1,g_2,g_3,g_4)$ implies the phase factor
$(-)^{g_4(g_1,g_2,g_3,g_4)}$ is a \emph{symmetric} LU transformation. However,
we still do not know whether the two wave functions with different patterns of fermion number $\{ \t
n_{ijk}\}$ and $\{ n_{ijk}\}$ describe the same fermionic SPT phase or not.

Indeed, we find that up to some \emph{symmetric} phase factors, the state
$|\Phi\rangle$ and $|\t\Phi\rangle$ can be related through a \emph{symmetric}
LU transformation $|\t \Phi\rangle\sim \hat U^\prime|\Phi\rangle$, where
\begin{widetext}
\begin{align}
 \hat U^\prime=&
\prod_{(0ij)}
 c_{(0ij)}^{m_1^\prime(g_i,g_j)}
\bar c_{(0ij)}^{m_1^\prime(g_i,g_j)}
\prod_{\bigtriangleup}
 c_{(0ij)}^{\dag m_1^\prime(g_i,g_j)}
 c_{(0jk)}^{\dag m_1^\prime(g_j,g_k)}
\bar c_{(0ik)}^{\dag m_1^\prime(g_i,g_k)}
 \left(c_{(ijk)}^\dag\right)^{\t n_2(g_i,g_j,g_k)-n_2(g_i,g_j,g_k)}
\times
\nonumber\\
&
\prod_{\bigtriangledown}
 \left(c_{(ijk)}^\dag\right)^{\t n_2(g_i,g_j,g_k)-n_2(g_i,g_j,g_k)}
 c_{(0ik)}^{\dag m_1^\prime(g_i,g_k)}
\bar c_{(0jk)}^{\dag m_1^\prime(g_j,g_k)}
\bar c_{(0ij)}^{\dag m_1^\prime(g_i,g_j)},
\end{align}
\end{widetext}
with $\t n_2(g_i,g_j,g_k)=
n_2(g_i,g_j,g_k)+m_1^\prime(g_i,g_j)+m_1^\prime(g_j,g_k)+m_1^\prime(g_k,g_i)$.
We note that both $n_2$ and $m_1^\prime$ are invariant under symmetry transformation.  Here
$\left(c_{(ijk)}^\dag\right)^{-1}$ is defined as
$\left(c_{(ijk)}^\dag\right)^{-1}=c_{(ijk)}$.
$\sim$ means equivalent up to some symmetric sign factors which will arise when
we reorder the fermion creation/annihilation operators in $\hat U'$. We also
note that $\hat U'$ is a generalized LU transformation which only has non-zero action
on the subspace with fixed fermion occupation pattern $n_2$. The above
discussions can be generalized into any dimension. Thus we have shown the wave
function constructed from the above $(\t \nu_{d+1},\t n_d, \t u_2^g)$ describes the same fermionic
SPT phase.

Finally, we can consider the more general case where $u^g_d$ also changes into $\t
u^g_d$. It is very easy to see that such changes in the wave function can be
realized by symmetric LU transformation since those $u^g_d$ evolving $g_0$
are all canceled in the wave function.

In conclusion, we have shown that $(\nu_{d+1},n_d,u^g_d)$ and $(\t\nu_{d+1},\t
n_d,\t u^g_d)$ belonging to the same group super-cohomology class will give rise to
fixed point wave functions describing the same fermionic SPT phase.

\subsection{Symmetry transformations and generic SPT states}

In Section \ref{notop}, we have constructed an ideal fermion SPT state $|\Psi\> = \hat U
|\Phi_0\>$ using the the fermionic LU transformation $\hat U$ constructed from
a cocycle $(\nu_3,n_2)$.  In this section, we are going to discuss how to
construct a more generic SPT state $|\Psi'\>$ that is in the same phase as the
ideal state $|\Psi\>$.

Clearly, $|\Psi'\>$ and $|\Psi\>$ have the same symmetry.  So to construct
$|\Psi'\>$, we need to first discuss how symmetry transformation changes under the
fermionic LU transformation $\hat U$.

The symmetry $G_b$ that act on $|\Psi\>$ is represented by the following (anti)
unitary operators
\begin{align}
\label{bsymm}
 \hat W(g)  = &\sum_{\{g_i\}}
\prod_{\bigtriangledown} [u_2^{g}(g_i,g_j,g_k)]^{-1}
\prod_{\bigtriangleup} u_2^{g}(g_i,g_j,g_k) \times
\nonumber\\
& \ \ \
\prod_i |g_i\>\<gg_i|\  K^{\frac{1-s(g)}{2}},
\end{align}
where $\hat W_i$ is the symmetry transformation acting on a single site $i$, and $K$ is the anti unitary operator
\begin{align}
K\imth K^{-1}=-\imth,\ \ \ \
K c K^{-1}=c, \ \ \
K c^\dag K^{-1}=c^\dag .
\end{align}
We find that
\begin{widetext}
\begin{align}
& \hat W^0(g) = \hat U^\dag \hat W(g) \hat U =
\sum_{\{g_i\}} \prod_i |g_i\>\<gg_i|
\prod_{\bigtriangledown} [u_2^{g}(g_i,g_j,g_k)]^{-1}
\prod_{\bigtriangleup} u_2^{g}(g_i,g_j,g_k)
\times
\\
&
\prod_{\bigtriangledown}
 c_{(ijk)}^{n_2(g_i,g_j,g_k)}
 c_{(0ik)}^{n_2(g_0,g_i,g_k)}
\bar c_{(0jk)}^{n_2(g_0,g_j,g_k)}
\bar c_{(0ij)}^{n_2(g_0,g_i,g_j)}
\prod_{\bigtriangleup}
 c_{(0ij)}^{n_2(g_0,g_i,g_j)}
 c_{(0jk)}^{n_2(g_0,g_j,g_k)}
\bar c_{(0ik)}^{n_2(g_0,g_i,g_k)}
\bar c_{(ijk)}^{n_2(g_i,g_j,g_k)}
\times
\nonumber\\
&
\prod_{\bigtriangleup} \nu_3(g_0,g_i, g_j, g_k)
\prod_{\bigtriangledown} \nu_3^{-1}(g_0,g_i, g_j, g_k)
\prod_{(0ij)}
 c_{(0ij)}^{\dag n_2(g_0,g_i,g_j)}
\bar c_{(0ij)}^{\dag n_2(g_0,g_i,g_j)}
\times
\nonumber\\
&  
\prod_{(0ij)}
 c_{(0ij)}^{n_2(g_0,gg_i,gg_j)}
\bar c_{(0ij)}^{n_2(g_0,gg_i,gg_j)}
\prod_{\bigtriangleup} [\nu_3^{-1}(g_0,gg_i, gg_j, gg_k)]^{s(g)}
\prod_{\bigtriangledown} [\nu_3(g_0,gg_i, gg_j,gg_k)]^{s(g)}
K^{\frac{1-s(g)}{2}} \times
\nonumber\\
&
\prod_{\bigtriangleup}
 c_{(0ij)}^{\dag n_2(g_0,gg_i,gg_j)}
 c_{(0jk)}^{\dag n_2(g_0,gg_j,gg_k)}
\bar c_{(0ik)}^{\dag n_2(g_0,gg_i,gg_k)}
\bar c_{(ijk)}^{\dag n_2(gg_i,gg_j,gg_k)}
\prod_{\bigtriangledown}
 c_{(ijk)}^{\dag n_2(gg_i,gg_j,gg_k)}
 c_{(0ik)}^{\dag n_2(g_0,gg_i,gg_k)}
\bar c_{(0jk)}^{\dag n_2(g_0,gg_j,gg_k)}
\bar c_{(0ij)}^{\dag n_2(g_0,gg_i,gg_j)}
\nonumber
\end{align}
\end{widetext}
Using the fact that $\hat U$ is independent of $g_0$, we can change the $g_0$
in the second half of the right-hand-side of the above expression to $gg_0$.
Then we can use the symmetry condition \eq{symmcnd} on $(\nu_3,n_2,u^g_2)$ to
show that the above is reduced to
\begin{align}
 \hat W^0(g)=\hat U^\dag \hat W(g) \hat U =\sum_{\{g_i\}} \prod_i |g_i\>\<gg_i| K^{\frac{1-s(g)}{2}}.
\end{align}
\emph{in the subspace with no fermions.}
Note that $ \hat W^0(g)$ acts on $|\Phi_0\>$. If we choose a new no-fermion
state $|\Psi'_0\>$
that is symmetric under the symmetry $G_b$: $ \hat W^0(g)
|\Psi'_0\>=|\Psi'_0\>$, then the resulting $|\Psi'\>=\hat U|\Psi'_0\>$ will
be symmetric under $\hat W(g)$: $ \hat W(g) |\Psi'\>=|\Psi'\>$.  Since $
\hat W^0(g)= \prod_i |g_i\>\<gg_i|K^{\frac{1-s(g)}{2}} $ has a simple form, it is easy to construct
the deformed $|\Phi_0\>$ that has the same symmetry under $\hat W^0(g)= \prod_i
|g_i\>\<gg_i|$.  Then after the fermion LU transformation $\hat U$, we can
obtain generic SPT states that are in the same phase as $|\Psi\>$.

Here we would like to stress that, only on a system without boundary, the total
symmetry transformation $\hat W$ is mapped into a simple on-site symmetry $\hat
W^0$ the LU transformation $\hat U$.  For system with boundary, under the LU
transformation $\hat U$, the total symmetry transformation $\hat W$ will be
mapped into a complicated symmetry transformation which does not have an on-site
form on the boundary.\cite{CLW1141}

\subsection{Entanglement density matrix}
\label{edgestate}

\begin{figure}[tb]
\begin{center}
\includegraphics[scale=1.0]{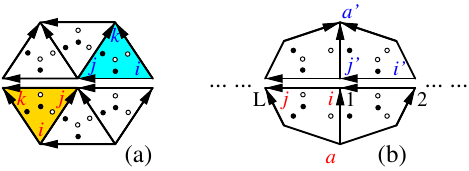}
\end{center}
\caption{
(Color online)
(a) We divide the system into two halve along a line by doubling the sites on the
line $|g_i\>\to |g_i\>\otimes |g_i\>$ and then cut between the pairs.  (b) The
interior of each region is deformed into a simpler lattice without affecting
the spectrum of the entanglement density matrix.
}
\label{edge}
\end{figure}


The non-trivialness of the SPT states is in their symmetry-protected gapless
boundary excitations.  To study the gapless boundary excitations, in this
section, we are going to study entanglement density matrix $\rho_E$ and its
entanglement Hamiltonian $H_E$: $\rho_E=\e^{-H_E}$, from the above constructed
ground state wave functions for SPT states.  The entanglement Hamiltonian $H_E$
can be viewed as the effective Hamiltonian for the gapless boundary excitations.

To calculate the entanglement density matrix, we first cut the system in to two
halves along a horizontal line.  To do cutting, we first split each site on the
cutting line into two sites: $|g_i\> \to |g_i\>\otimes|g_{i'}\>$ (see Fig.
\ref{edge}a).  The ground state lives in the subspace where $g_i=g_{i'}$.  We
then cut between the splitted sites.

Because the entanglement spectrum does not change if we perform LU
transformations within each half of the system, we can use the LU
transformations to deform the lattice into a simpler one in Fig. \ref{edge}b,
where the sites are labeled by $1$, $2$, .., $L$, $1'$, $2'$, .., $L'$, along
the edge which forms a ring, and the only two interior sites are labeled by
$a$ and $a'$.

On the deformed lattice, the ground state wave function can be written as
$\hat U|\Phi_0\>$, where
\begin{align}
|\Phi_0\>=
|\phi_{a}\>\<\phi_{a}|\otimes
|\phi_{a'}\>\<\phi_{a'}|
\otimes_{i=1}^L |\phi_i\>\<\phi_i|
\otimes_{i'=1'}^{L'} |\phi_{i'}\>\<\phi_{i'}|
\end{align}
and
\begin{widetext}
\begin{align}
& \hat U=
\prod_{i'} \nu_3^{-1}(g_0,g_{i'+1}, g_{i'}, g_{a'})
\prod_{i} \nu_3(g_0,g_a,g_{i+1}, g_i)
\prod_{i} \bar c_{(0,i+1,i)}^{n_2(g_0,g_{i+1},g_i)} c_{(0,{i'+1},i')}^{n_2(g_{0},g_{i'+1},g_{i'})}
\prod_{i} \bar c_{(0ai)}^{n_2(g_0,g_a,g_i)} c_{(0ai)}^{n_2(g_{0},g_{a},g_{i})}
\times \nonumber\\ &
\ \ \ \ \ \
\prod_{i'} \bar c_{(0i'a')}^{n_2(g_0,g_{i'},g_{a'})} c_{(0i'a')}^{n_2(g_{0},g_{i'},g_{a'})}
\prod_{i}
 c_{(a,{i+1},i)}^{\dag n_2(g_a,g_{i+1},g_i)}
 c_{(0ai)}^{\dag n_2(g_0,g_a,g_i)}
\bar c_{(0,i+1,i)}^{\dag n_2(g_0,g_{i+1},g_i)}
\bar c_{(0,a,i+1)}^{\dag n_2(g_0,g_a,g_{i+1})}
\times \nonumber \\ &
\ \ \ \ \ \
\prod_{i'}
 c_{(0,i'+1,i')}^{\dag n_2(g_{0},g_{i'+1},g_{i'})}
 c_{(0i'a')}^{\dag n_2(g_{0},g_{i'},g_{a'})}
\bar c_{(0,i'+1,a')}^{\dag n_2(g_{0},g_{i'+1},g_{a'})}
\bar c_{(i'+1,i',a')}^{\dag n_2(g_{i'+1},g_{i'},g_{a'})}
.
\end{align}
Now, let us rewrite $\hat U$ as
$\hat U=\hat U_\text{up}\hat U_\text{down}$,
where $\hat U_\text{up}$ acts on the upper half and
$\hat U_\text{down}$ acts on the lower half of the system:
\begin{align}
\label{Uupdown}
& \hat U= \Big[
(-)^{f(g_{1'},...,g_{L'})}
\prod_{i'} \nu_3^{-1}(g_0,g_{i'+1}, g_{i'}, g_{a'})
\prod_{i'} c_{(0,i'+1',i')}^{n_2(g_{0},g_{i'+1'},g_{i'})}
\prod_{i'} \bar c_{(0i'a')}^{n_2(g_0,g_{i'},g_{a'})} c_{(0i'a')}^{n_2(g_{0},g_{i'},g_{a'})}
\times \nonumber\\ &
\ \ \ \ \ \
\prod_{i'}
 c_{(0,i'+1,i')}^{\dag n_2(g_{0},g_{i'+1},g_{i'})}
 c_{(0i'a')}^{\dag n_2(g_{0},g_{i'},g_{a'})}
\bar c_{(0,i'+1,a')}^{\dag n_2(g_{0},g_{i'+1},g_{a'})}
\bar c_{(i'+1,i',a')}^{\dag n_2(g_{i'+1},g_{i'},g_{a'})}
\Big]
\times \nonumber\\ &
\ \ \ \
\Big[
\prod_{i} \nu_3(g_0,g_a,g_{i+1}, g_i)
\prod_{i} \bar c_{(0,i+1,i)}^{n_2(g_{0},g_{i+1},g_{i})}
\prod_{i} \bar c_{(0ai)}^{n_2(g_0,g_a,g_i)} c_{(0ai)}^{n_2(g_{0},g_{a},g_{i})}
\times \nonumber\\ &
\ \ \ \ \ \
\prod_{i}
 c_{(a,i+1,i)}^{\dag n_2(g_a,g_{i+1},g_i)}
 c_{(0ai)}^{\dag n_2(g_0,g_a,g_i)}
\bar c_{(0,i+1,i)}^{\dag n_2(g_0,g_{i+1},g_i)}
\bar c_{(0,a,i+1,)}^{\dag n_2(g_0,g_a,g_{i+1})}
\Big]
,
\nonumber
\end{align}
where the pure sign factor $(-)^{f(g_{1'},...,g_{L'})}$ arises from rewriting $
\prod_{i} \bar c_{(0,i+1,i)}^{n_2(g_0,g_{i+1},g_i)}
c_{(0,i'+1,i')}^{n_2(g_{0},g_{i'+1},g_{i'})} $ as $ \prod_{i'}
c_{(0,i'+1',i')}^{n_2(g_{0},g_{i'+1'},g_{i'})} \prod_{i}
c_{(0,i+1,i)}^{n_2(g_{0},g_{i+1},g_{i})} $, and we have used the fact that
$g_i=g_{i'}$ in the ground state subspace.  Thus the express \eqn{Uupdown} is
valid only when $\hat U$ acts within the ground state subspace.

Now the entanglement density matrix $\rho_E$ can be written as
\begin{align}
 \rho_E & =
\Tr_\text{upper}
\hat U_\text{up} \hat U_\text{down}
|\Phi_0\>\<\Phi_0|
\hat U^\dag_\text{down} \hat U^\dag_\text{up}
         = \Tr_\text{upper}
\hat U_\text{down}
|\Phi_0\>\<\Phi_0|
\hat U^\dag_\text{down}
,
\end{align}
where $\Tr_\text{upper}$ is the trace over the degrees of freedom on the upper
half of the system.  Since $g_i=g_{i'}$ along the edge in the ground state,
$\rho_E$ (as an operator) does not change $g_i$ along the edge (\ie $\rho_E$ is
diagonal in the $|g_i\>$ basis).  So it is sufficient to discuss $\rho_E$ in
the subspace of fixed $|g_i\>$'s:
\begin{align}
\rho_E(g_1,..)
\propto &
\sum_{g_a,g_a'}
\prod_{i}
\nu_3(g_0,g_a,g_{i+1}, g_i)
\nu_3^*(g_0,g_a',g_{i+1}, g_i)
\prod_{i=1}^L \bar c_{(0,i+1,i)}^{n_2(g_{0},g_{i+1},g_{i})}
\prod_{i} \bar c_{(0ai)}^{n_2(g_0,g_a,g_i)} c_{(0ai)}^{n_2(g_{0},g_{a},g_{i})}
\times \\ &
\prod_{i}
 c_{(a,i+1,i)}^{\dag n_2(g_a,g_{i+1},g_i)}
 c_{(0aj)}^{\dag n_2(g_0,g_a,g_j)}
\bar c_{(0,i+1,i)}^{\dag n_2(g_0,g_{i+1},g_i)}
\bar c_{(0,a,i+1)}^{\dag n_2(g_0,g_a,g_{i+1})}
|g_a\>\<g_a'|
\times \nonumber \\ &
\prod_{i}
\bar c_{(0,a,i+1)}^{n_2(g_0,g_a',g_{i+1})}
\bar c_{(0,i+1,i)}^{n_2(g_0,g_{i+1},g_i)}
 c_{(0ai)}^{n_2(g_0,g_a',g_i)}
 c_{(a,i+1,i)}^{n_2(g_a',g_{i+1},g_i)}
\prod_{i} c_{(0ai)}^{\dag n_2(g_{0},g_{a}',g_{i})} \bar c_{(0ai)}^{\dag n_2(g_0,g_a',g_i)}
\prod_{i=L}^1 \bar c_{(0,i+1,i)}^{\dag n_2(g_{0},g_{i+1},g_{i})}
\nonumber
\end{align}
We see that
$\rho_E$ has a form
\begin{align}
 \rho_E & =\sum_{\{g_i\}} \rho_E(g_1,...,g_L) \otimes
(\otimes_i |g_i\>\<g_i|)
  =\sum_{\{g_i\}} |\{g_i\}_\text{edge},g_0\>\<\{g_i\}_\text{edge},g_0|
\end{align}
\begin{align}
\label{edgeWav}
& |\{g_i\}_\text{edge},g_0\>  =
|G_b|^{-1/2}
\sum_{g_a}
\prod_{\bigtriangledown}
\nu_3(g_0,g_a,g_i, g_j)
\prod_{i=1}^L \bar c_{(0,i+1,i)}^{n_2(g_{0},g_{i+1},g_{i})}
\prod_{i} \bar c_{(0ai)}^{n_2(g_0,g_a,g_i)} c_{(0ai)}^{n_2(g_{0},g_{a},g_{i})}
\times \nonumber \\ &
\prod_i
 c_{(a,i+1,i)}^{\dag n_2(g_a,g_{i+1},g_i)}
 c_{(0ai)}^{\dag n_2(g_0,g_a,g_i)}
\bar c_{(0,i+1,i)}^{\dag n_2(g_0,g_{i+1},g_i)}
\bar c_{(0,a,i+1)}^{\dag n_2(g_0,g_a,g_{i+1})}
|g_a\>\otimes (\otimes_i |g_i\>)
\end{align}
This is a key result of this paper that allows us to understand gapless edge
excitations.  In fact $|\{g_i\}_\text{edge}\>$ is a basis of the low energy
subspace of the edge excitations.  We see that the  low energy edge excitations
are described by $g_i$ on the edge.

We can simplify the above expression \eq{edgeWav}.
First we note that
\begin{align}
&\ \ \ \
\prod_{i=1}^L \bar c_{(0,i+1,i)}^{n_2(g_{0},g_{i+1},g_{i})}
\prod_{i} \bar c_{(0ai)}^{n_2(g_0,g_a,g_i)} c_{(0ai)}^{n_2(g_{0},g_{a},g_{i})}
=
\prod_{i=1}^L
\bar c_{(0ai)}^{n_2(g_0,g_a,g_{i})}
c_{(0ai)}^{n_2(g_{0},g_{a},g_{i})}
\bar c_{(0,i+1,i)}^{n_2(g_{0},g_{i+1},g_{i})}
\nonumber\\
&= (-)^{n_2(g_{0},g_{a},g_{1})+n_2(g_{0},g_{a},g_{1})\sum_{i=1}^L
n_2(g_{0},g_{i+1},g_{i})
}
\prod_{i=1}^L
c_{(0ai)}^{n_2(g_{0},g_{a},g_{i})}
\bar c_{(0,i+1,i)}^{n_2(g_{0},g_{i+1},g_{i})}
\bar c_{(0,a,i+1)}^{n_2(g_0,g_a,g_{i+1})}
\end{align}
We also note that
\begin{align}
&\ \ \ \ c_{(a,i+1,i)}^{\dag n_2(g_a,g_{i+1},g_i)}
 c_{(0ai)}^{\dag n_2(g_0,g_a,g_i)}
\bar c_{(0,i+1,i)}^{\dag n_2(g_0,g_{i+1},g_i)}
\bar c_{(0,a,i+1)}^{\dag n_2(g_0,g_a,g_{i+1})}
\\
&=
\bar c_{(0,a,i+1)}^{\dag n_2(g_0,g_a,g_{i+1})}
\bar c_{(0,i+1,i)}^{\dag n_2(g_0,g_{i+1},g_i)}
 c_{(0ai)}^{\dag n_2(g_0,g_a,g_i)}
 c_{(a,i+1,i)}^{\dag n_2(g_a,g_{i+1},g_i)}
(-)^{[
 n_2(g_a,g_{i+1},g_i)
+n_2(g_0,g_a,g_i)
+n_2(g_0,g_{i+1},g_i)
+n_2(g_0,g_a,g_{i+1})
]/2}
\nonumber
\end{align}
Therefore,
\begin{align}
&\ \ \ \
\prod_{i=1}^L \bar c_{(0,i+1,i)}^{n_2(g_{0},g_{i+1},g_{i})}
\prod_{i} \bar c_{(0ai)}^{n_2(g_0,g_a,g_i)} c_{(0ai)}^{n_2(g_{0},g_{a},g_{i})}
\prod_i
 c_{(a,i+1,i)}^{\dag n_2(g_a,g_{i+1},g_i)}
 c_{(0ai)}^{\dag n_2(g_0,g_a,g_i)}
\bar c_{(0,i+1,i)}^{\dag n_2(g_0,g_{i+1},g_i)}
\bar c_{(0,a,i+1)}^{\dag n_2(g_0,g_a,g_{i+1})}
\nonumber \\
&= (-)^{n_2(g_{0},g_{a},g_{1})+n_2(g_{0},g_{a},g_{1})\sum_{i=1}^L
n_2(g_{0},g_{i+1},g_{i})
}
\prod_{i=1}^L
(-)^{
\frac{n_2(g_{0},g_{i+1},g_{i})+
n_2(g_{a},g_{i+1},g_{i})}{2}
+
n_2(g_0,g_a,g_i)
}
 c_{(a,i+1,i)}^{\dag n_2(g_a,g_{i+1},g_i)}
.
\nonumber
\end{align}
We find
\begin{align}
\label{edgeWavS}
 |\{g_i\}_\text{edge},g_0\>&  =
|G_b|^{-1/2}
\sum_{g_a}
\prod_{i}
\nu_3(g_0,g_a,g_{i+1}, g_i)
(-)^{n_2(g_{0},g_{a},g_{1}) [1+\sum_{i=1}^L
n_2(g_{0},g_{i+1},g_{i})]
}
\times \nonumber\\ &
\ \ \ \ \
\prod_{i=1}^L
(-)^{
\frac{n_2(g_{0},g_{i+1},g_{i})+
n_2(g_{a},g_{i+1},g_{i})}{2}
+
n_2(g_0,g_a,g_i)
}
 c_{(a,i+1,i)}^{\dag n_2(g_a,g_{i+1},g_i)}
|g_a\>\otimes (\otimes_i |g_i\>)
\end{align}
We see that the total number of fermions in $|\{g_i\}_\text{edge},g_0\>$ is
given by $N_F=\sum_i n_2(g_a,g_{i+1},g_i)$.  So the fermion number is not fixed
for fixed $g_i$'s (due to the $g_a$ dependence), Since $N_F = \sum_i
n_2(g_0,g_{i+1},g_i)$ mod 2,  the  fermion number parity is fixed for fixed
$g_i$'s.

To see how $|\{g_i\}_\text{edge},g_0\>$
depends on $g_0$, let us
use \eqn{edgeWavS} to calculate
$ \<\{g_i\}_\text{edge},g'_0 |\{g_i\}_\text{edge},g_0\>$:
\begin{align}
 \<\{g_i\}_\text{edge},g'_0 &|\{g_i\}_\text{edge},g_0\>
=
|G_b|^{-1}
\sum_{g_a}
(-)^{
\sum_{i=1}^L
\frac{
n_2(g'_{0},g_{i+1},g_{i})+
n_2(g_{0},g_{i+1},g_{i})
}{2}
+n_2(g_{a},g_{i+1},g_{i})
+n_2(g'_0,g_a,g_i)
+n_2(g_0,g_a,g_i)
}
\times \nonumber \\ &\ \ \ \
(-)^{
[n_2(g'_{0},g_{a},g_{1})
+n_2(g_{0},g_{a},g_{1})]
[1+\sum_{i=1}^L n_2(g_{0},g_{i+1},g_{i})] }
\prod_{i}
\nu^{-1}_3(g'_0,g_a,g_{i+1}, g_i)
\nu_3(g_0,g_a,g_{i+1}, g_i)
\nonumber\\
&=
|G_b|^{-1}
(-)^{
\sum_{i=1}^L
\frac{
n_2(g'_{0},g_{i+1},g_{i})+
n_2(g_{0},g_{i+1},g_{i})
}{2}
}
\sum_{g_a}
(-)^{
\sum_{i=1}^L
n_2(g_{a},g_{i+1},g_{i})
+n_2(g_0,g'_0,g_i)
+n_2(g_0,g'_0,g_a)
}
\times \nonumber \\ &\ \ \ \
(-)^{
[
n_2(g_0,g'_0,g_1)
+n_2(g_0,g'_0,g_a)
]
[1+\sum_{i=1}^L n_2(g_{0},g_{i+1},g_{i})] }
\prod_{i}
\nu^{-1}_3(g'_0,g_a,g_{i+1}, g_i)
\nu_3(g_0,g_a,g_{i+1}, g_i)
\end{align}
Using $ (-)^{\sum_{i=1}^L n_2(g_{0},g_{i+1},g_{i})}
=(-)^{\sum_{i=1}^L n_2(g_{a},g_{i+1},g_{i})}$
and the cocycle condition
\begin{align}
&\ \ \ \
\nu^{-1}_3(g'_0,g_a,g_{i+1}, g_i)
\nu_3(g_0,g_a,g_{i+1}, g_i)
\nonumber\\ &
=
\nu_3(g_0,g'_0,g_{i+1}, g_i)
\nu^{-1}_3(g_0,g'_0,g_a,g_i)
\nu_3(g_0,g'_0,g_a,g_{i+1})
(-)^{n_2(g_0,g'_0,g_a)n_2(g_a,g_{i+1},g_i) } ,
\end{align}
we can rewrite the above as
\begin{align}
&
 \<\{g_i\}_\text{edge},g'_0
|\{g_i\}_\text{edge},g_0\>
=
(-)^{
\sum_{i=1}^L
\frac{
n_2(g'_{0},g_{i+1},g_{i})+
n_2(g_{0},g_{i+1},g_{i})
}{2}
}
(-)^{
\sum_{i=1}^L
n_2(g_{0},g_{i+1},g_{i})
+n_2(g_0,g'_0,g_i)
}
\times \nonumber \\ &
\ \ \ \ \
\ \ \ \ \
(-)^{
n_2(g_0,g'_0,g_1)
[1+\sum_{i=1}^L n_2(g_{0},g_{i+1},g_{i})] }
\prod_{i}
\nu_3(g_0,g'_0,g_{i+1}, g_i)
|G_b|^{-1}
\sum_{g_a}
(-)^{(L+1) n_2(g_0,g'_0,g_a)}
\end{align}
We see that when
$L$ = odd,
$|\{g_i\}_\text{edge},g_0\>$ and
$|\{g_i\}_\text{edge},g_0'\>$ only differ by a phase
\begin{align}
|\{g_i\}_\text{edge},g_0\>
&=
(-)^{
\sum_{i=1}^L
\frac{
n_2(g'_{0},g_{i+1},g_{i})-
n_2(g_{0},g_{i+1},g_{i})
}{2}
+n_2(g_0,g'_0,g_i)
}
\times \nonumber \\ &
\ \ \ \
(-)^{
n_2(g_0,g'_0,g_1)
[1+\sum_{i=1}^L n_2(g_{0},g_{i+1},g_{i})] }
\prod_{i}
\nu_3(g_0,g'_0,g_{i+1}, g_i) \ \
 |\{g_i\}_\text{edge},g'_0 \>.
\end{align}

{}From the above results, and using the relation between the entanglement
density matrix and gapless edge excitations, we learn two things.  1) The low
energy edge degrees of freedom are labeled by $\{g_1,...,g_L\}$ on the boundary,
since $\rho_E(g_1,...,g_L)$ has one and only one non-zero eigenvalue for each
$\{g_1,...,g_L\}$.  2)  The low energy edge degrees of freedom are entangled
with the bulk degrees of freedom, since the states on the site-$a$ (in the
bulk) are different for different $\{g_1,...,g_L\}$.

Using the expression \eq{edgeWavS}, we calculate the low energy
effective Hamiltonian $H_\text{eff}$ on the edge
from a physical Hamiltonian $H_\text{edge}$ on the edge:
\begin{align}
\label{Heff}
 (H_\text{eff})_{ \{g_i'\}, \{g_i\}} =
\<\{g_i'\}_\text{edge},g_0| H_\text{edge} |\{g_i\}_\text{edge},g_0\>.
\end{align}
$H_\text{eff}$ has a short range interaction
and satisfies certain symmetry conditions
if $H_\text{edge}$ are symmetric.
In the following, we
are going to
study how
$|\{g_i\}_\text{edge},g_0\>$ transforms under the symmetry transformation.

\subsection{Symmetry transformation on edge states}
\label{symmedge}

Let us apply the symmetry operation $\hat W(g)$ \eq{bsymm}
to the edge state  $|\{gg_i\}_\text{edge},g_0\> $
\begin{align}
& \hat W(g) |\{gg_i\}_\text{edge},g_0\>
  =
|G_b|^{-1/2}
\sum_{g_a}
\prod_{i}
[u_2^{g}(g_a,g_{i+1},g_i)]^{-1}
\nu_3^{s(g)}(g_0,gg_a,gg_{i+1}, gg_i)
\prod_{i=1}^L \bar c_{(0,i+1,i)}^{n_2(g_{0},gg_{i+1},gg_{i})}
\times
\\ & \ \ \ \ \ \
\prod_{i} \bar c_{(0ai)}^{n_2(g_0,gg_a,gg_i)} c_{(0ai)}^{n_2(g_{0},gg_{a},gg_{i})}
\prod_i
 c_{(a,i+1,i)}^{\dag n_2(gg_a,gg_{i+1},gg_i)}
 c_{(0ai)}^{\dag n_2(g_0,gg_a,gg_i)}
\bar c_{(0,i+1,i)}^{\dag n_2(g_0,gg_{i+1},gg_i)}
\bar c_{(0,a,i+1)}^{\dag n_2(g_0,gg_a,gg_{i+1})}
|g_a\>\otimes (\otimes_i |g_i\>)
\nonumber\\
  &=
|G_b|^{-1/2}
\sum_{g_a}
\prod_{i}
[u_2^{g}(g^{-1}g_0,g_{i+1},g_i)]^{-1}
\nu_3(g^{-1}g_0,g_a,g_{i+1}, g_i)
\prod_{i=1}^L \bar c_{(0,i+1,i)}^{n_2(g^{-1}g_{0},g_{i+1},g_{i})}
\prod_{i} \bar c_{(0ai)}^{n_2(g^{-1}g_0,g_a,g_i)} c_{(0ai)}^{n_2(g^{-1}g_{0},g_{a},g_{i})}
\times \nonumber \\ & \ \ \ \ \ \
\prod_i
 c_{(a,i+1,i)}^{\dag n_2(g_a,g_{i+1},g_i)}
 c_{(0ai)}^{\dag n_2(g^{-1}g_0,g_a,g_i)}
\bar c_{(0,i+1,i)}^{\dag n_2(g^{-1}g_0,g_{i+1},g_i)}
\bar c_{(0,a,i+1)}^{\dag n_2(g^{-1}g_0,g_a,g_{i+1})}
|g_a\>\otimes (\otimes_i |g_i\>)
\propto |\{g_i\}_\text{edge},g^{-1}g_0\>
\nonumber
\end{align}
When
 $L$ = odd, $|\{g_i\}_\text{edge},g^{-1}g_0\> \propto
|\{g_i\}_\text{edge},g_0\>$ and we find
\begin{align}
 \hat W(g) |\{gg_i\}_\text{edge}&,g_0\>
 = w(\{g_i\},g) |\{g_i\}_\text{edge},g_0\>,
\\
w(\{g_i\},g) &= \frac{
 \<\{g_i\}_\text{edge},g_0 |\{g_i\}_\text{edge},g^{-1}g_0\>
\prod_{i=1}^L[u_2^{g}(g^{-1}g_0,g_{i+1},g_{i})]^{-1}
}{
 \<\{g_i\}_\text{edge},g_0 |\{g_i\}_\text{edge},g_0\>
}
\nonumber\\
&=
(-)^{
\sum_{i=1}^L
\frac{
n_2(g^{-1}g_{0},g_{i+1},g_{i})
-n_2(g_{0},g_{i+1},g_{i})
}{2}
+n_2(g^{-1}g_0,g_0,g_i)
}
\times \nonumber \\ &
\ \ \ \
(-)^{
n_2(g^{-1}g_0,g_0,g_1)
[L+\sum_{i=1}^L n_2(g^{-1}g_{0},g_{i+1},g_{i})] }
\prod_{i}
\nu_3(g^{-1}g_0,g_0,g_{i+1}, g_i)
\prod_{i=1}^L[u_2^{g}(g^{-1}g_0,g_{i+1},g_{i})]^{-1}
\nonumber
\end{align}
Note that we have rewritten $(-)^{ n_2(g^{-1}g_0,g_0,g_1) [1+\sum_{i=1}^L
n_2(g^{-1}g_{0},g_{i+1},g_{i})] } $ as $(-)^{ n_2(g^{-1}g_0,g_0,g_1)
[L+\sum_{i=1}^L n_2(g^{-1}g_{0},g_{i+1},g_{i})] } $ since $L$ is odd.

Although the above expression for the action of the edge effective symmetry is
obtain for $L$ = odd, we can show that the expression actually forms a
representation of $G_f$ for both $L$ = odd and $L$ = even.  Let us consider
\begin{align}
 \hat W(g') |\{g'g_i\}_\text{edge},g_0\>
=&
 \hat W(g^{-1}g') \hat W(g) |\{g g^{-1}g' g_i\}_\text{edge},g_0\>
=  \hat W(g^{-1}g') w(\{g^{-1}g' g_i\},g)  |\{g^{-1}g' g_i\}_\text{edge},g_0\>
\nonumber\\
=&   w(\{g^{-1}g' g_i\},g) w(\{g_i\},g^{-1}g')  |\{ g_i\}_\text{edge},g_0\>
=   w(\{g_i\},g') |\{ g_i\}_\text{edge},g_0\>
\end{align}
We see that in order for $\hat W(g)$ to form a representation, we require that $
w(\{g^{-1}g' g_i\},g) w(\{g_i\},g^{-1}g')= w(\{g_i\},g')$.  So let us examine
\begin{align}
w(\{g^{-1} & g'  g_i\},g)  w(\{g_i\},g^{-1}g')=
(-)^{
\sum_{i=1}^L
\frac{
n_2(g^{-1}g_{0},g^{-1}g'g_{i+1},g^{-1}g'g_{i})
-n_2(g_{0},g^{-1}g'g_{i+1},g^{-1}g'g_{i})
}{2}
+n_2(g^{-1}g_0,g_0,g^{-1}g'g_i)
}
\times \nonumber \\ &
(-)^{
n_2(g^{-1}g_0,g_0,g^{-1}g'g_1)
[L+\sum_{i=1}^L n_2(g^{-1}g_{0},g^{-1}g'g_{i+1},g^{-1}g'g_{i})] }
\prod_{i}
\nu_3(g^{-1}g_0,g_0,g^{-1}g'g_{i+1}, g^{-1}g'g_i)
\times \nonumber \\ &
\Big[
\prod_{i=1}^L[u_2^{g}(g^{-1}g_0,g^{-1}g'g_{i+1},g^{-1}g'g_{i})]^{-1}
\Big]
(-)^{
\sum_{i=1}^L
\frac{
n_2(g^{\prime -1}gg_{0},g_{i+1},g_{i})
-n_2(g_{0},g_{i+1},g_{i})
}{2}
+n_2(g^{\prime -1}gg_0,g_0,g_i)
}
\times \nonumber \\ &
(-)^{
n_2(g^{\prime -1}gg_0,g_0,g_1)
[L+\sum_{i=1}^L n_2(g^{\prime -1}gg_{0},g_{i+1},g_{i})] }
\prod_{i}
\nu_3(g^{\prime -1}gg_0,g_0,g_{i+1}, g_i)
\prod_{i=1}^L[u_2^{g^{-1}g'}(g^{\prime -1}gg_0,g_{i+1},g_{i})]^{-1}
.
\end{align}
Using
\begin{align}
& \nu_3(g^{-1}g_0,g_0,g^{-1}g'g_{i+1}, g^{-1}g'g_i)
=
\nu_3(g^{\prime -1}g_0,g^{\prime -1}gg_0,g_{i+1}, g_i)
\times \\
& \ \ \ \
u_2^{g^{-1}g'}(g^{\prime -1}gg_0,g_{i+1}, g_i)
[u_2^{g^{-1}g'}(g^{\prime -1}g_0,g_{i+1}, g_i)]^{-1}
u_2^{g^{-1}g'}(g^{\prime -1}g_0,g^{\prime -1}gg_0,g_i)
[u_2^{g^{-1}g'}(g^{\prime -1}g_0,g^{\prime -1}gg_0,g_{i+1}]^{-1})
\nonumber
\end{align}
we can simplify the above as
\begin{align}
w&(\{g^{-1}  g'  g_i\},g)  w(\{g_i\},g^{-1}g')=
(-)^{
[n_2(g_0,g'g_0,g'g_1)
+n_2(g_0,gg_0,g'g_0)]
[L+\sum_{i=1}^L n_2(g_{0},g_{i+1},g_{i})] }
\times \nonumber \\ &
(-)^{
\sum_{i=1}^L
\frac{
n_2(g^{\prime -1}g_{0},g_{i+1},g_{i})
-n_2(g^{\prime -1} gg_{0},g_{i+1},g_{i})
}{2}
+n_2(g^{\prime -1} g_0,g^{\prime -1} gg_0,g_i)
}
(-)^{
\sum_{i=1}^L
\frac{
n_2(g^{\prime -1}gg_{0},g_{i+1},g_{i})
-n_2(g_{0},g_{i+1},g_{i})
}{2}
+n_2(g^{\prime -1}gg_0,g_0,g_i)
}
\times \nonumber \\ &
\prod_{i}
\nu_3(g^{\prime -1}g_0,g^{\prime -1} g g_0,g_{i+1}, g_i)
\nu_3(g^{\prime -1}gg_0,g_0,g_{i+1}, g_i)
\prod_{i=1}^L[u_2^{g'}(g^{\prime -1}g_0,g_{i+1},g_{i})]^{-1}
\nonumber \\
= &
(-)^{
[n_2(g_0,g'g_0,g'g_1)
+n_2(g_0,gg_0,g'g_0)]
[L+\sum_{i=1}^L n_2(g_{0},g_{i+1},g_{i})] }
\times \nonumber \\ &
(-)^{
\sum_{i=1}^L
\frac{
n_2(g^{\prime -1}g_{0},g_{i+1},g_{i})
-n_2(g_{0},g_{i+1},g_{i})
}{2}
+n_2(g^{\prime -1} g_0,g_0,g_i)
+n_2(g^{\prime -1} g_0,g^{\prime -1} gg_0,g_0)
}
\times \nonumber \\ &
\prod_{i}
\nu_3(g^{\prime -1}g_0,g^{\prime -1} g g_0,g_{i+1}, g_i)
\nu_3(g^{\prime -1}gg_0,g_0,g_{i+1}, g_i)
\prod_{i=1}^L[u_2^{g'}(g^{\prime -1}g_0,g_{i+1},g_{i})]^{-1}
.
\end{align}
Using the cocycle condition
\begin{align}
&\ \ \ \
\nu_3(g^{\prime -1}gg_0,g_0,g_{i+1}, g_i)
\nu_3(g^{\prime -1}g_0,g^{\prime -1}gg_0,g_{i+1}, g_i)
\nonumber\\ &
=
\nu_3(g^{\prime -1}g_0,g_0,g_{i+1}, g_i)
\nu_3(g^{\prime -1}g_0,g^{\prime -1}gg_0,g_0,g_i)
\nu_3^{-1}(g^{\prime -1}g_0,g^{\prime -1}gg_0,g_0,g_{i+1})
(-)^{n_2(g^{\prime -1}g_0,g^{\prime -1}gg_0,g_0)n_2(g_0,g_{i+1},g_i) } ,
\end{align}
we can rewrite the above as
\begin{align}
w&(\{g^{-1}  g'  g_i\},g)  w(\{g_i\},g^{-1}g')=
(-)^{
n_2(g^{\prime -1}g_0,g_0,g_1)
[L+\sum_{i=1}^L n_2(g_{0},g_{i+1},g_{i})] }
\times \nonumber \\ &
(-)^{
\sum_{i=1}^L
\frac{
n_2(g^{\prime -1}g_{0},g_{i+1},g_{i})
-n_2(g_{0},g_{i+1},g_{i})
}{2}
+n_2(g^{\prime -1} g_0,g_0,g_i)
}
\prod_{i}
\nu_3(g^{\prime -1}g_0,g_0,g_{i+1}, g_i)
\prod_{i=1}^L[u_2^{g'}(g^{\prime -1}g_0,g_{i+1},g_{i})]^{-1}
\end{align}
\end{widetext}
We see that we indeed have $ w(\{g^{-1}g' g_i\},g)
w(\{g_i\},g^{-1}g')= w(\{g_i\},g')$.  The strange factor $(-)^{
n_2(g^{-1}g_0,g_0,g_1) [L+\sum_{i=1}^L n_2(g^{-1}g_{0},g_{i+1},g_{i})] } $ is
important to make $w(\{g_i\},g)$ to be consistent with $\hat W(g)$ being a
representation of $G_f$.

So if
$\hat W^\dag(g) H_\text{edge} \hat W(g)=H_\text{edge}$,
then $(H_\text{eff})_{ \{g_i'\}, \{g_i\}}$ satisfies
\begin{align}
\label{HeffSymm}
&\ \ \ \
 (H_\text{eff})_{ \{gg_i'\}, \{gg_i\}} =
\<\{gg_i'\}_\text{edge},g_0| H_\text{edge} |\{gg_i\}_\text{edge},g_0\>.
\nonumber\\ &
=
\<\{gg_i'\}_\text{edge},g_0|\hat W^\dag(g) H_\text{edge} \hat W(g) |\{gg_i\}_\text{edge},g_0\>
\nonumber\\ &
=
w^*(\{g_i'\},g)
w(\{g_i\},g)
\<\{g_i'\}_\text{edge},g_0|H_\text{edge} |\{g_i\}_\text{edge},g_0\>
\nonumber\\ &
=
w^*(\{g_i'\},g)
 (H_\text{eff})_{ \{g_i'\}, \{g_i\}}
w(\{g_i\},g)
.
\end{align}
In the operator form, the above can be rewritten as
\begin{align}
& W^\dag_\text{eff}(g) H_\text{eff} W_\text{eff}(g) = H_\text{eff}
\nonumber\\
& W_\text{eff}(g) |\{g_i\}_\text{edge},g_0\>
= w^*(\{g_i\},g) |\{g g_i\}_\text{edge},g_0\>.
\end{align}

We see that the symmetry transformation of $H_\text{eff}$ contains additional
phase factor $w(\{g_i\},g)$.  In particular, the phase factor cannot be written
as an on-site form $w(\{g_i\},g)=\prod_i f(g_i,g)$.  So the symmetry
transformation on the effective edge degrees of freedom is not an on-site
symmetry transformation.  The non-on-site symmetry of the edge state ensures the
gaplessness of the edge excitation if the symmetry is not broken.\cite{CLW1141}
Note that for bosonic case
$n_2=0$ and $u^g_2=1$. We have a simple
result
\begin{align}
 w(\{g_i\},g) &=
\prod_{i}
\nu_3(g^{-1}g_0,g_0,g_{i+1}, g_i)
.
\end{align}

Let us write $w(\{g_i\},g)$ as
\begin{align}
&
 w(\{g_i\},g)
\\
=&
(-)^{ n_2(g^{-1}g_0,g_0,g_1) [L+\sum_{i=1}^L
n_2(g^{-1}g_{0},g_{i+1},g_{i})] }
\prod_{i} w_{i,i+1},
\nonumber
\end{align}
where
\begin{align}
& w_{i,i+1} =
\imth^{
n_2(g^{-1}g_{0},g_{i+1},g_{i})
-n_2(g_{0},g_{i+1},g_{i})
+2n_2(g^{-1}g_0,g_0,g_i)
}
\times \nonumber \\ & \ \ \ \
\nu_3(g^{-1}g_0,g_0,g_{i+1}, g_i)[u_2^{g}(g^{-1}g_0,g_i,g_{i+1})]^{-1}.
\end{align}
We see that for fermion cases, we cannot write the phase factor $ w(\{g_i\},g)$
as a product of local non-on-site phase factors $w_{i,i+1}$.  The non-local phase
factor $(-)^{ n_2(g^{-1}g_0,g_0,g_1) [L+\sum_{i=1}^L
n_2(g^{-1}g_{0},g_{i+1},g_{i})] } $ seems must appear.

Each term in the
edge effective Hamiltonian
$H_\text{eff} =\sum_i H_\text{eff}(i)$ must satisfy
\begin{align}
W^\dag_\text{eff}(g) H_\text{eff}(i) W_\text{eff}(g) = H_\text{eff}(i)
,
\end{align}
where $H_\text{eff}(i)$ only acts on sites near site-$i$.  $ H_\text{eff}(i)$
must also preserve the fermion number parity $(-)^{N_F}=(-)^{\sum_{i=1}^L
n_2(g_0,g_{i+1},g_i)}$.  In other words $ H_\text{eff}(i)$ commutes with
$(-)^{\sum_{i=1}^L n_2(g_0,g_{i+1},g_i)}$.
So if
$H_\text{eff}(i)$ is far away from the site-1,
then $H_\text{eff}(i)$ commutes with the non-local phase
factor
$(-)^{ n_2(g^{-1}g_0,g_0,g_1) [L+\sum_{i=1}^L
n_2(g^{-1}g_{0},g_{i+1},g_{i})] }$.
Thus $H_\text{eff}(i)$ is invariant under
\begin{align}
\t W^\dag_\text{eff}(g) H_\text{eff}(i) \t W_\text{eff}(g) = H_\text{eff}
,
\end{align}
where
\begin{align}
\t W_\text{eff}(g) |\{g_i\}_\text{edge},g_0\>
=\t w^*(\{g_i\},g) |\{g g_i\}_\text{edge},g_0\>.
\end{align}
with
\begin{align}
\t w(\{g_i\},g) = \prod_{i} w_{i,i+1}.
\nonumber
\end{align}

\section{Ideal Hamiltonian from path integral}
\label{Hamiltonian}

\begin{figure}[h]
\begin{center}
\includegraphics[scale=0.7]{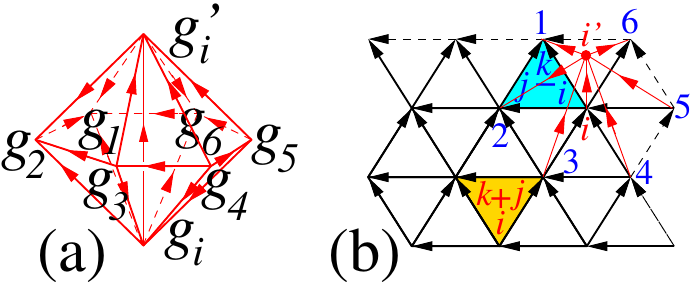}
\end{center}
\caption{(Color online)
(a) The matrix elements of the Hamiltonian term $H_i$ can be obtained from the
fermionic path integral of the fixed-point action amplitude $\cV_3$ on the
complex $\< g_1 g_2 g_3 g_4 g_5 g_6 g_ig_i'\> $.  The branching structure on the
complex is chosen to match that of the triangular lattice in Fig. \ref{trilatt}.
(b) The action of $H_i$ can be obtained by attaching the complex in (a)
to the triangular lattice in Fig. \ref{trilatt}.
The Grassmann numbers on the six overlapping triangles
are integrated out.
}
\label{Hi}
\end{figure}

After constructing the fermionic SPT state $|\Psi\>$ in the Section \ref{2dwav},
here we would like to construct a local Hamiltonian
\begin{align}
 H=-\sum_i H_i
\end{align}
that has the $G_f$ symmetry, such that $|\Psi\>$ is the unique ground state of
the Hamiltonian.

We note that the path integral of the fixed-point action amplitude not only
gives rise to an ideal ground state wave function (as discussed above), it also
gives rise to an ideal Hamiltonian.  From the structure of the fixed-point path
integral, we find that $H_i$ has a structure that it acts on a seven-spin cluster
labeled by $i$, 1 -- 6 in blue in Fig. \ref{trilatt}, and on the six-fermion
cluster on the six triangles inside the hexagon $\<123456\>$ in Fig.
\ref{trilatt}.

The matrix elements of $H_i$ can be obtained by evaluating the fixed-point
action amplitude $\cV_3^\pm$ on the complex in Fig. \ref{Hi}a, since the action
of $H_i$ can be realized by attaching the complex Fig. \ref{Hi}a to the
triangular lattice and then performing the fermionic path integral on the new
complex (see Fig. \ref{Hi}b).  The evaluation of $\cV_3^\pm$ on the complex
give us
\begin{widetext}
\begin{align}
\label{Hipath}
&
(-)^{m_1(g_4,g_i')+m_1(g_i',g_1)}
(-)^{ m_1(g_i,g_2)+m_1(g_3,g_i) +m_1(g_5,g_i) +m_1(g_i,g_6) }
\int \prod_\Si \cV_3
\nonumber\\
=&
(-)^{ m_1(g_4,g_i')+m_1(g_i',g_1)+m_1(g_i,g_2)+m_1(g_3,g_i) +m_1(g_5,g_i) +m_1(g_i,g_6) }
\int
\prod_{j=1,2,6}
\dd \th_{(ii'j)}^{n_2(g_i,g_i',g_j)} \dd \bar\th_{(ii'j)}^{n_2(g_i,g_i',g_j)}
\prod_{j=3,4,5}
\dd \th_{(jii')}^{n_2(g_j,g_i,g_i')}\dd  \bar\th_{(jii')}^{n_2(g_j,g_i,g_i')}
\nonumber\\ &
(-)^{
 m_1(g_4,g_i)
+m_1(g_5,g_i')
+m_1(g_i,g_6)
+m_1(g_i',g_1)
+m_1(g_i,g_2)
+m_1(g_3,g_i')
}
(-)^{
 m_1(g_i,g_i')
}
\times
\nonumber\\ &
\mathcal{V}^+_3(g_5,g_i,g_i',g_6)
\mathcal{V}^+_3(g_4,g_5,g_i,g_i')
\mathcal{V}^-_3(g_4,g_3,g_i,g_i')
\mathcal{V}^-_3(g_3,g_i,g_i',g_2)
\mathcal{V}^-_3( g_i,g_i',g_2,g_1)
\mathcal{V}^+_3(g_i,g_i',g_6,g_1)
\nonumber\\
=&
(-)^{ m_1(g_4,g_i')+m_1(g_i',g_1)+m_1(g_i,g_2)+m_1(g_3,g_i) +m_1(g_5,g_i) +m_1(g_i,g_6) }
\int
\prod_{j=1,2,6}
\dd \th_{(ii'j)}^{n_2(g_i,g_i',g_j)} \dd \bar\th_{(ii'j)}^{n_2(g_i,g_i',g_j)}
\prod_{j=3,4,5}
\dd \th_{(jii')}^{n_2(g_j,g_i,g_i')}\dd  \bar\th_{(jii')}^{n_2(g_j,g_i,g_i')}
\nonumber\\ &
(-)^{
 m_1(g_4,g_i)
+m_1(g_5,g_i')
+m_1(g_i,g_6)
+m_1(g_i',g_1)
+m_1(g_i,g_2)
+m_1(g_3,g_i')
}
(-)^{
 m_1(g_i,g_i')
}
\times
\nonumber\\ &
\nu_3(g_5,g_i,g_i',g_6)
\th_{(ii'6)}^{n_2(g_i,g_i',g_6)}
\th_{(5i6)}^{n_2(g_5,g_i,g_6)}
\bar\th_{(5i'6)}^{n_2(g_5,g_i',g_6)}
\bar\th_{(5ii')}^{n_2(g_5,g_i,g_i')}
\nu_3(g_4,g_5,g_i,g_i')
\th_{(5ii')}^{n_2(g_5,g_i,g_i')}
\th_{(45i')}^{n_2(g_4,g_5,g_i')}
\bar\th_{(4ii')}^{n_2(g_4,g_i,g_i')}
\bar\th_{(45i)}^{n_2(g_4,g_5,g_i)}
\nonumber\\ &
\nu_3^{-1}(g_4,g_3,g_i,g_i')
\th_{(43i)}^{n_2(g_4,g_3,g_i)}
\th_{(4ii')}^{n_2(g_4,g_i,g_i')}
\bar\th_{(43i')}^{n_2(g_4,g_3,g_i')}
\bar\th_{(3ii')}^{n_2(g_3,g_i,g_i')}
\nu^{-1}_3(g_3,g_i,g_i',g_2)
\th_{(3ii')}^{n_2(g_3,g_i,g_i')}
\th_{(3i'2)}^{n_2(g_3,g_i',g_2)}
\bar\th_{(3i2)}^{n_2(g_3,g_i,g_2)}
\bar\th_{(ii'2)}^{n_2(g_i,g_i',g_2)}
\nonumber\\ &
\nu^{-1}_3( g_i,g_i',g_2,g_1)
\th_{(ii'2)}^{n_2(g_i,g_i',g_2)}
\th_{(i21)}^{n_2(g_i,g_2,g_1)}
\bar\th_{(ii'1)}^{n_2(g_i,g_i',g_1)}
\bar\th_{(i'21)}^{n_2(g_i',g_2,g_1)}
\nu_3(g_i,g_i',g_6,g_1)
\th_{(i'61)}^{n_2(g_i',g_6,g_1)}
\th_{(ii'1)}^{n_2(g_i,g_i',g_1)}
\bar\th_{(i61)}^{n_2(g_i,g_6,g_1)}
\bar\th_{(ii'6)}^{n_2(g_i,g_i',g_6)}.
\end{align}
Just like the wave function, we have included an extra factor $
(-)^{m_1(g_4,g_i')+m_1(g_i',g_1)} (-)^{ m_1(g_i,g_2)+m_1(g_3,g_i) +m_1(g_5,g_i)
+m_1(g_i,g_6) } $ in the path integral.  With this extra factor, multiplying
\eqn{Hipath} to the wave function can then be viewed as attaching the complex
Fig. \ref{Hi}a to the triangular lattice (see Fig. \ref{Hi}b).  Note that the
factor $ (-)^{ m_1(g_i,g_2)+m_1(g_3,g_i) +m_1(g_5,g_i) +m_1(g_i,g_6) } $ plus
the factor that we included in the wave function $
(-)^{m_1(g_4,g_i)+m_1(g_i,g_1)} $ give us a factor
$(-)^{m_1(g_4,g_i)+m_1(g_i,g_1)+ m_1(g_i,g_2)+m_1(g_3,g_i) +m_1(g_5,g_i)
+m_1(g_i,g_6) } $ on all the six interior edges $\{i,1\}$, $\{i,2\}$,...,
$\{i,6\}$.  This factor is needed in the path integral on the complex in Fig.
\ref{Hi}b obtained by attaching the complex in Fig. \ref{Hi}a to the triangular
lattice.  Also note that the factor $ (-)^{m_1(g_4,g_i')+m_1(g_i',g_1)}$ is
what we need in the new wave function after the action of attaching the complex
Fig. \ref{Hi}a to the triangular lattice (see Fig. \ref{Hi}b).  By combining
all the $m_1$ factors on the right-hand-side of \eqn{Hipath}, we find that all
those $m_1$ factors become $(-)^{ n_2(g_3,g_i,g_i') +n_2(g_4,g_i,g_i')
+n_2(g_5,g_i,g_i')}$.  So we find
\begin{align}
\label{Hipath1}
& \ \ \ \
(-)^{m_1(g_4,g_i')+m_1(g_i',g_1)}
(-)^{ m_1(g_i,g_2)+m_1(g_3,g_i) +m_1(g_5,g_i) +m_1(g_i,g_6) }
\int \prod_\Si \cV_3
\\
&=(-)^{ n_2(g_3,g_i,g_i') +n_2(g_4,g_i,g_i') +n_2(g_5,g_i,g_i')}
\int
\prod_{j=1,2,6}
\dd \th_{(ii'j)}^{n_2(g_i,g_i',g_j)} \dd \bar\th_{(ii'j)}^{n_2(g_i,g_i',g_j)}
\prod_{j=3,4,5}
\dd \th_{(jii')}^{n_2(g_j,g_i,g_i')}\dd  \bar\th_{(jii')}^{n_2(g_j,g_i,g_i')}
\times
\nonumber\\ &
(-)^{
 n_2(g_4,g_5,g_i)  n_2(g_4,g_i,g_i')
+n_2(g_4,g_i,g_i') n_2(g_4,g_3,g_i)
+n_2(g_i',g_2,g_1) n_2(g_i,g_i',g_1)
+n_2(g_i,g_i',g_1) n_2(g_i',g_6,g_1)
}
\nonumber\\ &
\nu_3(g_5,g_i,g_i',g_6)
\th_{(ii'6)}^{n_2(g_i,g_i',g_6)}
\th_{(5i6)}^{n_2(g_5,g_i,g_6)}
\bar\th_{(5i'6)}^{n_2(g_5,g_i',g_6)}
\bar\th_{(5ii')}^{n_2(g_5,g_i,g_i')}
\nu_3(g_4,g_5,g_i,g_i')
\th_{(5ii')}^{n_2(g_5,g_i,g_i')}
\th_{(45i')}^{n_2(g_4,g_5,g_i')}
\bar\th_{(45i)}^{n_2(g_4,g_5,g_i)}
\bar\th_{(4ii')}^{n_2(g_4,g_i,g_i')}
\nonumber\\ &
\nu_3^{-1}(g_4,g_3,g_i,g_i')
\th_{(4ii')}^{n_2(g_4,g_i,g_i')}
\th_{(43i)}^{n_2(g_4,g_3,g_i)}
\bar\th_{(43i')}^{n_2(g_4,g_3,g_i')}
\bar\th_{(3ii')}^{n_2(g_3,g_i,g_i')}
\nu^{-1}_3(g_3,g_i,g_i',g_2)
\th_{(3ii')}^{n_2(g_3,g_i,g_i')}
\th_{(3i'2)}^{n_2(g_3,g_i',g_2)}
\bar\th_{(3i2)}^{n_2(g_3,g_i,g_2)}
\bar\th_{(ii'2)}^{n_2(g_i,g_i',g_2)}
\nonumber\\ &
\nu^{-1}_3( g_i,g_i',g_2,g_1)
\th_{(ii'2)}^{n_2(g_i,g_i',g_2)}
\th_{(i21)}^{n_2(g_i,g_2,g_1)}
\bar\th_{(i'21)}^{n_2(g_i',g_2,g_1)}
\bar\th_{(ii'1)}^{n_2(g_i,g_i',g_1)}
\nu_3(g_i,g_i',g_6,g_1)
\th_{(ii'1)}^{n_2(g_i,g_i',g_1)}
\th_{(i'61)}^{n_2(g_i',g_6,g_1)}
\bar\th_{(i61)}^{n_2(g_i,g_6,g_1)}
\bar\th_{(ii'6)}^{n_2(g_i,g_i',g_6)}.
\nonumber
\end{align}
In the above, we have also rearranged the order of the Grassmann numbers within
some simplexes to bring, say, $ \bar\th_{(4ii')}^{n_2(g_4,g_i,g_i')} $ next to
$ \th_{(4ii')}^{n_2(g_4,g_i,g_i')} $.  Now we can integrate out $\dd \th\dd
\bar\th$'s and obtain
\begin{align}
\label{Hipath2}
& \ \ \ \
(-)^{m_1(g_4,g_i')+m_1(g_i',g_1)}
(-)^{ m_1(g_i,g_2)+m_1(g_3,g_i) +m_1(g_5,g_i) +m_1(g_i,g_6) }
\int \prod_\Si \cV_3
=(-)^{n_2(g_i,g_i',g_6)} \times
\\
&(-)^{ n_2(g_3,g_i,g_i') +n_2(g_4,g_i,g_i') +n_2(g_5,g_i,g_i')
+ n_2(g_4,g_5,g_i)  n_2(g_4,g_i,g_i')
+n_2(g_4,g_i,g_i') n_2(g_4,g_3,g_i)
+n_2(g_i',g_2,g_1) n_2(g_i,g_i',g_1)
+n_2(g_i,g_i',g_1) n_2(g_i',g_6,g_1)
}
\nonumber\\ &
\nu_3(g_5,g_i,g_i',g_6)
\th_{(5i6)}^{n_2(g_5,g_i,g_6)}
\bar\th_{(5i'6)}^{n_2(g_5,g_i',g_6)}
\nu_3(g_4,g_5,g_i,g_i')
\th_{(45i')}^{n_2(g_4,g_5,g_i')}
\bar\th_{(45i)}^{n_2(g_4,g_5,g_i)}
\nu_3^{-1}(g_4,g_3,g_i,g_i')
\th_{(43i)}^{n_2(g_4,g_3,g_i)}
\bar\th_{(43i')}^{n_2(g_4,g_3,g_i')}
\nonumber\\ &
\nu^{-1}_3(g_3,g_i,g_i',g_2)
\th_{(3i'2)}^{n_2(g_3,g_i',g_2)}
\bar\th_{(3i2)}^{n_2(g_3,g_i,g_2)}
\nu^{-1}_3( g_i,g_i',g_2,g_1)
\th_{(i21)}^{n_2(g_i,g_2,g_1)}
\bar\th_{(i'21)}^{n_2(g_i',g_2,g_1)}
\nu_3(g_i,g_i',g_6,g_1)
\th_{(i'61)}^{n_2(g_i',g_6,g_1)}
\bar\th_{(i61)}^{n_2(g_i,g_6,g_1)}
\nonumber
\end{align}
where the factor $(-)^{n_2(g_i,g_i',g_6)}$ comes from bringing
$\bar\th_{(ii'6)}^{n_2(g_i,g_i',g_6)}$ in \eqn{Hipath1} all the way from the
back to the front.  Let us rearrange the order of the Grassmann numbers to
bring, say, $\bar\th_{(5i'6)}^{n_2(g_5,g_i',g_6)}$ in front of
$\th_{(5i6)}^{n_2(g_5,g_i,g_6)}$:
\begin{align}
\label{Hipath3}
& \ \ \ \
(-)^{m_1(g_4,g_i')+m_1(g_i',g_1)}
(-)^{ m_1(g_i,g_2)+m_1(g_3,g_i) +m_1(g_5,g_i) +m_1(g_i,g_6) }
\int \prod_\Si \cV_3
\nonumber\\ &
=(-)^{n_2(g_i,g_i',g_6)+n_2(g_3,g_i,g_i') +n_2(g_4,g_i,g_i') +n_2(g_5,g_i,g_i')} \times
\\
&\ \ \ \
(-)^{
 n_2(g_4,g_5,g_i)  n_2(g_4,g_i,g_i')
+n_2(g_4,g_i,g_i') n_2(g_4,g_3,g_i)
+n_2(g_i',g_2,g_1) n_2(g_i,g_i',g_1)
+n_2(g_i,g_i',g_1) n_2(g_i',g_6,g_1)
} \times
\nonumber\\ &\ \ \ \
\frac{
\nu_3(g_4,g_5,g_i,g_i')
\nu_3(g_5,g_i,g_i',g_6)
\nu_3(g_i,g_i',g_6,g_1)
}{
\nu_3( g_i,g_i',g_2,g_1)
\nu_3(g_3,g_i,g_i',g_2)
\nu_3(g_4,g_3,g_i,g_i')
}
(-)^{
 n_2(g_5,g_i',g_6) n_2(g_5,g_i,g_6)
+n_2(g_4,g_3,g_i') n_2(g_4,g_3,g_i)
+n_2(g_i',g_2,g_1) n_2(g_i,g_2,g_1)
}
\times
\nonumber\\ &\ \ \ \ \ \ \
\bar\th_{(5i'6)}^{n_2(g_5,g_i',g_6)}
\th_{(5i6)}^{n_2(g_5,g_i,g_6)}
\th_{(45i')}^{n_2(g_4,g_5,g_i')}
\bar\th_{(45i)}^{n_2(g_4,g_5,g_i)}
\bar\th_{(43i')}^{n_2(g_4,g_3,g_i')}
\th_{(43i)}^{n_2(g_4,g_3,g_i)} \times
\nonumber\\ &\ \ \ \ \ \ \
\th_{(3i'2)}^{n_2(g_3,g_i',g_2)}
\bar\th_{(3i2)}^{n_2(g_3,g_i,g_2)}
\bar\th_{(i'21)}^{n_2(g_i',g_2,g_1)}
\th_{(i21)}^{n_2(g_i,g_2,g_1)}
\th_{(i'61)}^{n_2(g_i',g_6,g_1)}
\bar\th_{(i61)}^{n_2(g_i,g_6,g_1)}
\nonumber
\end{align}

Indeed, the above expression can be regarded as the fermion coherent
state representation of the ideal Hamiltonian.  We can replace, for example,
$\bar\th_{(5i'6)}^{n_2(g_5,g_i',g_6)}\th_{(5i6)}^{n_2(g_5,g_i,g_6)}$ as:
\begin{align}
\hat C={\left(c^\dagger_{(5i6)}\right)}^{n_2(g_5,g_i',g_6)}
c_{(5i6)}^{n_2(g_5,g_i,g_6)}-\left(1-n_2(g_5,g_i',g_6)\right)\left(1-n_2(g_5,g_i,g_6)\right)c^\dagger_{(5i6)}c_{(5i6)},
\end{align}
We have
\begin{align}
 \hat C &= c_{(5i6)} c^\dagger_{(5i6)}, \text{ for }
[n_2(g_5,g_i',g_6), n_2(g_5,g_i,g_6)]=[0,0],
&
 \hat C &=c_{(5i6)}, \text{ for }
[n_2(g_5,g_i',g_6), n_2(g_5,g_i,g_6)]=[0,1],
\\
 \hat C &=c^\dag_{(5i6)}, \text{ for }
[n_2(g_5,g_i',g_6), n_2(g_5,g_i,g_6)]=[1,0],
&
 \hat C &=c^\dagger_{(5i6)} c_{(5i6)}, \text{ for }
[n_2(g_5,g_i',g_6), n_2(g_5,g_i,g_6)]=[1,1].
\nonumber
\end{align}
We note that the fermion coherent state is defined as
$|\th_{(5i6)}\rangle=|0\rangle-\th_{(5i6)}c^\dagger_{(5i6)}|0\rangle$. It is
easy to check
\begin{align}
\langle\bar\th_{(5i'6)}|{\left(c^\dagger_{(5i6)}\right)}^{n_2(g_5,g_i',g_6)}
c_{(5i6)}^{n_2(g_5,g_i,g_6)}-\left(1-n_2(g_5,g_i',g_6)\right)\left(1-n_2(g_5,g_i,g_6)\right)c^\dagger_{(5i6)}c_{(5i6)}|\th_{(5i6)}\rangle=
\bar\th_{(5i'6)}^{n_2(g_5,g_i',g_6)}\th_{(5i6)}^{n_2(g_5,g_i,g_6)}
\end{align}
However, a sign factor $(-)^{n_2(g_5,g_i,g_6)}$ is required when the
Hamiltonian acts on the coherent state since
$\dd\th_{(5i6)}^{n_2(g_5,g_i,g_6)}\dd\bar\th_{(5i6)}^{n_2(g_5,g_i,g_6)}$ should
be reordered as
$\dd\bar\th_{(5i6)}^{n_2(g_5,g_i,g_6)}\dd\th_{(5i6)}^{n_2(g_5,g_i,g_6)}$.
Similarly, for the face $(45i)$, the fermion coherent state is defined as
$|\bar\th_{(45i)}\rangle=|0\rangle-\bar\th_{(45i)}c^\dagger_{(45i)}|0\rangle$.
In this case
$\dd\th_{(45i)}^{n_2(g_4,g_5,g_i)}\dd\bar\th_{(45i)}^{n_2(g_4,g_5,g_i)}$ is the
correct ordering and we don't need to introduce the extra sign factor.

By applying the same discussions to other triangles, we find that $43i$ and $i12$ also contribute
sign factors $(-)^{n_2(g_4,g_3,g_i)}$ and $(-)^{n_2(g_i,g_2,g_1)}$. We also note that
\begin{align}
&\ \ \ \
(-)^{n_2(g_i,g_i',g_6)+n_2(g_3,g_i,g_i') +n_2(g_4,g_i,g_i') +n_2(g_5,g_i,g_i')}
(-)^{
 n_2(g_5,g_i,g_6)
+n_2(g_4,g_3,g_i)
+n_2(g_i,g_2,g_1)
}
\nonumber\\
&=(-)^{
 n_2(g_4,g_3,g_i')
+n_2(g_5,g_i',g_6)
+n_2(g_i,g_2,g_1)
}.
\end{align}
and finally obtain
\begin{align}
\label{Higi}
 H_i
&=\sum_{g_i,g_i'}
|g_i',g_1g_2g_3g_4g_5g_6\>\<g_i,g_1g_2g_3g_4g_5g_6| \
\frac{
\nu_3(g_4,g_5,g_i,g_i')
\nu_3(g_5,g_i,g_i',g_6)
\nu_3(g_i,g_i',g_6,g_1)
}{
\nu_3( g_i,g_i',g_2,g_1)
\nu_3(g_3,g_i,g_i',g_2)
\nu_3(g_4,g_3,g_i,g_i')
}  \times
\\
& \ \ \ \
(-)^{
 n_2(g_4,g_5,g_i)  n_2(g_4,g_i,g_i')
+n_2(g_4,g_i,g_i') n_2(g_4,g_3,g_i)
+n_2(g_i',g_2,g_1) n_2(g_i,g_i',g_1)
+n_2(g_i,g_i',g_1) n_2(g_i',g_6,g_1)
} \times
\nonumber\\ &\ \ \ \
(-)^{
 n_2(g_5,g_i',g_6) n_2(g_5,g_i,g_6)
+n_2(g_4,g_3,g_i') n_2(g_4,g_3,g_i)
+n_2(g_i',g_2,g_1) n_2(g_i,g_2,g_1)
}
(-)^{ n_2(g_4,g_3,g_i') +n_2(g_5,g_i',g_6) +n_2(g_i,g_2,g_1) }
\times
\nonumber\\ &\ \ \ \
\left[{\left(c^\dagger_{(5i6)}\right)}^{n_2(g_5,g_i',g_6)}
c_{(5i6)}^{n_2(g_5,g_i,g_6)}-\left(1-n_2(g_5,g_i',g_6)\right)\left(1-n_2(g_5,g_i,g_6)\right)c^\dagger_{(5i6)}c_{(5i6)}\right]\times
\nonumber\\ &\ \ \ \
\left[{\left(c^\dagger_{(45i)}\right)}^{n_2(g_4,g_5,g_i')}
c_{(45i)}^{n_2(g_4,g_5,g_i)}-\left(1-n_2(g_4,g_5,g_i')\right)\left(1-n_2(g_4,g_5,g_i)\right)c^\dagger_{(45i)}c_{(45i)}\right]\times
\nonumber\\ &\ \ \ \
\left[{\left(c^\dagger_{(43i)}\right)}^{n_2(g_4,g_3,g_i')}
c_{(43i)}^{n_2(g_4,g_3,g_i)}-\left(1-n_2(g_4,g_3,g_i')\right)\left(1-n_2(g_4,g_3,g_i)\right)c^\dagger_{(43i)}c_{(43i)}\right]\times
\nonumber\\ &\ \ \ \
\left[{\left(c^\dagger_{(3i2)}\right)}^{n_2(g_3,g_i',g_2)}
c_{(3i2)}^{n_2(g_3,g_i,g_2)}-\left(1-n_2(g_3,g_i',g_2)\right)\left(1-n_2(g_3,g_i,g_2)\right)c^\dagger_{(3i2)}c_{(3i2)}\right]\times
\nonumber\\ &\ \ \ \
\left[{\left(c^\dagger_{(i21)}\right)}^{n_2(g_i',g_2,g_1)}
c_{(i21)}^{n_2(g_i,g_2,g_1)}-\left(1-n_2(g_i',g_2,g_1)\right)\left(1-n_2(g_i,g_2,g_1)\right)c^\dagger_{(i21)}c_{(i21)}\right]\times
\nonumber\\ &\ \ \ \
\left[{\left(c^\dagger_{(i61)}\right)}^{n_2(g_i',g_6,g_1)}
c_{(i61)}^{n_2(g_i,g_6,g_1)}-\left(1-n_2(g_i',g_6,g_1)\right)\left(1-n_2(g_i,g_6,g_1)\right)c^\dagger_{(i61)}c_{(i61)}\right]
\end{align}
We can rewrite $H_i$ as
\begin{align}
\label{Higi6}
 H_i
&=\sum_{g_i,g_i'}
|g_i',g_1g_2g_3g_4g_5g_6\>\<g_i,g_1g_2g_3g_4g_5g_6|
O_{56;g_5g_6}^{g_{i'}g_i}
O_{45;g_4g_5}^{g_{i'}g_i}
O_{43;g_4g_3}^{g_{i'}g_i}
O_{32;g_3g_2}^{g_{i'}g_i}
O_{21;g_2g_1}^{g_{i'}g_i}
O_{61;g_6g_1}^{g_{i'}g_i}
\end{align}
where
$
O_{mn;g_mg_n}^{g_{i'}g_i}
$
are given by
\begin{align}
O_{21;g_2g_1}^{g_{i'}g_i} &=
\nu_3^{-1}( g_i,g_i',g_2,g_1)
(-)^{
 n_2(g_i',g_2,g_1) n_2(g_i,g_i',g_1)
+n_2(g_i',g_2,g_1) n_2(g_i,g_2,g_1)
+n_2(g_i,g_2,g_1) }
\times
\nonumber\\ &\ \ \ \
\left[{\left(c^\dagger_{(i21)}\right)}^{n_2(g_i',g_2,g_1)}
c_{(i21)}^{n_2(g_i,g_2,g_1)}-\left(1-n_2(g_i',g_2,g_1)\right)\left(1-n_2(g_i,g_2,g_1)\right)c^\dagger_{(i21)}c_{(i21)}\right]
\end{align}
\begin{align}
O_{32;g_3g_2}^{g_{i'}g_i}
&=
\nu_3^{-1}(g_3,g_i,g_i',g_2)
\left[{\left(c^\dagger_{(3i2)}\right)}^{n_2(g_3,g_i',g_2)}
c_{(3i2)}^{n_2(g_3,g_i,g_2)}-\left(1-n_2(g_3,g_i',g_2)\right)\left(1-n_2(g_3,g_i,g_2)\right)c^\dagger_{(3i2)}c_{(3i2)}\right]
\end{align}
\begin{align}
O_{43;g_4g_3}^{g_{i'}g_i}
&=
\nu_3^{-1}(g_4,g_3,g_i,g_i')
(-)^{
+n_2(g_4,g_i,g_i') n_2(g_4,g_3,g_i)
+n_2(g_4,g_3,g_i') n_2(g_4,g_3,g_i)
+ n_2(g_4,g_3,g_i') }
\times
\nonumber\\ &\ \ \ \
\left[{\left(c^\dagger_{(43i)}\right)}^{n_2(g_4,g_3,g_i')}
c_{(43i)}^{n_2(g_4,g_3,g_i)}-\left(1-n_2(g_4,g_3,g_i')\right)\left(1-n_2(g_4,g_3,g_i)\right)c^\dagger_{(43i)}c_{(43i)}\right]
\end{align}
\begin{align}
O_{45;g_4g_5}^{g_{i'}g_i}
&=
\nu_3(g_4,g_5,g_i,g_i')
(-)^{ n_2(g_4,g_5,g_i)  n_2(g_4,g_i,g_i')}
\times
\nonumber\\ &\ \ \ \
\left[{\left(c^\dagger_{(45i)}\right)}^{n_2(g_4,g_5,g_i')}
c_{(45i)}^{n_2(g_4,g_5,g_i)}-\left(1-n_2(g_4,g_5,g_i')\right)\left(1-n_2(g_4,g_5,g_i)\right)c^\dagger_{(45i)}c_{(45i)}\right]
\end{align}
\begin{align}
O_{56;g_5g_6}^{g_{i'}g_i}
&=
\nu_3(g_5,g_i,g_i',g_6)
(-)^{
 n_2(g_5,g_i',g_6) n_2(g_5,g_i,g_6)
+n_2(g_5,g_i',g_6) }
\times
\nonumber\\ &\ \ \ \
\left[{\left(c^\dagger_{(5i6)}\right)}^{n_2(g_5,g_i',g_6)}
c_{(5i6)}^{n_2(g_5,g_i,g_6)}-\left(1-n_2(g_5,g_i',g_6)\right)\left(1-n_2(g_5,g_i,g_6)\right)c^\dagger_{(5i6)}c_{(5i6)}\right]
\end{align}
\begin{align}
O_{61;g_6g_1}^{g_{i'}g_i}
&=
\nu_3(g_i,g_i',g_6,g_1)
(-)^{
+n_2(g_i,g_i',g_1) n_2(g_i',g_6,g_1)}
\times
\nonumber\\ &\ \ \ \
\left[{\left(c^\dagger_{(i61)}\right)}^{n_2(g_i',g_6,g_1)}
c_{(i61)}^{n_2(g_i,g_6,g_1)}-\left(1-n_2(g_i',g_6,g_1)\right)\left(1-n_2(g_i,g_6,g_1)\right)c^\dagger_{(i61)}c_{(i61)}\right]
\end{align}

The above expression for $H_i$ is valid in the subspace where the fermion
occupation number $n_{ijk}$ on each triangle $(ijk)$ satisfies $n_{ijk}=
n_2(g_i,g_j,g_k)$.
The ideal fermionic SPT state $|\Psi\>$ is also in this subspace.
We can add a term
\begin{align}
 H_{(ijk)}=U
|g_ig_jg_k\> \<g_ig_jg_k|
(c_{(ijk)}^\dag c_{(ijk)} - n_2(g_i,g_j,g_k))^2
\end{align}
with a large positive $U$
on each triangle to put the ground state in the subspace.

Next, we would like to show that $H_i$ is hermitian. This property is very important and it makes
the whole theory to be unitary. In the above, we express the matrix element $H_{i;g_i,g_i^\prime}(g_1g_2g_3g_4g_5g_6)$ as:
\begin{align}
(-)^{ m_1(g_4,g_i')+m_1(g_i',g_1)+m_1(g_i,g_2)+m_1(g_3,g_i) +m_1(g_5,g_i) +m_1(g_i,g_6) }
\int \prod_\Si \cV_3,
\end{align}
hence, its hermitian conjugate $H^*_{i;g_i^\prime,g_i}(g_1g_2g_3g_4g_5g_6)$ reads:
\begin{align}
&(-)^{ m_1(g_4,g_i)+m_1(g_i,g_1)+m_1(g_i',g_2)+m_1(g_3,g_i') +m_1(g_5,g_i') +m_1(g_i',g_6) }
\int
\prod_{j=1,2,6}
\dd \th_{(ii'j)}^{n_2(g_i',g_i,g_j)} \dd \bar\th_{(ii'j)}^{n_2(g_i',g_i,g_j)}
\prod_{j=3,4,5}
\dd \th_{(jii')}^{n_2(g_j,g_i',g_i)}\dd  \bar\th_{(jii')}^{n_2(g_j,g_i',g_i)}
\nonumber\\ &
(-)^{
 m_1(g_4,g_i')
+m_1(g_5,g_i)
+m_1(g_i',g_6)
+m_1(g_i,g_1)
+m_1(g_i',g_2)
+m_1(g_3,g_i)
}
(-)^{
 m_1(g_i',g_i)
}
\times
\nonumber\\ &
\nu_3^{-1}(g_5,g_i',g_i,g_6)
\th_{(5ii')}^{n_2(g_5,g_i',g_i)}
\th_{(5i'6)}^{n_2(g_5,g_i,g_6)}
\bar\th_{(5i6)}^{n_2(g_5,g_i',g_6)}
\bar\th_{(ii'6)}^{n_2(g_i',g_i,g_6)}
\nu_3^{-1}(g_4,g_5,g_i',g_i)
\th_{(45i)}^{n_2(g_4,g_5,g_i')}
\th_{(4ii')}^{n_2(g_4,g_i',g_i)}
\bar\th_{(45i')}^{n_2(g_4,g_5,g_i)}
\bar\th_{(5ii')}^{n_2(g_5,g_i',g_i)}
\nonumber\\ &
\nu_3(g_4,g_3,g_i',g_i)
\th_{(3ii')}^{n_2(g_3,g_i',g_i)}
\th_{(43i')}^{n_2(g_4,g_3,g_i)}
\bar\th_{(4ii')}^{n_2(g_4,g_i',g_i)}
\bar\th_{(43i)}^{n_2(g_4,g_3,g_i')}
\nu_3(g_3,g_i',g_i,g_2)
\th_{(ii'2)}^{n_2(g_i',g_i,g_2)}
\th_{(3i2)}^{n_2(g_3,g_i',g_2)}
\bar\th_{(3i'2)}^{n_2(g_3,g_i,g_2)}
\bar\th_{(3ii')}^{n_2(g_3,g_i',g_i)}
\nonumber\\ &
\nu_3( g_i',g_i,g_2,g_1)
\bar\th_{(i'21)}^{n_2(g_i,g_2,g_1)}
\bar\th_{(ii'1)}^{n_2(g_i',g_i,g_1)}
\th_{(i21)}^{n_2(g_i',g_2,g_1)}
\th_{(ii'2)}^{n_2(g_i',g_i,g_2)}
\nu^{-1}_3(g_i',g_i,g_6,g_1)
\th_{(ii'6)}^{n_2(g_i',g_i,g_6)}
\th_{(i61)}^{n_2(g_i',g_6,g_1)}
\bar\th_{(ii'1)}^{n_2(g_i',g_i,g_1)}
\bar\th_{(i'61)}^{n_2(g_i,g_6,g_1)}.
\end{align}
Note $\nu^*=\nu^{-1}$($\nu$ is a $U(1)$ phase factor) and $(\theta_1\theta_2\bar\theta_3\bar\theta_4)^*=\theta_4\theta_3\bar\theta_2\bar\theta_1$. However, we can not
directly compare the above expression with $H_{i;g_i,g_i^\prime}(g_1g_2g_3g_4g_5g_6)$, since its a function of new pairs of Grassmann
variable  $\th_{(5i'6)}^{n_2(g_5,g_i',g_6)}
\bar\th_{(5i6)}^{n_2(g_5,g_i,g_6)},
\bar\th_{(45i')}^{n_2(g_4,g_5,g_i')}
\th_{(45i)}^{n_2(g_4,g_5,g_i)},
\th_{(43i')}^{n_2(g_4,g_3,g_i')}
\bar\th_{(43i)}^{n_2(g_4,g_3,g_i)}
\bar\th_{(3i'2)}^{n_2(g_3,g_i',g_2)}
\th_{(3i2)}^{n_2(g_3,g_i,g_2)},
\th_{(i'21)}^{n_2(g_i',g_2,g_1)}
\bar\th_{(i21)}^{n_2(g_i,g_2,g_1)}$
and
$\bar\th_{(i'61)}^{n_2(g_i',g_6,g_1)}
\th_{(i61)}^{n_2(g_i,g_6,g_1)}$. Such a difference is simply because the action of $H_i$ maps, for example,
$\bar\th_{(5i6)}$ to $\bar\th_{(5i'6)}$ while the action of $H_i^\dagger$ maps $\bar\th_{(5i'6)}$ back to $\bar\th_{(5i6)}$. Thus, to see
whether they are the same mapping in the Hilbert space or not, we need to redefine $\bar\th_{(5i6)}(\th_{(5i6)})$ as $\bar\th_{(5i'6)}(\th_{(5i'6)})$ and $\bar\th_{(5i'6)}(\th_{(5i'6)})$ as $\bar\th_{(5i6)}(\th_{(5i6)})$. A simple way to do so is just replacing $i(i')$ by $i'(i)$ in the above expression:
\begin{align}
&(-)^{ m_1(g_4,g_i)+m_1(g_i,g_1)+m_1(g_i',g_2)+m_1(g_3,g_i') +m_1(g_5,g_i') +m_1(g_i',g_6) }
\int
\prod_{j=1,2,6}
\dd \th_{(i'ij)}^{n_2(g_i',g_i,g_j)} \dd \bar\th_{(i'ij)}^{n_2(g_i',g_i,g_j)}
\prod_{j=3,4,5}
\dd \th_{(ji'i)}^{n_2(g_j,g_i',g_i)}\dd  \bar\th_{(ji'i)}^{n_2(g_j,g_i',g_i)}
\nonumber\\ &
(-)^{
 m_1(g_4,g_i')
+m_1(g_5,g_i)
+m_1(g_i',g_6)
+m_1(g_i,g_1)
+m_1(g_i',g_2)
+m_1(g_3,g_i)
}
(-)^{
 m_1(g_i',g_i)
}
\times
\nonumber\\ &
\nu_3^{-1}(g_5,g_i',g_i,g_6)
\th_{(5i'i)}^{n_2(g_5,g_i',g_i)}
\th_{(5i6)}^{n_2(g_5,g_i,g_6)}
\bar\th_{(5i'6)}^{n_2(g_5,g_i',g_6)}
\bar\th_{(i'i6)}^{n_2(g_i',g_i,g_6)}
\nu_3^{-1}(g_4,g_5,g_i',g_i)
\th_{(45i')}^{n_2(g_4,g_5,g_i')}
\th_{(4i'i)}^{n_2(g_4,g_i',g_i)}
\bar\th_{(45i)}^{n_2(g_4,g_5,g_i)}
\bar\th_{(5i'i)}^{n_2(g_5,g_i',g_i)}
\nonumber\\ &
\nu_3(g_4,g_3,g_i',g_i)
\th_{(3i'i)}^{n_2(g_3,g_i',g_i)}
\th_{(43i)}^{n_2(g_4,g_3,g_i)}
\bar\th_{(4i'i)}^{n_2(g_4,g_i',g_i)}
\bar\th_{(43i')}^{n_2(g_4,g_3,g_i')}
\nu_3(g_3,g_i',g_i,g_2)
\th_{(i'i2)}^{n_2(g_i',g_i,g_2)}
\th_{(3i'2)}^{n_2(g_3,g_i',g_2)}
\bar\th_{(3i2)}^{n_2(g_3,g_i,g_2)}
\bar\th_{(3i'i)}^{n_2(g_3,g_i',g_i)}
\nonumber\\ &
\nu_3( g_i',g_i,g_2,g_1)
\bar\th_{(i21)}^{n_2(g_i,g_2,g_1)}
\bar\th_{(i'i1)}^{n_2(g_i',g_i,g_1)}
\th_{(i'21)}^{n_2(g_i',g_2,g_1)}
\th_{(i'i2)}^{n_2(g_i',g_i,g_2)}
\nu^{-1}_3(g_i',g_i,g_6,g_1)
\th_{(i'i6)}^{n_2(g_i',g_i,g_6)}
\th_{(i'61)}^{n_2(g_i',g_6,g_1)}
\bar\th_{(i'i1)}^{n_2(g_i',g_i,g_1)}
\bar\th_{(i61)}^{n_2(g_i,g_6,g_1)}\nonumber\\
&=(-)^{ m_1(g_4,g_i)+m_1(g_i,g_1)+m_1(g_i',g_2)+m_1(g_3,g_i') +m_1(g_5,g_i') +m_1(g_i',g_6) }
\int
\prod_{j=1,2,6}
\dd \th_{(i'ij)}^{n_2(g_i',g_i,g_j)} \dd \bar\th_{(i'ij)}^{n_2(g_i',g_i,g_j)}
\prod_{j=3,4,5}
\dd \th_{(ji'i)}^{n_2(g_j,g_i',g_i)}\dd  \bar\th_{(ji'i)}^{n_2(g_j,g_i',g_i)}
\nonumber\\ &
(-)^{
 m_1(g_4,g_i')
+m_1(g_5,g_i)
+m_1(g_i',g_6)
+m_1(g_i,g_1)
+m_1(g_i',g_2)
+m_1(g_3,g_i)
}
(-)^{
 m_1(g_i',g_i)
}
\times
\nonumber\\ &
\mathcal{V}_3^{-}(g_5,g_i',g_i,g_6)
\mathcal{V}_3^{-}(g_4,g_5,g_i',g_i)
\mathcal{V}_3^+(g_4,g_3,g_i',g_i)
\mathcal{V}_3^+(g_3,g_i',g_i,g_2)
\mathcal{V}_3^+( g_i',g_i,g_2,g_1)
\mathcal{V}^{-}_3(g_i',g_i,g_6,g_1)\nonumber\\
&=(-)^{ m_1(g_4,g_i)+m_1(g_i,g_1)+m_1(g_i',g_2)+m_1(g_3,g_i') +m_1(g_5,g_i') +m_1(g_i',g_6) }
\int \prod_\Si \cV_3.
\end{align}
We note that in the last line we evaluate the complex Fig.\ref{Hi}a in a different way(by choosing opposite
orientation for the internal link). The topological invariance of $\cV_3$ path integral implies the two different ways  must give out the same results. It is also not hard to see that the new sign factor $(-)^{ m_1(g_4,g_i)+m_1(g_i,g_1)+m_1(g_i',g_2)+m_1(g_3,g_i') +m_1(g_5,g_i') +m_1(g_i',g_6) }$ is equivalent to $(-)^{ m_1(g_4,g_i')+m_1(g_i',g_1)+m_1(g_i,g_2)+m_1(g_3,g_i) +m_1(g_5,g_i) +m_1(g_i,g_6) }$, since:
\begin{align}
&m_1(g_4,g_i)+m_1(g_i,g_1)+m_1(g_i',g_2)+m_1(g_3,g_i') +m_1(g_5,g_i') +m_1(g_i',g_6)+\nonumber\\
& m_1(g_4,g_i')+m_1(g_i',g_1)+m_1(g_i,g_2)+m_1(g_3,g_i) +m_1(g_5,g_i) +m_1(g_i,g_6)\nonumber\\
=&n_2(g_5,g_i',g_6)+n_2(g_5,g_i,g_6)+n_2(g_4,g_5,g_i')+n_2(g_4,g_5,g_i)+n_2(g_4,g_3,g_i')+n_2(g_4,g_3,g_i)+\nonumber\\
&n_2(g_3,g_i',g_2)+n_2(g_3,g_i,g_2)+n_2(g_i',g_2,g_1)+n_2(g_i,g_2,g_1)+n_2(g_i',g_6,g_1)+n_2(g_i,g_6,g_1)\nonumber\\
=&\text{$0$ mod $2$}.
\end{align}
\end{widetext}
Thus, we have proved that the matrix element $H_{i;g_i,g_i^\prime}(g_1g_2g_3g_4g_5g_6)$ is the same as $H^*_{i;g_i^\prime,g_i}(g_1g_2g_3g_4g_5g_6)$, implying
$H_i$ is a hermitian operator.

Finally, we note that the Hamiltonian only depends on $\nu_3$ and $n_2$.  It does not
depend on $m_1$. So the  Hamiltonian is symmetric under the $G_f$ symmetry.  By
construction, the ideal ground state wave function is an eigenstate of $H_i$
with eigenvalue 1.  The topological invariance of $\cV_3$ path integral implies
that $H_i^2=H_i$ and $H_iH_j=H_jH_i$.  Thus $H_i$ is a
hermitian projection operator, and the set $\{H_i\}$ is a set of commuting projectors.
Therefore $H=-\sum_i H_i$ is an exactly solvable Hamiltonian which realizes the
fermionic SPT state described by $(\nu_3,n_2,u_2^g) \in \fZ^3[G_f,U_T(1)]$.

\section{
The mapping $n_d \to f_{d+2}$ induces
a mapping $\cH^d(G_b,\Z_2) \to \cH^{d+2}(G_b,\Z_2)$}

\label{f5equivalent}

We note that \eqn{fdn} defines a mapping from a $d$-cochain $n_d \in
\cC^d(G_b,\Z_2)$ to a $(d+2)$-cochain $f_d \in \cC^{d+2}(G_b,\Z_2)$.  We can
show that, if $n_d$ is a cocycle $n_d \in \cZ^d(G_b,\Z_2)$, then the
corresponding $f_{d+2}$ is also a cocycle $f_d \in \cZ^{d+2}(G_b,\Z_2)$.  Thus
\eqn{fdn} defines a mapping from a $d$-cocycle $n_d \in \cZ^d(G_b,\Z_2)$ to a
$(d+2)$-cocycle $f_d \in \cZ^{d+2}(G_b,\Z_2)$.  In this appendix, we are going
to show that \eqn{fdn} actually defines a mapping from $d$-cohomology classes to
$(d+2)$-cohomology classes: $\cH^d(G_b,\Z_2)\to \cH^{d+2}(G_b,\Z_2)$.  This is
because if $n_d$ and $\t n_d$ differ by a coboundary, then the corresponding
$f_{d+2}$ and $\t f_{d+2}$ also differ by a coboundary.

To show this, let us first assume $d=2$.  Since the 4-cochain $f_4(g_0 ,g_1
,g_2 ,g_3 ,g_4)$ in $\cC^4(G_b,\Z_2)$ is the cup product to two 2-cocycles
$n_2(g_0,g_1,g_2)$ and $n_2(g_0,g_1,g_2)$ in $\cZ^2(G_b,\Z_2)$, so if we change
the two cocycles by a coboundary, the 4-cochain $f_4$ will also change by a
coboundary.  Thus there exists a vector $\v g_3$ such that
\begin{align}
\t{\v f}_4=\v f_4 +D_3 \v g_3 \text{ mod } 2.
\end{align}
and \eqn{ffg} is valid for $d=2$.

To show  \eqn{ffg} to be valid for $d=3$,
we need to show when $n_3(g_0,g_1,g_2,g_3)$ and $\t
n_3(g_0,g_1,g_2,g_3)$ are related by a coboundary
in $\cB^3(G_b,\Z_2)$:
\begin{align}
&\t
n_3(g_0,g_1,g_2,g_3)=n_3(g_0,g_1,g_2,g_3)+m^\prime_2(g_1,g_2,g_3)\nonumber\\&+ m^\prime_2(g_0,g_2,g_3)+ m^\prime_2(g_0,g_1,g_3)+ m^\prime_2(g_0,g_1,g_2)
 \text{ mod }  2,
\end{align}
the corresponding $\v f_5$ and $ \t{\v f}_5$ are also related by a
coboundary in $\cB^5(G_b,\Z_2)$:
\begin{align}
 \v  f_5=\t{\v f}_5+D_4 \v  g_4 \text{ mod }  2.
\end{align}
Let us introduce
\begin{align}
n_3'(g_0,g_1,g_2,g_3)
&= m^\prime_2(g_1,g_2,g_3)+ m^\prime_2(g_0,g_2,g_3)\nonumber\\&+ m^\prime_2(g_0,g_1,g_3)+ m^\prime_2(g_0,g_1,g_2).
\end{align}
Then we can express $\t{f}_5$ as $\t{ f}_5= f_5+f_5^I+ f_5^{II}+
f_5^{III}$, with:
\begin{widetext}
\begin{align}
&f_5^I(g_0,g_1,g_2,g_3,g_4,g_5)
\nonumber\\
=&n_3(g_0,g_1,g_2,g_3)n_3^\prime(g_0,g_3,g_4,g_5)+n_3(g_1,g_2,g_3,g_4)n_3^\prime(g_0,g_1,g_4,g_5)
+n_3(g_2,g_3,g_4,g_5)n_3^\prime(g_0,g_1,g_2,g_5),
 \nonumber\\
&f_5^{II}(g_0,g_1,g_2,g_3,g_4,g_5)
\nonumber\\
=&n_3^\prime(g_0,g_1,g_2,g_3)n_3(g_0,g_3,g_4,g_5)+n_3^\prime(g_1,g_2,g_3,g_4)n_3(g_0,g_1,g_4,g_5)
+n_3^\prime(g_2,g_3,g_4,g_5)n_3(g_0,g_1,g_2,g_5),
 \nonumber\\
&f_5^{III}(g_0,g_1,g_2,g_3,g_4,g_5)
\nonumber\\
=&n_3^\prime(g_0,g_1,g_2,g_3)n_3^\prime(g_0,g_3,g_4,g_5)+n_3^\prime(g_1,g_2,g_3,g_4)n_3^\prime(g_0,g_1,g_4,g_5)
+n_3^\prime(g_2,g_3,g_4,g_5)n_3^\prime(g_0,g_1,g_2,g_5).
\end{align}

In the following, we would like to show $g_4$ can be constructed as
$g_4=g_4^I+g_4^{II}+g_4^{III}$, with:
\begin{align}
\label{gd}
g_4^I(g_0,g_1,g_2,g_3,g_4)&=n_3(g_0,g_1,g_2,g_3) m^\prime_2(g_0,g_3,g_4)
+n_3(g_1,g_2,g_3,g_4) m^\prime_2(g_0,g_1,g_4),
\\
g_4^{II}(g_0,g_1,g_2,g_3,g_4)&=n_3(g_0,g_2,g_3,g_4) m^\prime_2(g_0,g_1,g_2)
+n_3(g_0,g_1,g_2,g_4) m_2^\prime(g_2,g_3,g_4)+n_3(g_0,g_1,g_3,g_4) m^\prime_2(g_1,g_2,g_3),
\nonumber \\
g_4^{III}(g_0,g_1,g_2,g_3,g_4)&
=n_3^\prime(g_0,g_1,g_2,g_3) m^\prime_2(g_0,g_3,g_4)
+n_3^\prime(g_1,g_2,g_3,g_4) m^\prime_2(g_0,g_1,g_4)
+ m^\prime_2(g_0,g_1,g_2) m^\prime_2(g_2,g_3,g_4).
\nonumber
\end{align}
Let us prove the above statement in three steps:

\noindent \textbf{First step}:
After plug-in, we find that $\v f_5^I -D_4 \v g_4^I$ is given by
\begin{align}
&f_5^I(g_0,g_1,g_2,g_3,g_4,g_5)-g_4^I(g_0,g_1,g_2,g_3,g_4)-g_4^I(g_0,g_1,g_2,g_3,g_5)
-g_4^I(g_0,g_1,g_2,g_4,g_5)
\nonumber\\
&
\ \ \ \ \ \ \ \ \ \
\ \ \ \ \ \ \ \ \ \
\ \ \ \ \ \ \ \ \
-g_4^I(g_0,g_1,g_3,g_4,g_5)
-g_4^I(g_0,g_2,g_3,g_4,g_5)-g_4^I(g_1,g_2,g_3,g_4,g_5)
\nonumber\\
=&\ \ \ n_3(g_0,g_1,g_2,g_3)\left[ m^\prime_2(g_0,g_3,g_4)+ m^\prime_2(g_0,g_3,g_5)+ m^\prime_2(g_0,g_4,g_5)+ m^\prime_2(g_3,g_4,g_5)\right]
\nonumber\\
& +n_3(g_1,g_2,g_3,g_4)\left[ m^\prime_2(g_0,g_1,g_4)+ m^\prime_2(g_0,g_1,g_5)+ m^\prime_2(g_0,g_4,g_5)+ m^\prime_2(g_1,g_4,g_5)\right]
\nonumber\\
&
+n_3(g_2,g_3,g_4,g_5)\left[ m^\prime_2(g_0,g_1,g_2)+ m^\prime_2(g_0,g_1,g_5)+ m^\prime_2(g_0,g_2,g_5)+ m^\prime_2(g_1,g_2,g_5)\right]
\nonumber\\ &
-n_3(g_0,g_1,g_2,g_3) m^\prime_2(g_0,g_3,g_4)
-n_3(g_1,g_2,g_3,g_4) m^\prime_2(g_0,g_1,g_4)-n_3(g_0,g_1,g_2,g_3) m^\prime_2(g_0,g_3,g_5)
\nonumber\\ &
-n_3(g_1,g_2,g_3,g_5) m^\prime_2(g_0,g_1,g_5)
-n_3(g_0,g_1,g_2,g_4) m^\prime_2(g_0,g_4,g_5)
-n_3(g_1,g_2,g_4,g_5) m^\prime_2(g_0,g_1,g_5)
\nonumber\\ &
-n_3(g_0,g_1,g_3,g_4) m^\prime_2(g_0,g_4,g_5)
-n_3(g_1,g_3,g_4,g_5) m^\prime_2(g_0,g_1,g_5)
-n_3(g_0,g_2,g_3,g_4) m^\prime_2(g_0,g_4,g_5)
\nonumber\\ &
-n_3(g_2,g_3,g_4,g_5) m^\prime_2(g_0,g_2,g_5)
-n_3(g_1,g_2,g_3,g_4) m^\prime_2(g_1,g_4,g_5)
-n_3(g_2,g_3,g_4,g_5) m^\prime_2(g_1,g_2,g_5)
\nonumber
\end{align}
We note that there are 5 terms containing $ m^\prime_2(g_0,g_4,g_5)$ and 5 terms
containing $ m^\prime_2(g_0,g_1,g_5)$. Using the condition that $n_3(g_0,g_1,g_2,g_3)$
is a 3 cocycle:
\begin{align}
&n_3(g_0,g_1,g_2,g_3)+n_3(g_0,g_1,g_2,g_3)+n_3(g_0,g_1,g_2,g_3)
+n_3(g_0,g_1,g_2,g_3)+n_3(g_0,g_1,g_2,g_3)=0
\text{ mode }2,
\end{align}
we find that those terms cancel out.  Also, there are two terms containing
$ m^\prime_2(g_0,g_3,g_4)$, two terms containing $ m^\prime_2(g_0,g_4,g_5)$, ..., etc.  Those
terms also cancel out under the mod 2 calculation.  But the $ m^\prime_2(g_0,g_1,g_2)$
term and the $ m^\prime_2(g_3,g_4,g_5)$ term appear only once. Thus we simplify the
above to
\begin{align}
&f_5^I(g_0,g_1,g_2,g_3,g_4,g_5)-g_4^I(g_0,g_1,g_2,g_3,g_4)-g_4^I(g_0,g_1,g_2,g_3,g_5)
-g_4^I(g_0,g_1,g_2,g_4,g_5)
\nonumber\\
&
\ \ \ \ \ \ \ \ \ \
\ \ \ \ \ \ \ \ \ \
\ \ \ \ \ \ \ \ \
-g_4^I(g_0,g_1,g_3,g_4,g_5)
-g_4^I(g_0,g_2,g_3,g_4,g_5)-g_4^I(g_1,g_2,g_3,g_4,g_5)
\nonumber\\
=&n_3(g_0,g_1,g_2,g_3) m^\prime_2(g_3,g_4,g_5)
+n_3(g_2,g_3,g_4,g_5) m^\prime_2(g_0,g_1,g_2)\text{ mod } 2
\end{align}

\noindent \textbf{Second step}: Similarly, we have
\begin{align}
&f_5^{II}(g_0,g_1,g_2,g_3,g_4,g_5)-g_4^{II}(g_0,g_1,g_2,g_3,g_4)-g_4^{II}(g_0,g_1,g_2,g_3,g_5)
-g_4^{II}(g_0,g_1,g_2,g_4,g_5)
\nonumber\\&
\ \ \ \ \ \ \ \ \ \
\ \ \ \ \ \ \ \ \ \
\ \ \ \ \ \ \ \ \ \ \
-g_4^{II}(g_0,g_1,g_3,g_4,g_5)
-g_4^{II}(g_0,g_2,g_3,g_4,g_5)-g_4^{II}(g_1,g_2,g_3,g_4,g_5)\nonumber\\
=&-n_3(g_0,g_1,g_2,g_3) m^\prime_2(g_3,g_4,g_5)-n_3(g_2,g_3,g_4,g_5) m^\prime_2(g_0,g_1,g_2)\text{
mod } 2
\end{align}

\noindent \textbf{Third step}: Finally, by using the results derived
in the first step, we have
\begin{align}
&f_5^{III}(g_0,g_1,g_2,g_3,g_4,g_5)-g_4^{III}(g_0,g_1,g_2,g_3,g_4)-g_4^{III}(g_0,g_1,g_2,g_3,g_5)
-g_4^{III}(g_0,g_1,g_2,g_4,g_5)
\nonumber\\&
\ \ \ \ \ \ \ \ \ \
\ \ \ \ \ \ \ \ \ \
\ \ \ \ \ \ \ \ \ \ \
-g_4^{III}(g_0,g_1,g_3,g_4,g_5)
-g_4^{III}(g_0,g_2,g_3,g_4,g_5)-g_4^{III}(g_1,g_2,g_3,g_4,g_5)\nonumber\\
=&n_3^\prime(g_0,g_1,g_2,g_3) m^\prime_2(g_3,g_4,g_5)+n_3^\prime(g_2,g_3,g_4,g_5) m^\prime_2(g_0,g_1,g_2)
\nonumber\\ & \ \ \ \ \ \
+ m^\prime_2(g_0,g_1,g_2) m^\prime_2(g_2,g_3,g_4)
+ m^\prime_2(g_0,g_1,g_2) m^\prime_2(g_2,g_3,g_5)
+ m^\prime_2(g_0,g_1,g_2) m^\prime_2(g_2,g_4,g_5)
\nonumber\\ & \ \ \ \ \ \
+ m^\prime_2(g_0,g_1,g_3) m^\prime_2(g_3,g_4,g_5)
+ m^\prime_2(g_0,g_2,g_3) m^\prime_2(g_3,g_4,g_5)+ m^\prime_2(g_1,g_2,g_3) m^\prime_2(g_3,g_4,g_5)\text{
mod }  2 \nonumber\\=& 0 \text{ mod } 2
\end{align}
\end{widetext}

Combining the results from the above three steps, we can show:
\begin{align}
D_4\v g_4=\v f_5^I+\v f_5^{II}+\v f_5^{III} \text{ mod }2
\end{align}
Thus, we prove \eqn{ffg} for $d=3$.

We would like to mention that the induced mapping $\cH^d(G_b,\Z_2)\to
\cH^{d+2}(G_b,\Z_2)$ by the $n_d\to f_{d+2}$ mapping \eq{fdn}, appears to be the
Steenrod square $Sq^2$.\cite{S4790,S5313}  This realization will allow us to
generalize \eqn{fdn} to higher dimensions.

\section{The mapping $n_d \to f_{d+2}$ preserves additivity}
\label{additive}

In this appendix, we will show that the operation \eqn{fadd} defines an Abelian
group structure in $\fH^d[G_f,U_T(1)]$.
The key step is to show that if $n_d$ maps to $f_{d+2}$, $n_d^\prime$ maps to
$f_{d+2}^\prime$, and $n_d^{\prime\prime}=n_d+n_d^\prime$ maps to
$f_{d+2}^{\prime\prime}$, then $f_{d+2}^{\prime\prime} - f_{d+2} -
f_{d+2}^\prime$ is a coboundary in $\cB^{d+1}(G_b,\Z_2)$. (It is also a
coboundary in $\cB^{d+1}(G_b,U_T(1))$).

Let us first consider the case $d=2$ and we have:
\begin{align}
f_4(g_0,g_1,g_2,g_3,g_4)=&n_2(g_0,g_1,g_2)n_2(g_2,g_3,g_4)
\\
f_4^\prime(g_0,g_1,g_2,g_3,g_4)=&n_2^\prime(g_0,g_1,g_2) n_2^\prime(g_2,g_3,g_4)\nonumber\\
f_4^{\prime\prime}(g_0,g_1,g_2,g_3,g_4)=&\left[n_2(g_0,g_1,g_2)+n_2^\prime(g_0,g_1,g_2)\right]
\nonumber\\
\times &\left[n_2(g_2,g_3,g_4)+n_2^\prime(g_2,g_3,g_4)\right]
\nonumber
\end{align}
So
\begin{align}
\label{2Dfactor}
&\ \ \ \
f_{4}^{\prime\prime}- f_{4}-f_{4}^\prime
\\
& =
n_2(g_0,g_1,g_2)n_2^\prime(g_2,g_3,g_4)
+n_2^\prime(g_0,g_1,g_2) n_2(g_2,g_3,g_4).
\nonumber
\end{align}
The above is indeed a coboundary:
\begin{align}
&
n_2(g_0,g_1,g_2)n_2^\prime(g_2,g_3,g_4)
+n_2^\prime(g_0,g_1,g_2) n_2(g_2,g_3,g_4)
\nonumber\\
& =
(\dd a_3)(g_0,...,g_4) ,
\\
&\ \ \ \
a_3(g_0,...,g_3)
\nonumber\\
&=
n_2(g_0,g_1,g_2)n_2^\prime(g_0,g_2,g_3)+n_2(g_1,g_2,g_3)n_2^\prime(g_0,g_1,g_3)
,
\nonumber
\end{align}
where $\dd$ is
a mapping from $(d+1)$-variable functions
$f_d(g_0,...,g_d)$ to $(d+2)$-variable functions
$(\dd f_d)(g_0,...,g_{d+1})$:
\begin{align}
\label{dddef}
(\dd f_d)(g_0,...,g_{d+1})
\equiv
\sum_{i=0}^{d+1}(-)^i
f_d(g_0,..,\hat g_i,..,g_{d+1}) .
\nonumber
\end{align}

It is easy to check that:
\begin{align}
&n_2(g_0,g_1,g_2)n_2^\prime(g_0,g_2,g_3)+n_2(g_1,g_2,g_3)n_2^\prime(g_0,g_1,g_3)
\nonumber\\&+n_2(g_0,g_1,g_2)n_2^\prime(g_0,g_2,g_4)+n_2(g_1,g_2,g_4)n_2^\prime(g_0,g_1,g_4)
\nonumber\\&+n_2(g_0,g_1,g_3)n_2^\prime(g_0,g_3,g_4)+n_2(g_1,g_3,g_4)n_2^\prime(g_0,g_1,g_4)
\nonumber\\&+n_2(g_0,g_2,g_3)n_2^\prime(g_0,g_3,g_4)+n_2(g_2,g_3,g_4)n_2^\prime(g_0,g_2,g_4)
\nonumber\\&+n_2(g_1,g_2,g_3)n_2^\prime(g_1,g_3,g_4)+n_2(g_2,g_3,g_4)n_2^\prime(g_1,g_2,g_4)
\nonumber
\end{align}
\begin{align}
=&n_2(g_0,g_1,g_2)\left[n_2^\prime(g_2,g_3,g_4)+n_2^\prime(g_0,g_3,g_4)\right]
\nonumber\\&+n_2(g_1,g_2,g_3)n_2^\prime(g_0,g_1,g_3)+n_2(g_1,g_2,g_4)n_2^\prime(g_0,g_1,g_4)
\nonumber\\&+n_2(g_0,g_1,g_3)n_2^\prime(g_0,g_3,g_4)+n_2(g_1,g_3,g_4)n_2^\prime(g_0,g_1,g_4)
\nonumber\\&+n_2(g_0,g_2,g_3)n_2^\prime(g_0,g_3,g_4)+n_2(g_2,g_3,g_4)n_2^\prime(g_0,g_2,g_4)
\nonumber\\&+n_2(g_1,g_2,g_3)n_2^\prime(g_1,g_3,g_4)+n_2(g_2,g_3,g_4)n_2^\prime(g_1,g_2,g_4)
\nonumber\\&\text{ mod 2 }\nonumber
\end{align}
\begin{align}
=&n_2(g_0,g_1,g_2)n_2^\prime(g_2,g_3,g_4)+n_2(g_1,g_2,g_3)n_2^\prime(g_0,g_3,g_4)
\nonumber\\&+n_2(g_1,g_2,g_3)n_2^\prime(g_0,g_1,g_3)+n_2(g_1,g_2,g_4)n_2^\prime(g_0,g_1,g_4)
\nonumber\\&+n_2(g_1,g_3,g_4)n_2^\prime(g_0,g_1,g_4)+n_2(g_2,g_3,g_4)n_2^\prime(g_0,g_2,g_4)
\nonumber\\&+n_2(g_1,g_2,g_3)n_2^\prime(g_1,g_3,g_4)+n_2(g_2,g_3,g_4)n_2^\prime(g_1,g_2,g_4)
\nonumber\\&\text{ mod 2 }\nonumber
\end{align}
\begin{align}
=&n_2(g_0,g_1,g_2)n_2^\prime(g_2,g_3,g_4)+n_2(g_1,g_2,g_3)n_2^\prime(g_0,g_1,g_4)
\nonumber\\&+n_2(g_1,g_3,g_4)n_2^\prime(g_0,g_1,g_4)+n_2(g_1,g_2,g_4)n_2^\prime(g_0,g_1,g_4)
\nonumber\\&+n_2(g_2,g_3,g_4)n_2^\prime(g_0,g_2,g_4)+n_2(g_2,g_3,g_4)n_2^\prime(g_1,g_2,g_4)
\nonumber\\&\text{ mod 2 }\nonumber
\end{align}
\begin{align}
=&n_2(g_0,g_1,g_2)n_2^\prime(g_2,g_3,g_4)+n_2(g_2,g_3,g_4)n_2^\prime(g_0,g_1,g_4)
\nonumber\\&+n_2(g_2,g_3,g_4)n_2^\prime(g_0,g_2,g_4)+n_2(g_2,g_3,g_4)n_2^\prime(g_1,g_2,g_4)
\nonumber\\&\text{ mod 2 }\nonumber
\end{align}
\begin{align}
=& n_2(g_0,g_1,g_2)n_2^\prime(g_2,g_3,g_4)+n_2(g_2,g_3,g_4)n_2^\prime(g_0,g_1,g_2)
\nonumber\\&\text{ mod 2 }
\end{align}
Here we make use of the fact that $n_2(g_0,g_1,g_2)$ and $n_2(g_0,g_1,g_2)^\prime$ satisfy the $2$-cocycle condition:
\begin{align}
&n_2(g_0,g_1,g_2)+n_2(g_0,g_1,g_3)+n_2(g_0,g_2,g_3)+n_2(g_1,g_2,g_3)\nonumber\\&=0 \text{ mod }2 \nonumber\\
&n_2(g_0,g_1,g_2)^\prime+n_2(g_0,g_1,g_3)^\prime+n_2(g_0,g_2,g_3)^\prime+n_2(g_1,g_2,g_3)^\prime \nonumber\\&=0 \text{ mod }2
\end{align}
Thus, we have shown that
$f_{4}^{\prime\prime}- f_{4} -f_{4}^\prime$ is a coboundary.

Next we consider the case $d=3$, we note that:
\begin{widetext}
\begin{align}
&f_5(g_0,g_1,g_2,g_3,g_4,g_5)
\nonumber\\
=&n_3(g_0,g_1,g_2,g_3)n_3(g_0,g_3,g_4,g_5)+n_3(g_1,g_2,g_3,g_4)n_3(g_0,g_1,g_4,g_5)
+n_3(g_2,g_3,g_4,g_5)n_3(g_0,g_1,g_2,g_5),
 \nonumber\\
&f_5^{\prime}(g_0,g_1,g_2,g_3,g_4,g_5)
\nonumber\\
=&n_3^\prime(g_0,g_1,g_2,g_3)n_3^\prime(g_0,g_3,g_4,g_5)+n_3^\prime(g_1,g_2,g_3,g_4)n_3^\prime(g_0,g_1,g_4,g_5)
+n_3^\prime(g_2,g_3,g_4,g_5)n_3^\prime(g_0,g_1,g_2,g_5),\nonumber\\
&f_5^{\prime\prime}(g_0,g_1,g_2,g_3,g_4,g_5)
\nonumber\\
=&[n_3(g_0,g_1,g_2,g_3)+n_3^\prime(g_0,g_1,g_2,g_3)][n_3(g_0,g_3,g_4,g_5)+n_3^\prime(g_0,g_3,g_4,g_5)]\nonumber\\
&+[n_3(g_1,g_2,g_3,g_4)+n_3^\prime(g_1,g_2,g_3,g_4)][n_3(g_0,g_1,g_4,g_5)+n_3^\prime(g_0,g_1,g_4,g_5)]\nonumber\\
&+[n_3(g_2,g_3,g_4,g_5)+n_3^\prime(g_2,g_3,g_4,g_5)][n_3(g_0,g_1,g_2,g_5)+n_3^\prime(g_0,g_1,g_2,g_5)].
\end{align}
So
\begin{align}
\label{3Dfactor},
&
f_{5}^{\prime\prime} - f_{5} - f_{5}^\prime =
\nonumber\\
&{n_3(g_0,g_1,g_2,g_3)n_3^\prime(g_0,g_3,g_4,g_5)+n_3(g_1,g_2,g_3,g_4)n_3^\prime(g_0,g_1,g_4,g_5)
+n_3(g_2,g_3,g_4,g_5)n_3^\prime(g_0,g_1,g_2,g_5)}\nonumber\\
&+{n_3^\prime(g_0,g_1,g_2,g_3)n_3(g_0,g_3,g_4,g_5)+n_3^\prime(g_1,g_2,g_3,g_4)n_3(g_0,g_1,g_4,g_5)
+n_3^\prime(g_2,g_3,g_4,g_5)n_3(g_0,g_1,g_2,g_5)}
\end{align}
which is indeed a coboundary:
\begin{align}
f_{5}^{\prime\prime} - f_{5} - f_{5}^\prime &= \dd a_4,
\nonumber\\
a_4(g_0,...,g_4) &=
n_3(g_0,g_1,g_2,g_3)n_3^\prime(g_0,g_1,g_3,g_4)+n_3^\prime(g_0,g_1,g_2,g_4)n_3(g_0,g_2,g_3,g_4)+
\nonumber\\
&\ \ \ \
 n_3(g_0,g_1,g_3,g_4)n_3^\prime(g_1,g_2,g_3,g_4)+n_3(g_0,g_1,g_2,g_3)n_3^\prime(g_1,g_2,g_3,g_4)
\end{align}

First, it is easy to check that:
\begin{align}
& n_3(g_0,g_1,g_2,g_3)n_3^\prime(g_0,g_1,g_3,g_4)+n_3(g_0,g_1,g_2,g_3)n_3^\prime(g_0,g_1,g_3,g_5)\nonumber\\
&+ n_3(g_0,g_1,g_2,g_4)n_3^\prime(g_0,g_1,g_4,g_5)+ n_3(g_0,g_1,g_3,g_4)n_3^\prime(g_0,g_1,g_4,g_5)\nonumber\\
&+ n_3(g_0,g_2,g_3,g_4)n_3^\prime(g_0,g_2,g_4,g_5)+ n_3(g_1,g_2,g_3,g_4)n_3^\prime(g_1,g_2,g_4,g_5)\nonumber\\
&+n_3^\prime(g_0,g_1,g_2,g_4)n_3(g_0,g_2,g_3,g_4)+n_3^\prime(g_0,g_1,g_2,g_5)n_3(g_0,g_2,g_3,g_5)\nonumber\\
&+n_3^\prime(g_0,g_1,g_2,g_5)n_3(g_0,g_2,g_4,g_5)+n_3^\prime(g_0,g_1,g_3,g_5)n_3(g_0,g_3,g_4,g_5)\nonumber\\
&+n_3^\prime(g_0,g_2,g_3,g_5)n_3(g_0,g_3,g_4,g_5)+n_3^\prime(g_1,g_2,g_3,g_5)n_3(g_1,g_3,g_4,g_5)\nonumber
\end{align}
\begin{align}
=& n_3(g_0,g_1,g_2,g_3)[n_3^\prime(g_0,g_3,g_4,g_5)+n_3^\prime(g_0,g_1,g_4,g_5)+n_3^\prime(g_1,g_3,g_4,g_5)] \nonumber\\
&+ n_3(g_0,g_1,g_2,g_4)n_3^\prime(g_0,g_1,g_4,g_5)+ n_3(g_0,g_1,g_3,g_4)n_3^\prime(g_0,g_1,g_4,g_5)\nonumber\\
&+ n_3(g_0,g_2,g_3,g_4)n_3^\prime(g_0,g_2,g_4,g_5)+ n_3(g_1,g_2,g_3,g_4)n_3^\prime(g_1,g_2,g_4,g_5)\nonumber\\
&+n_3^\prime(g_0,g_1,g_2,g_4)n_3(g_0,g_2,g_3,g_4)+n_3^\prime(g_0,g_1,g_2,g_5)n_3(g_0,g_2,g_3,g_5)\nonumber\\
&+n_3^\prime(g_0,g_1,g_2,g_5)n_3(g_0,g_2,g_4,g_5)+n_3^\prime(g_0,g_1,g_3,g_5)n_3(g_0,g_3,g_4,g_5)\nonumber\\
&+n_3^\prime(g_0,g_2,g_3,g_5)n_3(g_0,g_3,g_4,g_5)+n_3^\prime(g_1,g_2,g_3,g_5)n_3(g_1,g_3,g_4,g_5)\text{ mod }2\nonumber
\end{align}
\begin{align}
=& n_3(g_0,g_1,g_2,g_3)[n_3^\prime(g_0,g_3,g_4,g_5)+n_3^\prime(g_1,g_3,g_4,g_5)] \nonumber\\
&+ [n_3(g_0,g_2,g_3,g_4)+n_3(g_1,g_2,g_3,g_4)]n_3^\prime(g_0,g_1,g_4,g_5) \nonumber\\
&+ n_3(g_0,g_2,g_3,g_4)n_3^\prime(g_0,g_2,g_4,g_5)+ n_3(g_1,g_2,g_3,g_4)n_3^\prime(g_1,g_2,g_4,g_5)\nonumber\\
&+n_3^\prime(g_0,g_1,g_2,g_4)n_3(g_0,g_2,g_3,g_4)+n_3^\prime(g_0,g_1,g_2,g_5)n_3(g_0,g_2,g_3,g_5)\nonumber\\
&+n_3^\prime(g_0,g_1,g_2,g_5)n_3(g_0,g_2,g_4,g_5)+n_3^\prime(g_0,g_1,g_3,g_5)n_3(g_0,g_3,g_4,g_5)\nonumber\\
&+n_3^\prime(g_0,g_2,g_3,g_5)n_3(g_0,g_3,g_4,g_5)+n_3^\prime(g_1,g_2,g_3,g_5)n_3(g_1,g_3,g_4,g_5)\text{ mod }2\nonumber
\end{align}
\begin{align}
=& n_3(g_0,g_1,g_2,g_3)[n_3^\prime(g_0,g_3,g_4,g_5)+n_3^\prime(g_1,g_3,g_4,g_5)]+n_3(g_1,g_2,g_3,g_4)n_3^\prime(g_0,g_1,g_4,g_5) \nonumber\\
&+ n_3(g_0,g_2,g_3,g_4)[n_3^\prime(g_0,g_1,g_2,g_5)+n_3^\prime(g_1,g_2,g_4,g_5)]\nonumber\\
&+n_3^\prime(g_0,g_1,g_2,g_5)n_3(g_0,g_2,g_3,g_5)+n_3(g_1,g_2,g_3,g_4)n_3^\prime(g_1,g_2,g_4,g_5)\nonumber\\
&+n_3^\prime(g_0,g_1,g_2,g_5)n_3(g_0,g_2,g_4,g_5)+n_3^\prime(g_0,g_1,g_3,g_5)n_3(g_0,g_3,g_4,g_5)\nonumber\\
&+n_3^\prime(g_0,g_2,g_3,g_5)n_3(g_0,g_3,g_4,g_5)+n_3^\prime(g_1,g_2,g_3,g_5)n_3(g_1,g_3,g_4,g_5)\text{ mod }2\nonumber
\end{align}
\begin{align}
=& n_3(g_0,g_1,g_2,g_3)[n_3^\prime(g_0,g_3,g_4,g_5)+n_3^\prime(g_1,g_3,g_4,g_5)]+n_3(g_1,g_2,g_3,g_4)n_3^\prime(g_0,g_1,g_4,g_5) \nonumber\\
&+ [n_3(g_0,g_3,g_4,g_5)+n_3(g_2,g_3,g_4,g_5)]n_3^\prime(g_0,g_1,g_2,g_5)+n_3(g_0,g_2,g_3,g_4)n_3^\prime(g_1,g_2,g_4,g_5)\nonumber\\
&+n_3^\prime(g_0,g_1,g_3,g_5)n_3(g_0,g_3,g_4,g_5)+n_3(g_1,g_2,g_3,g_4)n_3^\prime(g_1,g_2,g_4,g_5)\nonumber\\
&+n_3^\prime(g_0,g_2,g_3,g_5)n_3(g_0,g_3,g_4,g_5)+n_3^\prime(g_1,g_2,g_3,g_5)n_3(g_1,g_3,g_4,g_5)\text{ mod }2\nonumber
\end{align}
\begin{align}
=& n_3(g_0,g_1,g_2,g_3)[n_3^\prime(g_0,g_3,g_4,g_5)+n_3^\prime(g_1,g_3,g_4,g_5)]+n_3(g_1,g_2,g_3,g_4)n_3^\prime(g_0,g_1,g_4,g_5) \nonumber\\
&+n_3(g_2,g_3,g_4,g_5)n_3^\prime(g_0,g_1,g_2,g_5)+ n_3(g_0,g_2,g_3,g_4)n_3^\prime(g_1,g_2,g_4,g_5) +n_3(g_1,g_2,g_3,g_4)n_3^\prime(g_1,g_2,g_4,g_5)\nonumber\\
&+[n_3^\prime(g_0,g_1,g_2,g_3)+n_3^\prime(g_1,g_2,g_3,g_5)]n_3(g_0,g_3,g_4,g_5)+n_3^\prime(g_1,g_2,g_3,g_5)n_3(g_1,g_3,g_4,g_5)\text{ mod }2\nonumber
\end{align}
\begin{align}
=& n_3(g_0,g_1,g_2,g_3)n_3^\prime(g_0,g_3,g_4,g_5)+n_3^\prime(g_0,g_1,g_2,g_3)n_3(g_0,g_3,g_4,g_5)
\nonumber\\&+n_3(g_1,g_2,g_3,g_4)n_3^\prime(g_0,g_1,g_4,g_5)+n_3(g_2,g_3,g_4,g_5)n_3^\prime(g_0,g_1,g_2,g_5) \nonumber\\
&+n_3(g_1,g_2,g_3,g_4)n_3^\prime(g_1,g_2,g_4,g_5)+n_3(g_0,g_2,g_3,g_4)n_3^\prime(g_1,g_2,g_4,g_5)+n_3(g_0,g_1,g_2,g_3)n_3^\prime(g_1,g_3,g_4,g_5)\nonumber\\
&+ n_3^\prime(g_1,g_2,g_3,g_5)n_3(g_1,g_3,g_4,g_5)+n_3^\prime(g_1,g_2,g_3,g_5)n_3(g_0,g_3,g_4,g_5) \text{ mod }2
\end{align}
In the above calculation, we make use of the fact that $n_3(g_0,g_1,g_2,g_3)$ and $n_3(g_0,g_1,g_2,g_3)^\prime$ satisfy the $4$-cocycle condition:
\begin{align}
&n_3(g_0,g_1,g_2,g_3)+n_3(g_0,g_1,g_2,g_4)+n_3(g_0,g_1,g_3,g_4)+n_3(g_0,g_2,g_3,g_4)+n_3(g_1,g_2,g_3,g_4) =0 \text{ mod }2 \nonumber\\
&n_3(g_0,g_1,g_2,g_3)^\prime+n_3(g_0,g_1,g_2,g_4)^\prime+n_3(g_0,g_1,g_3,g_4)^\prime+n_3(g_0,g_2,g_3,g_4)^\prime+n_3(g_1,g_2,g_3,g_4)^\prime=0 \text{ mod }2
\end{align}
Next, by using the same trick, we can see that:
\begin{align}
&n_3(g_0,g_1,g_2,g_3)n_3^\prime(g_0,g_1,g_3,g_4)+n_3(g_0,g_1,g_2,g_3)n_3^\prime(g_0,g_1,g_3,g_5)\nonumber\\
&+ n_3(g_0,g_1,g_2,g_4)n_3^\prime(g_0,g_1,g_4,g_5)+ n_3(g_0,g_1,g_3,g_4)n_3^\prime(g_0,g_1,g_4,g_5)\nonumber\\
&+ n_3(g_0,g_2,g_3,g_4)n_3^\prime(g_0,g_2,g_4,g_5)+ n_3(g_1,g_2,g_3,g_4)n_3^\prime(g_1,g_2,g_4,g_5)\nonumber\\
&+n_3^\prime(g_0,g_1,g_2,g_4)n_3(g_0,g_2,g_3,g_4)+n_3^\prime(g_0,g_1,g_2,g_5)n_3(g_0,g_2,g_3,g_5)\nonumber\\
&+n_3^\prime(g_0,g_1,g_2,g_5)n_3(g_0,g_2,g_4,g_5)+n_3^\prime(g_0,g_1,g_3,g_5)n_3(g_0,g_3,g_4,g_5)\nonumber\\
&+n_3^\prime(g_0,g_2,g_3,g_5)n_3(g_0,g_3,g_4,g_5)+n_3^\prime(g_1,g_2,g_3,g_5)n_3(g_1,g_3,g_4,g_5)\nonumber\\
&+n_3(g_0,g_1,g_3,g_4)n_3^\prime(g_1,g_2,g_3,g_4)+n_3(g_0,g_1,g_3,g_5)n_3^\prime(g_1,g_2,g_3,g_5)\nonumber\\
&+n_3(g_0,g_1,g_4,g_5)n_3^\prime(g_1,g_2,g_4,g_5)+n_3(g_0,g_1,g_4,g_5)n_3^\prime(g_1,g_3,g_4,g_5)\nonumber\\
&+n_3(g_0,g_2,g_4,g_5)n_3^\prime(g_2,g_3,g_4,g_5)+n_3(g_1,g_2,g_4,g_5)n_3^\prime(g_2,g_3,g_4,g_5)\nonumber
\end{align}
\begin{align}
=& n_3(g_0,g_1,g_2,g_3)n_3^\prime(g_0,g_3,g_4,g_5)+n_3^\prime(g_0,g_1,g_2,g_3)n_3(g_0,g_3,g_4,g_5)
\nonumber\\&+n_3(g_1,g_2,g_3,g_4)n_3^\prime(g_0,g_1,g_4,g_5)+n_3(g_2,g_3,g_4,g_5)n_3^\prime(g_0,g_1,g_2,g_5) \nonumber\\
&+n_3(g_1,g_2,g_3,g_4)n_3^\prime(g_1,g_2,g_4,g_5)+n_3(g_0,g_2,g_3,g_4)n_3^\prime(g_1,g_2,g_4,g_5)+n_3(g_0,g_1,g_2,g_3)n_3^\prime(g_1,g_3,g_4,g_5)\nonumber\\
&+ n_3^\prime(g_1,g_2,g_3,g_5)n_3(g_1,g_3,g_4,g_5)+n_3^\prime(g_1,g_2,g_3,g_5)n_3(g_0,g_3,g_4,g_5)\nonumber\\
&+ n_3(g_0,g_1,g_3,g_4)n_3^\prime(g_1,g_2,g_3,g_4)+n_3(g_0,g_1,g_3,g_5)n_3^\prime(g_1,g_2,g_3,g_5)\nonumber\\
&+n_3(g_0,g_1,g_4,g_5)n_3^\prime(g_1,g_2,g_4,g_5)+n_3(g_0,g_1,g_4,g_5)n_3^\prime(g_1,g_3,g_4,g_5)\nonumber\\
&+n_3(g_0,g_2,g_4,g_5)n_3^\prime(g_2,g_3,g_4,g_5)+n_3(g_1,g_2,g_4,g_5)n_3^\prime(g_2,g_3,g_4,g_5) \text{ mod }2\nonumber
\end{align}
\begin{align}
=& n_3(g_0,g_1,g_2,g_3)n_3^\prime(g_0,g_3,g_4,g_5)+n_3^\prime(g_0,g_1,g_2,g_3)n_3(g_0,g_3,g_4,g_5)
\nonumber\\&+n_3(g_1,g_2,g_3,g_4)n_3^\prime(g_0,g_1,g_4,g_5)+n_3(g_2,g_3,g_4,g_5)n_3^\prime(g_0,g_1,g_2,g_5) \nonumber\\
&+n_3(g_1,g_2,g_3,g_4)n_3^\prime(g_1,g_2,g_4,g_5)+n_3(g_0,g_2,g_3,g_4)n_3^\prime(g_1,g_2,g_4,g_5)+n_3(g_0,g_1,g_2,g_3)n_3^\prime(g_1,g_3,g_4,g_5)\nonumber\\
&+ n_3^\prime(g_1,g_2,g_3,g_5)[n_3(g_0,g_1,g_4,g_5)+n_3(g_0,g_1,g_3,g_4)]+ n_3(g_0,g_1,g_3,g_4)n_3^\prime(g_1,g_2,g_3,g_4) \nonumber\\
&+n_3(g_0,g_1,g_4,g_5)n_3^\prime(g_1,g_2,g_4,g_5)+n_3(g_0,g_1,g_4,g_5)n_3^\prime(g_1,g_3,g_4,g_5)\nonumber\\
&+n_3(g_0,g_2,g_4,g_5)n_3^\prime(g_2,g_3,g_4,g_5)+n_3(g_1,g_2,g_4,g_5)n_3^\prime(g_2,g_3,g_4,g_5) \text{ mod }2\nonumber
\end{align}
\begin{align}
=& n_3(g_0,g_1,g_2,g_3)n_3^\prime(g_0,g_3,g_4,g_5)+n_3^\prime(g_0,g_1,g_2,g_3)n_3(g_0,g_3,g_4,g_5)
\nonumber\\&+n_3(g_1,g_2,g_3,g_4)n_3^\prime(g_0,g_1,g_4,g_5)+n_3(g_2,g_3,g_4,g_5)n_3^\prime(g_0,g_1,g_2,g_5) \nonumber\\
&+n_3(g_1,g_2,g_3,g_4)n_3^\prime(g_1,g_2,g_4,g_5)+n_3(g_0,g_2,g_3,g_4)n_3^\prime(g_1,g_2,g_4,g_5)+n_3(g_0,g_1,g_2,g_3)n_3^\prime(g_1,g_3,g_4,g_5)\nonumber\\
&+ n_3^\prime(g_1,g_2,g_3,g_5)[n_3(g_0,g_1,g_4,g_5)+n_3(g_0,g_1,g_3,g_4)]+ n_3(g_0,g_1,g_3,g_4)n_3^\prime(g_1,g_2,g_3,g_4) \nonumber\\
&+n_3(g_0,g_1,g_4,g_5)n_3^\prime(g_1,g_2,g_4,g_5)+n_3(g_0,g_1,g_4,g_5)n_3^\prime(g_1,g_3,g_4,g_5)\nonumber\\
&+n_3(g_0,g_2,g_4,g_5)n_3^\prime(g_2,g_3,g_4,g_5)+n_3(g_1,g_2,g_4,g_5)n_3^\prime(g_2,g_3,g_4,g_5) \text{ mod }2\nonumber
\end{align}
\begin{align}
=& n_3(g_0,g_1,g_2,g_3)n_3^\prime(g_0,g_3,g_4,g_5)+n_3^\prime(g_0,g_1,g_2,g_3)n_3(g_0,g_3,g_4,g_5)
\nonumber\\&+n_3(g_1,g_2,g_3,g_4)n_3^\prime(g_0,g_1,g_4,g_5)+n_3(g_2,g_3,g_4,g_5)n_3^\prime(g_0,g_1,g_2,g_5) \nonumber\\
&+n_3(g_1,g_2,g_3,g_4)n_3^\prime(g_1,g_2,g_4,g_5)+n_3(g_0,g_2,g_3,g_4)n_3^\prime(g_1,g_2,g_4,g_5)+n_3(g_0,g_1,g_2,g_3)n_3^\prime(g_1,g_3,g_4,g_5)\nonumber\\
&+n_3(g_0,g_1,g_3,g_4)n_3^\prime(g_1,g_2,g_3,g_5)+ n_3(g_0,g_1,g_3,g_4)n_3^\prime(g_1,g_2,g_3,g_4) \nonumber\\
&+n_3(g_0,g_1,g_4,g_5)n_3^\prime(g_1,g_2,g_3,g_4)+[n_3(g_0,g_1,g_2,g_5)+n_3(g_0,g_1,g_2,g_4)]n_3^\prime(g_2,g_3,g_4,g_5)  \text{ mod }2\nonumber
\end{align}
\begin{align}
=& n_3(g_0,g_1,g_2,g_3)n_3^\prime(g_0,g_3,g_4,g_5)+n_3^\prime(g_0,g_1,g_2,g_3)n_3(g_0,g_3,g_4,g_5)
\nonumber\\&+n_3(g_1,g_2,g_3,g_4)n_3^\prime(g_0,g_1,g_4,g_5)+n_3(g_0,g_1,g_4,g_5)n_3^\prime(g_1,g_2,g_3,g_4)\nonumber\\
&+n_3(g_0,g_1,g_2,g_5)n_3^\prime(g_2,g_3,g_4,g_5)+n_3(g_2,g_3,g_4,g_5)n_3^\prime(g_0,g_1,g_2,g_5) \nonumber\\
&+[n_3(g_0,g_1,g_2,g_4)+n_3(g_0,g_1,g_2,g_3)+n_3(g_0,g_1,g_3,g_4)]n_3^\prime(g_1,g_2,g_4,g_5)
+n_3(g_0,g_1,g_2,g_3)n_3^\prime(g_1,g_3,g_4,g_5)\nonumber\\
&+n_3(g_0,g_1,g_3,g_4)n_3^\prime(g_1,g_2,g_3,g_5)+ n_3(g_0,g_1,g_3,g_4)n_3^\prime(g_1,g_2,g_3,g_4)
+n_3(g_0,g_1,g_2,g_4)n_3^\prime(g_2,g_3,g_4,g_5)  \text{ mod }2\nonumber
\end{align}
\begin{align}
=& n_3(g_0,g_1,g_2,g_3)n_3^\prime(g_0,g_3,g_4,g_5)+n_3^\prime(g_0,g_1,g_2,g_3)n_3(g_0,g_3,g_4,g_5)
\nonumber\\&+n_3(g_1,g_2,g_3,g_4)n_3^\prime(g_0,g_1,g_4,g_5)+n_3(g_0,g_1,g_4,g_5)n_3^\prime(g_1,g_2,g_3,g_4)\nonumber\\
&+n_3(g_0,g_1,g_2,g_5)n_3^\prime(g_2,g_3,g_4,g_5)+n_3(g_2,g_3,g_4,g_5)n_3^\prime(g_0,g_1,g_2,g_5) \nonumber\\
&+[n_3(g_0,g_1,g_2,g_4)+n_3(g_0,g_1,g_2,g_3)]n_3^\prime(g_1,g_2,g_4,g_5)
+n_3(g_0,g_1,g_2,g_3)n_3^\prime(g_1,g_3,g_4,g_5)\nonumber\\
&+n_3(g_0,g_1,g_3,g_4)[n_3^\prime(g_1,g_3,g_4,g_5)+n_3^\prime(g_2,g_3,g_4,g_5)]
+n_3(g_0,g_1,g_2,g_4)n_3^\prime(g_2,g_3,g_4,g_5)  \text{ mod }2\nonumber
\end{align}
\begin{align}
=& n_3(g_0,g_1,g_2,g_3)n_3^\prime(g_0,g_3,g_4,g_5)+n_3^\prime(g_0,g_1,g_2,g_3)n_3(g_0,g_3,g_4,g_5)
\nonumber\\&+n_3(g_1,g_2,g_3,g_4)n_3^\prime(g_0,g_1,g_4,g_5)+n_3(g_0,g_1,g_4,g_5)n_3^\prime(g_1,g_2,g_3,g_4)\nonumber\\
&+n_3(g_0,g_1,g_2,g_5)n_3^\prime(g_2,g_3,g_4,g_5)+n_3(g_2,g_3,g_4,g_5)n_3^\prime(g_0,g_1,g_2,g_5) \nonumber\\
&+n_3(g_0,g_1,g_2,g_4)n_3^\prime(g_1,g_2,g_4,g_5)+n_3(g_0,g_1,g_2,g_3)[n_3^\prime(g_1,g_2,g_3,g_4)+n_3^\prime(g_1,g_2,g_3,g_5)+n_3^\prime(g_2,g_3,g_4,g_5)]\nonumber\\
&+n_3(g_0,g_1,g_3,g_4)n_3^\prime(g_1,g_3,g_4,g_5)
+[n_3(g_0,g_1,g_2,g_3)+n_3(g_0,g_2,g_3,g_4)+n_3(g_1,g_2,g_3,g_4)]n_3^\prime(g_2,g_3,g_4,g_5)  \text{ mod }2\nonumber
\end{align}
\begin{align}
=& n_3(g_0,g_1,g_2,g_3)n_3^\prime(g_0,g_3,g_4,g_5)+n_3^\prime(g_0,g_1,g_2,g_3)n_3(g_0,g_3,g_4,g_5)
\nonumber\\&+n_3(g_1,g_2,g_3,g_4)n_3^\prime(g_0,g_1,g_4,g_5)+n_3(g_0,g_1,g_4,g_5)n_3^\prime(g_1,g_2,g_3,g_4)\nonumber\\
&+n_3(g_0,g_1,g_2,g_5)n_3^\prime(g_2,g_3,g_4,g_5)+n_3(g_2,g_3,g_4,g_5)n_3^\prime(g_0,g_1,g_2,g_5) \nonumber\\
&+n_3(g_0,g_1,g_2,g_4)n_3^\prime(g_1,g_2,g_4,g_5)+n_3(g_0,g_1,g_2,g_3)[n_3^\prime(g_1,g_2,g_3,g_4)+n_3^\prime(g_1,g_2,g_3,g_5)]\nonumber\\
&+n_3(g_0,g_1,g_3,g_4)n_3^\prime(g_1,g_3,g_4,g_5)
+[n_3(g_0,g_2,g_3,g_4)+n_3(g_1,g_2,g_3,g_4)]n_3^\prime(g_2,g_3,g_4,g_5)  \text{ mod }2
\end{align}
To this end, we can finally show that:
\begin{align}
&n_3(g_0,g_1,g_2,g_3)n_3^\prime(g_0,g_1,g_3,g_4)+n_3(g_0,g_1,g_2,g_3)n_3^\prime(g_0,g_1,g_3,g_5)\nonumber\\
&+ n_3(g_0,g_1,g_2,g_4)n_3^\prime(g_0,g_1,g_4,g_5)+ n_3(g_0,g_1,g_3,g_4)n_3^\prime(g_0,g_1,g_4,g_5)\nonumber\\
&+ n_3(g_0,g_2,g_3,g_4)n_3^\prime(g_0,g_2,g_4,g_5)+ n_3(g_1,g_2,g_3,g_4)n_3^\prime(g_1,g_2,g_4,g_5)\nonumber\\
&+n_3^\prime(g_0,g_1,g_2,g_4)n_3(g_0,g_2,g_3,g_4)+n_3^\prime(g_0,g_1,g_2,g_5)n_3(g_0,g_2,g_3,g_5)\nonumber\\
&+n_3^\prime(g_0,g_1,g_2,g_5)n_3(g_0,g_2,g_4,g_5)+n_3^\prime(g_0,g_1,g_3,g_5)n_3(g_0,g_3,g_4,g_5)\nonumber\\
&+n_3^\prime(g_0,g_2,g_3,g_5)n_3(g_0,g_3,g_4,g_5)+n_3^\prime(g_1,g_2,g_3,g_5)n_3(g_1,g_3,g_4,g_5)\nonumber\\
&+n_3(g_0,g_1,g_3,g_4)n_3^\prime(g_1,g_2,g_3,g_4)+n_3(g_0,g_1,g_3,g_5)n_3^\prime(g_1,g_2,g_3,g_5)\nonumber\\
&+n_3(g_0,g_1,g_4,g_5)n_3^\prime(g_1,g_2,g_4,g_5)+n_3(g_0,g_1,g_4,g_5)n_3^\prime(g_1,g_3,g_4,g_5)\nonumber\\
&+n_3(g_0,g_2,g_4,g_5)n_3^\prime(g_2,g_3,g_4,g_5)+n_3(g_1,g_2,g_4,g_5)n_3^\prime(g_2,g_3,g_4,g_5)\nonumber\\
&+n_3(g_0,g_1,g_2,g_3)n_3^\prime(g_1,g_2,g_3,g_4)+n_3(g_0,g_1,g_2,g_3)n_3^\prime(g_1,g_2,g_3,g_5)\nonumber\\
&+n_3(g_0,g_1,g_2,g_4)n_3^\prime(g_1,g_2,g_4,g_5)+n_3(g_0,g_1,g_3,g_4)n_3^\prime(g_1,g_3,g_4,g_5)\nonumber\\
&+n_3(g_0,g_2,g_3,g_4)n_3^\prime(g_2,g_3,g_4,g_5)+n_3(g_1,g_2,g_3,g_4)n_3^\prime(g_2,g_3,g_4,g_5)\nonumber\\
=& n_3(g_0,g_1,g_2,g_3)n_3^\prime(g_0,g_3,g_4,g_5)+n_3^\prime(g_0,g_1,g_2,g_3)n_3(g_0,g_3,g_4,g_5)
\nonumber\\&+n_3(g_1,g_2,g_3,g_4)n_3^\prime(g_0,g_1,g_4,g_5)+n_3(g_0,g_1,g_4,g_5)n_3^\prime(g_1,g_2,g_3,g_4)\nonumber\\
&+n_3(g_0,g_1,g_2,g_5)n_3^\prime(g_2,g_3,g_4,g_5)+n_3(g_2,g_3,g_4,g_5)n_3^\prime(g_0,g_1,g_2,g_5) \text{ mod }2
\end{align}
\end{widetext}
Thus, $f_{5}^{\prime\prime} - f_{5} - f_{5}^\prime = \dd a_4$ is a coboundary.
For general $d$, we believe such a statement is still correct, however, since
only the cases with $d=2$ and $d=3$ are relevant to physical reality, we are
not going to prove it for general $d$ and leave it as an open mathematical
problem.  Once we have $f_{d}^{\prime\prime} - f_{d} - f_{d}^\prime = \dd a_d$,
it is quite easy to show \eqn{fadd}.


%

\end{document}